\documentclass[a4paper,12pt,oneside]{memoir}

\usepackage[T1]{fontenc} 
\usepackage[utf8]{inputenc} 
\usepackage{microtype} 
\usepackage{lmodern} 
\usepackage{booktabs} 
\usepackage{mdframed} 
\usepackage{graphicx}
\usepackage{todonotes} 

\usepackage{float} 

\usepackage[breaklinks=true]{hyperref}
 \usepackage{color}
 \definecolor{dark-red}{rgb}{0.4,0.15,0.15}
 \definecolor{dark-blue}{rgb}{0.15,0.15,0.4}
 \definecolor{medium-blue}{rgb}{0,0,0.5}
 \hypersetup{colorlinks, linkcolor={dark-red}, citecolor={dark-blue}, urlcolor={medium-blue}}

\usepackage[hyphenbreaks]{breakurl}

\usepackage{cite}

\usepackage{tikz} 
\usepackage{tikz-cd}
\usepackage{tikz-3dplot}

\newcommand*{\shi}{1}
\newcommand*{\shih}{0.5}
\newcommand*{\shihh}{0.25}

\usepackage{amsmath,amsfonts,amssymb}
\usepackage{eucal} 
\usepackage{mleftright} \mleftright
\usepackage{tensor} 
\usepackage{braket} 
\usepackage{mathrsfs} 

\usepackage{cancel} %

\usepackage{amsthm} 
\theoremstyle{plain}
\newtheorem{theorem}{Theorem}[chapter]

\newtheorem{corollary}[theorem]{Corollary}

\theoremstyle{definition}
\newtheorem{definition}[theorem]{Definition}
\newtheorem{example}[theorem]{Example}

\theoremstyle{remark}

\usepackage{makeidx}
\usepackage{showidx}
\makeindex

\providecommand*{\dd}{\mathop{}\!d}
\renewcommand*{\dd}{\mathop{}\!d}
\providecommand*{\Dd}{\mathop{}\!D}
\renewcommand*{\Dd}{\mathop{}\!D}
\providecommand*{\pd}{\mathop{}\!\partial}
\renewcommand*{\pd}{\mathop{}\!\partial}
\providecommand*{\vd}{\mathop{}\!\delta}
\renewcommand*{\vd}{\mathop{}\!\delta}

\providecommand*{\dco}{\mathop{}\!\delta}
\renewcommand*{\dco}{\mathop{}\!\delta}

\providecommand*{\Ld}{\mathop{}\!\mathscr{L}}
\renewcommand*{\Ld}{\mathop{}\!\mathscr{L}}

\providecommand*{\ic}{\mathop{}\! i}
\renewcommand*{\ic}{\mathop{}\! i}

\providecommand*{\R}{{\mathbb{R}}}
\renewcommand*{\R}{{\mathbb{R}}}
\providecommand*{\C}{{\mathbb{C}}}
\renewcommand*{\C}{{\mathbb{C}}}

\providecommand*{\CS}{\mathrm{CS}}
\renewcommand*{\CS}{\mathrm{CS}}

\providecommand*{\os}{\overset{\mathrm{o.s.}}{=}}
\renewcommand*{\os}{\overset{\mathrm{o.s.}}{=}}

\usepackage{stmaryrd} 

\providecommand*{\Tb}{\mathtt{T}}
\renewcommand*{\Tb}{\mathtt{T}}
\providecommand*{\Zb}{\mathtt{Z}}
\renewcommand*{\Zb}{\mathtt{Z}}
\providecommand*{\Gb}{\mathtt{G}}
\renewcommand*{\Gb}{\mathtt{G}}
\providecommand*{\Hb}{\mathtt{H}}
\renewcommand*{\Hb}{\mathtt{H}}
\providecommand*{\Db}{\mathtt{D}}
\renewcommand*{\Db}{\mathtt{D}}
\providecommand*{\Xb}{\mathtt{X}}
\renewcommand*{\Xb}{\mathtt{X}}
\providecommand*{\Yb}{\mathtt{Y}}
\renewcommand*{\Yb}{\mathtt{Y}}

\providecommand*{\Hba}{\mathtt{H}}
\renewcommand*{\Hba}{\mathtt{H}}
\providecommand*{\Hbb}{\widetilde{\mathtt{H}}}
\renewcommand*{\Hbb}{\widetilde{\mathtt{H}}}
\providecommand*{\alphaa}{\alpha}
\renewcommand*{\alphaa}{\alpha}
\providecommand*{\betaa}{\beta}
\renewcommand*{\betaa}{\beta}
\providecommand*{\gammaa}{\gamma}
\renewcommand*{\gammaa}{\gamma}
\providecommand*{\alphab}{\widetilde{\alpha}}
\renewcommand*{\alphab}{\widetilde{\alpha}}
\providecommand*{\betab}{\widetilde{\beta}}
\renewcommand*{\betab}{\widetilde{\beta}}
\providecommand*{\gammab}{\widetilde{\gamma}}
\renewcommand*{\gammab}{\widetilde{\gamma}}

\providecommand*{\vdis}{\mathbin{\dot +}}
\renewcommand*{\vdis}{\mathbin{\dot +}}

\providecommand*{\dis}{\oplus}
\renewcommand*{\dis}{\oplus}

\providecommand*{\sdis}{\mathbin{\niplus}}
\renewcommand*{\sdis}{\mathbin{\niplus}}

\providecommand*{\inj}{\to}
\renewcommand*{\inj}{\to}
\providecommand*{\surj}{\to}
\renewcommand*{\surj}{\to}

\DeclareMathOperator{\ad}{ad}
\DeclareMathOperator{\der}{der}
\DeclareMathOperator{\End}{End}
\DeclareMathOperator{\spn}{span}

\DeclareMathOperator{\tr}{tr}

\DeclareMathOperator{\im}{im}

 \DeclareSymbolFont{bbold}{U}{bbold}{m}{n}
 \DeclareSymbolFontAlphabet{\mathbbold}{bbold}

\newcommand\parmp{\mathbin{\vcenter{\hbox{%
  \oalign{$\scriptstyle({-})$\cr
          \noalign{\kern-0.8ex}
          \hfil$\scriptscriptstyle+$\hfil\cr}%
}}}}
\def\comma{\,,\,}
\providecommand{\Gt}{{\mathtt{G}}}
\renewcommand{\Gt}{{\mathtt{G}}}
\providecommand{\Jt}{{\mathtt{J}}}
\renewcommand{\Jt}{{\mathtt{J}}}
\providecommand{\Ht}{{\mathtt{H}}}
\renewcommand{\Ht}{{\mathtt{H}}}

\providecommand{\Tt}{{\mathtt{T}}}
\renewcommand{\Tt}{{\mathtt{T}}}
\providecommand{\Pt}{{\mathtt{P}}}
\renewcommand{\Pt}{{\mathtt{P}}}
\providecommand{\Lt}{{\mathtt{L}}}
\renewcommand{\Lt}{{\mathtt{L}}}
\providecommand{\Wt}{{\mathtt{W}}}
\renewcommand{\Wt}{{\mathtt{W}}}

\newcommand{\hJ}{\hat{{\mathtt{J}}}}
\newcommand{\hP}{\hat{{\mathtt{P}}}}

\newcommand{\eq}[2]{\begin{equation} #1 \label{#2} \end{equation}}
\newcommand{\vp}{\varphi}
\newcommand{\eps}{\epsilon}

\newcommand{\extd}{\dd}

\providecommand{\At}{\ensuremath{\widetilde{A}}}
\renewcommand{\At}{\ensuremath{\widetilde{A}}}

\providecommand{\Ft}{\ensuremath{\widetilde{F}}}
\renewcommand{\Ft}{\ensuremath{\widetilde{F}}}

\providecommand{\ddt}{\ensuremath{\widetilde{\dd}}}
\renewcommand{\ddt}{\ensuremath{\widetilde{\dd}}}

\providecommand{\deltat}{\ensuremath{\widetilde{\delta}}}
\renewcommand{\deltat}{\ensuremath{\widetilde{\delta}}}

\newcommand{\overbar}[1]{\mkern 1.5mu\overline{\mkern-2mu#1\mkern-1mu}\mkern 1.5mu}
\newcommand{\overbarl}[1]{\mkern 1.5mu\overline{\mkern-1.5mu#1\mkern+0.5mu}\mkern 1.5mu}

\renewcommand{\O}[1]{\ensuremath{\mathcal{O}\left(#1\right)}}
\renewcommand{\L}{\ensuremath{\mathcal{L}}}

\newcommand{\W}{\ensuremath{\mathcal{W}}}
\newcommand{\Lba}{\ensuremath{\overbar{\mathcal{L}}}}
\newcommand{\Wba}{\ensuremath{\overbar{\mathcal{W}}}}

\newcommand{\Ab}{A^{-}}

\newcommand{\eb}{\ensuremath{\overbarl{\epsilon}}}

\begin{document}

\frontmatter





\title{\textbf{Chern--Simons Holography}\\
  {\normalsize
    Boundary Conditions, Contractions and Double Extensions \newline
    for a Journey Beyond Anti-de Sitter}}
\author{Stefan Prohazka\thanks{
    \href{mailto:prohazka@hep.itp.tuwien.ac.at}{prohazka@hep.itp.tuwien.ac.at},
    \href{mailto:stefan.prohazka@ulb.ac.be}{stefan.prohazka@ulb.ac.be},
    \href{https://orcid.org/0000-0002-3925-3983}{ORCID ID: 0000-0002-3925-3983}
  }
  \\
    Institute for Theoretical Physics\\
    TU Wien\\
    Wiedner Hauptstrasse 8-10/136\\
    Tower B (yellow), 10th floor\\
    A-1040 Wien, AUSTRIA
}

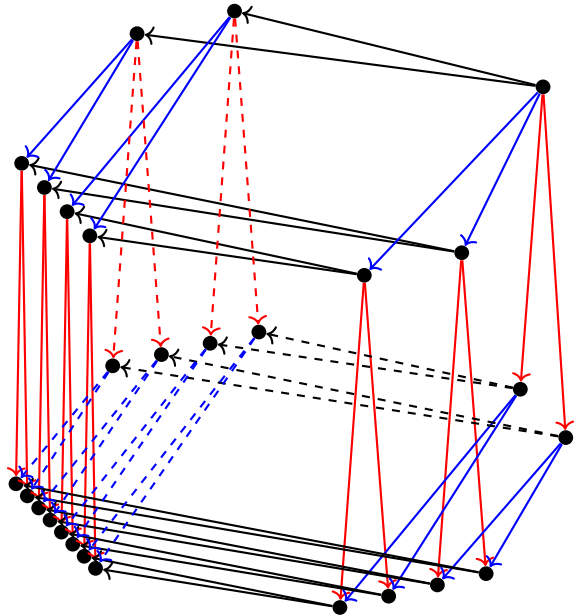
\date{}
\maketitle
\thispagestyle{empty}

\begin{figure}[h]
  \centering
\tdplotsetmaincoords{60}{110}
\begin{tikzpicture}[
tdplot_main_coords,
dot/.style={circle,fill,scale=0.5},
linf/.style={thick,->,blue},
cinf/.style={thick,->,red},
tinf/.style={thick,->},
stinf/.style={ultra thick,->,gray},
scale=0.5
]


\node (ads) at (0,10,10) [dot, 
] {};

\node (p) at (10,10,10) [
] {};

\node (p1) at (10+\shi,10-\shi,10) [dot] {};
\node (p2) at (10-\shi,10+\shi,10) [dot] {};


\node (nh1) at (-\shi,10-\shi,0) [dot] {};
\node (nh2) at (\shi,10+\shi,0) [dot, 
] {};

\node (pp) at (0,0,10) [
] {};

\node (pp1) at (-\shi,\shi,10) [dot] {};
\node (pp2) at (\shi,-\shi,10) [dot] {};

\node (g) at (10,10,0) [
] {};

\node (g3) at (10-\shi-\shih,10+\shi+\shih,0) [dot] {};
\node (g4) at (10-\shi+\shih,10+\shi-\shih,0) [dot] {};

\node (g1) at (10+\shi+\shih,10-\shi-\shih,0) [dot] {};
\node (g2) at (10+\shi-\shih,10-\shi+\shih,0) [dot] {};

\node (pg) at (0,0,0) [
] {};


\node (pg1) at (-\shi+\shih,\shi-\shih,0) [dot] {};
\node (pg2) at (-\shi-\shih,\shi+\shih,0) [dot] {};


\node (pg3) at (\shi+\shih,-\shi-\shih,0) [dot] {};
\node (pg4) at (\shi-\shih,-\shi+\shih,0) [dot] {};


\node (car1) at (10+\shi-\shih,\shi-\shih,10) [dot] {};
\node (car2) at (10+\shi+\shih,\shi+\shih,10) [dot] {};

\node (car3) at (10-\shi-\shih,-\shi-\shih,10) [dot, 
] {};
\node (car4) at (10-\shi+\shih,-\shi+\shih,10) [dot] {};

\node (st) at (10,0,0) [
] {};

\node (st3) at (10-\shi-\shih-\shihh,-\shi-\shih-\shihh,0) [dot] {};
\node (st4) at (10-\shi-\shih+\shihh,-\shi-\shih+\shihh,0) [dot] {};
\node (st7) at (10+\shi+\shih+\shihh,\shi+\shih+\shihh,0) [dot] {};
\node (st8) at (10+\shi+\shih-\shihh,\shi+\shih-\shihh,0) [dot] {};

\node (st5) at (10+\shi-\shih+\shihh,\shi-\shih+\shihh,0) [dot] {};
\node (st6) at (10+\shi-\shih-\shihh,\shi-\shih-\shihh,0) [dot] {};

\node (st1) at (10-\shi+\shih+\shihh,-\shi+\shih+\shihh,0) [dot] {};
\node (st2) at (10-\shi+\shih-\shihh,-\shi+\shih-\shihh,0) [dot] {};

\draw[tinf] (p1) -- (car2);
\draw[tinf] (p1) -- (car1);
\draw[tinf] (p2) -- (car3);
\draw[tinf] (p2) -- (car4);

\draw[cinf] (ads) -- node [left] {
} (nh1);
\draw[cinf] (ads) -- node [right] {
}(nh2);

\draw[tinf] (ads) -- node [above] {
}(pp1);
\draw[tinf] (ads) -- node [below] {
}(pp2);

\draw[cinf] (p1) -- (g1);
\draw[cinf] (p1) -- (g2);
\draw[cinf] (p2) -- (g3);
\draw[cinf] (p2) -- (g4);

\draw[linf] (ads) -- node [left] {
} (p1);
\draw[linf] (ads) -- node [left] {
} (p2);

\draw[linf] (nh1) -- (g1);
\draw[linf] (nh1) -- (g2);
\draw[linf] (nh2) -- (g3);
\draw[linf] (nh2) -- (g4);

\draw[linf] (pp1) -- (car1);
\draw[linf] (pp1) -- (car2);
\draw[linf] (pp2) -- (car3);
\draw[linf] (pp2) -- (car4);

\draw[cinf] (car2) -- (st8);
\draw[cinf] (car2) -- (st7);

\draw[cinf] (car1) -- (st6);
\draw[cinf] (car1) -- (st5);

\draw[cinf] (car3) -- (st4);
\draw[cinf] (car3) -- (st3);
\draw[cinf] (car4) -- (st2);
\draw[cinf] (car4) -- (st1);

\draw[tinf,dashed] (nh1) -- (pg1);
\draw[tinf,dashed] (nh1) -- (pg2);
\draw[tinf,dashed] (nh2) -- (pg3);
\draw[tinf,dashed] (nh2) -- (pg4);

\draw[cinf,dashed] (pp1) -- (pg1);
\draw[cinf,dashed] (pp1) -- (pg2);
\draw[cinf,dashed] (pp2) -- (pg3);
\draw[cinf,dashed] (pp2) -- (pg4);

\draw[tinf] (g1) -- (st8);
\draw[tinf] (g1) -- (st7);
\draw[tinf] (g2) -- (st6);
\draw[tinf] (g2) -- (st5);
\draw[tinf] (g3) -- (st4);
\draw[tinf] (g3) -- (st3);
\draw[tinf] (g4) -- (st2);
\draw[tinf] (g4) -- (st1);

\draw[linf,dashed] (pg1) -- (st6);
\draw[linf,dashed] (pg1) -- (st5);
\draw[linf,dashed] (pg2) -- (st8);
\draw[linf,dashed] (pg2) -- (st7);

\draw[linf,dashed] (pg3) -- (st3);
\draw[linf,dashed] (pg3) -- (st4);
\draw[linf,dashed] (pg4) -- (st1);
\draw[linf,dashed] (pg4) -- (st2);

\end{tikzpicture}
\end{figure}

\newpage


\chapter{Abstract}
\label{sec:abstract}


The holographic principle originates from
the observation that 
black hole entropy
is proportional to the horizon area
and not,
as expected from a quantum field theory perspective, to the volume.
This principle has found a concrete realization in the Anti-de Sitter/Conformal Field Theory (AdS/CFT) correspondence.
It is interesting to ponder whether the key insights about holography so far are specific to AdS/CFT or if they are general lessons for quantum gravity
and (non)relativistic field theories.

Relativistic and
nonrelativistic geometries
play a fundamental
role in advances of holography
beyond AdS spacetimes, e.g.,
for strongly coupled systems in
condensed matter physics.
Holography for higher spin theories
is comparably well understood
and they are therefore good candidates
to gain further insights.
In three spacetime dimensions
they are distinguished by
technical simplicity, the
possibility to write the
theory in Chern--Simons form
and the option to consistently
truncate the infinite higher spin
fields to any integer spin greater than two.

Here we will show progress
that has been made
to construct
relativistic and nonrelativistic theories
in spin-three gravity.
These theories describe
a coupled spin two and three field
and are based
on Chern--Simons theories
with
kinematical gauge algebras
of which the Poincar\'e,
Galilei and Carroll algebra
are prominent examples.
To have a spin-three theory where all fields
are dynamical it is sometimes necessary,
as will be shown,
to extend the gauge algebras
accordingly.

We will also discuss concepts
which are useful in these constructions.
Guidance is
provided by combining
Lie algebra contractions
and,
a procedure that
will be reviewed extensively,
double extensions.







\chapter{Acknowledgments}
\label{sec:acknowledgments}

It is unfortunate that I will not be able
to acknowledge every person that has shaped
and contributed to this work or
the enjoyable time I had in
preparing it.
However,
the following people deserve special
thanks for their direct or indirect
role in the development of this text. 

\subsection{Professional Community}
\label{sec:prof}

I am grateful to my supervisors Daniel Grumiller
and Mirah Gary for their continuous support
and encouragement.
Their guidance from my master's thesis
to the completion of this PhD thesis
was exceptional
and
the patience in answering my
never-ending questions
admirable.
It was always a pleasure to come
to ``work''.
This is  very much due to the
enjoyable and friendly atmosphere that was created in our
work group.
Daniel was available
for discussions
and advice
practically every time
and
as spontaneously as sometimes
necessary.
His enthusiasm,
creativity,
and
commitment for physics
and science in general
is something
I hope to be able
to emulate myself.

Without the ``office crew''
consisting of
Maria Irakleidou,
Jakob Salzer, and Friedrich Schöller
the last three years
would not have been as
gratifying as they were.
The
thorough
discussions
about physics and beyond 
shaped
my understanding
of nearly any aspect of life.

I am pleased to thank Jan Rosseel for fruitful collaborations
and for introducing me to double extensions
which play a fundamental role in this work.
Also his valuable advise
concerning social aspects
of theoretical physics
is very much appreciated.

During my studies I had the honor
to co-supervise works of 
Veronika Breunhölder
and
Raphaela Wutte.
This was 
an experience
that I hope
was as
valuable
 for
them as it was for me.

Of course the interesting and entertaining discussions
with 
Hernán Gon\-zález,
Iva Lovrekovic,
Wout Merbis,
and
Max Riegler
should not be unmentioned.

I would like to thank
Hamid Afshar,
Martin Ammon,
Arjun Bagchi,
Eric Bergshoeff,
St\'ephane Detournay,
Alfredo Perez,
Soo-Jong Rey,
David Tempo,
and
Ricardo Troncoso
for rewarding
and delightful
collaborations.

I thank Daniel Grumiller, Jakob Salzer and Raphaela Wutte
for lightning fast proof reading of this thesis.
Of course all remaining mistakes are solely mine.

Finally, I am thankful to Stefan Fredenhagen
and Radoslav Rashkov for being my external
experts and referees of this PhD thesis.

\subsection{Financial Support}
\label{sec:fincsupp}

The begin of my PhD studies was supported by the START project Y 435-N16 of the Austrian Science Fund (FWF) and the FWF project I 952-N16.
The main period was supported by the FWF project P 27396-N27.
Furthermore,
financial and logistical support by
all the conferences and workshop organizers
is very much appreciated.

\subsection{Financial Support -- Extended Version}
\label{sec:fincsuppext}

The extension of the original version
was supported by the ERC Advanced Grant ``High-Spin-Grav'' and
by FNRS-Belgium (convention FRFC PDR T.1025.14 and convention IISN 4.4503.15).

\subsection{Family and Friends}
\label{sec:family-friends}

Without the unfailing support and encouragement of 
my parents Christine and Wolfgang Prohazka
and my sister Ricarda Prohazka
I would have never come to this point.
Their influence on my life is obvious
and have formed me into the
person and scientist
that I am.
I also acknowledge the support provided by my grandparents,
uncles, aunts and cousins.

The biggest sacrifice during the write-up
of this thesis was
possibly made
by my
girlfriend
Lisa Schwarzbauer.
The long hours and
my sometimes mental absence
did not alter her
always supportive
nature
for which I am not
able
to fully formulate
my gratitude.

I am thankful to 
Paul Baumgarten for hysterical phone calls and for delaying this
thesis by making me his best man for his marriage with Claudia.
In addition I send my best regards to
Thomas Eigner for hard fights,
Bernhard Frühwirt for long runs,
Jakob Salzer for enjoyable swims,
Friedrich Schöller for support in basically every aspect,
Johannes Radl for beautiful sunsets,
and Thomas Wernhart for ghostly visits and medical support during the write up of this thesis.
Without you and all other friends this
life-changing
journey would not have been possible.

\clearpage
\tableofcontents

\chapter{Notes to the Reader}
\label{cha:notes-reader}
\epigraph{In der Kürze liegt die Würze.
Brevity is the soul of wit.}{}

\noindent
This PhD thesis is based,
some parts verbatim,
on my master's thesis and the following publications:
\begin{itemize}
\item[\cite{Afshar:2014rwa}]
  H.~Afshar, A.~Bagchi, S.~Detournay, D.~Grumiller, S.~Prohazka, and M.~Riegler,
  ``{Holographic Chern-Simons Theories},''\\
  \href{http://dx.doi.org/10.1007/978-3-319-10070-8_12}{{\em Lect. Notes Phys.}
    {\bfseries 892} (2015) 311--329},
  \href{http://arxiv.org/abs/1404.1919}{{\ttfamily arXiv:1404.1919 [hep-th]}}.

\item[\cite{Gary:2014mca}]
M.~Gary, D.~Grumiller, S.~Prohazka, and S.-J. Rey, ``{Lifshitz Holography with
  Isotropic Scale Invariance},''\\
  \href{http://dx.doi.org/10.1007/JHEP08(2014)001}{{\em JHEP} {\bfseries 1408}
  (2014) 001},
\href{http://arxiv.org/abs/1406.1468}{\ttfamily{arXiv:\hspace{0pt}1406.1468 [hep-th]}}.

\item[\cite{Breunhoelder:2015waa}]
  V.~Breunhölder, M.~Gary, D.~Grumiller, and S.~Prohazka, ``{Null warped AdS in
    higher spin gravity},''\\
  \href{http://dx.doi.org/10.1007/JHEP12(2015)021}{{\em
      JHEP} {\bfseries 12} (2015) 021},
  \href{http://arxiv.org/abs/1509.08487}{{\ttfamily arXiv:1509.08487 [hep-th]}}.

\item[\cite{Grumiller:2016kcp}]
  D.~Grumiller, A.~Perez, S.~Prohazka, D.~Tempo, and R.~Troncoso, ``{Higher Spin
  Black Holes with Soft Hair},''\\
  \href{http://dx.doi.org/10.1007/JHEP10(2016)119}{{\em JHEP} {\bfseries 10}
    (2016) 119},
  \href{http://arxiv.org/abs/1607.05360}{{\ttfamily arXiv:1607.05360 [hep-th]}}.

\item[\cite{Bergshoeff:2016soe}]
  E.~Bergshoeff, D.~Grumiller, S.~Prohazka, and J.~Rosseel, ``{Three-dimensional
    Spin-3 Theories Based on General Kinematical Algebras},''\\
  \href{http://dx.doi.org/10.1007/JHEP01(2017)114}{{\em JHEP} {\bfseries 01}
    (2017) 114},
  \href{http://arxiv.org/abs/1612.02277}{{\ttfamily arXiv:1612.02277 [hep-th]}}.

\item[\cite{Prohazka:2017equ}]
  S.~Prohazka, J.~Salzer, and F.~Schöller, ``{Linking Past and Future Null
  Infinity in Three Dimensions},''\\
  \href{http://dx.doi.org/10.1103/PhysRevD.95.086011}{{\em Phys. Rev.}
    {\bfseries D95} no.~8, (2017) 086011},
  \href{http://arxiv.org/abs/1701.06573}{{\ttfamily arXiv:1701.06573 [hep-th]}}.

\item[\cite{Ammon:2017vwt}]
  M.~Ammon, D.~Grumiller, S.~Prohazka, M.~Riegler, and R.~Wutte, ``{Higher-Spin
  Flat Space Cosmologies with Soft Hair},''\\
  \href{http://dx.doi.org/10.1007/JHEP05(2017)031}{{\em JHEP} {\bfseries 05}
    (2017) 031},
  \href{http://arxiv.org/abs/1703.02594}{{\ttfamily arXiv:1703.02594 [hep-th]}}.
\end{itemize}

I do not intend to reproduce all the
details of the aforementioned publications.
It was the goal of the publications to provide sufficient
information and repeating everything
without further insights felt to be an uninteresting and especially
useless endeavor.
So, I will only discuss the parts of these publications that seemed
necessary for my explanations or for other reasons.

My actual objective is to highlight and explain the underlying concepts used in
these publications.
At various stages I will review and combine information that
would usually not find its way into a publication.
The motivation is to get a deeper understanding and a bird eye's view
upon them, in such a way that the results become somehow obvious once
the fundamental theory is understood.

This led to some results that to my best knowledge are absent in the literature,
e.g., I am not aware of a place where contractions of double
extensions are studied.

I try to explain the abstract concepts and ideas first,
often with a preference for simple examples over lengthy explanations.
Furthermore, I did try to set my work into context with the literature
that I think might be useful for future investigations.

At first sight and in seeming opposition to the proverb stated in the beginning
of this note I will be redundant at various places and more explicit
than necessary.
The reason for this is that often it was useful for me.
For instance
having an abstract equation written explicitly in a basis, although an
easy exercise to an expert, takes time and leaves place for errors.
Of course this work is not exempt from errors, wrong and missing citations
or other misconceptions and I am therefore grateful for any e-mail to
\href{mailto:prohazka@hep.itp.tuwien.ac.at}{prohazka@hep.itp.tuwien.ac.at}
or
\href{mailto:stefan.prohazka@ulb.ac.be}{stefan.prohazka@ulb.ac.be}
that points them out.
I furthermore hope that these sometimes explicit calculations and
collections of formulas (mostly in the appendices) are useful to others.

\mainmatter

\chapter{Introduction}
\label{cha:introduction}

\begin{figure}[h]
  \centering
\tdplotsetmaincoords{80}{120}
\begin{tikzpicture}[
tdplot_main_coords,
dot/.style={circle,fill},
linf/.style={ultra thick,->,blue},
cinf/.style={ultra thick,->,red},
tinf/.style={ultra thick,->},
stinf/.style={ultra thick,->,gray},
scale=0.65
]

\node (ads) at (0,10,10) [dot, label=above:$\text{Anti-de Sitter}$] {};
\node (p) at (10,10,10) [dot,label=below left:Relativistic] {};
\node (nh) at (0,10,0) [dot
] {};
\node (pp) at (0,0,10) [dot
] {};
\node (g) at (10,10,0) [dot, label=below:Nonrelativistic] {};
\node (pg) at (0,0,0) [dot
] {};
\node (car) at (10,0,10) [dot
] {};
\node (st) at (10,0,0) [dot
] {};

\draw[linf] (ads) -- node [sloped,below] {$\ell \to \infty$ 
} (p);

\draw[linf] (nh) -- (g);
\draw[linf] (pp) -- (car);
\draw[linf,dashed] (pg) -- (st);

\draw[cinf] (ads) --  (nh);
\draw[cinf,dashed] (pp) -- (pg);
\draw[cinf] (p) -- node[left]  {$c \to \infty$
} (g);
\draw[cinf] (car) -- (st);

\draw[tinf] (ads) --  (pp);
\draw[tinf] (p) -- (car);
\draw[tinf] (g) -- (st);
\draw[tinf,dashed] (nh) -- (pg);

\end{tikzpicture}
\caption{This figure shows that
  the relativistic symmetries can be understood
  as a contraction from the Anti-de Sitter symmetries
  where the universe is negatively curved with radius $\ell$.
  From  relativistic systems we can send the speed of light to infinity
  to arrive at the nonrelativistic ones.}
  \label{fig:cubeint}
\end{figure}
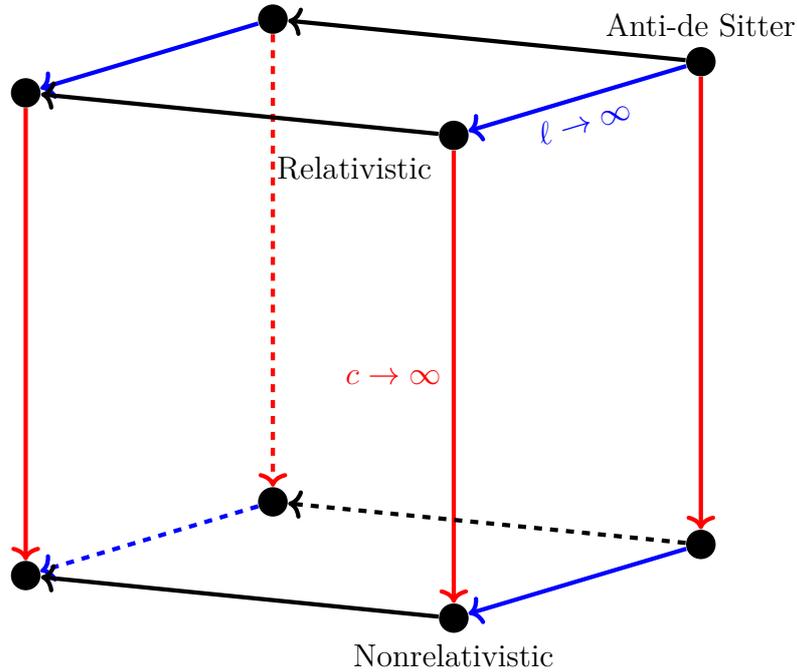

Symmetries have always been a successful guiding principle in
physics.
Already Galilei
realized that the everyday physical laws
are invariant
under transformations
like rotations, time and space translations but also
more general ones like the so called  inertial transformations.
Under
the assumptions~\cite{Bacry:1968zf}
that space is rotation invariant
and
boosts form a noncompact subgroup
(and another natural technical assumption)
and
sufficient knowledge of Lie algebras
Galilei would have seen that
his physical worldview might 
be an approximation
of a more fundamental one.
He would have arrived at the Poincar\'e algebra of which
the Galilei algebra emerges
as a contraction.
See Figure \ref{fig:cubeint},
where at each corner sits a so called ``kinematical algebra'' which
we will further discuss in Chapter \ref{cha:kinem-chern-simons}.


The difference between the laws of physics how Galilei
would have seen them and the
relativistic ones 
can be made obvious by
introducing the speed of light.
We know nowadays that the speed of light is a finite
constant approximately given by $c=3 \cdot 10^{8} \,\mathrm{m}/\mathrm{s}$.
For nonrelativistic theories there is no reason
for the speed of light to be finite
and they are often described as
approximate theories
in the 
$c \to \infty$ limit
of more fundamental relativistic ones (see Figure \ref{fig:cubeint}).

\subsection{Nonrelativistic Theories}
\label{sec:nonr-theor}

For many everyday phenomena the finiteness
of the speed of light is of no relevant 
consequence and can be safely ignored.
Interesting examples
with technological interest
are
strongly coupled condensed matter systems~(for a review see \cite{Hartnoll:2009sz})
or
the fractional quantum Hall effect~\cite{Tsui:1982yy,Laughlin:1983fy,Son:2013rqa,Geracie:2014nka}.
In both cases nonrelativistic
geometries play an influential role.




\subsection{AdS/CFT}
\label{sec:adscft}

Interestingly, new ways to analyze strongly coupled
systems have been found~\cite{Maldacena:1997re,Gubser:1998bc,Witten:1998qj,Aharony:1999ti}
and are best understood
on another place of the cube, to be specific,
at the Anti-de Sitter corner (see Figure \ref{fig:cubeint}).
Here one needs to introduce an additional constant which
equals the curvature of the universe.
These new techniques are due to the 
holographic principle~\cite{'tHooft:1993gx,Susskind:1994vu}
which states that a quantum gravitational
theory admits a dual description in terms
of a non-gravitational quantum field theory in lower spacetime dimension.
It is considered a key element of any approach to quantum gravity.

This principle found its realization in the famous AdS/CFT (Anti-de Sitter/Conformal field theory) correspondence~\cite{Maldacena:1997re,Gubser:1998bc,Witten:1998qj,Aharony:1999ti}.
But 
neither AdS spacetimes nor CFTs are strictly
necessary for the holographic principle to be true.
This begs the question if the tools used in AdS/CFT can lead to insights
at other corners of the cube.
For that it is useful to start
with theories where the duality has been tested
in detail and is comparably well understood.
For that higher spin gravity seems like a good candidate
which has passed various nontrivial checks
in different dimensions (for reviews see \cite{Giombi:2012ms,Gaberdiel:2012uj}).

\subsection{Higher Spin Theories}
\label{sec:test}

A very interesting class of theories where holography is realizable is higher spin theory.
Most of the work in this thesis is focused on  $2+1$ dimensions where the theory admits a Chern--Simons formulation~\cite{Blencowe:1988gj}.
Much of the simplicity comes then from the fact that there might be a two-dimensional conformal field
theory on the boundary.
Due to the large amount of symmetries in two-dimensions 
these conformal field theories
provide a high degree of analytic control and are therefore
distinguished theories for the exploration of conceptual questions
that seem far out of reach in higher dimensions.
The higher spin bulk theory can be understood as a generalization of
pure
$(2+1)$-dimensional Einstein gravity in the Chern--Simons formulation ~\cite{Achucarro:1987vz,Witten:1988hc}
accompanied by bosonic higher spin fields, or as a simplified version of the Fradkin--Vasiliev theory~\cite{Fradkin:1987ks}. 
These theories provide new insights with respect
to possible dualities~\cite{Henneaux:2010xg,Campoleoni:2010zq,Gaberdiel:2011wb,Campoleoni:2011hg},
higher spin generalizations
of black holes~\cite{Gutperle:2011kf}, singularity resolution thereof~\cite{Castro:2011fm}, thermodynamics~\cite{Banados:2012ue,deBoer:2013gz,Bunster:2014mua}, 
entanglement entropy~\cite{Ammon:2013hba,deBoer:2013vca}, holography~\cite{Gaberdiel:2010pz} and string theory~\cite{Gaberdiel:2014cha,Gaberdiel:2015wpo}.
Therefore, this seems like an interesting starting point to look for generalizations.

This work centers around which of the above mentioned features are specific to AdS and which can be generalized.
The discussion will be focused towards
spacetimes that have the possibility to describe boundary theories with applications in, e.g., condensed matter physics~\cite{Son:2008ye,Balasubramanian:2008dm}.

Two such spacetimes (Lifshitz and Null-warped) were realized explicitly in higher spin gravity, and consistent boundary conditions and the asymptotic
symmetry algebra were provided~\cite{Gary:2014mca,Breunhoelder:2015waa}. This showed that it is possible to realize spacetimes beyond AdS in higher spin gravity.



What was missing so far was a
systematic procedure to go from higher spin  Anti-de Sitter
to  (non)relativistic higher spin theories.
Concepts that will  provide this transition
will be investigated in this
thesis (see also~\cite{Bergshoeff:2016soe}).
It can be seen  on the cover of this thesis that symmetry was again a
useful guide in deriving these (non)relativistic
higher spin geometries.
Since nonrelativistic geometries play a central role in non-AdS holography~\cite{Christensen:2013lma,Christensen:2013rfa,Hartong:2014oma,Hartong:2016yrf}
the hope is that their higher spin geometry generalization lead to an equal important generalization.

\clearpage

\subsection{Outline}
\label{sec:outline}

The Chapters \ref{cha:introduction} to \ref{sec:contr-lie-algebr}
can be seen as introduction to
the main part
given by Chapters \ref{cha:ads-spin-3} to \ref{cha:kinem-high-spin}
after which conclusion, outlook and appendices follow.
The introductory chapters are without reference to a specific
gauge algebra and therefore of general interest.
Furthermore, various statements generalize
to any gauge theory that is based on a Lie algebra valued one-form.
In the main part we will focus on specific examples
of higher spin theories and follow closely the publications\cite{Afshar:2014rwa,Gary:2014mca,Breunhoelder:2015waa,Bergshoeff:2016soe,Grumiller:2016kcp,Prohazka:2017equ,Ammon:2017vwt}.
The appendices can in principle be omitted, but
they fix the notation (see also the Index at the end)
and provide useful additional information.

\begin{description}
\item[Chapter~\ref{cha:chern-simons-theory}]
  The theory that this work is
  centered around, the Chern--Simons theory, is introduced.
  It is usually based on a gauge algebra with a symmetric invariant
  nondegenerate bilinear form (invariant metric) and each of these requirements is
  examined for its importance.
\item[Chapter~\ref{sec:gauge-algebra}] Due to a structure theorem
  it is known how Lie algebras that posses such an invariant metric
  are constructed
  and it is therefore of interest to review the ingredients.
  Besides the direct sum of one-dimensional and simple Lie algebras,
  double extensions are
   introduced.
  This is  beneficial for later considerations of kinematical algebras,
  since they are based on these concepts.
\item[Chapter~\ref{sec:contr-lie-algebr}]
  For the study of approximate physical theories
  contractions are a useful tool since one
  is automatically guided by considerations of the original theory.
  Lie algebra contractions of different generality are discussed.
  Contractions are used later in Chapter \ref{cha:kinem-chern-simons} and \ref{cha:kinem-high-spin} for the classification
  of (spin-3) kinematical algebras.
\item[Chapter~\ref{sec:contr-invar-metr}]
  A contracted Lie algebra that is useful for gauge theories
  should be accompanied by an (also contracted) invariant metric.
  For self-dual algebras a special invariant metric preserving
  contraction is defined.
\item[Chapter~\ref{sec:boundary-conditions}]
  The global charges of Chern--Simons theories with boundary
  provide information concerning possible
  boundary theories and are therefore reviewed.
\item[Chapter~\ref{cha:ads-spin-3}]
  After a short review of higher spin theories
  the standard $\mathcal{W}_{3}$ boundary conditions
  are  introduced  as $\hat{\mathfrak{u}}(1)$ composite
  objects~\cite{Grumiller:2016kcp,Ammon:2017vwt}.
\item[Chapter~\ref{cha:non-ads-cs}]
  Consistent boundary conditions
  for Lifshitz~\cite{Gary:2014mca}
  and null-warped~\cite{Breunhoelder:2015waa} spin-3 gravity
  and difficulties concerning their interpretation
  are reviewed.
\item[Chapter~\ref{cha:kinem-chern-simons}]
  Kinematical algebras are analyzed 
  and boundary conditions for Carroll gravity~\cite{Bergshoeff:2016soe} are
  proposed.
\item[Chapter~\ref{cha:kinem-high-spin}]
  Using contractions spin-3 kinematical
  algebras are classified~\cite{Bergshoeff:2016soe}.
  For spin-3 Carroll gravity the invariant metric preserving contractions of Chapter
  \ref{sec:contr-invar-metr} show their usefulness
  whereas the considerations of Chapter~\ref{sec:gauge-algebra} concerning
  double extensions provide spin-3 Galilei gravity with an invariant metric.
\item[Chapter~\ref{cha:conclusions}]
  Conclusions and a discussion of interesting
  open problems and possible future projects are provided.
\item[Appendix~\ref{cha:conventions}]
  A summary of the conventions is provided in this appendix.
\item[Appendix~\ref{cha:lie-algebras}]
  A brief review of Lie algebra concepts that are
  used in the main part of this thesis is given,
  partially to fix the notation.
\item[Appendix~\ref{sec:useful}]
  Some explicit calculations for symmetry discussions
  for CS actions are provided.
\item[Appendix~\ref{cha:explicit-lie-algebra}]
  A useful and extensive
  overview of the various Lie algebras
  and their
  invariant metrics
  that underlie spin-2 and spin-3 gravity is given.
\end{description}

\chapter{Chern--Simons Theory}
\label{cha:chern-simons-theory}



We start by introducing the theory that forms the foundation of this
work, the Chern--Simons theory.
It is based on a Lie algebra with
an invariant metric.
The importance of each of the properties
of this symmetric nondegenerate invariant bilinear form will
be examined. 


\section{Chern--Simons Action}
\label{sec:what-chern-simons}

The Lagrange density of the three-dimensional Chern--Simons (CS)
theory~\cite{Chern:1974ft} (see also \cite{Deser:1982vy,Deser:1981wh} and \cite{Witten:1988hf})
is given by
\begin{align}
  \label{eq:CSLagrang}
  \CS[A]
  &=\langle \dd A \wedge A + \frac{2}{3} A \wedge A \wedge A \rangle
  \\
  &\equiv \langle \dd A \wedge A + \frac{1}{3} [A,A] \wedge A \rangle
  \\
  &=\langle \Tb_{a}\Tb_{b}\rangle
    (
    \dd A^{a} \wedge A^{b} + \frac{1}{3} f\indices{_{cd}^{a}} A^{c} \wedge A^{d} \wedge A^{b}
    ) 
\end{align}
\index{Chern--Simons theory!Action}%
\index{CS theory|see {Chern--Simons theory}}%
with some connection $A$. We also write
$A=A\indices{_{\mu}}\dd x^{\mu}=A^{a}\Tb_{a} =A\indices{^{a}_{\mu}}\,
\Tb_{a}\otimes \dd x^{\mu} \in \mathfrak{g} \otimes TM_{3}^{*}$,
which shows that $A$ is a Lie algebra valued one-form\footnote{
  By writing the Lagrangian density in this form we implicitly
  assume that the $G$ bundle is trivial.
  The connection can otherwise not be regarded as a Lie algebra
  valued one-form although a suitable generalized definition
  exists~(see, e.g., \cite{Dijkgraaf:1989pz}).
  For connected, simply connected Lie groups on a three manifold
  the $G$ bundle is necessarily trivial.
  So specifying the Lie group and not just the Lie algebra
  differentiates between the Lie groups whose Lie algebra is
  $\mathfrak{g}$
  and provides additional information.
  We will ignore this subtleties in the following and restrict mainly
  to discussions of the Lie algebra.
  For more information see~\cite{Dijkgraaf:1989pz}.
}.
We define the commutator between Lie algebra valued one-forms by $[A,B]\equiv A^{a} \wedge B^{b} \,[\Tb_{a},\Tb_{b}]$ where $[\Tb_{a},\Tb_{b}]=f\indices{_{ab}^{c}}\Tb_{c}$
is the Lie bracket.
The symmetric nondegenerate invariant bilinear form, also called invariant
metric, is denoted by $\langle \Tb_{a}\Tb_{b} \rangle$ (see Definition
\ref{def:invariant-metric} in Section \ref{sec:invariant-metric}).
Often this is written as $\tr(\Tb_{a}\Tb_{b})$, but as will become
clear this form can be defined without any reference to a matrix
representation and a trace thereof.
Therefore, this notation is reserved for places where the matrices are actually
defined.
Using the Lagrangian density the action is given by 
\begin{equation}
  \label{eq:CSAction}
  I_{\CS}[A]= \frac{k}{4 \pi} \int_{M_{3}} \CS[A] 
\end{equation}
where $M_{3}$ denotes an oriented three-dimensional manifold.

As we have just defined the Chern--Simons (bulk) theory this leaves still
some freedom:
\begin{enumerate}
\item The three-dimensional manifold is the spacetime and it is
  mostly assumed that we can decompose it as $M_{3}=\R \times \Sigma$.
  The time part $\R$ might get identified periodically when black holes are
  discussed in an Euclidean setup.
  The two-dimensional space part $\Sigma$ is for holographic purposes
  assumed to have an (asymptotic) boundary, see Figure \ref{fig:lor}.
  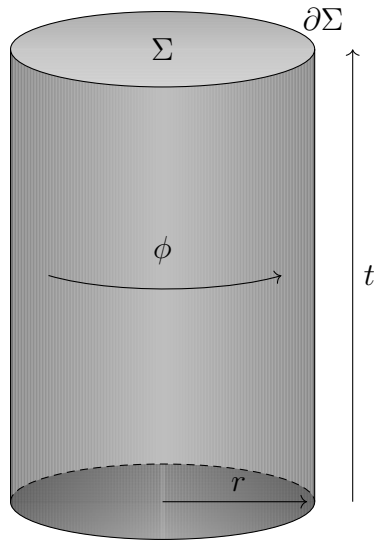
\begin{figure}[h]
  \centering
\begin{tikzpicture}
\fill[top color=gray!50!black,bottom color=gray!10,middle color=gray,shading=axis,opacity=0.25] (0,0) circle (2cm and 0.5cm);
\fill[left color=gray!50!black,right color=gray!50!black,middle color=gray!50,shading=axis,opacity=0.25] (2,0) -- (2,6) arc (360:180:2cm and 0.5cm) -- (-2,0) arc (180:360:2cm and 0.5cm);
\fill[top color=gray!90!,bottom color=gray!2,middle color=gray!30,shading=axis,opacity=0.25] (0,6) circle (2cm and 0.5cm);

\draw (-2,6) -- (-2,0) arc (180:360:2cm and 0.5cm) -- (2,6) ++ (-2,0) circle (2cm and 0.5cm);

\node [above] at (0,3) {$\phi$};
\draw [arrows=->] (-1.5,3) arc (220:320:2cm and 0.5cm) ;

\draw[densely dashed] (-2,0) arc (180:0:2cm and 0.5cm);
\node [above] at (1,0) {$r$};

\draw[arrows=->] (0,0) -- (1.9,0);

\draw[arrows=->] (2.5,0) -- (2.5,6);
\node [right] at (2.5,3) {$t$};

\node at (0,6) {$\Sigma$};
\node [right] at (1.7,6.4) {$\pd\Sigma$};
\end{tikzpicture}  
  \caption{The three-dimensional manifold $M_{3}$.}
  \label{fig:lor}
\end{figure}
\item Our goal is to describe three-dimensional gravitational theories
  using the  Chern--Simons description~\cite{Achucarro:1987vz,Witten:1988hc}.
  The Lie algebra $\mathfrak{g}$ specifies then which one.
  The Chern--Simons theory based on
  $\mathfrak{sl}(2,\R) \dis \mathfrak{sl}(2,\R)$ for example leads to
  three-dimensional gravity with negative cosmological constant,
  whereas $\mathfrak{sl}(2,\C)$ and $\mathfrak{isl}(2,\R)$ corresponds
  to positive and vanishing cosmological
  constant, respectively.
  Lie algebras that have these Lie algebras as a subalgebra can be understood
  as a generalization of Einstein gravity, e.g.,
  $\mathfrak{sl}(3,\R)\dis \mathfrak{sl}(3,\R)$ is a generalization
  including an additional spin $3$ field.
\item Additionally to the Lie algebra one needs to specify the invariant
  metric. Once a Lie algebra is chosen it might happen that the invariant metric
  has some freedom, outside of the overall scaling, that one needs to
  specify.
  Another possibility is that the Lie algebra might not posses an
  invariant metric.
  So, there is a tight connection between the Lie algebra and its
  possible invariant metric. To specify one without the
  other makes little sense for Chern--Simons theories.
  
  The importance of the various conditions of the
  invariant metric for a
  well defined Chern--Simons theory will be discussed in the next
  section.
  The kind of Lie algebras that have an invariant metric are reviewed
  in Section \ref{sec:gauge-algebra}.
\item One point that might not seem obvious from the definition of the
  action is the importance of boundary conditions. Without specifying
  these the action is not well defined and from a holographic point of
  view the boundary conditions determine the possible boundary
  theories. This will be discussed in Section
  \ref{sec:boundary-conditions}.
\end{enumerate}

One point that differentiates Chern--Simons theory from other
theories like electrodynamics and general relativity is that it is
independent of any spacetime metric.
It is thus a topological quantum field theory of Schwarz type, for a review see~\cite{Birmingham:1991ty}.

Another property of CS theories (in three dimensions) is that it has no
local degrees of freedom in the bulk or
in other words, there are no ``Chern--Simons waves'' propagating inside
the spacetime.
This fits nicely with the fact that pure three-dimensional gravity,
for any value of the cosmological constant,
also has no gravitational waves~\cite{Staruszkiewicz:1963zza}.

\subsection{Variation and Equations of Motion}
\label{sec:variation-action}

To get the equations of motion we vary the CS Lagrangian density
\begin{align}
  \vd \CS[A]
  &=\langle \dd\vd A \wedge A + (\dd A + 2 A \wedge A) \wedge \vd A \rangle \label{eq:variation}
  \\
  &=\langle  2  F \wedge \vd A \rangle - \dd \langle A \wedge \vd
    A\rangle \label{eq:variation2} \,.
\end{align}
Here we have defined the Lie algebra valued two-form $F = \dd A + A
\wedge A \equiv \dd A + \frac{1}{2} [A , A]$,
which is the curvature of the connection.
Given suitable boundary conditions, meaning that the boundary term in
\eqref{eq:variation2} vanishes when integrated,
leads to the equations of motion that the curvature is flat $F=0$,
or more explicitly,
\begin{align}
  F_{\mu\nu}^{a}
  =\pd_{\mu}A\indices{^{a}_{\nu}}-\pd_{\nu}A\indices{^{a}_{\mu}} + f\indices{^{a}_{bc}} A\indices{^{b}_{\mu}}A\indices{^{c}_{\nu}}
  =0 \,.
\end{align}
\index{Chern--Simons theory!Equations of motion}%
Solutions to the equations of motion can locally be written as $A=g^{-1}\dd g$ for a group element $g$, see Appendix \ref{sec:deta-solut-f=0}.
\index{Chern--Simons theory!Solutions}

\section{Invariant Metric}
\label{sec:invariant-metric}

We will now define what an invariant metric is.
Afterwards  will be  examined why
and to which extend each of its properties are really necessary for a
well defined CS theory.
This is of special importance since each part of the definition of
the invariant metric is an additional restriction on the possible Lie
algebras. For example, any Lie algebra would be possible for a well defined
CS theory if we
left out the condition of non-degeneracy.
So it is of interest if one could relax some conditions and still get a
well defined theory.

\begin{definition}
  \label{def:invariant-metric}
  An \textbf{invariant metric} is a bilinear form $\langle \cdot , \cdot \rangle: \mathfrak{g}\times \mathfrak{g}\rightarrow K$ on a Lie algebra $\mathfrak{g}$ with field $K$ which has the following three properties:
  \begin{enumerate}
  \item Symmetry
    \begin{equation}
      \langle X,Y \rangle
      = \langle Y,X\rangle
      \quad \text{ for all }\quad  X,Y \in \mathfrak{g} \,.
    \end{equation}
  \item Non-degeneracy
    \begin{equation}
      \text{If }
      \quad
      \langle X,Y \rangle =0
      \quad
      \text{ for all }
      \quad
      Y \in \mathfrak{g} 
      \quad
      \text{ then }
      \quad X=0 \,. 
    \end{equation}
  \item Invariance
    \begin{align}
      \label{eq:invar}
      \langle [Z,X], Y \rangle
      +\langle X,[Z,Y]\rangle=0
      \quad \text{ for all }
      \quad  X,Y,Z \in \mathfrak{g} \,.
    \end{align}
  \end{enumerate}
  A     \index{Lie algebra!Symmetric self-dual}%
  \textbf{symmetric self-dual Lie algebra}\footnote{
    We follow here \cite{FigueroaO'Farrill:1995cy} and for details concerning the nomenclature see Remark 2.2 therein.
    Other names are
    ``metric'', ``metrised'', ``orthogonal'', ``quadratic'',
    ``regular quadratic'',
    and ``self-dual'' Lie algebra~\cite{Ovando:Review}.
  }
  is a Lie algebra possessing an invariant metric.
\end{definition}
\index{Invariant Metric}
When there is no risk of confusion the comma between the two
arguments of the bilinear form will be omitted.
Given two symmetric self-dual algebras with their invariant metrics
$(\mathfrak{g}_{1},\langle \cdot,\cdot \rangle_{1})$ and $(\mathfrak{g}_{2},\langle \cdot,\cdot \rangle_{2})$
we can obtain a new symmetric self-dual algebra by using a direct sum of Lie algebras and the orthogonal direct product metric
$(\mathfrak{g}_{1} \dis \mathfrak{g}_{2},\langle \cdot,\cdot \rangle_{1} \vdis \langle \cdot,\cdot \rangle_{2})$.
A Lie algebra which can be written as such a direct sum is \textbf{decomposable}, if not it is \textbf{indecomposable}.
Examples for indecomposable symmetric self-dual Lie algebras are simple and one-dimensional Lie algebras,
whereas semisimple ones are decomposable.
That there are symmetric self-dual Lie algebras beyond these examples
will be shown in Section \ref{sec:gauge-algebra}.
\index{Invariant Metric!Decomposable}%
\index{Invariant Metric!Indecomposable}%

Using the basis $\Tb_{a}$ for the Lie algebra
$[\Tb_{a},\Tb_{b}]=f\indices{_{ab}^{c}}\Tb_{c}$ and
$\Tb_{ab}\equiv\langle \Tb_{a},\Tb_{b}\rangle$ the conditions
on the invariant metric
in
components are given by
\begin{align}
  \Tb_{ab}&=\Tb_{ba} & &\text{(Symmetry),}
                     \\
  \det (\Tb_{ab})&\neq 0 & &\text{(Non-degeneracy),}
\\
  f\indices{_{ab}^{d}}\Tb_{dc}+ f\indices{_{ac}^{d}}\Tb_{db}&=0 & &\text{(Invariance).}
\end{align}
Not every Lie algebra admits an invariant metric, e.g., the three-dimensional Galilei algebra
or the two-dimensional algebra
$[\Tb_{1},\Tb_{2}]=\Tb_{1}$, $[\Tb_{1},\Tb_{1}]= [\Tb_{2},\Tb_{2}]=0$
do not.
We will now analyze why symmetry, nondegeneracy and invariance
are important properties for CS theories.

\subsection{Symmetry}
\label{sec:symmetric}

We start with the CS Lagrangian and ignore the condition that
the invariant metric should be symmetric.
One is then always able to decompose the bilinear form into symmetric and antisymmetric parts
\begin{align}
  \langle \Tb_{a} \Tb_{b}\rangle
  &=\frac{1}{2}\left(\langle \Tb_{a} \Tb_{b}\rangle + \langle \Tb_{b} \Tb_{a}\rangle\right)
  +\frac{1}{2} (\langle \Tb_{a} \Tb_{b}\rangle - \langle \Tb_{b} \Tb_{a}\rangle)
  \\
  &= \langle \Tb_{a} \Tb_{b}\rangle_{\mathrm{S}}
  + \langle \Tb_{a} \Tb_{b}\rangle_{\mathrm{AS}}
   \,.
\end{align}
If we apply this to the CS Lagrangian
the symmetric part reduces to the well known CS Lagrangian \eqref{eq:CSLagrang},
the antisymmetric part reduces to a total derivative
\begin{align}
  \label{eq:CSas}
  \langle \dd A \wedge A + \frac{1}{3} [A,A] \wedge A \rangle_{\mathrm{AS}}
  =\frac{1}{2} \dd \langle A \wedge A \rangle_{\mathrm{AS}} \,.
\end{align}
The first term of the left hand side of \eqref{eq:CSas} leads to the total derivative
and the second one vanishes using the antisymmetry and the invariance of the bilinear form.

So, in principle, one could relax the symmetry condition,
but one would merely change the theory by a total derivative\footnote{
  This might have implications for boundary theories.
}
or equivalently, the equations of motion would stay unaltered.

\subsection{Non-degeneracy}
\label{sec:non-degenerate}


If we ignore the condition of non-degeneracy in the definition of the invariant metric
then there exists a vector subspace $V \subset \mathfrak{g}$ of the Lie algebra to which the whole Lie algebra is orthogonal,
i.e., $\langle V , \mathfrak{g} \rangle =0$.
An immediate consequence is that the fields that are part of $V$ have no kinetic term $\langle A \wedge \dd A \rangle$
and are therefore not dynamical.

So non-degeneracy is necessary if we want a theory where all fields
have a kinetic term.

\subsection{Invariance}
\label{sec:invariant}

We will illustrate the importance of the invariance of the metric
for nonabelian gauge theories by applying (part of) a gauge transformation $g=e^{Z}$ to $\langle X,Y \rangle$,
\begin{align}
  \langle g^{-1}X g ,g^{-1}Y g\rangle
    &=\langle X,Y \rangle -\langle [Z,X],Y\rangle - \langle X ,[Z,Y]\rangle + \mathcal{O}(Z^{2}) \,.
\end{align}
The invariance of the metric 
$\langle [Z,X],Y\rangle + \langle X ,[Z,Y]\rangle=0$
(equation \eqref{eq:invar})
is sufficient that
these kind of ``gauge transformations'' vanish.
Not having this invariance property might lead to additional constraints for the possible Lie algebras.

If one inserts for $X$ and $Y$ the curvature and the Hodge dual curvature of a connection this
calculation basically shows also the importance of the invariance of the
 metric for the gauge invariance of the Yang--Mills action.
Invariance is also important for similar calculations concerning the CS theory in Section \ref{sec:boundary-conditions},
as well as for other gauge theories like, e.g., the Wess--Zumino--Witten (WZW) model.

\subsection{Summary}
\label{sec:summary}

For a well defined Chern--Simons theory where all fields have
a kinetic term it seems reasonable to search for Lie algebras with
invariant metric, i.e., for symmetric self-dual Lie algebras.
In the next chapter we will discuss what kind of Lie algebras posses
such an invariant metric.

\chapter{Symmetric Self-dual Lie Algebras}
\label{sec:gauge-algebra}








Lie algebras that posses an invariant metric play a fundamental role
for gauge theories in physics,
e.g., as possible gauge algebras for Yang--Mills, CS and WZW theories.

We will now examine what kind of Lie algebras admit such a metric.
The discussions in this chapter are general
and independent of any specific gauge theory.
Appendix \ref{cha:lie-algebras} contains further details and definitions.

\section{Reductive Lie Algebras and the Killing Form}
\label{sec:simple-lie-algebras}

Given a Lie algebra $\mathfrak{g}$ over a real or complex field
$K$ one can always construct
the \textbf{Killing form} $\kappa:\mathfrak{g}\times \mathfrak{g} \to K$ by defining
\begin{align}
  \label{eq:killingf}
  \kappa(X,Y)\equiv \tr (\ad_{X} \circ \ad_{Y})
  \quad \text{ or using a basis } \quad
  \kappa(\Tb_{a},\Tb_{b})= f\indices{_{ac}^{d}} f\indices{_{bd}^{c}}\,.
\end{align}
\index{Lie algebra!Killing form}%
The definition of the adjoint action (see appendix \ref{sec:basic-concepts})
and the invariance of the trace under cyclic permutations
shows that the Killing form is a symmetric invariant bilinear form on the Lie algebra.
However, as stated by Cartan's criterion, in general the Killing form might be degenerate.
\begin{theorem}[Cartan's criterion]
  A Lie algebra is semisimple if and only if its Killing form is non-degenerate.
\end{theorem}
So it follows that only for the semisimple Lie algebras the Killing
form automatically provides us with an invariant metric.
For simple Lie algebras this invariant metric is even unique up to overall normalization.
\begin{example}[$\mathfrak{sl}(2,\R)$]
  An example for a simple Lie algebra is $\mathfrak{sl}(2,\R)$ given
  by the commutation relations
  \begin{align}
    \label{eq:sl2ex1}
    [\Lt_{a},\Lt_{b}]=(a-b) \Lt_{a+b}
  \end{align}
  where $a=-1,0,+1$.
  As just discussed, since the Lie algebra is simple we can construct
  the Killing form that then automatically provides us with an
  invariant metric.
  An explicit calculation gives the Killing form (for further details see Section \ref{sec:sl2-r})
  \begin{align}
    \kappa (\Lt_{a}\,\Lt_{b})=
    \left(
    \begin{array}{c|ccc}
                 & \Lt_{-1}       & \Lt_0       & \Lt_{+1}    \\
      \hline
      \Lt_{-1}   & 0              & 0           & -4          \\
      \Lt_0      & 0              & 2 & 0           \\
      \Lt_{+1}   & -4             & 0           & 0           \\
    \end{array}
  \right)
  \end{align}
  which indeed fulfills all requirements of a well defined invariant
  metric. So does any invariant metric proportional to it.

  For a semisimple Lie algebra one could now add a second
  $\mathfrak{sl}(2,\R)$ Lie algebra as a direct sum.
  So additionally to \eqref{eq:sl2ex1} we have now
  \begin{align}
    [\widetilde \Lt_{a},\widetilde \Lt_{b}]&=(a-b)\widetilde \Lt_{a+b}
    &
    [\widetilde \Lt_{a}, \Lt_{b}]&=0
  \end{align}
  for which we get the Killing form $\kappa (\widetilde \Lt_{a}\,
  \widetilde \Lt_{b})=\kappa (\Lt_{a}\,\Lt_{b})$ and $\kappa
  (\Lt_{a}\,\widetilde \Lt_{b})=0$.
  For the direct sum we have two parameters in the invariant metric,
  one for each factor,
  that we can freely choose.
\end{example}

A generalization of semisimple Lie algebras is given by
\textbf{reductive} Lie algebras which are direct sums of simple and abelian Lie algebras.
\index{Lie algebra!Reductive}
Since the commutator of an abelian Lie algebra vanishes,
so does their Killing form.
Nevertheless is it possible to construct an invariant metric for reductive Lie algebras.

\begin{example}[$\mathfrak{u}(1)$]
\label{sec:dimension-1}
  The abelian Lie algebra $\mathfrak{u}(1)$, which is the unique one-dimensional algebra,
  is given by the commutation relation
  $[\Tb,\Tb]=0$.
  Even though the Killing form is $\kappa(\Tb,\Tb)=0$ we can define
  an invariant metric by $\langle \Tb,\Tb \rangle = \mu$,
  where $\mu$ is a nonzero real constant.

  A reductive Lie algebra would then be for example the direct sum
  $\mathfrak{sl}(2,\R) \dis \mathfrak{u}(1)$, with the same invariant
  metrics as on their factors and $\langle \Lt_{a}, \Tb \rangle = 0$.
\end{example}

One might ask if one could take direct sums of
Lie algebras and find an invariant metric that makes it
indecomposable.
This would mean for Example \ref{eq:sl2ex1} that $\kappa
(\Lt_{a}\,\widetilde \Lt_{b})\neq 0$
or for Example \ref{sec:dimension-1} that $\langle \Lt_{a}, \Tb \rangle \neq 0$.

That this is not possible for the direct sum with a simple Lie
algebra can be easily shown.
Suppose we have a symmetric self-dual Lie algebra $\mathfrak{s} \dis
\mathfrak{g}$ which is a direct sum of a simple one $\mathfrak{s}$
with another arbitrary Lie algebra $\mathfrak{g}$. Then the invariant
metric is orthogonal since
$
\langle \mathfrak{s}, \mathfrak{g} \rangle
=
\langle [\mathfrak{s},\mathfrak{s}], \mathfrak{g} \rangle
=
\langle \mathfrak{s}, [\mathfrak{g},\mathfrak{s}] \rangle
=
0
$~\cite{FigueroaO'Farrill:1995cy}.
We have used that simple Lie algebras are perfect
($[\mathfrak{s},\mathfrak{s}]=\mathfrak{s}$), that the metric is
invariant and afterwards that the Lie algebras are a direct sum.
For abelian Lie algebras one can always find an isomorphism that also
diagonalizes the invariant metric and therefore makes it decomposable.
So, reductive Lie algebras are also always decomposable.

Many important gauge theories are based on reductive gauge algebras,
e.g.,
electrodynamics with $\mathfrak{u}(1)$
and the Standard Model of particle physics with $\mathfrak{su}(3)\oplus \mathfrak{su}(2) \oplus \mathfrak{u}(1)$.


\section{Double Extensions}
\label{sec:double-extensions}

An interesting question is of course if there are Lie algebras
possessing an invariant metric besides the reductive ones.
We answer this in the affirmative, via the construction of double
extensions,
and discuss the construction
of any such symmetric self-dual Lie algebra in the next section.
This section is based on the work of Medina and Revoy~\cite{Medina1985},
but we will follow closely~\cite{FigueroaO'Farrill:1995cy}.
For the notation see Appendix \ref{cha:lie-algebras} or the Index
(the symbol $\vdis$ means direct sum as vector space).

\begin{definition}[Double extension~\cite{Medina1985}]
  \label{def:double-ext}
  Let $(\mathfrak{g}, \langle \cdot , \cdot \rangle_{\mathfrak{g}})$ be a Lie algebra with an invariant metric 
  on which a Lie algebra $\mathfrak{h}$ acts via antisymmetric derivations, i.e.,
  \begin{align}
    h \cdot [x,y]_{\mathfrak{g}} = [h \cdot x,y]_{\mathfrak{g}}+ [x, h \cdot y]_{\mathfrak{g}}
    \quad \text{and} \quad
      \langle h \cdot x,y\rangle_{\mathfrak{g}} + \langle x, h \cdot y \rangle_{\mathfrak{g}}= 0 \,.
  \end{align}
  Then we can define on the vector space $\mathfrak{g} \vdis \mathfrak{h} \vdis \mathfrak{h}^{*}$
  the Lie algebra $\mathfrak{d}$, called the \textbf{double extension} of $\mathfrak{g}$ by $\mathfrak{h}$,
  by
  \begin{multline}
    \label{eq:doublext}
    [(x,h,\alpha), (x',h',\alpha')]=
     \\
     (
    [x,x']_{\mathfrak{g}}+ h \cdot x' - h' \cdot x
    ,
    [h,h']_{\mathfrak{h}}
    ,
    \beta(x,x') + \ad_{h}^{*}\cdot\, \alpha' - \ad_{h'}^{*} \cdot \,\alpha
    )    
  \end{multline}
  where $x,x' \in \mathfrak{g}$, $h,h' \in \mathfrak{h}$, $\alpha, \alpha' \in \mathfrak{h}^{*}$.
  The skew-symmetric bilinear form $\beta: \mathfrak{g} \times \mathfrak{g} \to \mathfrak{h}^{*}$ fulfills
  \begin{align}
    \langle h \cdot x,y \rangle_{\mathfrak{g}}
    =
    \langle h , \beta(x,y) \rangle \,.
  \end{align}
\end{definition}
\index{Lie algebra!Double extension}%
Double extensions are symmetric self-dual Lie algebras since they
always carry an invariant metric defined by
\begin{align}
  \langle (x,h,\alpha) , (x',h',\alpha')\rangle
  =
  \langle x, x'\rangle_{\mathfrak{g}}
  +
  \langle h, h' \rangle_{\mathfrak{h}}
  +
  \alpha(h') + \alpha'(h)
\end{align}
where $\langle \cdot, \cdot \rangle_{\mathfrak{h}}$ is a
(possibly degenerate)
invariant symmetric bilinear form on $\mathfrak{h}$.

The stars following $\mathfrak{h}^{*}$ and $\ad^{*}$ denote dual space and
coadjoint representation, respectively.
We will denote double extensions by $D(\mathfrak{g},\mathfrak{h})$ or
\index{$D(\mathfrak{g},\mathfrak{h})$}
the mnemonic $ (\mathfrak{g} \dis_{c} \mathfrak{h}^{*})\sdis
\mathfrak{h}$.
This also explains the name double extension since $\mathfrak{g}$ is
centrally extended by $\mathfrak{h}^{*}$ which then split extends $\mathfrak{h}$.

Any nontrivial double extension, meaning that $\mathfrak{h}$ is
nontrivial, is non-semisimple.
This is due to the abelian ideal $[\alpha,\alpha']=0$.
If $\mathfrak{h}$ is also nonabelian we have a new class
of symmetric self-dual Lie algebras.

Before we discuss further details of double extensions we will write it in a basis.
For $\mathfrak{g}$ we fix the basis $\{\Gb_{i}\}$ in which the invariant metric of $\mathfrak{g}$ is given by $\Omega^{\mathfrak{g}}_{ij}$
and the commutation relations by $[\Gb_{i},\Gb_{j}]_{\mathfrak{g}}=f\indices{_{ij}^{k}}\Gb_{k}$.
For the Lie algebra $\mathfrak{h}$ the basis  $\{\Hb_{\alpha}\}$ has the Lie bracket $[\Hb_{\alpha},\Hb_{\beta}]_{\mathfrak{h}}= f\indices{_{\alpha\beta}^{\gamma}} \Hb_{\gamma}$
which acts via antisymmetric derivations
\begin{align}
  \Hb_{\alpha} \cdot \Gb_{i}
  &= f\indices{_{\alpha i}^{j}} \Gb_{j}
  \qquad  (f\indices{_{\alpha i}^{j}}=-f\indices{_{i \alpha }^{j}})\,.
\end{align}
The fact that it is a derivation can be read of from
$
\Hb_{\alpha} \cdot [\Gb_{i},\Gb_{j}]
=[ \Hb_{\alpha} \cdot \Gb_{i},\Gb_{j}]
+[ \Gb_{i},  \Hb_{\alpha} \cdot \Gb_{j}],
$
and is equivalent to
\begin{align}
  \label{eq:derivexp}
  f\indices{_{\alpha k}^{l}} f\indices{_{ij}^{k}}= f\indices{_{\alpha i}^{k}} f\indices{_{kj}^{l}} + f\indices{_{i k}^{l}} f\indices{_{\alpha j}^{k}}
\end{align}
whereas the antisymmetry condition
$
\langle [\Hb_{\alpha}, \Gb_{i}],\Gb_{j}\rangle_{\mathfrak{g}}
+
\langle  \Gb_{i},[\Hb_{\alpha},\Gb_{j}]\rangle_{\mathfrak{g}}
= 0
$
leads to
\begin{align}
  \label{eq:antisymexp}
  f\indices{_{\alpha i}^{k}}\Omega_{kj}^{\mathfrak{g}}+f\indices{_{\alpha j}^{k}}\Omega_{ki}^{\mathfrak{g}}=0 \,.
\end{align}
Its canonical dual basis is given by $\{\Hb^{\alpha}\}$.

Then the Lie algebra $\mathfrak{d}=D(\mathfrak{g},\mathfrak{h})$ defined on the vector space $\mathfrak{g} \dot + \mathfrak{h} \dot +\mathfrak{h}^{*}$ by 
\begin{align}
 [\Gb_{i},\Gb_{j}]                            & =f\indices{_{ij}^{k}} \Gb_{k}+f\indices{_{\alpha i}^{k}} \Omega_{kj}^{\mathfrak{g}} \Hb^{\alpha} 
                    \label{eq:GG}           \\
 [\Hb_{\alpha},\Gb_{i}]                       & =f\indices{_{\alpha i}^{j}} \Gb_{j}
                         \label{eq:HG}      \\
 [\Hb_{\alpha},\Hb_{\beta}]                   & =f\indices{_{\alpha \beta}^{\gamma}} \Hb_{\gamma}
                             \label{eq:HH}  \\
 [\Hb_{\alpha},\Hb^{\beta}]                   & =-f\indices{_{\alpha \gamma}^{\beta}} \Hb^{\gamma}
                             \label{eq:HHd} \\
  [\Hb^{\alpha},\Gb_{j}]                      & =0
                          \label{eq:HdG}    \\
  [\Hb^{\alpha},\Hb^{\beta}]                  & =0
                              \label{eq:HdHd}
\end{align}
is a double extension of $\mathfrak{g}$ by $\mathfrak{h}$. It has the invariant metric
\begin{align}
  \Omega_{ab}^{\mathfrak{d}}= \bordermatrix{~ & \Gb_{j}                    & \Hb_{\beta}                       & \Hb^{\beta} \cr
                            \Gb_{i}           & \Omega_{ij}^{\mathfrak{g}} & 0                                 & 0  \cr
                             \Hb_{\alpha}     & 0                          & h_{\alpha\beta}                   & \delta\indices{_{\alpha}^{\beta}} \cr
                             \Hb^{\alpha}     & 0                          & \delta\indices{^{\alpha}_{\beta}} & 0 \cr}
\end{align}
where $h_{\alpha\beta}$ is some arbitrary (possibly degenerate) invariant symmetric bilinear form on $\mathfrak{h}$.

Even though the notation $D(\mathfrak{g},\mathfrak{h})$ might suggest otherwise further information is necessary to fully define the double extension.
Since that can be clearly illustrated via the explicit Lie bracket realizations
\eqref{eq:GG} to \eqref{eq:HdHd}, the necessary expression is provided in the parentheses:
\begin{enumerate}
\item The Lie algebra $\mathfrak{g}$ ($f\indices{_{ij}^{k}}$).
\item An invariant metric on $\mathfrak{g}$ ($\Omega^{\mathfrak{g}}_{ij}$).
\item The Lie algebra $\mathfrak{h}$ ($f\indices{_{\alpha \beta}^{\gamma}}$).
\item The action (antisymmetric derivation) of $\mathfrak{h}$ on $\mathfrak{g}$  ($f\indices{_{\alpha i}^{j}}$).
\end{enumerate}
As can be seen the remaining structure is mandated by the given one.
  
To fix the invariant metric of the double extension
one has to additionally provide the invariant symmetric bilinear form on $\mathfrak{h}$ ($h_{\alpha\beta}$).
This part of the bilinear form can be freely rescaled
without disturbing the properties of the full invariant metric.

It might be illuminating to check that the double extended Lie algebra
is indeed well defined.
Antisymmetry for the right hand side of the $[\Gb_{i},\Gb_{j}]$
commutator, see equation \eqref{eq:GG}, follows from
the definition of $\mathfrak{g}$ and the antisymmetry condition
\eqref{eq:antisymexp}.
Otherwise antisymmetry follows from the definition of the derivation
and of the Lie algebra $\mathfrak{h}$.

To verify that the Jacobi identity is satisfied
we will use that
$f_{ijk} \equiv f\indices{_{ij}^{l}} \Omega_{lk}^{\mathfrak{g}}$
and
$f_{ij\alpha} \equiv f\indices{_{\alpha i}^{k}}
\Omega_{kj}^{\mathfrak{g}}$
are totally antisymmetric in $ijk$ and $ij\alpha$, respectively.
Then
\begin{align}
  \label{eq:GGGjac}
  \underset{ijk}{\circlearrowleft} [[\Gb_{i},\Gb_{j}],\Gb_{k}]
  &=
  \underset{ijk}{\circlearrowleft} f\indices{_{ij}^{l}}
    f\indices{_{\alpha l}^{m}} \Omega_{mk}^{\mathfrak{g}} \Hb^{\alpha}
  \\
    &=
    (
    f\indices{_{ljk}}
    f\indices{_{\alpha i}^{l}}
    +
    f\indices{_{ilk}}
    f\indices{_{\alpha j}^{l}}
    +
    f\indices{_{ki}^{l}}
    f\indices{_{lj\alpha }}
    +
    f\indices{_{jk}^{l}}
    f\indices{_{li \alpha }}
      )\Hb^{\alpha}
  \\
  &=0
\end{align}
where we used in the first line that $\mathfrak{g}$ itself satisfies
Jacobi's identity, which leaves us with the remaining terms.
In the second line the sum is expanded and the derivation condition
\eqref{eq:derivexp} is used on the first term. The total antisymmetry
of $f_{ijk}$ and $f_{ij\alpha}$ can then be used to show that the
first (second) and last (third) term cancel.
The other Jacobi identities can be verified in a similar manner.



\section{Indecomposable Symmetric Self-dual Lie Algebras}
\label{sec:selfd-symm-lie}

In the last section we have seen that double extensions provide an
additional way to construct Lie algebras with invariant metrics.
The proof that all symmetric self-dual Lie algebras can be obtained by
direct sums and double extensions of simple and one-dimensional Lie
algebras is due to the structure theorem of Medina and Revoy
\cite{Medina1985}.
Here, we add two additional refinements
(\ref{item:MRg} and \ref{item:MRhout})
which,
to my best knowledge, were first presented in~\cite{FigueroaO'Farrill:1995cy}.
\begin{theorem}
  \label{thm:MR}
  Every indecomposable Lie algebra which permits an invariant metric,
  i.e., every indecomposable symmetric self-dual Lie algebra
  is either:
  \begin{enumerate}
  \item A simple Lie algebra.
  \item A one-dimensional Lie algebra.
  \item A double extended Lie algebra $D(\mathfrak{g},\mathfrak{h})$ where:
    \begin{enumerate}
    \item $\mathfrak{g}$ has no factor $\mathfrak{p}$ for which
      $H^{1}(\mathfrak{p},\R) = H^{2}(\mathfrak{p},\R) = 0$.
      \label{item:MRg}
    \item $\mathfrak{h}$ is either simple or one-dimensional.
      \label{item:MRhsimp}
    \item $\mathfrak{h}$ acts on $\mathfrak{g}$ via outer derivations.
      \label{item:MRhout}
    \end{enumerate}
  \end{enumerate}
\end{theorem}
Since every decomposable Lie algebra can be obtained from the
indecomposable ones this theorem describes how all of them can be generated,
see Figure \ref{fig:selfdual}.

In Theorem \ref{thm:MR} we have presented further restrictions on double extensions that
are necessary to make them indecomposable. They are not sufficient as
will be shown in Example~\ref{ex:D0u1}.
First, the restrictions on indecomposable double extensions of Theorem \ref{thm:MR} will be further discussed.

\newcommand*{\dist}{4}

\begin{figure}
  \centering
  \begin{tikzpicture}[
  txt/.style={draw,rectangle, minimum height=18},
  fund/.style={draw,rectangle, minimum height=18,rounded corners=3mm,
    top color=white, bottom color=black!20},
  indec/.style={draw,rectangle, minimum height=18, top color=white, bottom color=black!20},
  nonsuf/.style={->,densely dashed}
]

\node (simple) at (-\dist,0) [fund] {Simple};
\node (onedim) at (0,0) [fund] {One-dimensional};
\node (de) at (\dist,0) [indec] {$D(\mathfrak{g},\mathfrak{h})$};

\node (sdis) at (-\dist,-1) [label=center:$\oplus$] {};
\node (onedis) at (0,-1) [label=center:$\oplus$] {};
\node (dedis) at (\dist,-1) [label=center:$\oplus$] {};

\draw[->] (simple) -- (sdis);
\draw[->] ($(simple.south) + (4mm,0)$) -- (sdis);
\draw[->] ($(simple.south) - (4mm,0)$) -- (sdis);
\draw[->] (onedim) -- (onedis);
\draw[->] ($(onedim.south) + (4mm,0) $)-- (onedis);
\draw[->] ($(onedim.south) - (4mm,0) $)-- (onedis);
\draw[->] (de) -- (dedis);
\draw[->] ($(de.south) + (4mm,0) $) -- (dedis);
\draw[->] ($(de.south) - (4mm,0) $) -- (dedis);

\node (semisimple) at (-\dist,-2) [txt] {Semisimple};
\draw[->] (sdis) -- (semisimple);

\node (abelian) at (0,-2) [txt] {Abelian};
\draw[->] (onedis) -- (abelian);

\node (doubleext) at (\dist,-2) [txt] {Double extensions};
\draw[->] (dedis) -- (doubleext);

\node (redsum) at (-\dist/2,-3) [label=center:$\oplus$] {};
\draw[->] (abelian.south) -- (redsum.east);
\draw[->] (semisimple.south) -- (redsum.west);

\node (red) at (-\dist/2,-4) [txt] {Reductive};
\draw[->] (redsum) -- (red);

\node (or) at (0,1) [circle,draw,inner sep=1pt] {or};
\draw[nonsuf] (simple.north) -- (-\dist,1) -- (or.west);
\draw[nonsuf] (onedim) -- (or);
\draw[nonsuf] (or.east) -- ($(\dist,1) + (3mm,0)$) -- ($(de.north) + (3mm,-1.5mm)$);

\node (abdesum) at (1.5,-2) [label=center:$\oplus$] {};
\draw[nonsuf] (abelian.east) -- (abdesum);
\draw[nonsuf] (doubleext.west) -- (abdesum);
\draw[nonsuf]  (abdesum) -- (1.5,-1) -- (2.5,-1) -- (2.5,0.75) -- ($(\dist,0.75)+(-0.3mm,0)$) -- ($(de.north) + (-0.3mm,-1.5mm)$);

\node (redsssdis) at (0,-5) [label=center:$\oplus$] {};
\draw[->] (red.south) -- (redsssdis.west) ;
\draw[->] (doubleext.south) -- (redsssdis.east) ;
\node (ssd) at (0,-6) [txt] {Symmetric self-dual};
\draw[->] (redsssdis) -- (ssd);
\end{tikzpicture}
\caption{This diagram shows how all symmetric self-dual Lie algebras can be constructed.
  The simple and one-dimensional Lie algebras are the fundamental indecomposable building blocks.
  Direct sums of these algebras lead to the decomposable semisimple
  and abelian Lie algebras, which in turn can be summed to the reductive ones.
  Other indecomposable Lie algebras with invariant metric can be obtained
  by double extending an abelian or already double extended Lie algebra, denoted by $\mathfrak{g}$,
  via outer derivations by a simple or one-dimensional one, denoted by $\mathfrak{h}$ (see the dashed lines, Definition \ref{def:double-ext} and Theorem \ref{thm:MR}).
  \label{fig:selfdual}
}
\end{figure}
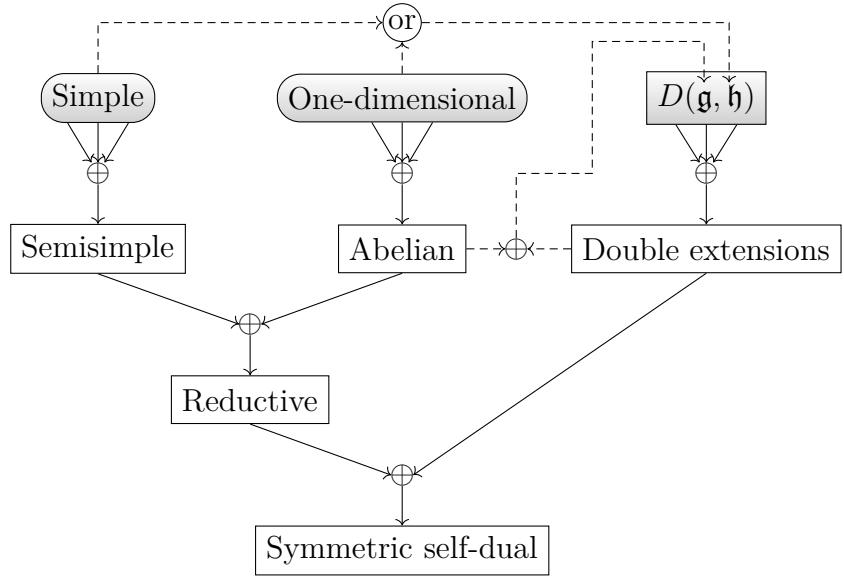
The condition \ref{item:MRg} is necessary since otherwise the factor
$\mathfrak{p}$ would also factor out of the double extension,
i.e.,
$D(\mathfrak{p} \dis \mathfrak{g},\mathfrak{h})
=
\mathfrak{p} \dis D(\mathfrak{g},\mathfrak{h})$
which makes it decomposable.
This is basically due to the restriction \ref{item:MRhout}, since for such
a factor $\mathfrak{p}$ all derivations are inner
and these also factor out of double extensions~\cite{FigueroaO'Farrill:1995cy}.
Lie algebras $\mathfrak{p}$ with $H^{1}(\mathfrak{p},\R) = H^{2}(\mathfrak{p},\R) = 0$
are sometimes called pluperfect~\cite{FigueroaO'Farrill:1995cy}.
Partly because $H^{1}(\mathfrak{p},\R)=0$ is equivalent to the
condition that $\mathfrak{p}$ is perfect ($[\mathfrak{p},\mathfrak{p}]=\mathfrak{p}$).
The second condition $H^{2}(\mathfrak{p},\R) = 0$ is equivalent to the
condition that $\mathfrak{p}$ does not admit any nontrivial
one-dimensional central extensions, see Appendix \ref{sec:central-extension-1}.
Semisimple Lie algebras are pluperfect and are therefore not allowed
as factors if the resulting double extension should be indecomposable~\cite{FigueroaO'Farrill:1995cy}.
This restricts the class of Lie algebras that one could double extend
to the abelian and the ones that have already been double extended, see the dashed lines in Figure \ref{fig:selfdual}.

It should be emphasized that there is no restriction concerning decomposability on $\mathfrak{g}$.
Example \ref{ex:u1u1u1} shows the double extension of a degenerate
symmetric self-dual Lie algebra to an indecomposable one.

One special case are double extensions of trivial Lie algebras.
For these the resulting symmetric self-dual Lie algebra is a
semidirect sum since
$
D(0,\mathfrak{h})
=
(0 \dis_{c} \mathfrak{h}^{*})\sdis\mathfrak{h}
=
\mathfrak{h}^{*} \sdis \mathfrak{h}
$.

\begin{example}[Poincar\'e]
  The three-dimensional Poincar\'e algebra
  is a double extension of a trivial Lie algebra by
  $\mathfrak{so}(2,1)$, i.e., $D(0,\mathfrak{so}(2,1))$.
  This can be seen explicitly from the commutation relations
  \begin{align} 
    \left[ \Jt_A ,  \Jt_B \right]
    &=
     \epsilon\indices{_{AB}^{C}}  \Jt_{C} 
    &
      \left[ \Jt_A ,  \Pt^{B}\right]
    &= -\epsilon\indices{_{AC}^{B}} \Pt^C
    &
      \left[ \Pt^{A} , \Pt^{B} \right]
    & = 0
  \end{align}
  and the invariant metric
  \begin{align}
    \langle \Jt_{A}, \Jt_{B}\rangle
    &= \eta_{AB}
    &
      \langle \Jt_{A},\Pt^{B}\rangle
    &= \delta\indices{_{A}^{B}}
    &
      \langle \Pt^{A},\Pt^{B}\rangle
    &= 0
      \,.
  \end{align}
\end{example}

\section{Low-Dimensional Symmetric Self-dual Lie Algebras}
\label{sec:low-dimens-symm}
To get more familiar with the above mentioned constructions the lowest
dimensional symmetric self-dual Lie algebras will be discussed.
Since the dimension $1$ was already discussed in Example
\ref{sec:dimension-1} we proceed with dimension $2$.

\subsection{Dimension $2$}
\label{sec:dimension-2}

\begin{example}[$\mathfrak{u}(1) \dis \mathfrak{u}(1)$]
  \label{ex:u12}
  For the direct sum of two $\mathfrak{u}(1)$ algebras one has the
  commutation relations $[\Gb_{i},\Gb_{j}]=0$ for $i,j=1,2$ and the
  most general invariant metric is given by
  \begin{align}
    \label{eq:u12invm}
    \Omega=
    \begin{pmatrix}
      a & c
      \\
      c & b
    \end{pmatrix}
          \quad \text{ for }\quad
          a b - c^{2} \neq 0
          \,.
  \end{align}
  Since it is a symmetric matrix we can always find an invertible
  matrix that diagonalizes it.
  This in turn shows that there exists an isomorphism that makes it
  decomposable.
  Depending on the three parameters $a$, $b$ and $c$ the metric might be positive definite.

  Similar reasoning generalizes to higher-dimensional abelian Lie
  algebras,
  which always admit an invariant metric and are decomposable.
\end{example}

\begin{example}[$D(0,\mathfrak{u}(1))$]
  \label{ex:D0u1}
  We will now double extend a trivial Lie algebra $\mathfrak{g}$
  by an abelian algebra
$\mathfrak{u}(1)$.
This example fulfills all the necessary requirements of Theorem~\ref{thm:MR}
for a double extension to be indecomposable.
In the end it will fail to be so.
Of course, in the case at hand the conditions \ref{item:MRg} and
\ref{item:MRhout} are rather trivial.

Since $\mathfrak{u}(1)$ is abelian it follows that the double
extension is also abelian, $[\Hb,\Hb]=[\Hb,\Hb^{*}]=[\Hb^{*},\Hb^{*}]=0$.
This is the same Lie algebra as discussed in Example \ref{ex:u12}.
As discussed, double extensions admit an invariant metric 
given by $\langle \Hb, \Hb^{*} \rangle=c$.
In this example we could also add $\langle \Hb , \Hb \rangle=a$
and, what is more uncommon for double extensions,
the  $\langle \Hb^{*} , \Hb^{*} \rangle=b$ term.
We will ignore these two terms subsequently.
Using the isomorphism $\Hb^{\pm}=1/2(\Hb \pm \Hb^{*})$ we can show that the
double extension is decomposable. 
The commutation relations remain that of an abelian algebra and the
invariant metric is given by $\langle \Hb^{\pm}, \Hb^{\pm}\rangle= \pm
1$ and $\langle \Hb^{+}, \Hb^{-}\rangle= 0$.
So the double extension is decomposable
$D(0,\mathfrak{u}(1)) \simeq (\mathfrak{u}(1) \dis \mathfrak{u}(1), 
\langle -,- \rangle_{+} \vdis \langle -,- \rangle_{-})$.

As for generic double extensions,
with possibly $a \neq 0$ but $b=0$,
we see that the invariant metric is not positive definite.
\end{example}

Even though two Lie algebras might be isomorphic this might not
be true when in addition their invariant metrics
as additional structure
are taken into consideration.
An example for this phenomena would be the just mentioned abelian Lie algebras,
where once we take the invariant metric \eqref{eq:u12invm} with $c=0$
and once with $a=b=0$.
Even though the Lie algebras are isomorphic the metric is positive
definite and indefinite, respectively\footnote{
This leads naturally to the following notion:
Two symmetric self dual Lie algebras
$(g,\langle \cdot  , \cdot \rangle_{\mathfrak{g}})$
and
$(\widetilde{g},\langle \cdot  , \cdot \rangle_{\widetilde{\mathfrak{g}}})$
are isometrically isomorph (short i-isomorph)
if the Lie algebra isomorphism $\phi: \mathfrak{g} \to \widetilde{\mathfrak{g}}$
satisfies
$\langle\phi(X),\phi(Y)\rangle_{\widetilde{\mathfrak{g}}} = \langle X,Y \rangle_{\mathfrak{g}}$ for all $X,Y \in \mathfrak{g}$.
}.

\begin{example}[Nonabelian]
In two dimensions up to isomorphism there is exactly one nonabelian
Lie algebra.
The nonzero commutator is 
$[\Gb_{1},\Gb_{2}]=c_{1} \Gb_{1} + c_{2} \Gb_{2}$ with the restriction that not both $c_{1}$ and $c_{2}$ are
allowed to vanish.
The most general invariant metric is then proportional to
\begin{align}
  \label{eq:nonab2met}
  \Omega= \begin{pmatrix}
    c_{2}^{2} & - c_{1} c_{2}
    \\
    -c_{1} c_{2} & c_{1}^{2}
  \end{pmatrix}
\end{align}
which is always degenerate.

That it is the only nonabelian Lie algebra of dimension two can be
shown with the isomorphism 
$\Gb_{i}=T\indices{_{i}^{j}}\widetilde\Gb_{j}$ (see
\ref{sec:basic-concepts}), which leads with
\begin{align}
  \label{eq:na21}
 \widetilde c_{2}
  T\indices{_{1}^{1}} +  \widetilde c_{1} T\indices{_{1}^{2}} &= -c_{2} 
  \\
  \label{eq:na22}
  \widetilde c_{2}
  T\indices{_{2}^{1}}+ \widetilde c_{1} T\indices{_{2}^{2}}  &= c_{1} 
\end{align}
to
$
[\widetilde\Gb_{1},\widetilde\Gb_{2}] = \widetilde c_{1} \widetilde\Gb_{1} + \widetilde c_{2} \widetilde\Gb_{2}
$.
Except for $\widetilde c_{1}= \widetilde c_{2}=0$ we can always find
a invertible $T\indices{_{i}^{j}}$ that fulfills \eqref{eq:na21} and
\eqref{eq:na22}.
\end{example}

\subsection{Dimension $3$}
\label{sec:dimension-3}

The smallest simple Lie algebras have dimension three
and are either isomorphic to
$\mathfrak{sl}(2,\R) \simeq \mathfrak{so}(2,1) \simeq \mathfrak{su}(1,1)$
or
$\mathfrak{so}(3) \simeq \mathfrak{su}(2)$.
They have an invariant metric that is proportional to the Killing form.
Furthermore, we have the three-dimensional abelian Lie algebra
which was discussed in Example \ref{ex:u12}.

\begin{example}[$D(\mathfrak{u}(1),\mathfrak{u}(1)$] This is the only
  possible three-dimensional double extension, but it leads to an
  abelian Lie algebra as we will show. We start with
  $\mathfrak{g}=\mathfrak{u}(1)$ since it has an invariant metric, as
  is required for double extensions, see Example
  \ref{sec:dimension-1}. It is not perfect and therefore
  $H^{1}(\mathfrak{u}(1),\R) \neq 0$, which is another requirement for
  a possibly indecomposable double extension. However, antisymmetric
  outer derivations do not exist since the most general derivation
  $\Db \cdot \Gb = c\, \Gb$ (which is outer for $c \neq 0$), using the
  antisymmetry condition, leads to
  $\langle \Db \cdot \Gb, \Gb\rangle + \langle \Gb,\Db \cdot
  \Gb\rangle = 2 c\langle \Gb, \Gb\rangle = 2 c a $, which is nonzero
  for outer derivations and thus not
  antisymmetric. \end{example}
\subsection{Dimension $4$} \label{sec:dimension-4}

There exist no four-dimensional simple
Lie algebras. Again there is the abelian algebra given by the direct
sum of four $\mathfrak{u}(1)$ algebras. Additionally there are
reductive Lie algebras given by the direct sum of the
three-dimensional simple Lie algebras with $\mathfrak{u}(1)$. In four
dimensions exist the first indecomposable nontrivial double
extensions. The Lie algebra $\mathfrak{u}(1) \dis \mathfrak{u}(1)$ is
actually the only option for which an indecomposable double extension
is possible.
\begin{example}[$D(\mathfrak{\mathfrak{u}}(1) \dis
  \mathfrak{u}(1),\mathfrak{u}(1)$] \label{ex:u1u1u1} We start with
  the decomposable direct sum
  $\mathfrak{g}=\mathfrak{u}(1) \dis \mathfrak{u}(1)$ explicitly given
  by the commutators $[\Gb_{i},\Gb_{j}]=0$ and the invariant metric
  $\langle \Gb_{i}, \Gb_{j}\rangle = \delta_{ij}$\footnote{ We
    restrict the invariant metric to specific values. }. All
  conditions for a possibly indecomposable double extension are
  fulfilled since there exists an outer antisymmetric derivation given
  by $[\Hb, \Gb_{i}]= \epsilon\indices{_{i}^{j}} \Gb_{j}$, where
  $\epsilon\indices{_{1}^{2}}=-\epsilon\indices{_{2}^{1}}=1$. The
  double extension is then given by

  \begin{align}
    [\Gb_{i}, \Gb_{j}]&=
                        \epsilon\indices{_{i}^{k}} \delta_{kj} \Hb^{*}
    \\
    [\Hb, \Gb_{i}]&=
                    \epsilon\indices{_{i}^{j}}
                    \Gb_{j}
    \\
    [\Hb, \Hb^{*}]&=0
  \end{align}
  with the
  invariant
  metric
  \begin{align}
    \Omega_{ab}^{\mathfrak{d}}=
    \bordermatrix{
    ~ &
        \Gb_{j}
    & \Hb &
            \Hb^{*}
            \cr
            \Gb_{i}
    &
      \delta_{ij}
    & 0 & 0
          \cr
          \Hb
    & 0 & h
    & 1 \cr
      \Hb^{*}
    & 0 & 1
    & 0 \cr
      }
      \,.
  \end{align}
  This algebra has been used by Nappi and
  Witten to construct a nonsemisimple WZW
  model~\cite{Nappi:1993ie}.
\end{example}

There is another double extension of $\mathfrak{u}(1) \dis \mathfrak{u}(1)$
leading to a  four-dimensional algebra that is not isomorphic to the just mentioned one, 
for a review see \cite{Ovando:Review}.

\subsection{Summary}
\label{sec:summary-1}

All symmetric self-dual Lie algebras are given by application of direct
sums and double extensions to simple and $\mathfrak{u}(1)$ Lie
algebras, see Figure \ref{fig:selfdual}.
The indecomposable ones are of the type described in Theorem \ref{thm:MR}.

\chapter{Contractions of Lie Algebras}
\label{sec:contr-lie-algebr}


Contractions go back to the works of Segal~\cite{segal1951}
and Inönü and Wigner~\cite{Inonu:1953sp}.
While also mathematically interesting,
in physics their importance comes from the fact that they are related to
approximations.
The probably most famous example is the contraction from the
Poincar\'e group to the Galilei group~\cite{Inonu:1953sp},
i.e., going from relativistic to nonrelativistic physics.

We will discuss contractions on the level of Lie algebras
and in relation to invariant metrics,
but often useful insights for other interesting structures like the Lie group
and representations follow.

We start by introducing contractions,
generalized and simple Inönü--Wigner contractions
and briefly discuss their relations.
Afterwards we show how contractions
can make trivial central extensions to nontrivial ones.
In the next section the effect of
contractions on invariant metrics will be
investigated.

We will follow partially
\cite{06NesterenkoLieContr,00WeimarWoods}\footnote{%
  Lemma 2.2.\ in \cite{00WeimarWoods} is not correct
  and therefore the proof of Theorem 3.1.\ does not work~\cite{06NesterenkoLieContr,08WeimarWoods}.
}
where further details can be found.

\section{Contractions}
\label{sec:contractions}

We will start with the most general Lie algebra contraction definition.
For that we start with a Lie algebra $\mathfrak{g}$ with an underlying
vector space $V$ over $\R$.
\begin{definition}[Contraction]
  \index{Contraction}
  \index{Lie algebra!Contraction|see {Contraction}}
  Let $T(\epsilon)$, with $0 < \epsilon \leq 1$, be a family of
  continuous non-singular linear maps on $V$.
  Then the Lie algebras
  \begin{align}
    \mathfrak{g}_{T(\epsilon)}=(V,[ \cdot, \cdot]_{T(\epsilon)})
    \quad \text{ for } \quad
    \epsilon >0,
  \end{align}
  where
  \begin{align}
    [x,y]_{T(\epsilon)} = T^{-1}(\epsilon) [T(\epsilon)x, T(\epsilon)y
    ]
    \quad \text{ with } \quad
    x,y \in V
  \end{align}
  are isomorphic to
  $\mathfrak{g}=(V,[ \cdot , \cdot])$.
  If the limit
  \begin{align}
    [x,y]_{T} \equiv \lim_{\epsilon \to 0}\, [x,y]_{T(\epsilon)}
  \end{align}
  exists for all $x,y \in V$, then $[\cdot, \cdot]_{T}$ is a Lie
  product and the Lie algebra $\mathfrak{g}_{T}=(V,[\cdot,\cdot]_{T})$
  is called the \textbf{contraction} of $\mathfrak{g}$ by
  $T(\epsilon)$,
  in short,
  \begin{align}
    \mathfrak{g} \overset{T(\epsilon)}{\longrightarrow} \mathfrak{g}_{T} \,.
  \end{align}
\end{definition}
When a basis is fixed $T(\epsilon)$ is a matrix and we can define
the limit on the structure constants by
\begin{align}
  (f_{T})\indices{_{ab}^{c}}
  \equiv
  \lim_{\epsilon \to 0}
  T(\epsilon)\indices{_{a}^{d}} T(\epsilon)\indices{_{b}^{e}} T^{-1}(\epsilon)\indices{_{f}^{c}}
  f\indices{_{de}^{f}} \,.
\end{align}
When the specific contraction is clear we will sometimes leave out the
$T(\epsilon)$ or just write an $\epsilon$.

Two  contractions always exist:
\begin{enumerate}
\item $\mathfrak{g}_{T} \simeq \mathfrak{g}$:
  Contractions where the contracted Lie algebra is isomorphic to the original one are
  called \textbf{improper}.
  \index{Contraction! Improper}%
  Such a contraction can be defined using just an identity
  matrix for $T(\epsilon)$.
\item Abelian $\mathfrak{g}_{T}$: This \textbf{trivial} contraction
  also always exists.
  \index{Contraction! Trivial}%
  One just has to set $T(\epsilon)=\mathrm{diag} (\epsilon, \dotsc,\epsilon)$, which leads to
  \begin{align}
    [x,y]_{T}
    = \lim_{\epsilon \to 0} \epsilon^{-1}[ \epsilon x,\epsilon y]
    = \lim_{\epsilon \to 0} \epsilon[ x, y]
    = 0 \,.
  \end{align}
\end{enumerate}

This definition of a contraction is more general than the one for
Inönü--Wigner contractions~\cite{Inonu:1953sp} (IW-contractions) as well as for
Saletan contractions \cite{Saletan:61},
because we do not restrict to the existence of the limit
\begin{align}
  \label{eq:Twelldef}
  T(0) \equiv \lim_{\epsilon \to 0} T(\epsilon) \,.
\end{align}
In all cases the dimension of the Lie algebra stays unaltered.

One justification for this generalization is that this restriction is
not necessary for the existence of the contracted Lie algebra.
But it might be useful to have an equivalent contraction where $T(0)$
is well defined.
Because there might arise situations where one wants to
interpret the quantities $T(0)x$ and not only
the contracted Lie algebra bracket.
Therefore, it might be preferential if all components of $T(0)$ are finite,
or are composed in such a way that the structure of interest
is finite.
One such situation where this is the case
will be seen in Section \ref{sec:contr-invar-metr}.
Here by \textbf{equivalent}
\index{Contraction!Equivalence}%
we mean contractions that lead from two
isomorphic Lie algebras $\mathfrak{g} \simeq \mathfrak{g}'$ again to
two isomorphic Lie algebras $\mathfrak{g}_{T} \simeq
\mathfrak{g}'_{S}$, i.e., 
\begin{equation*}
\begin{tikzcd}
  \mathfrak{g} \arrow[d, "\simeq" left ,leftrightarrow]
  \arrow[r,"T(\epsilon)"] & \arrow[d, "\simeq",leftrightarrow] \mathfrak{g}_{T}
  \\
  \mathfrak{g}' \arrow[r,"S(\epsilon)"] & \mathfrak{g}'_{S}
\end{tikzcd}\, .
\end{equation*}




\section{Generalized IW-contractions}
\label{sec:gener-iw-contr}

A subclass of contractions are the generalized Inönü--Wigner contractions.
Due to their diagonal form they are very useful for explicit calculations.
They go back to work of Doebner and Melsheimer~\cite{67DoebnerGener}.
\begin{definition}
  A contraction $\mathfrak{g} \overset{T(\epsilon)}{\longrightarrow}
  \mathfrak{g}_T$
  is called a \textbf{generalized Inönü--Wigner contraction} (gIW-contraction)
  \index{Contraction!Generalized IW-contraction}
  if the matrix $T(\epsilon)$ has the form
  \begin{align}
    \label{eq:TgIW}
    T(\epsilon)\indices{_{a}^{b}}= \delta\indices{_{a}^{b}}
    \epsilon^{n_{b}}
    \quad \text{where} \quad
    n_{b} \in \R; \, \epsilon >0; \, a,b= 1,2,\dotsc,\dim (\mathfrak{g})
  \end{align}
  for some basis $\Gb_{1},\dotsc,\Gb_{N}$.
\end{definition}
Another way to write gIW-contractions is
$T(\epsilon)
=\mathrm{diag}(\epsilon^{n_{1}}, \dotsc ,\epsilon^{n_{\dim\mathfrak{g}}})$.
There are no sums over the exponents $n_{a}$, which can
be restricted to integer values
without loss
of generality.
This includes negative exponents which, as already discussed,
render the $\epsilon \to 0$ limit of $T(\epsilon)$ (see \eqref{eq:TgIW})
non-existent.
Furthermore, the matrices $T(\epsilon)$ for gIW-contractions are not
necessarily linear in $\epsilon$, which differentiates them from
IW-contractions~\cite{Inonu:1953sp} and Saletan contractions~\cite{Saletan:61}.

For a generic gIW-contraction our definition leads to
\begin{align}
  \label{eq:combefore}
  [\Gb_{a},\Gb_{b}]_{T(\epsilon)}
  =
  \epsilon^{n_{a} + n_{b}- n_{c}} f\indices{_{ab}^{c}} \Gb_{c} \,.
\end{align}
It is a contraction, i.e., well defined in the $\epsilon \to 0$ limit,
if and only if $n_{a} + n_{b}- n_{c} \geq 0$ for nonzero $f\indices{_{ab}^{c}}$.
For such a well defined contraction we then get
\begin{align}
  (f_{T})\indices{_{ab}^{c}} =
  \begin{cases}
    f\indices{_{ab}^{c}} & \text{if }  n_{a}+n_{b}=n_{c}
    \\
    0 & \text{otherwise.}  
  \end{cases}
\end{align}

\section{Simple IW-contractions}
\label{sec:simple-iw-contr}


A special class of gIW-contractions are the ones originally defined by
Inönü and Wigner~\cite{Inonu:1953sp}.

\begin{definition}
  A \textbf{(simple) Inönü--Wigner contraction}
  \index{Contraction!(Simple) IW-contraction}%
  ((s)IW-con\-trac\-tion) is a
  generalized Inönü--Wigner contraction where all $n_{b}$ in
  \begin{align}
    T(\epsilon)\indices{_{a}^{b}}= \delta\indices{_{a}^{b}}
    \epsilon^{n_{b}}
    \quad 
    (\epsilon >0)
  \end{align}
   are either $0$ or $1$.
\end{definition}
An immediate consequence of this definition is that $T(0)$ always
exists for sIW-contractions.
This is of course always true when all $n_{a} \geq 0$.

The condition for the existence of sIW-contractions can be translated
into conditions for Lie subalgebras.
Suppose we start with a Lie algebra $\mathfrak{g}$ that is a (non-intersecting)
vector space direct sum $\mathfrak{g}= \mathfrak{h} \vdis \mathfrak{i}$.
We then set for $\mathfrak{h}$ all $n_{a}=0$
and for $\mathfrak{i}$ all $n_{a}=1$.
We see, using \eqref{eq:combefore},  that the commutator of two elements of $\mathfrak{h}$ is not allowed to close into $\mathfrak{i}$ for the contraction $(\mathfrak{h} \vdis \mathfrak{i})
\overset{T(\epsilon)}{\longrightarrow}
(\mathfrak{h} \vdis \mathfrak{i})_{T}$ to be well defined because
\begin{align}
  [\mathfrak{h},\mathfrak{h}]_{T(\epsilon)}
  = \mathfrak{h} \vdis \epsilon^{-1} \mathfrak{i}
\end{align}
is not well defined in the $\epsilon \to 0$ limit.
So a Lie algebra can be contracted with a sIW-contraction with respect to a Lie subalgebra (e.g., $\mathfrak{h}$ above) and only with respect to a Lie subalgebra~\cite{Inonu:1953sp}.
This subalgebra specifies the contracted Lie algebra (e.g., $(\mathfrak{h} \vdis \mathfrak{i})_{T}$) uniquely up to isomorphism~\cite{Saletan:61}.
This property makes sIW-contractions, although less general than
arbitrary contractions and gIW-contractions,
conceptually much easier and gives an easy
criterion for when the contraction exists.

So explicitly the whole contraction $(\mathfrak{h} \vdis \mathfrak{i})
\overset{T(\epsilon)}{\longrightarrow}
(\mathfrak{h} \vdis \mathfrak{i})_{T}$ with respect to the subalgebra $\mathfrak{h}$ is given by
\begin{align}
  [\mathfrak{h},\mathfrak{h}]_{T(\epsilon)} & = \phantom{\epsilon} \mathfrak{h}  &  &                                       & [\mathfrak{h},\mathfrak{h}]_{T} & =  \mathfrak{h} 
\\                                                                                                             
  [\mathfrak{h},\mathfrak{i}]_{T(\epsilon)} & = \epsilon \mathfrak{h} \vdis \mathfrak{i}                         &  & \overset{T(\epsilon)}{\longrightarrow} & [\mathfrak{h},\mathfrak{i}]_{T} & =  \mathfrak{i}                        
\\                                                                                                             
  [\mathfrak{i},\mathfrak{i}]_{T(\epsilon)} & =\epsilon \mathfrak{h} \vdis \epsilon^{2} \mathfrak{i}             &  &                                       & [\mathfrak{i},\mathfrak{i}]_{T} & =0 \,.
\end{align}

The ideal $\mathfrak{i}$ of $(\mathfrak{h} \vdis \mathfrak{i})_{T}$ is abelian
and
therefore any proper sIW-contraction leads to a nonsemisimple Lie algebra.
The subalgebra $\mathfrak{h}$ stays unaltered under the contraction
and is isomorphic to the quotient algebra
$(\mathfrak{h} \vdis \mathfrak{i})_{T}/\mathfrak{i}$.

It should be noted, that the sIW-contractions do not exhaust all
possible contractions. A Saletan contraction where no equivalent 
sIW-contraction exists was already constructed by Saletan in \cite{Saletan:61}.
Even for the wider class of Saletan contractions, which are still
linear in $\epsilon$ and include the sIW-contractions, no contraction
from $\mathfrak{so}(3)$ to the Heisenberg algebra exists.
On the other hand a gIW-contraction from $\mathfrak{so}(3)$ to the
Heisenberg algebra exists, but there are other contractions where no
equivalent diagonal gIW-contraction is possible.
This phenomena starts with four-dimensional Lie algebras
\cite{06NesterenkoLieContr}.
One might hope that every gIW-contraction is decomposable in
sIW-contractions.
In full generality this is not the case (for more details see,
e.g., \cite{06NesterenkoLieContr}).

Leaving this very general considerations aside a lot of physically
interesting contractions, see e.g., \cite{Inonu:1953sp,Bacry:1968zf} are
given by sIW and gIW-contractions
and we will in the following restrict to these cases.

One remark should be added concerning this general discussion,
especially because it will become important later.
Here we have discussed Lie algebras merely on the level of an abstract
mathematical structure without any reference to physics.
When a physical Lie algebra is discussed specific Lie algebra generators
have an interpretation, e.g., as generator of time translations.
So just because two Lie algebras are isomorphic does not mean
they are
physically the same.
Exchanging the interpretation of a rotation and a time translation
leads to the same Lie algebras, but obviously our physical
interpretation would change drastically.
Suddenly elements that commuted with time translations do not
commute anymore and are therefore not conserved.
One such example are the Poincar\'e and
para-Poincar\'e algebras that will be discussed later.

\section{Contractions and Central Extensions}
\label{sec:contr-centr-extens}

There is an interesting interplay between contractions and
central extensions.
The Lie algebra $\mathfrak{a}$ will be abelian and more details concerning
Lie algebra cohomology are given in Appendix \ref{cha:lie-algebras}.
One consequence is the following diagram
\begin{equation*}
  \begin{tikzcd}
    \mathfrak{g} \dis \mathfrak{a} \arrow[rr,"\text{Contraction}" above] \arrow[dd,"\text{Change of coboundary}" left ]  && \mathfrak{g}_{T} \dis \mathfrak{a} \arrow[dd,"\cancel{\phantom{X}}" description, equal] \\
    \\
    \mathfrak{g} \dis \mathfrak{a}  \arrow[rr,"\text{Contraction}" below]                                   && \mathfrak{g}_{T} \dis_{c} \mathfrak{a}
  \end{tikzcd}
\end{equation*}
which means that trivial central extensions ($\mathfrak{g} \dis \mathfrak{a}$)
might lead after contraction to nontrivial ones
($\mathfrak{g} \dis_{c} \mathfrak{a}$)~\cite{Saletan:61,Hermann1966}.
The reason for that is that the coboundary that makes a central extension trivial might be gone after the contraction.
One famous example for this effect is the contraction of the
trivial centrally extended Poincar\'e algebra
to the nontrivial centrally extended Galilei algebra, the Bargmann algebra.
In that case the central term is of importance since it is the mass of the system.

\begin{example}
  We start by trivially centrally extending the two-dimensional
  Lie algebra with the nonzero commutator $[\Xb,\Yb]=\Xb$,
  i.e., we add the commutators $[\Xb,\Zb]=[\Yb,\Zb]=0$.
  We now change by a coboundary, which means that the central
  extension $\Zb$ is still trivial, to get
  \begin{align}
    [\Xb,\Yb]=\Xb +\Zb \,.
  \end{align}
  This can be implemented by shifting $\Xb$ by $\Zb$.
  The contraction with $n_{\Xb}=0$ and $n_{\Yb}=n_{\Zb}=1$,
  which leads to
  \begin{align}
    [\Xb,\Yb]_{\epsilon}= \epsilon \Xb +\Zb
  \end{align}
  shows now that that the central extension is
  not trivial anymore
  in
  the $\epsilon \to 0$ limit.
  The reason for this is that the necessary coboundary is gone
  or in other words that the shift by $\Xb$ is not possible anymore.
\end{example}

These considerations also have an influence on invariant metrics
since nontrivial central extensions can then render a degenerate invariant metric
nondegenerate.
An example for this is the three-dimensional
Galilei algebra, which can be centrally extended to the Extended Bargmann algebra.
Similar to the above considerations contractions from a trivially extended (Anti)-de Sitter algebra or Poincar\'e algebra are possible, see Section \ref{sec:extend-kinem}.

\chapter{Contractions and Invariant Metrics}
\label{sec:contr-invar-metr}






We now want to set Lie algebra contractions in relation to
existence of invariant metrics.
Instead of trying to give a complete discussion we will focus on
examples that are of relevance for our later considerations.

\section{Contraction of Invariant Metric}
\label{sec:invar-metr-contr}

We start with a Lie algebra $\mathfrak{g}$ and a contraction
$T(\epsilon)$.
Given this contraction of the Lie algebra
we can induce one on the invariant metric by
\begin{align}
  \langle x,y \rangle_{T(\epsilon)}
  =
  \langle T(\epsilon) x, T(\epsilon) y \rangle \, .
\end{align}
We see that a divergent $T(0)$ might
lead to a divergent contracted invariant metric.
Of course one could always ignore that an invariant metric exists
for the original Lie algebra,
contract it, and look afterwards for invariant metrics.
While this is certainly an option it might, e.g., for a theory given by an action,
be beneficial to also have a contraction on the level of the invariant metric.
For CS theories the contraction on the level of the invariant
metric basically corresponds to the limit on the level of the action.
Therefore, it might lead to additional insights and input for the contracted
theory.

The considerations of Sections \ref{sec:gener-iw-contr}
and \ref{sec:simple-iw-contr} can be adapted in a straightforward manner
for invariant metrics
and will therefore not be explicitly carried out.

\section{Contraction to Inhomogeneous Lie Algebras}
\label{sec:cont-inh}

Here we will show how the invariant metric of the inhomogeneous Lie
algebras can be derived via contractions from a direct sum of
two simple Lie algebras $\mathfrak{g} \dis \tilde{\mathfrak{g}}$.
This is of special importance since this is how the Poincar\'e algebra
and its higher spin generalizations in three spacetime dimensions are contracted.
The Lie algebras $\mathfrak{g}$ and $\tilde{\mathfrak{g}}$ are isomorphic
and since they are simple automatically admit an invariant metric.
A basis for the first and second summand is given by $\Gb_{a}$ and
$\tilde\Gb_{a}$, respectively.
The commutation relations are of the form
\begin{align}
  [\Gb_{a},\Gb_{b}]&=f\indices{_{ab}^{c}} \Gb_{c}
&
  [\Gb_{a},\tilde \Gb_{b}]&=0
&
  [\tilde\Gb_{a},\tilde\Gb_{b}]&=f\indices{_{ab}^{c}} \tilde \Gb_{c} 
\end{align}
with the most general invariant metric
\begin{align}
  \langle \Gb_{a} \Gb_{b} \rangle &= \mu \, \Omega_{ab}
&
  \langle \Gb_{a} \tilde \Gb_{b} \rangle &= 0
&
  \langle \tilde\Gb_{a} \tilde\Gb_{b} \rangle = \tilde\mu  \, \Omega_{ab}
\end{align}
where $\mu \tilde \mu \neq 0$.
Defining
\begin{align}
  \Gb^{\pm}_{a}&=\Gb_{a} \pm \tilde \Gb_{a}
\end{align}
leads to
\begin{align}
  [\Gb^{+}_{a},\Gb^{+}_{b}]&=f\indices{_{ab}^{c}} \Gb^{+}_{c}
&
  [\Gb^{+}_{a}, \Gb^{-}_{b}]&=f\indices{_{ab}^{c}} \Gb^{-}_{c}
&
  [\Gb^{-}_{a},\Gb^{-}_{b}]&=f\indices{_{ab}^{c}} \Gb^{+}_{c}
\end{align}
with the invariant metric
\begin{align}
  \langle \Gb^{+}_{a} \Gb^{+}_{b} \rangle &= \mu^{+} \, \Omega_{ab}
&
  \langle \Gb^{+}_{a} \Gb^{-}_{b} \rangle &= \mu^{-} \, \Omega_{ab}
&
  \langle \Gb^{-}_{a} \Gb^{-}_{b} \rangle =  \mu^{+} \, \Omega_{ab} \, .
\end{align}
where $\mu^{\pm} = \mu \pm \tilde \mu$.
The generators $\Gt^{+}_{a}$ span a Lie subalgebra with respect
to which we now make a sIW-contraction.
%
This leads to the Lie algebra $\mathfrak{g}_{\epsilon}$
\begin{align}
  [\Gb^{+}_{a},\Gb^{+}_{b}]_{\epsilon}
  &=
    f\indices{_{ab}^{c}} \Gb^{+}_{c}
&
  [\Gb^{+}_{a}, \Gb^{-}_{b}]_{\epsilon}
  &=
    f\indices{_{ab}^{c}} \Gb^{-}_{c}
&
  [\Gb^{-}_{a},\Gb^{-}_{b}]_{\epsilon}
  &=
    \epsilon^{2}
    f\indices{_{ab}^{c}} \Gb^{+}_{c}
\end{align}
and the, for $\epsilon \to 0$ degenerate, bilinear form
\begin{align}
  \langle \Gb^{+}_{a} \Gb^{+}_{b} \rangle_{\epsilon}
  &=
    \mu^{+} \, \Omega_{ab}
&
  \langle \Gb^{+}_{a} \Gb^{-}_{b} \rangle_{\epsilon}
  &=
    \epsilon
    \mu^{-} \, \Omega_{ab}
&
  \langle  \Gb^{-}_{a} \Gb^{-}_{b} \rangle_{\epsilon}
  &=
    \epsilon^{2}
    \mu^{+} \, \Omega_{ab} \, .
    \label{eq:inhcontr1}
\end{align}
This degeneracy is to be expected since we have basically contracted the
Killing form which for nonsemisimple Lie algebras should be degenerate.
We know on the other hand that this can not be the most general
invariant metric since the contracted algebra is a trivial double
extension $D(0,\mathfrak{g}^{+}_{0})=\mathfrak{g}^{-}_{0} \sdis
\mathfrak{g}^{+}_{0}$ and therefore
symmetric self-dual.
We know from our earlier considerations that we can add
$\langle \Gb^{+}_{a}  \Gb^{-}_{b} \rangle_{0} =\mu^{-} \, \Omega_{ab}$
to make it nondegenerate.
One might ask, if it is possible to also get this term using our
current discussion.
It works if one recognizes that one could rescale
$\mu^{-} \mapsto \epsilon^{-1}\mu^{-}$ 
to cancel the $\epsilon$ term in
\eqref{eq:inhcontr1} leading to the final contracted Lie algebra
with an invariant metric
\begin{align}
  [\Gb^{+}_{a},\Gb^{+}_{b}]_{0}
  &=
    f\indices{_{ab}^{c}} \Gb^{+}_{c}
&
  [\Gb^{+}_{a}, \Gb^{-}_{b}]_{0}
  &=
    f\indices{_{ab}^{c}} \Gb^{-}_{c}
&
  [\Gb^{-}_{a},\Gb^{-}_{b}]_{0}
  &=
    0
    \\
  \langle \Gb^{+}_{a} \Gb^{+}_{b} \rangle_{0}
  &=
    \mu^{+} \, \Omega_{ab}
&
  \langle \Gb^{+}_{a} \Gb^{-}_{b} \rangle_{0}
  &=
    \mu^{-} \, \Omega_{ab}
&
  \langle  \Gb^{-}_{a} \Gb^{-}_{b} \rangle_{0}
  &=0\, .
    \label{eq:inhcontr2}
\end{align}

\section{Invariant Metric Preserving Contraction}
\label{sec:double-extens-pres}

There exists a special class of contractions,
which we will call invariant metric preserving,
that lead from a double extended Lie algebra to another one.
Therefore, it leaves the properties of the invariant metric untouched.
This is done in a fashion that is naturally adapted to double extensions.
and it is not just of theoretical importance.
As we will see in Example \ref{exp:poitocar}, these contractions explain
why the contraction of Poincar\'e to Carroll (higher spin)
algebras in $2+1$ dimensions leaves the degeneracy of the invariant metric untouched.
This gives another explanation for the algebras discussed in \cite{Bergshoeff:2016soe,Grumiller:2017sjh}.
To the best of my knowledge this special kind of contraction has not
yet been discussed in the literature.

One starts with a double extended Lie algebra
$D(\mathfrak{g}, \mathfrak{h} \vdis \widetilde{\mathfrak{h}})$,
where $\mathfrak{h}$ should be a Lie subalgebra,
in analogy to the simple IW-contractions.
This allows us to perform a sIW-contraction on
$\mathfrak{h} \vdis \widetilde{\mathfrak{h}}$ with respect to the
subalgebra $\mathfrak{h}$, since this is a subalgebra
of the whole double extension.
Now, this would not leave the invariant metric invariant since the
important part for
nondegeneracy 
$\langle \widetilde{\mathfrak{h}} , \widetilde{\mathfrak{h}}^{*}\rangle_{T(\epsilon)}
=
\epsilon \langle \widetilde{\mathfrak{h}} , \widetilde{\mathfrak{h}}^{*}\rangle $
would degenerate.
But, this already hints towards the solution that we have to do the
``dual'', i.e., the inverse transformation
on the dual space $\widetilde{\mathfrak{h}}^{*}$.
Given the knowledge of double extensions this seems a very
natural thing to do.
We will now write this contraction explicitly in a basis.

Using the contraction
$T(\epsilon)\widetilde{\mathfrak{h}}=\epsilon
\widetilde{\mathfrak{h}}$
and
$T(\epsilon)\widetilde{\mathfrak{h}}^{*}=\epsilon^{-1}
\widetilde{\mathfrak{h}}^{*}$
where the remaining parts stay unaltered we write it,
in hopefully obvious notation (we omit the subscript $T(\epsilon)$ for
the Lie brackets and vanishing commutators)
\begin{align}
 [\Gb_{i},\Gb_{j}]              & =f\indices{_{ij}^{k}} \Gb_{k}+f\indices{_{\alphaa i}^{k}} \Omega_{kj}^{\mathfrak{g}} \Hba^{\alphaa} + \epsilon f\indices{_{\alphab i}^{k}} \Omega_{kj}^{\mathfrak{g}} \Hbb^{\alphab} 
                    \label{eq:TGGT}             \\
 [\Hba_{\alphaa},\Gb_{i}]       & =f\indices{_{\alphaa i}^{j}} \Gb_{j}
                         \label{eq:THGTa}       \\
 [\Hbb_{\alphab},\Gb_{i}]       & =\epsilon f\indices{_{\alphab i}^{j}} \Gb_{j}
                         \label{eq:THGTb}       \\
 [\Hba_{\alphaa},\Hba_{\betaa}] & =f\indices{_{\alphaa \betaa}^{\gammaa}} \Hba_{\gammaa} + \xcancel{\epsilon^{-1} f\indices{_{\alphaa \betaa}^{\gammab}} \Hbb_{\gammab}}
                             \label{eq:THHTaa}  \\
 [\Hba_{\alphaa},\Hbb_{\betab}] & =\epsilon f\indices{_{\alphaa \betab}^{\gammaa}} \Hba_{\gammaa} + f\indices{_{\alphaa \betab}^{\gammab}} \Hbb_{\gammab}
                             \label{eq:THHTab}  \\
 [\Hbb_{\alphab},\Hbb_{\betab}] & =\epsilon^{2} f\indices{_{\alphab \betab}^{\gammaa}} \Hba_{\gammaa} +\epsilon f\indices{_{\alphab \betab}^{\gammab}} \Hbb_{\gammab}
                             \label{eq:THHTbb}  \\
 [\Hba_{\alphaa},\Hba^{\betaa}] & =-f\indices{_{\alphaa \gammaa}^{\betaa}} \Hba^{\gammaa} - \epsilon f\indices{_{\alphaa \gammab}^{\betaa}} \Hbb^{\gammab}
                             \label{eq:THHdTaa} \\
 [\Hba_{\alphaa},\Hbb^{\betab}] & =-\xcancel{\epsilon^{-1} f\indices{_{\alphaa \gammaa}^{\betab}} \Hba^{\gammaa}} -f\indices{_{\alphaa \gammab}^{\betab}} \Hbb^{\gammab}
                             \label{eq:THHdTab} \\
 [\Hbb_{\alphab},\Hba^{\betaa}] & =- \epsilon f\indices{_{\alphab \gammaa}^{\betaa}} \Hba^{\gammaa} - \epsilon^{2} f\indices{_{\alphab \gammab}^{\betaa}} \Hbb^{\gammab}
                             \label{eq:THHdTba} \\
 [\Hbb_{\alphab},\Hbb^{\betab}] & =-f\indices{_{\alphab \gammaa}^{\betab}} \Hba^{\gammaa} - \epsilon f\indices{_{\alphab \gammab}^{\betab}} \Hbb^{\gammab}\,.
                             \label{eq:THHdTbb} 
\end{align}
The two crossed terms indicate elements that would render the
contraction not well defined.
But the first term
in \eqref{eq:THHTaa} is no obstruction, because we require
$\mathfrak{h}$ to be a subalgebra.
This is just the usual condition for sIW-contractions.
It is nice that this property automatically also renders the second
crossed term nonexistent and therefore the whole contraction is well defined.

The corresponding invariant metric is given by
\begin{align}
  \Omega_{ab}^{\mathfrak{d}}=
  \bordermatrix{
  ~                             & \Gb_{j}                    & \Hba_{\betaa}                       & \Hbb_{\betab}                       & \Hba^{\betaa}                       & \Hbb^{\betab} \cr
   \Gb_{i}                      & \Omega_{ij}^{\mathfrak{g}} & 0                                   & 0                                   & 0                                   & 0  \cr
   \Hba_{\alphaa}               & 0                          & h_{\alphaa \betaa}                  & \epsilon h_{\alphaa \betab }        & \delta\indices{_{\alphaa}^{\betaa}} & 0\cr
   \Hbb_{\alphab}               & 0                          & \epsilon h_{\alphab \betaa}         & \epsilon^{2} h_{\alphab \betab}     & 0                                   & \delta\indices{_{\alphab}^{\betab}}\cr
   \Hba^{\alphaa}               & 0                          & \delta\indices{^{\alphaa}_{\betaa}} & 0                                   & 0                                   & 0 \cr
   \Hbb^{\alphab}               & 0                          & 0                                   & \delta\indices{^{\alphab}_{\betab}} & 0                                   & 0 \cr
             }
\end{align}
and one can see that this special kind of contraction leaves it
nondegenerate and well defined.

Given that after the $\epsilon \to 0$ limit we have again a symmetric
self-dual Lie algebra one might ask what kind of double extension this contraction leads.
It is of the form $D(\mathfrak{g}\dis D(0,\widetilde{\mathfrak{h}}),\mathfrak{h})$.
Notice that according to Theorem \ref{thm:MR} the decomposability of
$\mathfrak{g}\dis D(0,\widetilde{\mathfrak{h}})$ is in principle
no problem for
the indecomposability of the new double extension.
One nice feature of this contraction is that,
like for sIW-contractions,
just the specification of a subalgebra gives a very easy criterion for
a well defined contraction.
So we have proven the following theorem
(commutators that are not explicitly given vanish).

\begin{theorem}[Invariant metric preserving contraction]
  \label{thm:invar-metr-pres}
  Let the double extended Lie algebra $D(\mathfrak{g}, \mathfrak{h}
  \vdis \widetilde{\mathfrak{h}})$
  have a Lie subalgebra $\mathfrak{h}$.
  Then a contraction of the form
  $T(\epsilon)\widetilde{\mathfrak{h}}=\epsilon \widetilde{\mathfrak{h}}$
  and
  $T(\epsilon)\widetilde{\mathfrak{h}}^{*}=\epsilon^{-1}\widetilde{\mathfrak{h}}^{*}$
  with the remaining elements unaltered,
  see \eqref{eq:TGGT} to \eqref{eq:THHdTbb},
  is a contraction that leads to a double extension
  $D(\mathfrak{g}\dis D(0,\widetilde{\mathfrak{h}}_{T}),\mathfrak{h}_{T})$
  explicitly given by
  \begin{align}
 [\Gb_{i},\Gb_{j}]              & =f\indices{_{ij}^{k}} \Gb_{k}+f\indices{_{\alphaa i}^{k}} \Omega_{kj}^{\mathfrak{g}} \Hba^{\alphaa} 
                    \label{eq:GGT}             \\
 [\Hbb_{\alphab},\Gb_{i}]       & =0
                         \label{eq:HGTb}       \\
 [\Hbb_{\alphab},\Hbb_{\betab}] & =0
                             \label{eq:HHTbb}  \\
 [\Hbb_{\alphab},\Hbb^{\betab}] & =-f\indices{_{\alphab \gammaa}^{\betab}} \Hba^{\gammaa} 
                                  \label{eq:HHdTbb} \\
 [\Hba_{\alphaa},\Gb_{i}]       & =f\indices{_{\alphaa i}^{j}} \Gb_{j}
                         \label{eq:HGTa}       \\
 [\Hba_{\alphaa},\Hbb_{\betab}] & = f\indices{_{\alphaa \betab}^{\gammab}} \Hbb_{\gammab}
                             \label{eq:HHTab}  \\
 [\Hba_{\alphaa},\Hbb^{\betab}] & = -f\indices{_{\alphaa \gammab}^{\betab}} \Hbb^{\gammab}
                             \label{eq:HHdTab} \\
 [\Hba_{\alphaa},\Hba_{\betaa}] & =f\indices{_{\alphaa \betaa}^{\gammaa}} \Hba_{\gammaa}
                             \label{eq:HHTaa}  \\
 [\Hba_{\alphaa},\Hba^{\betaa}] & =-f\indices{_{\alphaa \gammaa}^{\betaa}} \Hba^{\gammaa} 
                             \label{eq:HHdTaa} \\
  \end{align}
  with the invariant metric
  \begin{align}
  \Omega_{ab}=
  \bordermatrix{
  ~                             & \Gb_{j}                    & \Hba_{\betaa}                       & \Hbb_{\betab}                       & \Hba^{\betaa}                       & \Hbb^{\betab} \cr
   \Gb_{i}                      & \Omega_{ij}^{\mathfrak{g}} & 0                                   & 0                                   & 0                                   & 0  \cr
   \Hba_{\alphaa}               & 0                          & h_{\alphaa \betaa}                  & 0                                   & \delta\indices{_{\alphaa}^{\betaa}} & 0\cr
   \Hbb_{\alphab}               & 0                          & 0                                   & 0                                   & 0                                   & \delta\indices{_{\alphab}^{\betab}}\cr
   \Hba^{\alphaa}               & 0                          & \delta\indices{^{\alphaa}_{\betaa}} & 0                                   & 0                                   & 0 \cr
   \Hbb^{\alphab}               & 0                          & 0                                   & \delta\indices{^{\alphab}_{\betab}} & 0                                   & 0 \cr
             } \,.
\end{align}
\end{theorem}
We will call this type of contractions \textbf{invariant metric preserving}.
The contracted Lie algebra $\widetilde{\mathfrak{h}}_{T}$
is abelian.
Equation \eqref{eq:HHdTbb} can be written in a,
from the point of view of the double extension,
more suggestive way
\begin{align}
  [\Hbb_{\alphab},\Hbb^{\betab}]  =
  f\indices{_{\gammaa \alphab}^{\gammab}}
  \delta\indices{_{\gammab}^{\betab}}
  \Hba^{\gammaa}
  =-f\indices{_{\alphab \gammaa}^{\betab}} \Hba^{\gammaa}  \,.
\end{align}

Ignoring the double extension structure and since we rescale
$\widetilde{\mathfrak{h}}^{*}$ by inverse powers,
this is a gIW-contraction.
Taking the full structure into account one could see this contraction
as a sIW-contraction and its dual,
which makes it the natural generalization of sIW-contractions
for double extensions.


One observation will be useful, when we want to explain why the
sIW-contractions from the Poincar\'e to Carroll algebras
lead to such a contraction.
\begin{corollary}
  For trivial double extensions, i.e.,
  $D(0,\mathfrak{h} \vdis \widetilde{\mathfrak{h}})$
  the contraction
  described in Theorem \ref{sec:double-extens-pres} equals to a
  sIW-contraction with respect to the subalgebra
  $\mathfrak{h} \vdis \widetilde{\mathfrak{h}}^{*}$.
\end{corollary}
This explains why, even though sIW-contractions were done in \cite{Bergshoeff:2016soe} the invariant metric stayed nondegenerate.

\begin{example}[Poincar\'e to Carroll]
  \label{exp:poitocar}
  The Poincar\'e algebra in $2+1$ dimension is a trivial double
  extension $D(0,\mathfrak{h})$ where   
  $\mathfrak{h}=\{\Jt,\Gt_{a} \}$
  and
  $\mathfrak{h}^{*}=\{\Ht,\Pt_{a} \}$, see Table \ref{tab:PoiandCar}.
  There exists a sIW-contraction, with respect to the subalgebra
  $\{ \Jt, \Pt_{a}\}$ to the Carroll algebra~\cite{Bacry:1968zf}.
  Similar to considerations of Section \ref{sec:cont-inh} we could have
  found the invariant metric of the Carroll algebra.
  But in this case it is equivalent to an invariant metric preserving contraction,
  with the notation of before
  $\Hba_{\alphaa}=\Jt$,
  $\Hbb_{\alphab}=\Gt_{a}$
  and
  $\Hba^{\alphaa}=\Ht$,
  $\Hbb^{\alphab}=\Pt_{a}$.

  This discussion generalizes to contractions of the higher spin versions of Poincar\'e and Carroll.
  \begin{table}[H]
    \centering
    $
    \begin{array}{l r r r r r}
      \toprule %
                                                & \mathfrak{poi}          & \mathfrak{car}         \\ \midrule
      \left[\,\Jt  \comma \Gt_{a} \,\right]     & \epsilon_{am}  \Gt_{m}  & \epsilon_{am}  \Gt_{m} \\
      \left[\, \Jt \comma \Pt_{a} \,\right]     & \epsilon_{am}  \Pt_{m}  & \epsilon_{am}  \Pt_{m} \\ 
      \left[\,\Gt_{a}  \comma \Gt_{b} \,\right] & - \epsilon_{ab}  \Jt    & 0                      \\
      \left[\, \Gt_{a} \comma \Ht \,\right]     & -\epsilon_{am}  \Pt_{m} & 0                      \\
      \left[\, \Gt_{a} \comma \Pt_{b} \,\right] & -\epsilon_{ab} \Ht      & -\epsilon_{ab} \Ht     \\ \midrule
      \langle \Ht \,, \Jt \rangle               & -\mu^{-}                & -\mu^{-}               \\
      \langle \Pt_{a} \,, \Gt_{b} \rangle       & \mu^{-} \delta_{ab}     & \mu^{-} \delta_{ab}    \\
      \langle \Jt \,, \Jt \rangle               & -\mu^{+}                & -\mu^{+}               \\
      \langle \Gt_{a} \,, \Gt_{b} \rangle       & \mu^{+} \delta_{ab}     & 0    \\
      \bottomrule
    \end{array}
    $
    \caption{Poincar\'e and Carroll algebra and their invariant metrics.}
    \label{tab:PoiandCar}
  \end{table}
\end{example}

\chapter{Charges and Boundary Conditions}
\label{sec:boundary-conditions}


According to the AdS/CFT dictionary asymptotic symmetries of the bulk theory
correspond to global symmetries of the boundary theory.
So to get information concerning possible
boundary theories the asymptotic symmetry algebra
is a very useful tool.
To construct it one first needs to define
differentiable gauge transformations.
From there global charges can be defined,
which one then quotients by the true (proper)
gauge transformations.


Although they are
not of direct importance to the
considerations of these sections
some possibly useful and explicit calculations in relation
to symmetries of CS theories
are summarized in Appendix \ref{sec:useful}.

\section{Global Charges}
\label{sec:global-charges}

To construct global charges for CS theories we follow the approach
pioneered by Regge and Teitelboim~\cite{Regge:1974zd} (see also \cite{Benguria:1976in})
and first applied to CS theories by Ba\~nados~\cite{Banados:1994tn}.
I will follow Section 3 of \cite{Campoleoni:2010zq} (which is based on \cite{Banados:1994tn,Banados:1998ta,Banados:1998gg}),\cite{Bunster:2014mua} and \cite{Troessaert:2013fba} where more information can be found.

We start by $2+1$ decomposing\footnote{
 We ignore terms at $t = \pm \infty$. 
}
the CS action (for the notation see Appendix \ref{sec:2+1-decomposition})
\begin{align}
  I_{\mathrm{CS}}[A]&=\frac{k}{4 \pi}\int_{M_{3}}
                 \langle
                 A \wedge \dd A + \frac{2}{3} A \wedge A \wedge A
                 \rangle
  \\
  &=\frac{k}{4 \pi}\int_{\R \times \Sigma} 
    \langle
    d_{N} \At \At + 2 A_{N} \Ft +\ddt (A_{N} \At)
    \rangle
   \\
  &=\frac{k}{4 \pi}\int_{\R \times \Sigma} 
  \langle
    \dot{A}_{i}A_{j} +  A_{t} F_{ij}
    \rangle \dd t \dd x^{i} \dd x^{j}+\frac{k}{4 \pi}\int_{\R \times\partial \Sigma} \tr (A_t \At_{i}) \dd t \dd x^{i} \,.
    \label{eq:lagrsplit}
\end{align}
This action principle is that of a constrained system in Hamiltonian form, i.e.,
it has the form $\int( \dot q p - u \gamma(q,p))\dd t$.
The $\dim(\mathfrak{g})$ Lagrange multipliers $A_{0}$ enforce the first-class constraints
and the (bulk) Hamiltonian consists only of these.
There are $2 \cdot \dim(\mathfrak{g})$ canonical/dynamical fields $A_{i}$ and via the standard formula (e.g., \cite{Henneaux:1992ig}), and since there are no second-class constraints, we get
\begin{align}
  2 \cdot
  \begin{pmatrix}
      \text{Number of physical} \\
  \text{degrees of freedom}
  \end{pmatrix}
&= \begin{pmatrix}
  \text{Canonical}
  \\
  \text{variables}
\end{pmatrix}
    - 2 \cdot
  \begin{pmatrix}
    \text{First-class}
    \\
     \text{constraints}
\end{pmatrix}
  \\
&= 2 \dim(\mathfrak{g}) - 2  \dim(\mathfrak{g})
  \\
  &=0 \,.
\end{align}
So there are no (local) degrees of freedom (in the bulk).

The equal-time Poisson bracket for two differentiable functionals
$M[A_{i}]$ and $N[A_{i}]$
is defined by
\begin{align}
  \{M,N\}&=\frac{2 \pi}{k}
           \int_{\Sigma}\dd x^{i}\wedge \dd x^{j}
           \left\langle
           \frac{\delta M}{\vd A_{i}(x)} \frac{\delta N}{\vd A_{j}(x)}
           \right\rangle \,.
\end{align}
Using the Poisson bracket
first-class constraints generate gauge transformations (if they are differentiable)
by defining the gauge generator
\begin{align}
  \label{eq:Gauge}
  \bar G[\lambda]= \frac{k}{2 \pi} \int_{\Sigma}
               \langle
               \lambda \Ft
               \rangle \,.
\end{align}
The variation of this gauge transformation shows that gauge generator is not differentiable
as can be seen from the nonvanishing boundary term in
\begin{align}
  \delta \bar G[\lambda]&=
                      \frac{k}{2 \pi}\int_{\Sigma}
                      \langle
                      \delta \lambda \,\Ft-\deltat_{\lambda}\At \wedge \delta \At
                      \rangle
                      +
                      \frac{k}{2 \pi}\int_{\partial \Sigma}
                      \langle
                      \lambda \delta \At
                      \rangle 
\end{align}
where \begin{equation}
  \label{eq:tilgauge}
  \deltat_{\lambda}\bullet \equiv \widetilde{\dd}+[\bullet,\lambda] \,.
\end{equation}
Only differentiable gauge transformations
are allowed to enter the
Poisson bracket so
 one needs to add
the boundary
term $\vd Q[\lambda]$.
Assuming that $\lambda$ is independent of
dynamical fields\footnote{
  Here might appear a problem with integrability if this is not the case.
} $\vd Q[\lambda]$ can be integrated in field space and leads to
\begin{align}
  \label{eq:gaugetrafo}
  G[\lambda]   &=\bar G[\lambda]+Q[\lambda]
  \\
               &=  \frac{k}{2 \pi}\int_{\Sigma}
              \langle
              \lambda \Ft
              \rangle
              - \frac{k}{2 \pi}
              \int_{\partial \Sigma}
              \langle
              \lambda  \At
              \rangle
              \\
               &= \frac{k}{4 \pi}\int_{\Sigma} \dd x^{i}\wedge \dd x^{j}
              \langle
              \lambda F_{ij}
              \rangle
              - \frac{k}{2 \pi}
              \int_{\partial \Sigma} \dd x^{i}
              \langle
              \lambda  A_{i}
              \rangle \,.
\end{align}
We can now plug this differentiable gauge generator
into the Poisson algebra
\begin{align}
  \{ G[\lambda],G[\sigma]\}=G\left[[\lambda,\sigma]\right] + \frac{k}{2\pi} \int_{\pd \Sigma}\dd x^{i} \langle \lambda \pd_{i}\sigma\rangle \, .
\end{align}
One has to differentiate between two categories of
differentiable gauge transformations~\cite{Regge:1974zd,Benguria:1976in}:
\begin{itemize}
\item \textbf{Proper} or \textbf{true} gauge transformations are defined by $G[\lambda] = 0$ on the constrained surface $\tilde F=0$.
  \index{Gauge transformation}%
  \index{Gauge transformation!Proper}%
  \index{Gauge transformation!True}%
  This implies that generically $Q[\lambda] = 0$ since this term
  does not automatically vanish on-shell.
  On the other hand $\bar G[\lambda]= 0$ vanishes automatically since
  this is the part that consists of the constraints.
  These are the true gauge symmetries of the system
  in the sense that they are a redundancy of the description.
  Or said in a more drastic fashion,
  proper gauge transformations do physically nothing.
  They form an ideal subalgebra of the
  differentiable gauge transformations.
\item \textbf{Improper} gauge transformations
  \index{Gauge transformation!Improper}
  are nonzero on the constraint surface
  and therefore $G[\lambda] = Q[\lambda]\neq 0$.
  These are no true gauge transformations
  and they lead
  to the global symmetries of the theory.
  They change the physical state of the system
  and
  are the origin of the boundary degrees of freedom.
\end{itemize}

When the constraints are solved and the gauge is fixed, the
$Q[\lambda]$ give the global charges of the theory,
which in turn generate the asymptotic symmetry algebra
(when the quotient by the proper gauge symmetries is taken)\footnote{
  This is possible since the proper gauge symmetries are
  an ideal.
}.
The global symmetries are then  generated by
\begin{align}
  \label{eq:globgen}
  \delta_{\lambda}M
  = \{Q(\lambda),M \}
\end{align}
and on the reduced phase space lead to
\begin{align}
  \{ Q[\lambda],Q[\sigma]\}=Q\left[[\lambda,\sigma]\right] + \frac{k}{2\pi} \int_{\pd \sigma}\dd x^{i} \langle \lambda \pd_{i}\sigma\rangle \,.
\end{align}





\section{Boundary Conditions}
\label{sec:boundary-conditions-1}

Once an action principle is fixed
the procedure to establish boundary conditions
``is one of trial and error''~\cite{Bunster:2014mua}.
This means no bullet proof recipe is known,
but one minimum requirement is
that the extremized action
gives the desired equations of motion
up to surface terms at (spatial) infinity
\begin{align}
  \label{eq:Ivar}
  \delta I_{CS}[A]
  &=\frac{k}{2\pi}\int_{M_{3}}
    \langle
    F \wedge \delta A
    \rangle
    -\frac{k}{4\pi}
    \int_{\partial M_{3}}
    \langle
    A \wedge \delta A
    \rangle
  \\ 
  &=\frac{k}{2 \pi}\int_{\R \times \Sigma} \dd t \wedge 
    \langle
    (\partial_{t} \At - \deltat_{A_{t}}\At)\delta \At + \Ft \delta A_{t}
    \rangle
    \nonumber\\
  &\quad +\frac{k}{4 \pi}\int_{\R \times\partial \Sigma}\dd t \wedge
    \langle \At\delta A_{t}-A_{t}\delta \At \rangle \,.
\end{align}
Using the boundary conditions the final action should be differentiable,
i.e., extremized without additional boundary terms.
Furthermore, the  boundary conditions should allow for all solutions
of interest.

For CS theories and ignoring any specific physical requirements one might have,
there exist always (up to topological obstructions) boundary conditions that are related to the WZW model~\cite{Witten:1988hf,Elitzur:1989nr}.






\chapter{AdS Higher Spin Gravity}
\label{cha:ads-spin-3}

We will first review higher spin theories\footnote{We partially follow \cite{Prohazka:2013,Afshar:2014rwa}.
},
with emphasis towards $(2+1)$-dimensional spacetimes.
A nice and more complete review can be
found in \cite{Campoleoni:2010zq}.
Afterwards we discuss,
following closely \cite{Grumiller:2016kcp}, the $\mathfrak{u}(1)$
higher spin boundary conditions.

\section{Higher Spin Theories}
\label{sec:hs-theories}

The equations for non interacting massless particles of integer spin
in $(3+1)$ dimensions on a flat background were found by
Frondsdal~\cite{Fronsdal:1978rb}\footnote{We will restrict our explanations to integer spin for the sake of simplicity.}.
For $s=0,1,2$ they reduce to the well known Klein--Gordon equation, Maxwell equation and to linearized general relativity.
It is comparably easy to write down these free higher spin fields.
But coupling these for $s > 2$ to gravity leads to various no-go theorems (for a review see \cite{Bekaert:2010hw}).
Fradkin and Vasiliev \cite{Fradkin:1987ks} showed that consistent higher spin gauge theories involving gravity need to be defined on a curved background.
They were first formulated by Vasiliev \cite{Vasiliev:1992av} (and are reviewed in \cite{Bekaert:2005vh,Vasiliev:2012vf,Didenko:2014dwa}).
These theories involve an infinite tower of massless fields and can be constructed on (A)dS spaces.

One interesting aspect of higher spin gauge fields is that they
might be connected to string theory in the tensionless limit in which
the massive excitations of string theory become massless.
It is conjectured that string theory is a broken phase of a
higher spin gauge theory.
For more details see \cite{Sagnotti:2011qp} and references therein.

Furthermore the holographic principle finds a realization
in the form of the proposal made by Klebanov and Polyakov~\cite{Klebanov:2002ja}
and Sezgin and Sundell~\cite{Sezgin:2002rt,Sezgin:2003pt}.
They conjectured that there exists a duality in the large $N$
limit of the critical 3-dimensional $O(N)$ model and the  minimal bosonic higher spin theory in $\mathrm{AdS}_{4}$.
This holographic proposal got supported by calculations of Giombi and Yin \cite{Giombi:2009wh} and is reviewed in \cite{Giombi:2012ms}.

In $2+1$ spacetime dimensions the situation changes significantly.
Massless gauge fields with ``spin''\footnote{``Spin'' in $D=3$ refers to the transformation properties of the field under Lorentz transformations.} $s>1$ posses no local degrees of freedom anymore.
This makes theories in $2+1$ dimensions interesting in various aspects.
While there is still enough structure to be nontrivial the
technical difficulties that arise in the higher-dimensional cases are often circumvented.

This is already the case in the famous result by
Brown and Henneaux \cite{Brown:1986nw} which can be seen as a
precursor of the $\mathrm{AdS}_{3}/\mathrm{CFT}_{2}$ correspondence.
They showed that three-dimensional Einstein--Hilbert gravity with a
negative cosmological constant and Brown--Henneaux boundary conditions
leads to asymptotic symmetries given by the infinite-dimensional
conformal algebra in two dimensions.
These are two copies of the Virasoro algebra (see Section \ref{app:Walg})
with a nonvanishing central charge.
Equivalent results were derived in the CS formulation,
based on $\mathfrak{sl}(2,\R)\dis \mathfrak{sl}(2,\R)$~\cite{Banados:1994tn}.

The central charge appears again in the analysis of another
unexpected result, the Ba\~nados--Teitelboim--Zanelli (BTZ)
black hole \cite{Banados:1992wn,Banados:1992gq}.
Even though there are no local degrees of freedom in three-dimensional gravity,
for the case of negative cosmological constant these black holes exist.
Using the central charge it was shown that it is possible to calculate
the asymptotic density of states and the entropy~\cite{Strominger:1997eq}.
So a microscopic interpretation for the states of the black hole is possible and the holographic principle is realized.

To add interacting fields with spin $s>2$, in contrast to the higher-dimensional case in $2+1$ dimension, no infinite number of higher spin fields are needed
(at least in the classical theory) \cite{Aragone:1983sz}.
The Brown--Henneaux analysis has been generalized to higher spin fields \cite{Henneaux:2010xg,Campoleoni:2010zq,Gaberdiel:2011wb,Campoleoni:2011hg}.
In the case of the coupling of a spin-$3$ field to gravity the asymptotic symmetries are given by $\mathcal{W}_{3} \dis \mathcal{W}_{3}$ algebras~\cite{Henneaux:2010xg,Campoleoni:2010zq}.
For a review of $\mathcal{W}$ algebras see \cite{Bouwknegt:1992wg} and for
the explicit commutation relations see Section \ref{app:Walg}.
Fields of spin $s=3,4,\ldots,N$ coupled to gravity are given by a Chern-Simons theory with gauge algebra $\mathfrak{sl}(n,\R) \dis \mathfrak{sl}(n,\R)$ (see Appendix~\ref{cha:explicit-lie-algebra} for the commutation relations) and have  in the case of an $\mathrm{AdS_{3}}$ background the  asymptotic symmetries $\mathcal{W}_{N} \dis \mathcal{W}_{N}$~\cite{Gaberdiel:2011wb,Campoleoni:2011hg}. Using the infinite-dimensional higher spin algebras $\mathfrak{hs}[\lambda] \dis \mathfrak{hs}[\lambda]$ as gauge algebra
we get gravity coupled to spin fields $s=3,4,\ldots,\infty$ and again for $\mathrm{AdS_{3}}$ asymptotic symmetries  $\mathcal{W}_{\infty}[\lambda] \dis \mathcal{W}_{\infty}[\lambda]$~\cite{Gaberdiel:2011wb}.
The $\mathfrak{hs}[\lambda]$ algebra can be truncated to $\mathfrak{sl}(N,\R)$
for integer $N$, see Appendix \ref{sec:hs}.

Another aspect that is advantageous in $2+1$ dimensions
is that the dual to $\mathrm{AdS}_{3}$ is given by $\mathrm{CFT}_{2}$
and extensions thereof.
Two-dimensional conformal field theories are well understood and offer a high degree  of analytic control. It was proposed by Gaberdiel and Gopakumar \cite{Gaberdiel:2010pz} that the $\mathfrak{hs}[\lambda]$ theory on $\mathrm{AdS}_{3}$ is dual to the large-$N$ limit of $\mathcal{W}_{N}$ minimal models on the $\mathrm{CFT}$ side. As a hint for the validity of this proposal can be seen that this limit on the CFT side leads, like in the bulk theory, also to a $\mathcal{W}_{\infty}$ algebra. The duality is reviewed in \cite{Gaberdiel:2012uj} and new developments can be found in \cite{Gaberdiel:2014cha,Gaberdiel:2015wpo}.

The BTZ black hole can also be generalized to higher spin black holes\cite{Gutperle:2011kf}.
Since higher spin gauge theories
have an extended gauge symmetry with respect to general relativity
new questions
concerning gauge invariant characterization and black hole thermodynamics arise
(for a review of the proposed answers see \cite{Ammon:2012wc,Perez:2014pya}).

Before background and boundary conditions
beyond AdS$_{3}$ will be discussed
it is useful to review the standard
spin-3 ones.
There exist excellent resources
where they are derived from first principles~\cite{Campoleoni:2010zq,Bunster:2014mua,deBoer:2014fra}
and therefore we will choose a different route.
We will construct them following~\cite{Grumiller:2016kcp}
where they are composed out of $\mathfrak{u}(1)$ boundary conditions~\cite{Afshar:2016wfy}.

\section{\texorpdfstring{$\mathcal{W}_3$}{W3} via \texorpdfstring{$\hat{\mathfrak{u}}(1)$}{u(1)} Boundary Conditions }
\label{sec:w_3-bound-cond}

Higher spin gravity in $2+1$ dimensions can be
generically described in terms of the difference of two Chern--Simons
actions for independent gauge fields $A^{\pm}$ that take values in
$\mathfrak{sl}\left(N,\mathbb{R}\right)$, so that the action reads
\begin{equation}
I=I_{\mathrm{CS}}\left[A^{+}\right]-I_{\mathrm{CS}}\left[A^{-}\right]\;,\label{eq:HS-Action}
\end{equation}
with
\begin{equation}
I_{\mathrm{CS}}[A]=\frac{k_{N}}{4\pi}\int_{M_{3}}\text{tr}\left(A\wedge dA+\frac{2}{3}A\wedge A\wedge A\right)\;,\label{CS-Action}
\end{equation}
where $\text{tr}\left(\cdots\right)$ stands for the trace in the
fundamental representation of $\mathfrak{sl}\left(N,\mathbb{R}\right)$ (see
Appendix \ref{sec:slN-r}).
The level in \eqref{CS-Action} relates
to the Newton constant and the AdS radius according to $k_{N}=\frac{k}{2\epsilon_{N}}=\frac{\ell}{8G\epsilon_{N}}$,
whose normalization is determined by $\epsilon_{N}=\frac{N(N^{2}-1)}{12}$.

The gauge fields are related to a suitable generalization of the zuvielbein
and the spin connection, defined through 
\begin{equation}
A^{\pm}=\omega\pm\frac{e}{\ell}
\end{equation}
and hence, the spacetime metric and the higher spin fields can be
reconstructed from
\begin{equation}
g_{\mu\nu}=\frac{1}{\epsilon_{N}}\text{tr}(e_{\mu}e_{\nu}) \qquad\qquad \Phi_{\mu_{1}\dots\mu_{s}}=\frac{1}{\epsilon_{N}^{(s)}}\text{tr}(e_{(\mu_{1}}\dots e_{\mu_{s})}) \,. \label{MetricAndSpin-S-Fields}
\end{equation}

\subsection{Asymptotic Structure}
\label{Asymptotic-structure}

The asymptotic structure of AdS gravity coupled to higher spin fields
in three-dimensional spacetimes was investigated in \cite{Henneaux:2010xg,Campoleoni:2010zq},
where it was shown that the asymptotic symmetries are spanned by two
chiral copies of $\mathcal{W}$ algebras (see also \cite{Gaberdiel:2011wb,Campoleoni:2011hg}).
In order to accommodate the different higher spin black hole solutions
in \cite{Gutperle:2011kf,Castro:2011fm}, and \cite{Henneaux:2013dra,Bunster:2014mua},
the asymptotic behavior has to be extended so as to incorporate chemical
potentials associated to the global charges.  The one in \cite{Gutperle:2011kf,Ammon:2011nk}
successfully accommodates the black hole solution with higher spin
fields of \cite{Gutperle:2011kf}, while the set of boundary conditions
in \cite{Henneaux:2013dra,Bunster:2014mua} do for the higher spin black holes described
therein. It is worth pointing out that the asymptotic symmetries of
both sets are different.

Here we construct an inequivalent set of boundary conditions, which
reduces to the one recently introduced in \cite{Afshar:2016wfy} when
the higher spin fields are switched off. The asymptotic behavior
of the $\mathfrak{sl}\left(3,\R\right)$ gauge fields is proposed to
be given by 
\begin{equation}
A^{\pm}=b_{\pm}^{-1}\left(d+a^{\pm}\right)b_{\pm}\label{Amn}
\end{equation}
so that the dependence on the radial coordinate is completely contained
in the group elements
\begin{equation}
b_{\pm}=\exp\left(\pm\frac{1}{\ell\zeta^{\pm}}\Lt_{1}\right)\cdot\exp\left(\pm\frac{\rho}{2}\Lt_{-1}\right)\;.\label{b-null}
\end{equation}
The auxiliary connection reads
\begin{equation}
a^{\pm}=\left(\pm\mathcal{J}^{\pm}\ d\varphi+\zeta^{\pm}\ dt\right) \Lt_{0} + \left(\pm\mathcal{J}_{\left(3\right)}^{\pm}\ d\varphi+\zeta_{\left(3\right)}^{\pm}\ dt\right)\Wt_{0} \label{amn}
\end{equation}
where $\Lt_{i}, \Wt_{n}$, with $i=-1,0,1$, and $n=-2,-1,0,1,2$, span
the $sl\left(3,\mathbb{R}\right)$ algebra (see Appendix \ref{sec:sl3-r}). Following~\cite{Henneaux:2013dra},
it can be seen that $\mathcal{J}^{\pm}$ and $\mathcal{J}_{\left(3\right)}^{\pm}$
stand for arbitrary functions of (advanced) time and the angular coordinate
that correspond to the dynamical fields, while $\zeta^{\pm}$ and
$\zeta_{\left(3\right)}^{\pm}$ describe their associated Lagrange
multipliers that can be assumed to be fixed at the boundary without
variation ($\delta\zeta^{\pm}=\delta\zeta_{\left(3\right)}^{\pm}=0$). We shall refer to $\zeta^\pm$, $\zeta^\pm_{(3)}$ as chemical potentials.

The field equations, implying the local flatness of the gauge fields,
then reduce to 
\begin{equation}
\dot{\mathcal{J}}^{\pm}=\pm\zeta^{\prime} \qquad\qquad \dot{\mathcal{J}}_{\left(3\right)}^{\pm}=\pm\zeta_{\left(3\right)}^{\prime}\;,\label{FE-DG}
\end{equation}
where dot and prime denote derivatives with respect to $t$ and $\varphi$,
respectively.

\subsection{Asymptotic Symmetries and Canonical Generators}

In the canonical approach \cite{Regge:1974zd}, the variation of the conserved charges 
\begin{equation}
Q[\epsilon^{+},\epsilon^{-}]={\mathcal{Q}}^{+}[\epsilon^{+}]-{\mathcal Q}^{-}[\epsilon^{-}]
\end{equation}
associated to gauge symmetries spanned by $\epsilon^{\pm}=\epsilon_{i}^{\pm} \Lt_{i}+\epsilon_{\left(3\right)n}^{\pm} \Wt_{n}$,
that maintain the asymptotic form of the gauge fields, is determined
by
\begin{equation}
\delta{\mathcal Q}^{\pm}\left[\epsilon^{\pm}\right]=\mp\frac{k}{4\pi}\int d\varphi\left(\eta^{\pm}\delta\mathcal{J}^{\pm}+\frac{4}{3}\eta_{\left(3\right)}^{\pm}\delta\mathcal{J}_{\left(3\right)}^{\pm}\right)\;,
\end{equation}
with $\eta^{\pm}=\epsilon_{0}^{\pm}$, and $\eta_{\left(3\right)}^{\pm}=\epsilon_{\left(3\right)0}^{\pm}$.
According to \eqref{amn}, the asymptotic symmetries fulfill $\delta_{\epsilon^{\pm}}a^{\pm}=d\epsilon^{\pm}+[a^{\pm},\,\epsilon^{\pm}]={\mathcal O}(\delta a^{\pm})$,
provided that the transformation law of the dynamical fields reads
\begin{equation}
\delta\mathcal{J}^{\pm}=\pm\eta^{\pm\prime} \quad\quad \delta\mathcal{J}_{\left(3\right)}^{\pm}=\pm\eta_{\left(3\right)}^{\pm\prime} \label{TransfLaw JmnJ3mn}
\end{equation}
and the parameters are time-independent ($\dot{\eta}^{\pm}=\dot{\eta}_{\left(3\right)}^{\pm}=0$).
One has to take the quotient over the remaining
components of $\epsilon^{\pm}$, since they just span trivial gauge
transformations that neither appear in the variation of the global
charges nor in the transformation law of the dynamical fields. 

The surface integrals that correspond to the conserved charges associated
with the asymptotic symmetries then readily integrate as
\begin{equation}
{\mathcal Q}^{\pm}\left[\eta^{\pm},\eta_{\left(3\right)}^{\pm}\right]=\mp\frac{k}{4\pi}\int d\varphi\left(\eta^{\pm}\left(\varphi\right)\mathcal{J}^{\pm}\left(\varphi\right)+\frac{4}{3}\eta_{\left(3\right)}^{\pm}\left(\varphi\right)\mathcal{J}_{\left(3\right)}^{\pm}\left(\varphi\right)\right)\;,\label{Qmn}
\end{equation}
which are manifestly independent of the radial coordinate $\rho$.
Consequently, the boundary could be located at any fixed value $\rho=\rho_{0}$.
Hereafter, we assume that $\rho_{0}\rightarrow\infty$, since this
choice has the clear advantage of making our analysis to cover the
entire spacetime in bulk.

The algebra of the global charges can then be obtained directly from
the computation of their Poisson brackets; or as a shortcut, by virtue
of $\delta_{Y}Q\left[X\right]=\{Q\left[X\right],Q\left[Y\right]\}$,
from the variation of the dynamical fields in \eqref{TransfLaw JmnJ3mn}.
Expanding in Fourier modes
\begin{equation}
\label{eq:Jmodes}
\mathcal{J}^{\pm}\left(\varphi\right)=\frac{2}{k}\sum_{n=-\infty}^{\infty}J_{n}^{\pm}e^{\pm in\varphi} \qquad \qquad \mathcal{J}_{\left(3\right)}^{\pm}\left(\varphi\right)=\frac{3}{2k}\sum_{n=-\infty}^{\infty}J_{n}^{\left(3\right)\pm}e^{\pm in\varphi}
\end{equation}
leads to the asymptotic symmetry algebra which is described by a set
of $\hat{\mathfrak{u}}\left(1\right)$ currents whose nonvanishing brackets are
given by 
\begin{equation}
i\left\{ J_{n}^{\pm},J_{m}^{\pm}\right\} =\frac{1}{2}kn\delta_{m+n,0} \qquad \qquad i\left\{ J_{n}^{\left(3\right)\pm},J_{m}^{\left(3\right)\pm}\right\} =\frac{2}{3}kn\delta_{m+n,0}\;,\label{algebra3}
\end{equation}
with levels $\frac{1}{2}k$, and $\frac{2}{3}k$, respectively. 

\subsection{(Higher Spin) Soft Hair}
\label{soft hair}

Following the spin-2 construction \cite{Afshar:2016wfy}, we consider now all vacuum descendants $|\psi(q)\rangle$ labeled by a set $q$ of non-negative integers $N^\pm$, $N_{(3)}^\pm$, $n_i^\pm$, $n_i^{(3)\,\pm}$, $m_i^\pm$ and $m_i^{(3)\,\pm}$
  \begin{equation}
\label{eq:angelinajolie}
\big|\psi(q)\rangle = N(q)\prod_{i=1}^{N^\pm}\Big(J^\pm_{-n_i^\pm}\Big)^{m_i^\pm}\prod_{i=1}^{N^\pm_{(3)}}\Big(J^{(3)\,\pm}_{-n_i^{(3)\,\pm}}\Big)^{m_i^{(3)\,\pm}}\big|0\rangle \,.
  \end{equation}
Here $N(q)$ is some normalization constant such that $\langle\psi(q)|\psi(q)\rangle=1$ and the vacuum state\footnote{%
The vacuum state considered here resembles Poincar\'e-AdS rather than
global AdS. The state corresponding to global AdS is gapped by an
imaginary amount of the zero mode charges from the vacuum state
considered here.
} is defined through highest weight conditions, $J_n^\pm|0\rangle=J_n^{(3)\,\pm}|0\rangle=0$ for non-negative $n$.

We want to check now if all vacuum descendants $|\psi(q)\rangle$ have the same energy as the vacuum and are thus soft hair (our discussion easily generalizes from soft hair descendants of the vacuum to soft hair descendants of any higher spin black hole state). To this end we consider the surface integral associated with the generator in time, given by
\begin{equation}
H:=Q\left(\partial_{t}\right)=\frac{k}{4\pi}\int d\varphi\left(\zeta^{+}\mathcal{J}^{+}+\zeta^{-}\mathcal{J}^{-}+\frac{4}{3}\zeta_{\left(3\right)}^{+}\mathcal{J}_{\left(3\right)}^{+}+\frac{4}{3}\zeta_{\left(3\right)}^{-}\mathcal{J}_{\left(3\right)}^{-}\right)\;.
\end{equation}
For constant chemical potentials $\zeta^{\pm}$,
$\zeta_{\left(3\right)}^{\pm}$ the field equations \eqref{FE-DG}
imply that the dynamical fields become time-independent, and the total
Hamiltonian reduces to
\begin{equation}
H=\zeta^{+}J_{0}^{+}+\zeta^{-}J_{0}^{-}+\zeta_{\left(3\right)}^{+}J_{0}^{\left(3\right)+}+\zeta_{\left(3\right)}^{-}J_{0}^{\left(3\right)-}\;,
\end{equation}
which clearly commutes with the whole set of asymptotic symmetry generators
spanned by $J_{n}^{\pm}$ and $J_{m}^{\left(3\right)\pm}$. One then
concludes that for an arbitrary fixed value of the total energy, configurations
endowed with different sets of nonvanishing $\hat{\mathfrak{u}}\left(1\right)$
charges turn out to be inequivalent, because they can not be related
to each other through a pure gauge transformation. Since excitations \eqref{eq:angelinajolie} associated with the generators $J_{n}^{\pm}$, $J_{m}^{\left(3\right)\pm}$ preserve the total energy and cannot be gauged away, they are (higher spin) soft hair in the sense of Hawking, Perry and Strominger \cite{Hawking:2016msc}.

\subsection{Highest Weight Gauge and the Emergence
of Composite \texorpdfstring{$\mathcal{W}_{3}$}{W_3} Symmetries}
\label{Mapping}

Quite remarkably, it can be seen that spin-2 and spin-3 charges naturally
emerge as composite currents constructed out from the $\hat{\mathfrak{u}}(1)$
ones. 
Actually, the full set of generators of the $\mathcal{W}_{3}$ algebra
arises from suitable composite operators of the $\hat{\mathfrak{u}}(1)$ charges
through a twisted Sugawara construction. Here we show this explicitly
through the comparison of the new set of boundary conditions proposed
in the previous section with the ones that accommodate the higher
spin black holes in \cite{Henneaux:2013dra,Bunster:2014mua}, whose asymptotic symmetries
are described by two copies of the $\mathcal{W}_3$ algebra. In order to carry
out this task it is necessary to express both sets in terms of the
same variables. The asymptotic behavior
described by \eqref{Amn} and \eqref{amn} is formulated so that the
auxiliary connections $a^{\pm}$ are written in the diagonal gauge,
while the set in \cite{Henneaux:2013dra,Bunster:2014mua} was formulated in the so-called
highest weight gauge. Consequently, what we look for can be unveiled
once the gauge fields in \eqref{Amn} and \eqref{amn} are expressed
in terms of the variables that are naturally adapted to the gauge
fields $\hat{A}^{\pm}$ in the highest weight gauge. 

For a generic choice of Lagrange multipliers, which are still unspecified,
the asymptotic form of the gauge fields in the highest weight gauge
reads \cite{Henneaux:2013dra,Bunster:2014mua}
\begin{equation}
\hat{A}^{\pm}=\hat{b}_{\pm}^{-1}(d+\hat{a}^{\pm})\hat{b}_{\pm}\;,
\end{equation}
where the radial dependence can be captured by the choice $\hat{b}_{\pm}=e^{\pm\rho \Lt_{0}}$,
and 
\begin{align}
\hat{a}_{\varphi}^{\pm} & =\Lt_{\pm1}-\frac{2\pi}{k}\mathcal{L}_{\pm} \Lt_{\mp1} - \frac{\pi}{2k}\mathcal{W}_{\pm}\Wt _{\mp2} \qquad \qquad \hat{a}_{t}^{\pm}=\Lambda^{\pm}\left[\mu_{\pm},\nu_{\pm}\right]\;,\label{amnHW}
\end{align}
with
\begin{align}
\Lambda^{\pm}
  &\! =\! \pm\left[\mu_{\pm}\Lt_{\pm1}+\nu_{\pm}\Wt_{\pm2}\mp\mu_{\pm}^{\prime}\Lt_{0}\mp\nu_{\pm}^{\prime}\Wt_{\pm1}+\tfrac{1}{2}\left(\mu_{\pm}^{\prime\prime}-\tfrac{4\pi}{k}\mu_{\pm}\mathcal{L}_{\pm}+\tfrac{8\pi}{k}\mathcal{W}_{\pm}\nu_{\pm}\right)\Lt_{\mp1}\right.\nonumber \\
 & \quad -\left(\tfrac{\pi}{2k}\mathcal{W}_{\pm}\mu_{\pm}+\tfrac{7\pi}{6k}\mathcal{L}_{\pm}^{\prime}\nu_{\pm}^{\prime}+\tfrac{\pi}{3k}\nu_{\pm}\mathcal{L}_{\pm}^{\prime\prime}+\tfrac{4\pi}{3\kappa}\mathcal{L}_{\pm}\nu_{\pm}^{\prime\prime}\right.\left.-\tfrac{4\pi^{2}}{k^{2}}\mathcal{L}_{\pm}^{2}\nu_{\pm}-\tfrac{1}{24}\nu_{\pm}^{\prime\prime\prime\prime}\right)\Wt_{\mp2}\nonumber \\
 & \quad +\left.\tfrac{1}{2}\left(\nu_{\pm}^{\prime\prime}-\tfrac{8\pi}{k}\mathcal{L}_{\pm}\nu_{\pm}\right)\Wt_{0}\mp\tfrac{1}{6}\left(\nu_{\pm}^{\prime\prime\prime}-\tfrac{8\pi}{k}\nu_{\pm}\mathcal{L}_{\pm}^{\prime}-\tfrac{20\pi}{k}\mathcal{L}_{\pm}\nu_{\pm}^{\prime}\right)\Wt_{\mp 1}\right]\ ,\label{Lambda-HW}
\end{align}
where ${\mathcal L}_{\pm}$, ${\mathcal W}_{\pm}$ and $\mu_{\pm}$, $\nu_{\pm}$
stand for arbitrary functions of $t,\varphi$.

One then needs to find suitable permissible gauge transformations
span\-ned by group elements $g_{\pm}$, for which $\hat{a}^{\pm}=g_{\pm}^{-1}\left(d+a^{\pm}\right)g_{\pm}$.
These group elements indeed exist and are given,
as well as necessary consistency conditions, explicitly
in \cite{Grumiller:2016kcp}.
The gauge fields $a^{\pm}$ and $\hat{a}^{\pm}$ are then mapped to
each other provided
\begin{align}
{\mathcal L}_{\pm} & =\pm\frac{k}{4\pi}\left(\frac{1}{2}\left({\mathcal J}^{\pm}\right)^{2}+\frac{2}{3}\left({\mathcal J}_{\left(3\right)}^{\pm}\right)^{2}+{\mathcal J}^{\pm\prime}\right) \label{eq:MiuraL}\\
{\mathcal W}_{\pm} & =\mp\frac{k}{6\pi}\left(-\frac{8}{9}\left({\mathcal J}_{\left(3\right)}^{\pm}\right)^{3}+2\left({\mathcal J}^{\pm}\right)^{2}{\mathcal J}_{\left(3\right)}^{\pm}+{\mathcal J}_{\left(3\right)}^{\pm}{\mathcal J}^{\pm\prime}+3{\mathcal J}^{\pm}{\mathcal J}_{\left(3\right)}^{\pm\prime}+{\mathcal J}_{\left(3\right)}^{\pm\prime\prime}\right) \label{eq:MiuraW}
\end{align}
from which one recognizes the Miura transformation between the variables,
see e.g. \cite{Bouwknegt:1992wg}.

Note that the functions ${\mathcal L}_{\pm}$, ${\mathcal W}_{\pm}$, that
are naturally defined in the highest weight gauge, depend on the global
charges ${\mathcal J}^{\pm}$, ${\mathcal J}_{\left(3\right)}^{\pm}$ as in
eqs. \eqref{eq:MiuraL}, \eqref{eq:MiuraW}. 
In sum, our proposal for boundary conditions once expressed in the
highest weight gauge, is such that the Lagrange multipliers $\mu_{\pm}$
and $\nu_{\pm}$ depend on the dynamical variables. 

Indeed, for a generic choice of Lagrange multipliers in the highest
weight gauge, the field equations read \cite{Bunster:2014mua}
\begin{align}
\dot{\mathcal{L}}_{\pm} & =\pm2\mathcal{L}_{\pm}\mu_{\mathcal{\pm}}^{\prime}\pm\mu_{\pm}\mathcal{L}_{\pm}^{\prime}\mp\frac{k}{4\pi}\mu_{\mathcal{\pm}}^{\prime\prime\prime}\mp2\nu_{\pm}\mathcal{W}_{\pm}^{\prime}\mp3\mathcal{W}_{\pm}\nu_{\pm}^{\prime}\label{eq:LPuntoW3-1}\\
\dot{\mathcal{W}}_{\pm} & =\pm3\mathcal{W}_{\pm}\mu_{\mathcal{\pm}}^{\prime}\pm\mu_{\pm}\mathcal{W}_{\pm}^{\prime}\pm\frac{2}{3}\nu_{\pm}\left(\mathcal{L}_{\pm}^{\prime\prime\prime}-\frac{16\pi}{k}\mathcal{L}_{\pm}^{2\prime}\right)\pm3\left(\mathcal{L}_{\pm}^{\prime\prime}-\frac{64\pi}{9k}\mathcal{L}_{\pm}^{2}\right)\nu_{\pm}^{\prime}\nonumber \\
 &\quad \pm5\nu_{\pm}^{\prime\prime}\mathcal{L}_{\pm}^{\prime}\pm\frac{10}{3}\mathcal{L}_{\pm}\nu_{\pm}^{\prime\prime\prime}\mp\frac{k}{12\pi}\nu_{\pm}^{\left(5\right)}\ ,\label{eq:WPuntoW3}
\end{align}
which by virtue of the definition of our boundary conditions
reduce to the remarkably simple ones, given by $\dot{\mathcal{J}}^{\pm}=\pm\zeta^{\prime}$,
$\dot{\mathcal{J}}_{\left(3\right)}^{\pm}=\pm\zeta_{\left(3\right)}^{\prime}$,
which were directly obtained in the diagonal gauge (see eq. \eqref{FE-DG}).

It is also worth highlighting that eqs. \eqref{eq:MiuraL}, \eqref{eq:MiuraW}
can be regarded as the higher spin gravity version of the twisted
Sugawara construction. In fact, as show in \cite{Grumiller:2016kcp} the currents ${\mathcal L}_\pm$, ${\mathcal W}_\pm$ fulfill the $\mathcal{W}_3$ algebra.
\begin{align}
\delta\mathcal{L}_{\pm} & =\pm2\mathcal{L}_{\pm}\varepsilon_{\mathcal{\pm}}^{\prime}\pm\varepsilon_{\pm}\mathcal{L}_{\pm}^{\prime}\mp\frac{k}{4\pi}\varepsilon_{\mathcal{\pm}}^{\prime\prime\prime}\mp2\chi_{\pm}\mathcal{W}_{\pm}^{\prime}\mp3\mathcal{W}_{\pm}\chi_{\pm}^{\prime} \\
\delta\mathcal{W}_{\pm} & =\pm3\mathcal{W}_{\pm}\varepsilon_{\mathcal{\pm}}^{\prime}\pm\varepsilon_{\pm}\mathcal{W}_{\pm}^{\prime}\pm\frac{2}{3}\chi_{\pm}\left(\mathcal{L}_{\pm}^{\prime\prime\prime}-\frac{16\pi}{k}\mathcal{L}_{\pm}^{2\prime}\right)\pm3\left(\mathcal{L}_{\pm}^{\prime\prime}-\frac{64\pi}{9k}\mathcal{L}_{\pm}^{2}\right)\chi_{\pm}^{\prime}\nonumber \\
 &\quad \pm 5 \chi_{\pm}^{\prime\prime}\mathcal{L}_{\pm}^{\prime}\pm\frac{10}{3}\mathcal{L}_{\pm}\chi_{\pm}^{\prime\prime\prime}\mp\frac{k}{12\pi}\chi_{\pm}^{\left(5\right)}\ .
\end{align}
It is then apparent that ${\mathcal L}_{\pm}$ and ${\mathcal W}_{\pm}$ turn
out to be composite anomalous spin-2 and spin-3 currents, respectively.
In other words, the asymptotic $\mathcal{W}_3$ algebra obtained in \cite{Henneaux:2013dra,Bunster:2014mua}
for a different set of boundary conditions, being defined through
requiring the Lagrange multipliers in the highest weight gauge to
be fixed without variation ($\delta\mu_{\pm}=\delta\nu_{\pm}=0$),
is recovered as a composite one that emerges from the $\hat{\mathfrak{u}}\left(1\right)$
currents.

Despite of the fact that the spin-2 and spin-3 currents
${\mathcal L}_{\pm}$, ${\mathcal W}_{\pm}$ fulfill the $\mathcal{W}_3$ algebra,
their associated global charges generate the $\hat{\mathfrak{u}}\left(1\right)$
current algebras discussed in section \eqref{Asymptotic-structure}.
This is so because, by virtue of the consistency conditions
and \eqref{eq:MiuraL}, \eqref{eq:MiuraW} the variation of the global
charges readily reduces to 
\begin{equation}
\delta{\mathcal Q}^{\pm}=\mp\int d\varphi\left(\varepsilon_{\pm}\delta{\mathcal L}_{\pm}-\chi_{\pm}\delta{\mathcal W}_{\pm}\right)=\mp\frac{k}{4\pi}\int d\varphi\left({\mathcal \eta}^{\pm}\delta{\mathcal J}^{\pm}+\frac{4}{3}{\mathcal \eta}_{\left(3\right)}^{\pm}\delta{\mathcal J}_{\left(3\right)}^{\pm}\right)\;,
\end{equation}
so that they satisfy the current algebras in \eqref{algebra3}. Indeed,
this result just reflects the fact that the gauge transformation that
maps our asymptotic conditions in the highest weight and diagonal
gauges
is a permissible one in the sense of \cite{Bunster:2014mua}. Therefore, the
global charges associated with our asymptotic conditions, although written
in the highest weight gauge 
manifestly do not fulfill
the $\mathcal{W}_3$ algebra. This is because the Lagrange multipliers $\mu_{\pm}$,
$\nu_{\pm}$, are not chosen to be fixed at infinity without variation
as in \cite{Henneaux:2013dra,Bunster:2014mua}, but instead, here they explicitly depend
on the global charges. What is actually kept fixed at the boundary
without variation is the set of Lagrange multipliers that is naturally
defined in the diagonal gauge ($\delta\zeta^{\pm}=\delta\zeta_{\left(3\right)}^{\pm}=0$).

\subsection{Higher Spin Black Holes with Soft Hair}
\label{higher spin black hole}


As shown  the simpler subset of our boundary
conditions, obtained by choosing the Lagrange multipliers $\zeta^{\pm}$,
$\zeta_{\left(3\right)}^{\pm}$ to be constants, possesses the noticeable
property of making the global charges $J_{n}^{\pm}$, $J_{m}^{\left(3\right)\pm}$
to behave as (higher spin) soft hair.
An additional
remarkable feature that also occurs in this case is the fact
that regularity of the whole spectrum of Euclidean solutions that
fulfill our boundary conditions holds everywhere, regardless the value
of the global charges.

An interesting effect occurs for the
branch of higher spin black holes that is continuously connected
to the BTZ black hole \cite{Banados:1992wn,Banados:1992gq}, corresponding to $m=0$, $n=1$. Indeed, for
this branch the entropy 
is found to depend just on the
zero modes of the electric-like $\hat{\mathfrak{u}}\left(1\right)$ charges of
the purely gravitational sector, i.e.,
\begin{equation}
S=2\pi\left(J_{0}^{+}+J_{0}^{-}\right)\;.\label{SbtzBranch}
\end{equation}
Nonetheless, the information about the presence of the higher spin
fields is subtle hidden within the purely gravitational global charges,
as can be seen from the map between the $\hat{\mathfrak{u}}\left(1\right)$
and $\mathcal{W}_3$ currents. In fact, for the spherically symmetric higher
spin black hole, by virtue of \eqref{eq:MiuraL}, \eqref{eq:MiuraW},
the relationship between the zero modes of the purely gravitational
$\hat{\mathfrak{u}}\left(1\right)$ charges and the zero modes of the $\mathcal{W}_3$
ones reads
\begin{equation}
J_{0}^{\pm}=\sqrt{2\pi k\mathcal{L}_{\pm}}\cos\left[\frac{1}{3}\arcsin\left(\frac{3}{8}\sqrt{\frac{3k}{2\pi\mathcal{L}_{\pm}^{3}}}\mathcal{W}_{\pm}\right)\right]\;.\label{Jo-LW}
\end{equation}
Therefore, replacing \eqref{Jo-LW} into \eqref{SbtzBranch} one recovers
the following expression for the higher spin black hole entropy in
terms of the spin-2 and spin-3 charges, which reads
\begin{align}
S & =2\pi\sqrt{2\pi k}\left(\sqrt{\mathcal{L}_{+}}\cos\left[\frac{1}{3}\arcsin\left(\frac{3}{8}\sqrt{\frac{3k}{2\pi\mathcal{L}_{+}^{3}}}\mathcal{W}_{+}\right)\right]\right.\nonumber \\
 & \quad \left.+\sqrt{\mathcal{L}_{-}}\cos\left[\frac{1}{3}\arcsin\left(\frac{3}{8}\sqrt{\frac{3k}{2\pi\mathcal{L}_{-}^{3}}}\mathcal{W}_{-}\right)\right]\right)\;,
\end{align}
in full agreement with the result obtained in \cite{Bunster:2014mua}.

This analysis was generalized to arbitrary spin~\cite{Grumiller:2016kcp}
as well as to the case of flat space~\cite{Afshar:2016kjj} and flat space higher spin~\cite{Ammon:2017vwt}.
For more on ``Black Hole Horizon Fluff'' see~\cite{Afshar:2016uax}.

\chapter{Non-AdS Higher Spin Gravity}
\label{cha:non-ads-cs}

In Chapter \ref{cha:ads-spin-3} we have discussed
higher spin theories based on
$\mathfrak{sl}(3,\R) \dis \mathfrak{sl}(3,\R)$
algebras and (higher spin generalized) AdS spacetimes.
We  want to stick to the same underlying Lie algebra (this will be changed in the following chapters),
but we want to generalize to backgrounds and boundary conditions beyond AdS.

In many applications it is necessary to generalize holography to spacetimes more general than asymptotic AdS, for a review see e.g., \cite{Hartnoll:2009sz}.
Examples for which the spacetime can be constructed
in higher spin theories are~\cite{Gary:2012ms}:
\begin{description}
\item[Null-warped AdS]  spacetimes which arise in proposed holographic duals of nonrelativistic CFTs describing cold atoms \cite{Son:2008ye,Balasubramanian:2008dm}.
\item[Schr\"{o}dinger]  spacetimes, which generalize null warped AdS by introducing an arbitrary scaling exponent \cite{Adams:2008wt}.
\item[Lifshitz] spacetimes, which arise in gravity duals of Lifshitz-like fixed points \cite{Kachru:2008yh} and also have a scaling exponent parametrizing spacetime an\-iso\-tropy.
\end{description}



A variational principle for 3-dimensional higher spin gravity that accommodates spacetimes like asymptotically $\mathrm{AdS}_{2}\times \mathbb{R}$, $\mathbb{H}_{2}\times \mathbb{R}$, Schr\"{o}dinger, Lifshitz or warped AdS spacetimes was proposed and the connections that generate this backgrounds presented \cite{Gary:2012ms}. 
For the case of  $\mathbb{H}_{2}\times \mathbb{R}$ realized in $\mathfrak{sl}(3,\mathbb{R})$ HS gravity in the non-principal embedding the asymptotic symmetry algebra turned out to be the direct sum
of the $\mathcal{W}^{(2)}_{3} \oplus \hat{\mathfrak{u}}(1)$ \cite{Afshar:2012nk}.
We now want to investigate following~\cite{Gary:2014mca} the case of
Lifshitz higher spin theories.
Since the situation for null-warped AdS~\cite{Breunhoelder:2015waa} follows similar
considerations we will only provide a short overview.


\section{Lifshitz Higher Spin}
\label{sec:lifshitz-higher-spin}



A variety of condensed matter systems exhibits anisotropic scaling near a renormalization group fixed point.  Classical Lifshitz fixed points, in which the system scales anisotropically in different spatial directions, are extensively explored. Quantum Lifshitz fixed points, in which time and space scale anisotropically,  with relative scaling ratio $z$, are particularly common in strongly correlated systems     \cite{Grinstein:1981rbe,Hornreich:1975zz,Rokhsar:1988zz,Henley:1997,Sachdev:1999,Ardonne:2004,Ghaemi:2005,Varma:1997,Si:2003,Sachdev:1996,Yang:2004}.
Many-body field theories describing such anisotropic fixed points were proposed to be holographically dual to gravity in the background of Lifshitz geometries, where time and space scale asymptotically with the same ratio $z$ \cite{Kachru:2008yh}.

\subsection{Lifshitz Spacetime in Three Dimensions}\label{Lifshitz}

The $(2+1)$-dimensional Lifshitz spacetime \cite{Kachru:2008yh} is described by the line element
\begin{equation}\label{lineElr}
\extd s_{\mathrm{Lif}_{z}}^2 = \ell^2 \big(-r^{2z}\,\extd t^2 + \frac{\extd r^2}{r^2} + r^2\,\extd x^2 \big)\, .
 \end{equation}
The Lifshitz spacetime (\ref{lineElr}) is invariant under the anisotropic scaling ($z\in\mathbb{R}$):
\begin{equation}
  \label{anisotrop}
  t \rightarrow \lambda^z t \qquad x \rightarrow \lambda x \qquad r\rightarrow \lambda^{-1} r \, .
\end{equation}
For $z=1$, the scaling is isotropic and the spacetime (\ref{lineElr}) reduces to Poincar\'e patch AdS$_3$.

It is often useful to consider a change of coordinates to the radial variable $\rho = \ln  r$.
The spacetime (\ref{lineElr}) now becomes
\begin{equation}\label{Lifshitz3}
\extd s_{\mathrm{Lif}_{z}}^2 = \ell^{2} \left(- e^{2z\rho}\extd t^2 + \extd\rho^2 + e^{2\rho}\extd x^2\right)\, .
\end{equation}
The asymptotic region is approached for $\rho \rightarrow \infty$.

The Lifshitz spacetime (\ref{Lifshitz3}) possesses spacetime isometries. These Lifshitz isometries are generated by the Killing vector fields
\begin{equation}
 \xi_{\mathbb{H}} = \partial_t \qquad
 \xi_{\mathbb{P}} = \partial_x \qquad
 \xi_{\mathbb{D}} = -zt\,\partial_t+\partial_\rho -x\,\partial_x
\label{eq:iso}
\end{equation}
whose isometry algebra is the Lifshitz algebra $\mathfrak{lif}(z,\mathbb{R})$
\begin{equation}
 [\xi_{\mathbb{H}},\,\xi_{\mathbb{P}}] = 0 \qquad [\xi_{\mathbb{D}},\,\xi_{\mathbb{H}}] = z\,\xi_{\mathbb{H}}\qquad [\xi_{\mathbb{D}},\,\xi_{\mathbb{P}}] = \xi_{\mathbb{P}}
\label{eq:Lalgebra}
\end{equation}
The Killing vector $\xi_{\mathbb{H}}$ ($\xi_{\mathbb{P}}$) [$\xi_{\mathbb{D}}$] generates time translations (spatial translations) [anisotropic dilatations]. The Lifshitz spacetime with $z=1$ corresponds to the Poincar\'e patch of the isotropic AdS$_3$ spacetime. With enhanced $(1+1)$-dimensional Lorentz (boost) invariance, the isometry algebra gets enlarged to $\mathfrak{sl}(2, \mathbb{R}) \oplus \mathfrak{sl}(2, \mathbb{R})$ associated with two copies of chiral and anti-chiral excitations. 
Conversely, the Lifshitz algebra $\mathfrak{lif}(1, \mathbb{R})$ is a subalgebra of the $\mathfrak{sl}(2, \mathbb{R}) \oplus \mathfrak{sl}(2, \mathbb{R})$ isometry algebra of the AdS$_3$ spacetime.

Since the Lifshitz spacetime does not fulfill the vacuum Einstein equations, matter contributions are necessary. Known realizations so far involve, e.g., $p$-form gauge fields \cite{Kachru:2008yh}. For example, AdS Einstein gravity coupled to two 1-form abelian gauge fields $F_2 = \extd A_1, \ G_2 = \extd C_1$,
\begin{equation}\label{quadratic}
I = \frac{1}{16\pi G_3} \int \extd^3 x \sqrt{-g} \left[ R(g) + \frac{2}{\ell^{2}} + \frac{1}{4} ||F_2||^2 + \frac{1}{4\alpha} ||G_{2}||^2 + \frac{1}{2} \ast( A_{1} \wedge G_{2}) \right]\,,
\end{equation}
admits the Lifshitz spacetime as a classical solution, where the scaling ratio $z$ is determined by
\begin{equation}
z = \alpha \pm \sqrt{\alpha^2-1} \qquad (\alpha \ge 1)\,.
\end{equation}
Some other constructions require either a massive gauge field \cite{Taylor:2008tg}, a massive graviton \cite{AyonBeato:2009nh,Lu:2012xu} or Ho\v{r}ava--Lifshitz gravity \cite{Griffin:2012qx}.

Here, we take a different route and realize the Lifshitz spacetime by coupling AdS$_3$ Einstein gravity to a spin-3 field with full higher-spin gauge symmetry. In the next section, we construct an explicit example of  $(2+1)$-dimensional $z=2$ Lifshitz spacetime \eqref{Lifshitz3} with non-trivial spin-3 background field. We shall then carefully examine boundary conditions for the gravitational and spin-3 excitations over this Lifshitz spacetime.

\subsection{Lifshitz Boundary Conditions}
\label{strict}

In order to find the Lifshitz spacetime, we decompose again as
in \eqref{Amn} but with the group element $b_{\pm}=e^{\pm\rho \Lt_{0}}$.

To fix a variational principle, we take $\delta A^{+}_t=0=\delta\Ab_{t}$ at asymptotic infinity $\rho \rightarrow \infty$, where this time we denote our boundary
coordinates  by $t$ and $x$. With the boundary term
\begin{equation}
\frac{k}{4\pi}\int_{\R \times \partial\Sigma}\!\!\tr\left(  A_{t} A_{x} - \Ab_{t} \Ab_{x} \right) \extd t  \extd x
\label{eq:bt}
\end{equation}
added to the bulk action \eqref{CS-Action}, such a variational principle is well-posed \cite{Gary:2012ms}.
We take as a background that leads to the Lifshitz spacetime the connections
\begin{subequations}
\label{background}
  \begin{align}
    \hat{a}^{+} &=\tfrac{4}{9}\Wt_{+2} \dd t +  \Lt_{+1}\dd x \\
    \hat{a}^{-} &= \Wt_{-2}\dd t + \Lt_{-1}\dd x \, .
  \end{align}
  \end{subequations}
The specific numerical coefficients are chosen to cancel factors arising from traces.

Using the standard definition of the metric in terms of the zuvielbein
\eqref{MetricAndSpin-S-Fields}
leads to the geometry
\begin{equation}\label{backgroundLineEl}
\extd s_{\mathrm{Lif_{2}}}^2 =\ell^2 \left(- e^{4\rho}\extd t^2+\extd\rho^2  + e^{2\rho}\extd x^2 \right)\, .
\end{equation}
We thus obtain as a classical configuration the $(2+1)$-dimensional Lifshitz spacetime \eqref{Lifshitz3} with $z=2$. The classical solution also involves the totally symmetric spin-3 gauge field.
For our configuration, we find that the Lifshitz spacetime is supported by a nontrivial  spin-3 background gauge field
\begin{equation}\label{backgroundSpin3}
\phi_{\mu\nu\lambda}\,\extd x^\mu\extd x^\nu\extd x^\lambda = -\frac{5 \ell^{3}}{4}\, e^{4\rho}\,\extd t\,(\extd x)^2 \, .
\end{equation}
From now on we set $\ell=1$ to reduce clutter. The spin-3 gauge field is invariant under the transformations generated by the Killing vector fields (\ref{eq:iso}). We conclude that the classical configuration (\ref{backgroundLineEl}),  (\ref{backgroundSpin3}) respects the Lifshitz algebra $\mathfrak{lif}(2, \mathbb{R})$.
The above construction of the Lifshitz spacetime is quite elementary and simple.

Let us next examine the algebra of the symmetry currents for the Lifshitz system we have constructed. To this end, we first need to impose boundary conditions consistent with the background Lifshitz spacetime geometry. Note that we take the ansatz used in \cite{Afshar:2012nk,Prohazka:2013}, which differs from the asymptotic behavior $A-\hat{A}=\O{1}$ used in \cite{Campoleoni:2010zq,Gutperle:2013oxa}, where $\hat{A}$ was a fixed background connection. The fluctuations, which are already on-shell, turn out to take the following form
  \begin{align}
    a^{+} &= \big(\tfrac{8\pi}{9k}t\W(x)\Lt_0 - \tfrac{\pi}{2k}\L(x)\Lt_{-1}\big)\,\extd x\nonumber\\
    &\quad + \big(-\tfrac{32\pi}{81k}t^2\W(x)\Wt_{+2} + \tfrac{8\pi}{9k}t\L(x)\Wt_{+1} + \tfrac{2\pi}{9k}\W(x)\Wt_{-2}\big)\,\extd x\label{strictBC1}\\
    a^{-} &= \big(- \tfrac{2\pi}{k}t\Wba(x)\Lt_0-\tfrac{\pi}{2k}\Lba(x)\Lt_{+1}\big)\,\extd x\nonumber\\
    &\quad + \big(- \tfrac{2\pi}{k}t^2\Wba(x)\Wt_{-2} - \tfrac{2\pi}{k}t\Lba(x)\Wt_{-1} +\tfrac{2\pi}{9k}\Wba(x)\Wt_{+2}\big)\,\extd x\label{strictBC3}
  \end{align}
The set of all boundary functions $\L$, $\Lba$, $\W$ and $\Wba$ specify the set of all admissible fluctuations about the Lifshitz background.
 
A interesting and possibly disturbing feature of these boundary conditions is the polynomial time dependence. In general, time-dependent boundary conditions lead to non-conservation of canonical charges.
However, due to the specific form of the boundary conditions,
and as can be seen explicitly in \cite{Gary:2014mca}
all $t$-dependence is canceled in the boundary charge density and hence the canonical charges are conserved.

Below, we address some immediate consequences of the above boundary conditions, which all point to the fact that consistency of the boundary conditions is a highly non-trivial result.

Using \eqref{MetricAndSpin-S-Fields}, we also extract fluctuations of spin-2 and spin-3 fields. Up to the sub-leading terms, fluctuations of the spin-2 field take the form (for notational simplification, we suppress the $x$-dependence of all component functions hereafter)
\begin{subequations}
    \label{metricfluc}
  \begin{align}
    g_{t t}&= \boldsymbol{-e^{4 \rho }} \\
    g_{t \rho}&=0 \\
    g_{t x}&=t^2 e^{4 \rho } \big(\pi \Wba+\tfrac{4\pi}{9} \W\big)+\tfrac{\pi}{4}  \Wba+\tfrac{\pi}{9} \W \\
    g_{\rho \rho}&=\boldsymbol{1} \\
    g_{\rho x}&=t\,\big(\tfrac{\pi}{2} \Wba+\tfrac{2\pi}{9} \W\big) \\
    g_{x x}&= \boldsymbol{e^{2 \rho }}- t^4e^{4 \rho }\tfrac{16\pi^2}{81} \Wba \W - t^2e^{2 \rho }\tfrac{\pi^2}{9} \Lba \L \nonumber\\
    &\quad+\tfrac{\pi}{6} \Lba+\tfrac{\pi}{6} \L+t^2\tfrac{8\pi^2}{81}
     \Wba \W+\tfrac{\pi^2}{36} e^{-2 \rho } \Lba \L-\tfrac{\pi^2}{81}
     e^{-4 \rho } \Wba \W \ ,
  \end{align}
\end{subequations}
while fluctuations of the spin-3 field take the form
\begin{subequations}
    \label{spin3fluc}
  \begin{align}
    \phi_{txx}&=\boldsymbol{-\tfrac{5}{12}e^{4 \rho }}+ t^2e^{4 \rho }\, \big(\tfrac{\pi ^2}{3 k^2}\L^2-\tfrac{3 \pi ^2}{4 k^2}\Lba^2\big)
    \nonumber\\
    &\quad+e^{2 \rho}\, \big(\tfrac{\pi}{3 k}\L-\tfrac{3 \pi}{4 k}\Lba\big)+\tfrac{\pi ^2}{12 k^2}\L^2-\tfrac{3 \pi ^2}{16 k^2}\Lba^2 \\
    \phi_{\rho xx}&=te^{2 \rho }\, \big(\tfrac{2 \pi}{3 k}\L-\tfrac{3 \pi}{2 k}\Lba\big) + t\,\big(\tfrac{\pi ^2}{3 k^2}\L^2 - \tfrac{3 \pi ^2}{4 k^2}\Lba^2\big) \nonumber    \\
    \phi_{xxx}&=t^4 e^{4 \rho }\, \big(\tfrac{2 \pi ^3}{k^3}\Lba^2 \W-\tfrac{2 \pi ^3}{k^3}\L^2 \Wba\big)+t^2 e^{4 \rho }\, \big(\tfrac{9 \pi}{2 k}\Wba-\tfrac{8 \pi }{9 k}\W\big)\nonumber\\
    &\quad+t^2 e^{2 \rho }\, \big(\tfrac{2 \pi ^2}{k^2} \L \Wba-\tfrac{2 \pi ^2 }{k^2}\Lba \W\big)+t^2\, \big(\tfrac{\pi^3}{k^3}\L^2 \Wba-\tfrac{\pi ^3  }{k^3}\Lba^2 \W\big)-\tfrac{\pi}{2 k}\Wba+\tfrac{\pi}{2 k} \W \nonumber\\
    &\quad+e^{-2 \rho }\, \big(\tfrac{\pi^2}{2 k^2}\Lba \W-\tfrac{\pi^2}{2 k^2}\L \Wba\big)+e^{-4 \rho }\, \big(\tfrac{\pi^3}{8 k^3}\Lba^2\W-\tfrac{\pi^3}{8 k^3}\L^2 \Wba\big) \\
    \phi_{\mu\nu\lambda}&= 0 \quad \textrm{otherwise}\, .
  \end{align}
\end{subequations}
The boldfaced terms denote background geometry, while the remaining terms correspond to state-dependent contributions to the spin-2 and spin-3 fields.

It is also interesting to observe that, although the background geometry is Lifshitz, the boundary conditions also admit spin-2 field configurations that have asymptotically stronger divergent contributions in $\rho$ than the background geometry. For example, it is possible to have configurations whose $g_{tt}$ and $g_{xx}$ have the same asymptotic growth, $\sim e^{4\rho}$. Nevertheless, as we are going to show below, all the configurations allowed by our boundary conditions correspond to finite energy excitations, in the sense that all the canonical charges associated with these configurations are finite (as well as integrable and conserved).  It should be stressed that this feature crucially relies on higher-spin gauge symmetry that acts nontrivially on the spin-2 metric field: the would-be infinite energy density in Einstein-gravity for configurations of $\sim e^{4 \rho}$ asymptotic growth is canceled off by the spin-3 gauge transformations in higher-spin gravity.

The computation of the asymptotic symmetries and the canonical
charges is quite lengthy and cumbersome.
Therefore we refer to \cite{Gary:2014mca} for the details
and jump directly to the answers.
The canonical charges are conserved and well defined and
the asymptotic symmetry algebra is given by two commuting
$\mathcal{W}_{3}$ algebras (see Appendix \ref{app:Walg}),
i.e., $\mathcal{W}_{3} \dis \mathcal{W}_{3}$
with the same central charge as Brown--Henneaux~\cite{Brown:1986nw}.
This are not the symmetries one might expect of a
nonrelativistic Lifshitz system
and one might therefore ask what to the
aforementioned Lifshitz symmetries happened.
But, as remarked in~\cite{Gary:2014mca}
and
by virtue of the relation between gauge symmetries and diffeomorphisms, $\epsilon = \xi^\mu A_\mu$, $\eb=\xi^\mu \Ab_\mu$ \cite{Witten:1988hc}
(see Section~\ref{sec:diff-as-gauge})
one can see
that the Lifshitz symmetries~\eqref{eq:Lalgebra} get enhanced.
With the identification $\W_{-2}\leftrightarrow \mathbb{H}$, $\L_{-1}\leftrightarrow \mathbb{P}$, $\L_0\leftrightarrow \mathbb{D}$ and the use of \eqref{eq:W3}, it becomes obvious that we have the isometry subalgebra $\mathfrak{lif}(2,\R)$
as a subalgebra of $\mathcal{W}_{3}$.

Another work focusing on aspects of Lifshitz black holes~\cite{Gutperle:2013oxa}
also found boundary conditions that lead  to a $\W_3$ algebra, as pointed out in \cite{Perez:2014pya}. 
In fact, their field configurations turn out to be a special case of a general class of solutions of spin-3 gravity in the presence of chemical potentials \cite{Compere:2013nba,Henneaux:2013dra}.

Built upon their work and ours, we
put forth the conjecture that for generic higher-spin Lifshitz holography the asymptotic symmetry algebra gets ubiquitously enhanced to a class of $\W$-algebras.

It was pointed out in \cite{Lei:2015ika}, that
when considering gravitational theories in the first order formalism it can sometimes happen that the spin connection is not uniquely determined by the zuvielbein. In such cases the second order formulation is difficult to interpret as a gravitational theory in the traditional sense. While this is not an obstruction to studying such theories, it can make the interpretation more difficult
and our Lifshitz theory is plagued by this issues.
Further remarks concerning the degeneracy
of the nonrelativistic solutions can be found in \cite{Lei:2015ika}.

\section{Null-warped Higher Spin}
\label{sec:null-warped-higher}

In \cite{Breunhoelder:2015waa}
three-dimensional spin-3 gravity
was equipped with a set of boundary conditions called
``asymptotically null warped AdS''.
Null warped AdS is a special case of a large class of geometries studied by a number of researchers mainly in the context of topologically massive gravity \cite{Deser:1982vy,Deser:1981wh,Deser:1982a}, see e.g.~\cite{Clement:1994sb,Deser:2004wd,Detournay:2005fz,Anninos:2008fx,Gibbons:2008vi,Ertl:2010dh,Anninos:2010pm}.
The asymptotic symmetry algebra for the higher spin
generalization found in \cite{Breunhoelder:2015waa}
was found to be a chiral copy of the  $\mathcal{W}^{(2)}_{3}$ Polyakov--Bershadsky algebra
reminiscent of the situation in topologically massive gravity with strict null warped AdS boundary conditions (see \cite{Anninos:2010pm}).
Again, the ``usual'' null warped isometry algebra
get enhanced to a much bigger one.

Furthermore, was it shown that the invertibility issues~\cite{Lei:2015ika}
are not a problem for the null warped AdS case.
Given the asymptotic symmetries it seemed natural to check if
our boundary conditions can be mapped to
asymptotically AdS boundary conditions that also lead to a $\mathcal{W}^{(2)}_{3}$ algebra\cite{Bunster:2014mua}
which was indeed the case.
We refer to \cite{Breunhoelder:2015waa} for further details
concerning the introduction 
of chemical potentials,
the derivation of  entropy, free energy, and the holographic response functions.

\subsection{Summary}
\label{sec:summary-2}

As seen in this chapter,
it is nontrivial
to get boundary conditions where the
asymptotic symmetry algebra does
not get enhanced to one that
could be considered as relativistic.
It can be observed that
in both cases the resulting asymptotic
symmetry algebra is related to the gauge algebra.
The $\mathcal{W}_{3}$ algebras as well
as the $\mathcal{W}^{(2)}_{3}$ algebra
arise naturally in connection to the
two inequivalent embedding of $\mathfrak{sl}(2,\R)$
into $\mathfrak{sl}(3,\R)$
and their highest weight boundary conditions
(for details concerning this differentiation see, e.g., \cite{Bunster:2014mua}).

This considerations already
hint towards a way
to make other symmetries than (A)dS
manifest.
A change of gauge algebra
seems like a reasonable starting point
to get asymptotic symmetry algebras
of different kinematics
and will be discussed in the next chapters.

\chapter{Kinematical Spin-2 Theories}
\label{cha:kinem-chern-simons}

Due to the principle of relativity,
the notion of kinematical or spacetime symmetry algebras, which contain all symmetries that relate different inertial frames, is a crucial ingredient in the construction of physical theories. Bacry and L\'evy-Leblond have classified all possibilities for kinematical algebras \cite{Bacry:1968zf}, consisting of spacetime translations, spatial rotations and boosts, under some reasonable assumptions. Apart from the relativistic Poincar\'e and (A)dS algebras, this classification also contains the Galilei and Carroll algebras (and generalizations thereof that include a cosmological constant), that appear as kinematical algebras in the nonrelativistic ($c \rightarrow \infty$) and ultra-relativistic ($c\rightarrow 0$) limit. Even though fundamental theories are relativistic, the Galilei and Carroll algebras continue to play an important role in current explorations of string theory, holography and also phenomenology.

For instance, nonrelativistic symmetries  underlie Newton--Cartan geometry, a differential geometric framework for nonrelativistic spacetimes that has found recent applications in holography \cite{Son:2008ye, Kachru:2008yh, Balasubramanian:2008dm, Christensen:2013lma, Christensen:2013rfa, Hartong:2014oma, Hartong:2014pma, Bergshoeff:2014uea, Hartong:2015wxa}, Ho\v{r}\-ava--Lifshitz gravity \cite{Horava:2009uw, Hartong:2015zia, Hartong:2016yrf} and in the construction of effective field theories for strongly interacting condensed matter systems \cite{Son:2005rv, Hoyos:2011ez, Son:2013rqa, Abanov:2014ula, Gromov:2014vla, Gromov:2015fda, Geracie:2014nka, Festuccia:2016awg}.

On the other hand, ultra-relativistic Carroll symmetries have recently been studied in relation to their connection \cite{Duval:2014uva} with the Bondi--Metzner--Sachs (BMS) algebra of asymptotic symmetries of flat spacetime \cite{Bondi:1962px, Sachs:1962zza}. As such, Carroll symmetries play a role in attempts to construct holographic dualities in asymptotically flat spacetimes \cite{Barnich:2010eb, Barnich:2012aw, Bagchi:2009my, Bagchi:2010eg, Bagchi:2014iea, Hartong:2015xda, Bagchi:2015wna, Hartong:2015usd, Bagchi:2016bcd}, as symmetries of the $S$-matrix in gravitational scattering \cite{Strominger:2013jfa} and in the recent notion of soft hair on black hole horizons \cite{Hawking:2016msc, Hawking:2016sgy}.

The kinematical algebras that have been classified by Bacry and L\'evy-Leblond pertain to theories that contain bosonic fields with spins up to 2. One can also consider theories in which massless higher spin fields are coupled to gravity \cite{Vasiliev:1990en}. These so-called higher spin gauge theories have been formulated in (A)dS spacetimes (see \cite{Bekaert:2005vh, Vasiliev:2012vf, Didenko:2014dwa} for reviews) and have featured prominently in the AdS/CFT literature, as a class of theories for which holographic dualities can be constructed rigorously \cite{Giombi:2009wh, Giombi:2010vg, Giombi:2012ms, Gaberdiel:2010pz, Gaberdiel:2011zw, Candu:2012jq, Gaberdiel:2012uj, Candu:2012ne, Beccaria:2013wqa}, essentially because they are a weak-weak type of duality, i.e., CFTs with unbroken higher spin currents are free \cite{Maldacena:2011jn}. They typically contain an infinite number of higher spin fields. As a consequence, their spacetime symmetries are extended to infinite-dimensional algebras that include higher spin generalizations of spacetime translations, spatial rotations and boosts. Higher spin gauge theories have thus far mostly been considered in relativistic (A)dS spacetimes, with relativistic CFT duals\footnote{See however \cite{Gary:2012ms, Afshar:2012nk, Gutperle:2013oxa, Gary:2014mca, Breunhoelder:2015waa, Lei:2015ika,Lei:2015gza} for attempts to consider higher spin theories in non-AdS backgrounds with nonrelativistic CFT duals.}.

Since both higher spin gauge theories as well as non- and ultra-rela\-tivi\-stic spacetime symmetries have played an important role in recent developments in holography, it is natural to ask whether one can combine the two. In order to answer this question, one needs to know which non- and ultra-relativistic kinematical algebras can appear as symmetries of higher spin theories.
This will first be discussed without the additional higher spin symmetries
and in Chapter \ref{cha:kinem-high-spin} including them.

Chapter \ref{cha:kinem-chern-simons}
and Chapter \ref{cha:kinem-high-spin}
are based on \cite{Bergshoeff:2016soe}.
There is a slight change of terminology,
which hopefully does not lead to confusion.
In order to be consistent with the
introductory material presented in the beginning
the term ``contraction procedure'' is substituted
by just ``contraction'' or special cases thereof.

\section{Kinematical Algebras}
\label{sec:kinematical-algebras}

Before discussing spin-3, it is convenient to start with giving  a short review of the spin-2 case~\cite{Bacry:1968zf}.
Since both the spin-2 and spin-3 cases make use of the sIW-contractions
thoroughly reviewed in Section \ref{sec:simple-iw-contr}
we just fix the notation that will be used throughout the next sections.

Starting from a Lie algebra $\mathfrak{g}$, one can choose a subalgebra $\mathfrak{h}$ and consider the decomposition
$\mathfrak{g} = \mathfrak{h} \vdis \mathfrak{i}$.
As already discussed $\mathfrak{h}$ will be the subalgebra
with respect to which we will sIW-contract the original Lie algebra leading to
\begin{align} \label{eq:IWstructure}
  [\,\mathfrak{h} \comma \mathfrak{h} \,] & \subset \mathfrak{h}  &
  [\,\mathfrak{h} \comma \mathfrak{i}\,] & \subset  \mathfrak{i}  &
  [\,\mathfrak{i} \comma \mathfrak{i}\,] & = 0 \,.
\end{align}
Remember that a nontrivial sIW-contraction is uniquely specified by a suitable choice of the subalgebra $\mathfrak{h} \subset \mathfrak{g}$.

\begin{figure}[h]
  \centering
\tdplotsetmaincoords{80}{120}
\begin{tikzpicture}[
tdplot_main_coords,
dot/.style={circle,fill},
linf/.style={ultra thick,->,blue},
cinf/.style={ultra thick,->,red},
tinf/.style={ultra thick,->},
stinf/.style={ultra thick,->,gray},
scale=0.7
]

\node (ads) at (0,10,10) [dot, label=above:$\mathfrak{(A)dS}$] {};
\node (p) at (10,10,10) [dot, label=above:$\mathfrak{poi}$] {};
\node (nh) at (0,10,0) [dot, label=below:$\mathfrak{nh}$] {};
\node (pp) at (0,0,10) [dot, label=above:$\mathfrak{ppoi}$] {};
\node (g) at (10,10,0) [dot, label=below:$\mathfrak{gal}$] {};
\node (pg) at (0,0,0) [dot, label=below:$\mathfrak{pgal}$] {};
\node (car) at (10,0,10) [dot, label=above:$\mathfrak{car}$] {};
\node (st) at (10,0,0) [dot, label=below:$\mathfrak{st}$] {};

\draw[linf] (ads) -- node [sloped,above,xshift=-4mm] {Space-time
}  (p);

\draw[linf] (nh) -- (g);
\draw[linf] (pp) -- (car);
\draw[linf,dashed] (pg) -- (st);

\draw[cinf,double] (ads) -- node [sloped,above] {Speed-space 
} (nh);
\draw[cinf,double,dashed] (pp) -- (pg);
\draw[cinf] (p) -- (g);
\draw[cinf] (car) -- (st);

\draw[tinf,double] (ads) -- node [sloped,above] {Speed-time
} (pp);
\draw[tinf] (p) -- (car);
\draw[tinf] (g) -- (st);
\draw[tinf,double,dashed] (nh) -- (pg);

\draw[stinf,dashed] (ads) -- node [sloped,above] {General}(st);
\end{tikzpicture}
  \caption{This cube summarizes the contractions starting from $\mathfrak{(A)dS}$. The lines represent contractions and the dots represent the resulting contracted Lie algebra. We consider contractions starting from AdS and dS simultaneously. Each dot can therefore represent one Lie algebra, if the contractions from AdS and dS lead to the same algebra, or two Lie algebras, if the contractions from AdS and dS lead to two different results. We have indicated this in the cube by using single lines, for contraction that lead to the same contraction, and double lines otherwise. Dashed lines have no specific meaning except that they should convey the feeling of a three-dimensional cube.}
  \label{fig:cube}
\end{figure}
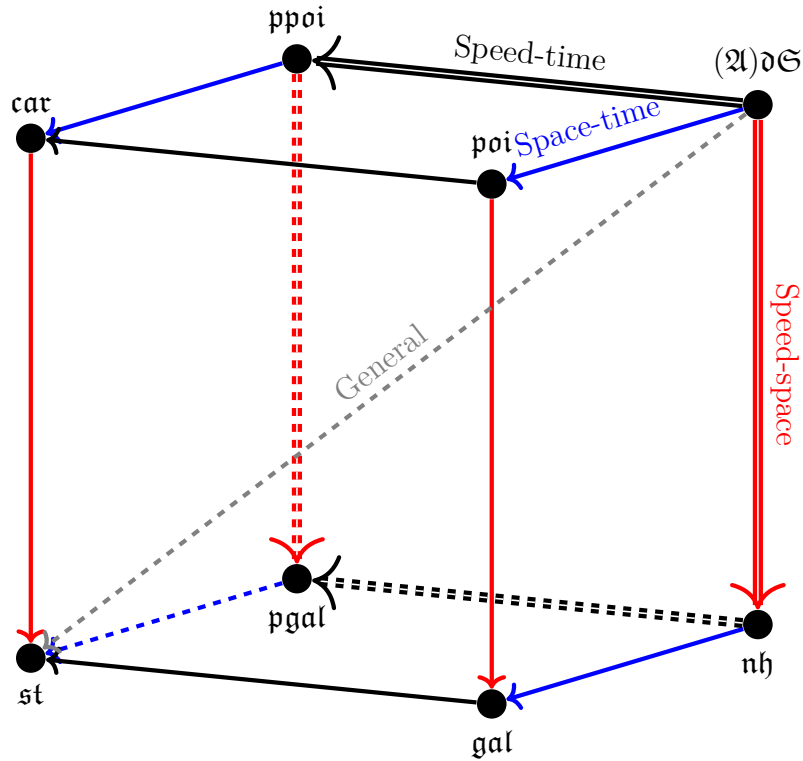

Not all possible subalgebras, however, lead to interesting contractions
that can, e.g., be interpreted as kinematical algebras.
For spin-2, the question which contractions of the isometry algebras of
AdS or dS lead to kinematical algebras,
has been addressed by Bacry and L\'evy--Leblond~\cite{Bacry:1968zf}.
In particular, they have shown that there are only four different
sIW-contractions of the AdS or dS algebras that lead to kinematical algebras.
These have been called ``space-time'', ``speed-space'', ``speed-time''
and ``general'' in \cite{Bacry:1968zf}.
Effectively, the first three of these contractions can be described by either taking a limit of the (A)dS radius $\ell$ or the speed of light $c$. Specifically, the space-time contraction corresponds to $\ell \to \infty$, the speed-time contraction corresponds to $c \rightarrow 0$ and the speed-space contraction corresponds to $c\rightarrow \infty$. However, in this work we suppress factors of $\ell$ and $c$.
The general contraction can also be obtained as consecutive sIW-contractions of the other three and therefore does not provide us with a new algebra.
Moreover, it has been shown that there are in total 8 possible kinematical algebras\footnote{The possible kinematical algebras considered in \cite{Bacry:1968zf} are all possible spacetime symmetry algebras that obey the assumptions that space is isotropic and therefore their generators have the correct ($\Ht$ is a scalar, $\Pt,\Jt,\Gt$ are vectors) transformation behavior under rotations. Furthermore, parity and time-reversal are automorphisms and  boosts are non-compact.} that can be obtained by combining different sIW-contractions of the AdS or dS isometry algebras. We have summarized the four sIW-contractions in the following Table \ref{tab:spin2contr}, by indicating the subalgebra $\mathfrak{h}$ with respect to which the contraction is taken, as well as the generators that form the abelian ideal $\mathfrak{i}$.
\begin{table}[H]
  \centering
$
  \begin{array}{l l l l l }
\toprule
    \text{Contraction } & \phantom{aa} & \multicolumn{1}{c}{\mathfrak{h}} & \phantom{aa} & \multicolumn{1}{c}{\mathfrak{i}} \\ \midrule
    \text{Space-time}   &              & \{\Jt,\Gt_{a} \}                 &              & \{\Ht,\Pt_{a} \}                 \\
    \text{Speed-space}  &              & \{\Jt,\Ht \}                     &              & \{\Gt_{a},\Pt_{a} \}             \\
    \text{Speed-time}   &              & \{\Jt, \Pt_{a} \}                &              & \{\Gt_{a}, \Ht \}                \\
     \text{General}     &              & \{\Jt \}                         &              & \{\Ht, \Pt_{a}, \Gt_{a} \}       \\ \bottomrule
  \end{array}
$
\caption{The four different IW contractions classified in~\cite{Bacry:1968zf}.}
  \label{tab:spin2contr}
\end{table}
\noindent The names of the eight kinematical algebras of \cite{Bacry:1968zf}, along with the symbols we will use to denote them, are given in Table \ref{tab:kinsp2}.
\begin{table}[h]
  \centering
  \begin{tabular}{l  l} \toprule
    Name & Symbol\\ \midrule
    (Anti)-de Sitter & $\mathfrak{(A)dS}$\\
    Poincar\'e & $\mathfrak{poi}$\\
    Para-Poincar\'e & $\mathfrak{ppoi}$\\
    Newton--Hooke & $\mathfrak{nh}$\\
    Galilei & $\mathfrak{gal}$\\
    Para-Galilei & $\mathfrak{pgal}$\\
    Carroll & $\mathfrak{car}$\\
    Static & $\mathfrak{st}$\\ \bottomrule
  \end{tabular}
  \caption{Names of the kinematical algebras and the symbols that denote them.}
  \label{tab:kinsp2}
\end{table}
The sIW-contractions and the resulting Lie algebras
that we have discussed so far can be conveniently summarized as a cube,
see Figure \ref{fig:cube}
and all the commutation relations of the resulting Lie algebras,
together with the most general invariant metric of (A)dS,
are collected in Appendix \ref{cha:kinematical-spin-2}.

For discussions concerning the invariant metric we have copied
some of them in Table \ref{tab:Spincopy}.
The $\mathfrak{(A)dS}_{\parmp}$ Lie algebras
are real (semi)simple Lie algebras
are have therefore an invariant metric proportional to
the Killing form.
The contraction to the $\mathfrak{poi}$ algebra
is of the form discussed in Section \ref{sec:gener-iw-contr}
and leads therefore also to a Lie algebra with invariant metric.
Another kinematical algebra
that is automatically equipped with an
invariant metric is given by the $\mathfrak{car}$
algebra.
This is due to the
invariant metric preserving contraction
of $\mathfrak{poi}$ to $\mathfrak{car}$
(see Section \ref{sec:double-extens-pres})
and was shown explicitly in Example \ref{exp:poitocar}.
So, the (A)dS, Poincar\'e and Carroll algebra
permit an invariant metric, but the Newton--Hooke
and Galilei algebra do not.

\begin{table}[H]
  \centering
$
\begin{array}{l r r r r r}
\toprule %
                                            & \mathfrak{(A)dS}_{\parmp} & \mathfrak{poi}          & \mathfrak{nh}             & \mathfrak{gal}          & \mathfrak{ebarg}        \\ \midrule
  \left[\,\Jt  \comma \Gt_{a} \,\right]     & \epsilon_{am}  \Gt_{m}    & \epsilon_{am}  \Gt_{m}  & \epsilon_{am}  \Gt_{m}    & \epsilon_{am}  \Gt_{m}  & \epsilon_{am}  \Gt_{m}  \\
  \left[\, \Jt \comma \Pt_{a} \,\right]     & \epsilon_{am}  \Pt_{m}    & \epsilon_{am}  \Pt_{m}  & \epsilon_{am}  \Pt_{m}    & \epsilon_{am}  \Pt_{m}  & \epsilon_{am}  \Pt_{m}  \\ 
  \left[\,\Gt_{a}  \comma \Gt_{b} \,\right] & - \epsilon_{ab}  \Jt      & - \epsilon_{ab}  \Jt    & 0                         & 0                       & \epsilon_{ab} \Ht^{*}   \\
  \left[\, \Gt_{a} \comma \Ht \,\right]     & -\epsilon_{am}  \Pt_{m}   & -\epsilon_{am}  \Pt_{m} & -\epsilon_{am}  \Pt_{m}   & -\epsilon_{am}  \Pt_{m} & -\epsilon_{am}  \Pt_{m} \\
  \left[\, \Gt_{a} \comma \Pt_{b} \,\right] & -\epsilon_{ab} \Ht        & -\epsilon_{ab} \Ht      & 0                         & 0                       & \epsilon_{ab} \Jt^{*}   \\
  \left[\,\Ht  \comma \Pt_{a} \,\right]     & \pm \epsilon_{am} \Gt_{m} & 0                       & \pm \epsilon_{am} \Gt_{m} & 0                       & 0                       \\
  \left[\,\Pt_{a}  \comma \Pt_{b} \,\right] & \mp  \epsilon_{ab}\Jt     & 0                       & 0                         & 0                       & 0                       \\ \bottomrule
\end{array}                                                                                                                                                                
$
\label{tab:Spincopy}
\caption{The commutation relations of the (Anti)-de Sitter, Poincar\'e, Newton--Hooke, Galilei and Extended Bargmann algebras.}
\end{table}

We will now analyze what needs to be done to
get an extension of the Galilei algebra that is symmetric
self-dual (see Table \ref{tab:Spincopy}).
Here knowledge about double extensions is useful.
Restricting to $\Pt_{a}$ and $\Gt_{b}$ and recognizing that
$\langle \Pt_{a} \,, \Gt_{b} \rangle = \delta_{ab} $
is an invariant metric on this restricted Lie algebra
leads to the insight
that we can double extend $\mathfrak{g}=\mathfrak{u}(1)^{4}= \{\Pt_{a},\Gt_{b}\}$
by $\Ht$ and $\Jt$ which leads to two nontrivial central extensions $\Ht^{*}$
and $\Jt^{*}$, respectively.
This algebra will be called Extended Bargmann algebra or $\mathfrak{ebarg}$.
``Bargmann algebra'' because the importance of the
central extension $\Jt^{*}$, which  is possible in 
any spacetime dimension and is interpreted as mass,
has been emphasized by Bargmann~\cite{Bargmann:1954gh}.
``Extended Bargmann algebra'' because of the second central extension,
which is not possible for higher dimensions
(for a discussion and references concerning possible
interpretations see, e.g., the introduction of \cite{Papageorgiou:2009zc}).
In three spacetime dimension
there actually is a third nontrivial central extension possible.
Since it is not necessary to get an invariant metric and
does not correspond to a central extension of the group \cite{levygalgr}
we will ignore it in the following.

The projective unitary irreducible representations
of this extended Ga\-li\-lei group were analyzed in \cite{Grigore:1993fz}.
The invariant metric that the Extended Bargmann algebra possesses
was used in \cite{Papageorgiou:2009zc}
to define ``Galilean quantum gravity'' using a CS formulation.
Furthermore, the coadjoint orbits of the group were discussed.
In \cite{Hartong:2016yrf} is was shown that this theory is
related to projectable
Ho\v{r}ava--Lifshitz gravity with a local $\mathfrak{u}(1)$ gauge symmetry
and without a cosmological constant.
There also exists
an extension to Extended Bargmann supergravity~\cite{Bergshoeff:2016lwr}.
We will now study if we can arrive at the Extended Bargmann algebra
using contractions.

\section{Extended Kinematical Algebras}
\label{sec:extend-kinem}

We have already discussed in Section \ref{sec:contr-centr-extens}
that trivial central extensions
can lead to nontrivial ones upon contraction.
Since we want to start our investigations from
$\mathfrak{(A)dS}$ algebras which are
(semi)simple our only option is to
centrally extend trivially.
With hindsight we shift 
$\Jt \to \Jt- \Ht^{*}$ and  $\Ht \to \Ht- \Jt^{*}$
where the starred generators denote the trivial
central extensions.
The shift applied to the commutation relations
and to the invariant metric, also normalized with hindsight,
can be seen in Table \ref{tab:extendedkin}.

The contraction that leads from (A)dS to the Poincar\'e algebra
is given by a sIW-contraction with
respect to the subalgebra spanned by $\{\Jt,\Gt_{a},\Ht^{*}\}$.
Or with the notation of \eqref{eq:combefore} where we just denote
the subscript of $a$ of each $n_{a}$ (e.g., $\Ht=1$ means that $n_{\Ht}=1$):
$\Jt=\Gt_{a}=\Ht^{*}=0$, $\Ht=\Pt_{a}=\Jt^{*}=1$ and $\mu^{-}=-1$.
The $\mathfrak{poi}$ algebra does still not allow for nontrivial
central extension.

Now the interesting contractions
are the ones from the centrally extended
relativistic algebras (A)dS and Poincar\'e
to the extended nonrelativistic Newton--Hooke and Galilei algebra.
They indeed lead to nontrivial central extended ones
which posses an invariant metric.
The gIW-contraction is in both cases given by:
  $\Jt=\Ht=0$, 
  $\Gt_{a}=\Pt_{a}=1$, 
  $\Jt^{*}=\Ht^{*}=2$ and
  $\mu^{-}=-2$.

  For completeness we also provide the
  gIW-contraction $\mathfrak{nh}\dis_{c} \mathfrak{u}(1)^{2} \to \mathfrak{ebarg}$:
  $\Jt=\Jt^{*}=0$,
  $\Pt_{a}=-\Gt_{a}=1$,
  $\Ht=-\Ht^{*}=2$ and
  $\mu^{-}=0$.



\begin{table}[H]
  \centering
$
\begin{array}{l r r r r r}
\toprule %
                                            & \mathfrak{(A)dS}_{\parmp}         & \mathfrak{poi}\quad\quad                 & \mathfrak{nh} \quad\quad             & \mathfrak{ebarg}        \\ 
                                            & \dis \,\mathfrak{u}(1)^{2}          & \dis \, \mathfrak{u}(1)^{2}       & \dis_{c} \, \mathfrak{u}(1)^{2}                         \\ \midrule
  \left[\,\Jt  \comma \Gt_{a} \,\right]     & \epsilon_{am}  \Gt_{m}            & \epsilon_{am}  \Gt_{m}         & \epsilon_{am}  \Gt_{m}     & \epsilon_{am}  \Gt_{m}  \\
  \left[\, \Jt \comma \Pt_{a} \,\right]     & \epsilon_{am}  \Pt_{m}            & \epsilon_{am}  \Pt_{m}         & \epsilon_{am}  \Pt_{m}     & \epsilon_{am}  \Pt_{m}  \\ 
  \left[\,\Gt_{a}  \comma \Gt_{b} \,\right] & - \epsilon_{ab}  (\Jt -\Ht^{*})   & - \epsilon_{ab} (\Jt -\Ht^{*}) & \epsilon_{ab}  \Ht^{*}     & \epsilon_{ab} \Ht^{*}   \\
  \left[\, \Gt_{a} \comma \Ht \,\right]     & -\epsilon_{am}  \Pt_{m}           & -\epsilon_{am}  \Pt_{m}        & -\epsilon_{am}  \Pt_{m}    & -\epsilon_{am}  \Pt_{m} \\
  \left[\, \Gt_{a} \comma \Pt_{b} \,\right] & -\epsilon_{ab} (\Ht -\Jt^{*})     & -\epsilon_{ab} (\Ht -\Jt^{*})  & \epsilon_{ab}  \Jt^{*}     & \epsilon_{ab} \Jt^{*}   \\
  \left[\,\Ht  \comma \Pt_{a} \,\right]     & \pm \epsilon_{am} \Gt_{m}         & 0                              & \pm \epsilon_{am} \Gt_{m}  & 0                       \\
  \left[\,\Pt_{a}  \comma \Pt_{b} \,\right] & \mp  \epsilon_{ab}(\Jt - \Ht^{*}) & 0                              & \pm  \epsilon_{ab} \Ht^{*} & 0                       \\ \midrule
  \langle \Pt_{a} \,, \Gt_{b} \rangle       & \mu^{-} \delta_{ab}               & \mu^{-} \delta_{ab}            & \mu^{-} \delta_{ab}        & \mu^{-} \delta_{ab}     \\
  \langle \Jt^{*} \,, \Ht^{*} \rangle       & \mu^{-}                           & \mu^{-}                        & 0                          & 0                       \\
  \langle \Jt \,, \Jt^{*} \rangle           & \mu^{-}                           & \mu^{-}                        & \mu^{-}                    & \mu^{-}                 \\
  \langle \Ht \,, \Ht^{*} \rangle           & \mu^{-}                           & \mu^{-}                        & \mu^{-}                    & \mu^{-}                 \\ \bottomrule
\end{array}
$
\caption{The central extended Lie algebras of (A)dS, Poincar\'e, Newton--Hooke and Galilei
  and their invariant metrics.
  The central extension of $\mathfrak{(A)dS}$ and $\mathfrak{poi}$ are trivial.
  For $\mathfrak{nh}$ and $\mathfrak{ebarg}$ they are nontrivial
  and necessary to permit an invariant metric.
  Nondegeneracy of the invariant metric demands that $\mu^{-}\neq 0$.}
\label{tab:extendedkin}
\end{table}

\section{Carroll Gravity}\label{sec:4}

\newcommand{\boost}{B}

In this section we address whether there are interesting infinite extensions of the algebras discussed above, in the same way that the global conformal algebra in two dimensions gets extended to the Virasoro algebra by imposing Brown--Henneaux boundary conditions \cite{Brown:1986nw}.
We will focus here on a specific simple example. In fact, as a first step we consider spin-2 Carroll gravity, defined by a Chern--Simons gauge theory with the connection
\eq{
A =  \tau\, \Ht + e^a\, \Pt_a  + \omega\, \Jt + \boost^a\, \Gt_a
}{eq:car1}
takes values in the spin-2 Carroll algebra ($a=1,2$), whose non-vanishing commutation relations read
\begin{subequations}
\label{eq:car2}
\begin{align}
    [\Jt,\,\Pt_a] &= \epsilon_{ab}\, \Pt_b\,, \\
    [\Jt,\,\Gt_a] &= \epsilon_{ab}\, \Gt_b\,, \\
    [\Pt_a,\,\Gt_b] &= -\epsilon_{ab}\, \Ht\,,
\end{align}
\end{subequations}
where we use the convention $\epsilon_{12}=+1$ for the antisymmetric $\epsilon$-symbol. The invariant metric has the non-vanishing entries
\eq{
\langle \Ht,\Jt\rangle = -1 \qquad \langle \Pt_a,\Gt_b\rangle = \delta_{ab} \,.
}{eq:car3} 

Our main goal is not just to find some infinite extension of the algebra \eqref{eq:car2} (this always exists at least in the form of the loop algebra of the underlying gauge algebra, see e.g.~\cite{Elitzur:1989nr}; for AdS$_3$ gravity such boundary conditions were investigated recently in \cite{Grumiller:2016pqb}), but rather to find an extension that has a ``nice'' geometric interpretation along the lines of the Brown--Henneaux boundary conditions. This means that we want to achieve a suitable Drinfeld--Sokolov type of reduction where not all algebraic components of the connection are allowed to fluctuate. The words ``nice'' and ``suitable'' here mean that,
in particular,  we want that the appropriate Carroll background geometry as part of our spectrum of physical states is allowed by our boundary conditions, and that all additional states are fluctuations around this background. First, we recall  some basic aspects of Carroll geometry.

The Carroll-zweibein for the flat background geometry in some Feffer\-man--Graham like coordinates should take the form
\eq{
e^1_\vp = \rho\qquad e^2_\rho = 1\qquad e^1_\rho=e^2_\vp = 0
}{eq:car4}
so that the corresponding two-dimensional line-element reads
\eq{
\extd s^2_{(2)} = e^a e^b \delta_{ab} = \rho^2\extd\vp^2 + \extd\rho^2\,.
}{eq:car5}
We shall refer to $\rho$ as radial coordinate and to $\vp$ as angular coordinate, assuming $\vp\sim\vp+2\pi$. Moreover, on the background the time-component should be fixed as
\eq{
\tau = \extd t\,.
}{eq:car6}
Below we shall allow subleading (in $\rho$) fluctuations in the two-dimensional line-element \eqref{eq:car5} and leading fluctuations in the time-component \eqref{eq:car6}.

We proceed now by stating the result for the boundary conditions that define our example of Carroll gravity and discuss afterwards the rationale behind our choices as well as the consistency of the boundary conditions by proving the finiteness, integrability, non-triviality and conservation of the canonical boundary charges. We follow the general recipe reviewed e.g.~in \cite{Afshar:2012nk, Riegler:2016hah}. First, we bring the connection \eqref{eq:car1} into a convenient gauge~\cite{Banados:1994tn}
\eq{
A = b^{-1}(\rho)\,\big(\extd+a(t,\,\vp)\big)\,b(\rho)
}{eq:car7}
where the group element
\eq{
b(\rho) = e^{\rho \Pt_2}
}{eq:car8}
is fixed as part of the specification of our boundary conditions, $\delta b=0$. The boundary connection $a$ does not depend on the radial coordinate $\rho$ and is given by
\begin{subequations}
\label{eq:car9}
\begin{align}
a_\vp &= -\Jt + h(t,\,\vp)\, \Ht + p_a(t,\,\vp)\, \Pt_a + g_a(t,\,\vp)\, \Gt_a\,,  \label{eq:car9a} \\
a_t &= \mu(t,\,\vp)\, \Ht\,, \label{eq:car9b}
\end{align}
\end{subequations}
where $\mu$ is arbitrary but fixed, $\delta\mu=0$, while all other functions are arbitrary and can vary. This means that the allowed variations of the boundary connection are given by
\eq{
\delta a = \delta a_\vp\,\extd\vp = \big(\delta h\, \Ht + \delta p_a \,\Pt_a + \delta g_a\, \Gt_a \big)\,\extd\vp\,.
}{eq:car10}
The full connection in terms of the boundary connection is then given by
\eq{
A = a + \Pt_2 \extd\rho + \rho\, [a,\,\Pt_2]
}{eq:car7too}
and acquires its non-trivial radial dependence through the last term,
\begin{equation}
\rho\, [a,\,\Pt_2] = \rho\,(\Pt_1 - g_1(t,\,\vp)\,\Ht)\,\extd\vp.  
\end{equation}
Only the $\vp$-component of the connection is then allowed to vary.
\eq{
\delta A = \delta a + \rho\, [\delta a,\,\Pt_2] = \big(\delta h\, \Ht  + \delta p_a\, \Pt_a + \delta g_a\, \Gt_a - \rho\,\delta g_1\, \Ht \big)\,\extd\vp
}{eq:car10too}

The above boundary conditions lead to Carroll-geometries of the form
\eq{
\extd s^2_{(2)} = \big[\big(\rho + p_1(t,\,\vp)\big)^2 + p_2(t,\,\vp)^2\big]\,\extd\vp^2 + 2p_2(t,\,\vp)\,\extd\vp\extd\rho + \extd\rho^2
}{eq:car11}
and
\eq{
\tau = \mu(t,\,\vp)\,\extd t + \big(h(t,\,\vp) - \rho\, g_1(t,\,\vp)\big)\,\extd\vp\,.
}{eq:car12}
Thus, we see that to leading order in $\rho$ the background line-element \eqref{eq:car5} is recovered from \eqref{eq:car11}, plus subleading (state-dependent) fluctuations captured by the functions $p_a(t,\,\vp)$. As we shall see in the next paragraph the functions $p_a$ and $g_a$ are $t$-independent on-shell. In the metric-formulation our boundary conditions can be phrased as
\eq{
  \extd s^2_{(2)} = \big(\rho^2 + {\mathcal{O}}(\rho)\big)\,\extd\vp^2
  + {\mathcal{O}}(1)\,\extd\rho\extd\vp + \extd\rho^2
}{eq:car27}
and
\eq{
\tau = \mu(t,\,\vp)\,\extd t + {\mathcal{O}}(\rho)\,\extd\vp\,.
}{eq:car29}
Note that while the asymptotic form of the two-dimensional line-element \eqref{eq:car27} may have been guessed easily, the specific form of the time-component \eqref{eq:car29} is much harder to guess, particularly the existence of a shift-com\-po\-nent proportional to $\extd\vp$ that grows linearly in $\rho$. Fortunately, the Chern--Simons formulation together with the gauge choice \eqref{eq:car7} minimizes the amount of guesswork needed to come up with meaningful boundary conditions.

We consider now the impact of the equations of motion on the free functions in the boundary connection \eqref{eq:car9}. Gauge-flatness $F=0$ implies
\eq{
\partial_t a_\vp - \partial_\vp a_t + [a_t,\,a_\vp]=\partial_t a_\vp - \partial_\vp a_t = 0\,.
}{eq:car13}
As a consequence, we get the on-shell conditions (which also could be called ``holographic Ward identities'')
\eq{
\partial_t p_a = \partial_t g_a = 0\qquad \partial_t h = \partial_\vp\mu\,.
}{eq:car14}
Thus, most of the functions in the boundary connection \eqref{eq:car9} are time-independent, with the possible exception of $h$ and $\mu$.

The boundary-condition preserving transformations, $\delta_{\hat\lambda} A = \extd\hat\lambda + [A,\,\hat\lambda] = {\mathcal{O}}(\delta A)$, generated by $\hat\lambda=b^{-1}\lambda b$ have to obey the relations
\begin{subequations}
\label{eq:car15}
\begin{align}
  \delta_\lambda a_t &= \partial_t\lambda + [a_t,\,\lambda] = \partial_t \lambda = 0\,,\ \label{eq:car15a} \\
  \delta_\lambda a_\vp &= \partial_\varphi\lambda + [a_\vp,\,\lambda] =
                         {\mathcal{O}}(\delta a_\vp)\,, \label{eq:car15b}
\end{align}
\end{subequations}
where $ {\mathcal{O}}(\delta a_\vp)$ denotes all the allowed variations displayed in \eqref{eq:car10}. It is useful to decompose $\lambda$ with respect to the algebra \eqref{eq:car1}.
\eq{
\lambda = \lambda^\Ht\, \Ht + \lambda^{\Pt_a}\, \Pt_a + \lambda^\Jt\,\Jt + \lambda^{\Gt_a}\, \Gt_a\,.
}{eq:car18}
The first line in \eqref{eq:car15} establishes the time-independence of $\lambda$, while the second line yields the consistency condition
\eq{
\partial_\vp \lambda^\Jt = 0
}{eq:car16}
as well as the transformations rules
\begin{subequations}
\label{eq:car17}
\begin{align}
 \delta_\lambda h &= \partial_\vp \lambda^\Ht - \big(p_1 \lambda^{\Gt_2} - p_2 \lambda^{\Gt_1} + g_1 \lambda^{\Pt_2} -  g_2 \lambda^{\Pt_1}\big)\,,\\
 \delta_\lambda p_a &= \partial_\vp \lambda^{\Pt_a} - \eps_{ab} \big(\lambda^{\Pt_b} - p_b\lambda^\Jt\big)\,,\\
 \delta_\lambda g_a &= \partial_\vp \lambda^{\Gt_a} - \eps_{ab} \big(\lambda^{\Gt_b} - g_b\lambda^\Jt\big)\,.
\end{align}
\end{subequations}

Applying the Regge--Teitelboim approach \cite{Regge:1974zd} to Chern--Simons theories yields the following background-independent result for the variation of the canonical boundary charges
\eq{
\delta Q[\lambda] = \frac{k}{2\pi}\,\oint\langle\hat\lambda\, \delta A\rangle = \frac{k}{2\pi}\,\oint\langle\lambda\, \delta a_\vp\rangle\,\extd\vp
}{eq:car19}
which in our case expands to
\eq{
\delta Q[\lambda] = \frac{k}{2\pi}\,\oint\big(-\lambda^\Jt \delta h + \lambda^{\Pt_a} \delta g_a + \lambda^{\Gt_a} \delta p_a\big)\,\extd\vp\,.
}{eq:car20}
The canonical boundary charges are manifestly finite since the $\rho$-dependence drops out in \eqref{eq:car19}; they are also integrable in field-space since our $\lambda$ is state-independent.
\eq{
Q[\lambda] = \frac{k}{2\pi}\,\oint\big(-\lambda^\Jt h + \lambda^{\Pt_a} g_a + \lambda^{\Gt_a} p_a\big)\,\extd\vp\,.
}{eq:car21}
The result \eqref{eq:car21} clearly is non-trivial in general. To conclude the proof that we have meaningful boundary conditions we finally check conservation in time, using the on-shell relations \eqref{eq:car14} as well as the time-independence of $\lambda$, see \eqref{eq:car15a}:
\eq{
\partial_t Q[\lambda]\big|_{\textrm{\tiny EOM}} = -\frac{k}{2\pi}\,\oint \lambda^\Jt \partial_t h\,\extd\vp = -\frac{k}{2\pi}\,\oint \lambda^\Jt \partial_\vp \mu\,\extd\vp = \frac{k}{2\pi}\,\oint \mu \partial_\vp \lambda^\Jt \,\extd\vp\,.
}{eq:car22}
By virtue of \eqref{eq:car16} we see that the last integrand vanishes and thus we have established charge conservation on-shell:
\eq{
\partial_t Q[\lambda]\big|_{\textrm{\tiny EOM}}=0\,.
}{eq:car23}
Since our canonical boundary charges \eqref{eq:car21} are finite, integrable in field space, non-trivial and conserved in time the boundary conditions \eqref{eq:car7}-\eqref{eq:car10too} are consistent and lead to a non-trivial theory. For later purposes, it is useful to note that due to the constancy of $\lambda^\Jt$ only the zero mode charge associated with the function $h$ can be non-trivial. This means that we can gauge-fix our connection using proper gauge transformations such that $h=\mathrm{const.}$

 We now introduce Fourier modes in order to be able to present the asymptotic symmetry algebra in a convenient form.\footnote{%
There is no meaning to the index positions in this section. The only reason why we write $\Pt^a_n$ and $\Gt^a_n$ instead of corresponding quantities with lower indices is that our current convention is easier to read.
}
\begin{subequations}
\label{eq:car24}
\begin{align}
  \Pt^a_n &:= \frac{1}{2\pi}\,\oint \extd\vp\, e^{in\vp}g_a(t,\,\vp)\big|_{\textrm{\tiny EOM}}\,, \\
  \Gt^a_n &:= \frac{1}{2\pi}\,\oint \extd\vp\, e^{in\vp}p_a(t,\,\vp)\big|_{\textrm{\tiny EOM}}\,, \\
  \Jt &:= -\frac{1}{2\pi}\,\oint \extd\vp\, h(t,\,\vp)\big|_{\textrm{\tiny EOM}}\,.
\end{align}
\end{subequations}
A few explanations are in order. Due to our off-diagonal bilinear form \eqref{eq:car3} we associate the $n^{\textrm{th}}$ Fourier mode of the functions $g_a$ ($p_a$) with the generator $\Pt^a_n$ ($\Gt^a_n$). For the same reason we associate $\Jt$ with minus the zero mode of $h$. Finally, the subscript EOM means that all integrals are evaluated on-shell, in which case all $t$-dependence drops out (and in the last integral also all $\vp$-dependence).

We make a similar Fourier decomposition of the gauge parameters $\lambda^i$, where $i$ refers to the generators $\Pt_a$, $\Gt_a$ and $\Jt$; the parameter $\lambda^\Ht$ is not needed since it does not appear in the canonical boundary charges \eqref{eq:car21}, so all gauge transformations associated with it are proper ones and can be used to make $h$ constant.
\begin{subequations}
\label{eq:car25}
\begin{align}
  \lambda^{\Pt_a}_n &:= \frac{1}{2\pi}\,\oint \extd\vp\, e^{in\vp}\lambda^{\Pt_a}(\vp)\,, \\
  \lambda^{\Gt_a}_n &:= \frac{1}{2\pi}\,\oint \extd\vp\, e^{in\vp}\lambda^{\Gt_a}(\vp)\,.
\end{align}
\end{subequations}
Note that we have used \eqref{eq:car15} to eliminate all time-dependence and that $\lambda^\Jt$ is a constant according to \eqref{eq:car16} thus requiring no Fourier decomposition.

The variations \eqref{eq:car17} of the state-dependent functions then establish corresponding variations in terms of the Fourier components \eqref{eq:car24}, \eqref{eq:car25}.
\begin{subequations}
\label{eq:car26}
\begin{align}
  \delta \Pt_n^a &= -in \lambda^{\Gt_a}_n - \eps_{ab} \lambda^{\Gt_b}_n + \eps_{ab}\lambda^\Jt \Pt^b_n\,,\\
  \delta \Gt_n^a &= -in \lambda^{\Pt_a}_n - \eps_{ab} \lambda^{\Pt_b}_n + \eps_{ab}\lambda^\Jt \Gt^b_n\,,\\
  \delta \Jt &= \sum_{n\in\mathbb{Z}} \eps_{ab}\Big(\Gt^a_n\lambda^{\Gt_b}_{-n} + \Pt^a_n \lambda_{-n}^{\Pt_b}\Big)\,.
\end{align}
\end{subequations}
From the variations \eqref{eq:car26} we can read off the asymptotic symmetry algebra, using the  fact that the canonical generators generate gauge transformations via the Dirac bracket $\delta_{\lambda_1} Q[\lambda_2]=\{Q[\lambda_1],\,Q[\lambda_2]\}$.

Converting Dirac brackets into commutators then establishes the asymptotic symmetry algebra as the commutator algebra of the infinite set of generators $\Pt^a_n$, $\Gt^a_n$ and $\Jt$. The central element of this algebra will be associated with (minus) $\Ht$, concurrent with the notation of \eqref{eq:car2}. Evaluating \eqref{eq:car26} yields\footnote{%
Note that our definitions of Fourier-components \eqref{eq:car24}, \eqref{eq:car25} require that we associate the negative Fourier components of the $\lambda$ with the positive Fourier components of the generators so that, for instance, $[\Pt^b_n,\,\Jt]=\delta_{\lambda^{\Pt_b}_{-n}} \Jt$.
}
\begin{subequations}
\label{eq:car28}
\begin{align}
  [\Jt,\,\Pt^a_n] &= \eps_{ab}\,\Pt^b_n\,, \\
  [\Jt,\,\Gt^a_n] &= \eps_{ab}\,\Gt^b_n\,, \\
  [\Pt^a_n,\,\Gt^b_m] &= -\big(\eps_{ab} + in \delta_{ab}\big)\,\Ht\,\delta_{n+m,\,0}\,,
\end{align}
\end{subequations}
where all commutators not displayed vanish. We have thus succeeded in providing an infinite lift of the Carroll algebra \eqref{eq:car2}, which is contained as a subalgebra of our asymptotic symmetry algebra \eqref{eq:car28} by restricting to the zero-mode generators $\Pt_a = \Pt^a_0$, $\Gt_a= \Gt^a_0$ in addition to $\Jt$ and $\Ht$. As a simple consistency check one may verify that the Jacobi identities indeed hold. The only non-trivial one to be checked is the identity $[[\Jt,\,\Pt^a_n],\,\Gt_m^b]+\textrm{cycl.} = 0$.

We conclude this section with a couple of remarks. The boundary conditions \eqref{eq:car7}-\eqref{eq:car9} by no means are unique and can be either generalized or specialized to looser or stricter ones, respectively.
Another set of well defined boundary
condition has been proposed in~\cite{Grumiller:2017sjh}.
In particular, we have switched off nearly all chemical potentials in our specification of the time-component of the connection \eqref{eq:car9b}, and it could be of interest to allow arbitrary chemical potentials. Apart from this issue there is only one substantial generalization of our boundary conditions, namely to allow for a state-dependent function in front of the generator $\Jt$ in the angular component of the connection \eqref{eq:car9a}. As mentioned in the opening paragraph of this section, in that case the expected asymptotic symmetry algebra is the loop algebra of the Carroll algebra \eqref{eq:car2}. In principle, it is possible to make our boundary conditions stricter, but that would potentially eliminate interesting physical states like some of the Carroll geometries \eqref{eq:car11}, \eqref{eq:car12}.
Thus, while our choice \eqref{eq:car7}-\eqref{eq:car9} is not unique it
provides an interesting set of boundary conditions for spin-2 Carroll gravity. Using the same techniques it should be straightforward to extend the discussion of this section to higher spin Carroll gravity and related theories discussed in this thesis.

\chapter{Kinematical Spin-3 Theories}
\label{cha:kinem-high-spin}

The reason for restricting ourselves to three spacetime dimensions stems from the fact that, as far as higher spin gauge theory is concerned, this case is a lot simpler than its higher-dimensional counterpart. For instance, in three dimensions it is  possible to consider higher spin gauge theory in flat spacetimes \cite{Afshar:2013vka, Gonzalez:2013oaa, Grumiller:2014lna, Gary:2014ppa, Matulich:2014hea}, unlike the situation in higher dimensions where a non-zero cosmological constant is required\footnote{See however \cite{Sleight:2016dba,Sleight:2016xqq,Ponomarev:2016lrm} for recent progress concerning higher spin theories in four-dimensional flat space.}.
Moreover, and as already discussed, in three dimensions higher spin gauge theories with only a finite number of higher spin fields can be constructed \cite{Aragone:1983sz}. In the relativistic case, such theories assume the form of Chern--Simons theories, for a gauge group that is a suitable finite-dimensional extension of the three-dimensional (A)dS and Poincar\'e algebras. For theories with integer spins ranging from 2 to $N$ in AdS spacetime, the gauge algebra is given by $\mathfrak{sl}(N,\R) \dis \mathfrak{sl}(N,\R)$. Here, we will restrict ourselves for simplicity to ``spin-3 theory'' for which $N=3$, although our analysis can straightforwardly be generalized to arbitrary $N$.

We will thus extend the discussion of kinematical algebras of \cite{Bacry:1968zf} and review in Section~\ref{cha:kinem-chern-simons} to theories in three spacetime dimensions that include a spin-3 field coupled to gravity. In particular, we will start from the observation made in \cite{Bacry:1968zf} that all kinematical algebras can be obtained by taking sequential In\"on\"u-Wigner (IW) contractions of the (A)dS algebras. We will then classify all possible sIW contractions\footnote{
  It should be emphasized that this does not classify the Lie algebras
  that result from the contraction.
} of the kinematical algebra of spin-3 theory in (A)dS$_3$, as well as all possible kinematical algebras that can be obtained by sequential contractions.
Some of the kinematical algebras that are obtained in this way can be interpreted as spin-3 extensions of the Galilei and Carroll algebras. We will show that one can construct Chern--Simons theories for (suitable extensions of) these algebras. These can then be interpreted as non- and ultra-relativistic three-dimensional spin-3 theories. We will in particular argue that these theories can be viewed as higher spin generalizations of Extended Bargmann gravity \cite{Papageorgiou:2009zc,Papageorgiou:2010ud,Bergshoeff:2016lwr,Hartong:2016yrf} and Carroll gravity \cite{Bergshoeff:2017btm}, two examples of non- and ultra-relativistic gravity theories that have been considered in the literature recently.

The kinematical algebras of spin-3 theories that we obtain are finite-dimensional. Relativistic three-dimensional kinematical algebras have infinite-dimensional extensions that are obtained as asymptotic symmetry algebras upon imposing suitable boundary conditions on metric and higher spin fields, such as the Virasoro algebra (for the AdS algebra) \cite{Brown:1986nw}, the BMS algebra (for the Poincar\'e algebra) \cite{Ashtekar:1996cd, Barnich:2006av} or $\mathcal{W}$-algebras (for their higher spin generalizations) \cite{Henneaux:2010xg, Campoleoni:2010zq}.
One such example for the Carroll algebra was discussed in Section \ref{sec:4}.
It is interesting to ask whether the found non- and ultra-relativistic algebras also have infinite-dimensional extensions that correspond to asymptotic symmetry algebras of their corresponding higher spin gravity theories.

This chapter is based on Section 2 and 3 of \cite{Bergshoeff:2016soe}.
We will first, in section \ref{sec:algebras}, classify all sIW contractions of the kinematical algebra of spin-3 theory in (A)dS$_3$.
We then classify all kinematical algebras that can be obtained by combining these various contractions.
In section \ref{sec:spin-3}, we restrict ourselves to the algebras that can be interpreted as non- and ultra-relativistic ones, for zero cosmological constant.
We argue that in the ultra-relativistic cases, a Chern--Simons theory can be constructed in a straightforward manner.
This is due to the considerations of Section \ref{sec:double-extens-pres}.
This is not true for the nonrelativistic cases.
However, we demonstrate that the nonrelativistic kinematical algebras can be suitably extended in such a way that a Chern--Simons action can be written down.
Here the knowledge of  double extension will be useful.
We then show via a linearized analysis that the non- and ultra-relativistic spin-3 Chern--Simons theories thus obtained can be viewed as spin-3 generalizations of Extended Bargmann gravity and Carroll gravity, respectively.

\section{Kinematical Spin-3 Algebras}
\label{sec:algebras}

In this section, we will be concerned with three-dimensional kinematical spin-3 algebras, i.e., generalized spacetime symmetry algebras of theories of interacting, massless spin-2 and spin-3 fields. In particular, following Bacry and L\'evy--Leblond~\cite{Bacry:1968zf} we will classify all such algebras that can be obtained by combining different sIW-contractions from the algebras that underlie spin-3 gravity in AdS$_3$ and dS$_3$.
After recalling the latter, we will present all possible ways of contracting them, such that non-trivial kinematical spin-3 algebras are obtained, via a classification theorem.
Combining different of these sIW-contractions leads to various kinematical spin-3 algebras,
some of which will be discussed in the next section as a starting point for considering Carroll and Galilei spin-3 gravity Chern--Simons theories.

\subsection{AdS$_3$ and dS$_3$ Spin-3 Algebras}
\label{ssec:AdSdS}

Remember
that Spin-3 gravity in (A)dS$_3$ \cite{Henneaux:2010xg,Campoleoni:2010zq} can be written as a Chern--Simons theory for the Lie algebra $\mathfrak{sl}(3,\R) \oplus \mathfrak{sl}(3,\R)$ for AdS$_3$ or $\mathfrak{sl}(3,\mathbb{C})$ (viewed as a real Lie algebra) for dS$_3$.
In the following we will often denote the higher spin algebra  $\mathfrak{sl}(3,\R) \oplus \mathfrak{sl}(3,\R)$, realizing Spin-3 gravity in AdS$_{3}$, by $\mathfrak{hs}_{3}\mathfrak{AdS}$.
Similarly, we indicate  the higher spin algebra $\mathfrak{sl}(3,\mathbb{C})$, realizing Spin-3 gravity in dS$_{3}$, by $\mathfrak{hs}_{3}\mathfrak{dS}$. In both cases, the algebra consists of the generators of Lorentz transformations $\hat \Jt_A$ and translations $\hat \Pt_A$ along with ``spin-3 rotations'' $\hat \Jt_{AB}$ and ``spin-3 translations'' $\hat \Pt_{AB}$, that are traceless-symmetric in the $(AB)$ indices ($A=0,1,2$) \footnote{We refer to Appendix \ref{cha:conventions} for index and other conventions used in this and upcoming sections.}:
\begin{align} \label{eq:JPsymmtraceless}
   & \hat\Jt_{AB} =\hat\Jt_{BA}     & \eta^{AB}\hat \Jt_{AB} = 0 
\\
   & \hat\Pt_{AB}=\hat\Pt_{BA}      & \eta^{AB}\hat \Pt_{AB} = 0 \,.
\end{align}
Here, $\eta^{AB}$ is the three-dimensional Minkowski metric.
We will often refer to $\{\hat \Jt_A,\hat \Pt_A\}$ as the ``spin-2 generators'' or the ``spin-2 part'' and similarly to $\{\hat \Jt_{AB}, \hat \Pt_{AB}\}$ as the ``spin-3 generators'' or ``spin-3 part''.
Their commutation relations are given by~\cite{Henneaux:2010xg,Campoleoni:2010zq}
\begin{align} \label{eq:AdSdSalgebra}
 \left[\,\hat \Jt_A \comma \hat \Jt_B \,\right]  &=  \epsilon_{ABC} \,\hat \Jt^C \,,  & \left[\,\hat \Jt_A \comma \hat \Pt_B \,\right]  &=  \epsilon_{ABC} \,\hat \Pt^C\,, \nonumber \\
 \left[\,\hat \Pt_A \comma \hat \Pt_B \,\right] & = \pm  \epsilon_{ABC} \,\hat \Jt^C \,,  & & \nonumber \\
 \left[\, \hat \Jt_A \comma \hat \Jt_{BC} \,\right] & =  \epsilon\indices{^M_{A(B}} \, \hat \Jt_{C)M} \,,
      & \left[\, \hat \Pt_A \comma \hat \Pt_{BC} \,\right]  &=  \pm 
        \, \epsilon\indices{^M_{A(B}} \, \hat \Jt_{C)M} \,, \nonumber
    \\
 \left[\, \hat \Jt_A \comma \hat \Pt_{BC} \,\right]  &=  \epsilon\indices{^M_{A(B}}\, \hat \Pt_{C)M} \,,
 &  \left[\, \hat \Pt_A \comma \hat \Jt_{BC} \,\right] & =  \epsilon\indices{^M_{A(B}} \, \hat \Pt_{C)M} \,, \nonumber \\
\left[\, \hat \Jt_{AB} \comma\hat \Jt_{CD} \,\right]  &=  -  \eta_{(A(C} \epsilon_{D)B)M}  \, \hat \Jt^M \,,
   & \left[\,\hat \Jt_{AB} \comma \hat \Pt_{CD} \,\right]  &=  -  \eta_{(A(C} \epsilon_{D)B)M} \, \hat \Pt^M \,, \nonumber
    \\
    \left[\, \hat \Pt_{AB} \comma \hat \Pt_{CD} \,\right] & =  \mp  
      \eta_{(A(C} \epsilon_{D)B)M} \, \hat \Jt^M  \,,
\end{align}
where the upper sign refers to $\mathfrak{hs}_{3}\mathfrak{AdS}$ and the lower sign to $\mathfrak{hs}_{3}\mathfrak{dS}$. Note that the first two lines constitute the isometry algebra of (A)dS$_3$, i.e., $\mathfrak{sl}(2,\R) \oplus \mathfrak{sl}(2,\R)$ for AdS$_3$ and $\mathfrak{sl}(2,\mathbb{C})$, viewed as a real Lie algebra, for dS$_3$.

The above mentioned algebra is (semi)simple
and therefore has an invariant metric.
The most general one is given in Section \ref{cha:invar-metr}
but we will restrict here to
\begin{equation} \label{eq:bilformAdSdS}
  \langle \hP_A\,, \hJ_B\rangle = \eta_{AB}  \qquad \langle \hP_{AB}\,, \hJ_{CD}\rangle = \eta_{A(C} \eta_{D)B} - \frac23 \eta_{AB} \eta_{CD} \,.
\end{equation}
Note that this represents an invariant metric for both $\mathfrak{hs}_{3}\mathfrak{AdS}$ and $\mathfrak{hs}_{3}\mathfrak{dS}$. The existence of this metric allows one to construct Chern--Simons actions for the algebras $\mathfrak{hs}_{3}\mathfrak{AdS}$ and $\mathfrak{hs}_{3}\mathfrak{dS}$, that correspond to the actions for spin-3 gravity in (A)dS$_3$ \cite{Henneaux:2010xg,Campoleoni:2010zq}.

In the following, it will prove convenient to introduce a time-space splitting of the indices $A = \{0,a; a=1,2\}$. We will thereby   use the following notation:
\begin{align}
  \label{eq:nothsalgebras}
\Jt &= \hat \Jt_0  & \Gt_a &= \hat \Jt_a  & \Ht &= \hat \Pt_0  & \Pt_a &= \hat \Pt_a\\
\Jt_a &= \hat \Jt_{0a}  & \Gt_{ab} &= \hat \Jt_{ab}  & \Ht_a &= \hat \Pt_{0a}  & \Pt_{ab} &= \hat \Pt_{ab}\,.
\end{align}
Note that we have left out the generators $\hat{\Pt}_{00}$ and $\hat{\Gt}_{00}$ here. These generators are not independent, due to the tracelessness constraint (\ref{eq:JPsymmtraceless}) and in the following we will eliminate them in favor of $\Pt_{ab}$ and $\Gt_{ab}$. After these substitutions, the commutation relations of $\mathfrak{hs}_{3}\mathfrak{(A)dS}$ in this new basis are given in the first column of Table~\ref{tab:adspoin}.

\subsection{All Kinematical Spin-3 Algebras by Contracting $\mathfrak{hs}_{3}\mathfrak{(A)dS}$}
\label{sec:allkinalgs}

We now consider the spin-3 case where,
following the spin-2 case (see Section \ref{sec:kinematical-algebras}), we will obtain a classification of all possible contractions\footnote{
  Here, we will classify different contractions, in the sense defined above as different choices of subalgebra $\mathfrak{h}$, i.e., we restrict to sIW-contractions.
  This does not mean that all these contractions lead to non-isomorphic Lie algebras. Indeed, in the analysis of \cite{Bacry:1968zf}, e.g.,  one can see that the space-time and speed-time contractions applied to the AdS$_3$ isometry algebra lead to two Lie algebras that are both isomorphic to the Poincar\'e algebra.
  We should however mention that these algebras are isomorphic in the mathematical sense; physically they can be regarded as non-equivalent as the isomorphism that relates them corresponds to an interchange of boost and translation generators.
  Note also that the different contractions that are classified here are not necessarily independent.
  As an example, one can check that the general sIW-contraction of Table \ref{tab:spin2contr} can be obtained by sequential space-time, speed-space and speed-time sIW-contractions in an arbitrary order.} of $\mathfrak{hs}_{3}\mathfrak{AdS}$ and $\mathfrak{hs}_{3}\mathfrak{dS}$ by listing all their possible subalgebras. We start from $\mathfrak{hs}_3\mathfrak{(A)dS}$ since these are semisimple algebras and can therefore not be viewed as a result of a sIW-contraction (since proper sIW-contractions always lead to algebras with an abelian ideal that are thus not semisimple). Now, in order to obtain contractions that can be identified as interesting kinematical spin-3 algebras, we will impose two restrictions:
\begin{itemize}
\item When restricted to the spin-2 part of the algebra, the sIW-contraction should correspond to those considered in Table \ref{tab:spin2contr}. This ensures that the spin-2 parts of the algebras obtained by various combinations of these contractions correspond to the kinematical algebras of \cite{Bacry:1968zf}.
\item Furthermore, we will also demand that in the resulting Lie algebra not all commutators of the spin-3 part are vanishing. This requirement is motivated by the fact that we are interested in using these contractions to describe fully interacting theories of massless spin-2 and spin-3 fields. Indeed, as we will show later on, for some of the algebras obtained here, one can construct a Chern--Simons action for spin-2 and spin-3 fields. Only when the commutators of the spin-3 part are not all vanishing, do the spin-3 fields contribute to the equations of motion of the spin-2 fields.
\end{itemize}
All ways of sIW-contracting $\mathfrak{hs}_{3}\mathfrak{AdS}$ and $\mathfrak{hs}_{3}\mathfrak{dS}$ that obey these two restrictions can then be summarized by the following theorem:
\begin{theorem}
  \label{THM:KINEMATICAL}
  All possible sIW-contractions, that reduce to those considered in Table \ref{tab:spin2contr} when restricted to the spin-2 part and that are nonabelian on the subspace spanned by the spin-3 generators $\{\Jt_{a},\Ht_{a},\Gt_{ab},\Pt_{ab} \}$, are given by 10 ``democratic'' contractions that are specified in Table \ref{tab:contr} and 7 ``traceless'' contractions, given in Table \ref{tab:excontr}. As in Table \ref{tab:spin2contr}, we have specified these contractions
  by indicating the subalgebra $\mathfrak{h}$ with respect to which $\mathfrak{hs}_{3}\mathfrak{(A)dS}$ is contracted, as well as by giving the resulting abelian ideal $\mathfrak{i}$.

  \begin{table}[H]
  \centering
$
  \begin{array}{l r l l }
\toprule
    \text{Contraction }                                                                                                & \# & \multicolumn{1}{c}{\mathfrak{h}}    & \multicolumn{1}{c}{\mathfrak{i}}                     \\ \midrule
    \text{Space-time}                                                                                                  & 1  & 
                                           \{\Jt, \Gt_{a},\Jt_{a},\Gt_{ab}\}                                           & 
                                                                                \{\Ht, \Pt_{a},\Ht_{a}, \Pt_{ab} \}                                                                                                      \\
                                                                                                                       & 2  & 
                                                                                    \{\Jt, \Gt_{a},\Ht_{a}, \Pt_{ab}\} & 
                                                                                                                         \{\Ht, \Pt_{a},\Jt_{a},\Gt_{ab} \}                                                              \\ \midrule
    \text{Speed-space}                                                                                                 & 3  & \{\Jt,\Ht,\Jt_{a}, \Ht_{a} \}       & \{\Gt_{a},\Pt_{a},\Gt_{ab},\Pt_{ab} \}               \\
                                                                                                                       & 4  & \{\Jt,\Ht,\Gt_{ab},\Pt_{ab} \}      & \{\Gt_{a},\Pt_{a},\Jt_{a}, \Ht_{a} \}                \\ \midrule
    \text{Speed-time}                                                                                                  & 5  & \{\Jt, \Pt_{a},\Jt_{a},\Pt_{ab} \}  & \{\Gt_{a}, \Ht,\Ht_{a},\Gt_{ab} \}                   \\
                                                                                                                       & 6  & \{\Jt, \Pt_{a},\Ht_{a},\Gt_{ab}  \} & \{\Gt_{a}, \Ht,\Jt_{a},\Pt_{ab} \}                   \\ \midrule
                                                                                                                       & 7  & \{\Jt,\Jt_{a} \}                    & \{\Ht, \Pt_{a}, \Gt_{a},\Ht_{a},\Gt_{ab},\Pt_{ab} \} \\
     \text{General}                                                                                                    & 8  & \{\Jt,\Gt_{ab} \}                   & \{\Ht, \Pt_{a}, \Gt_{a},\Jt_{a},\Ht_{a} ,\Pt_{ab} \} \\
                                                                                                                       & 9  & \{\Jt,\Ht_{a} \}                    & \{\Ht, \Pt_{a}, \Gt_{a},\Jt_{a},\Gt_{ab},\Pt_{ab} \} \\
                                                                                                                       & 10 & \{\Jt,\Pt_{ab} \}                   & \{\Ht, \Pt_{a}, \Gt_{a},\Jt_{a},\Ht_{a} ,\Gt_{ab} \} \\ \bottomrule
  \end{array}
$
  \caption{All democratic sIW-contractions.}
  \label{tab:contr}
\end{table}

\begin{table}[H]
  \centering
$
  \begin{array}{l r l l l }
\toprule
    \text{Contr.}   & \#  & \multicolumn{1}{c}{\mathfrak{h}}                                     & \multicolumn{1}{c}{\mathfrak{i}}                    \\ \midrule
    \text{Speed}    & 4a  & \{\Jt,\Ht,\Gt_{ab},\Pt_{12}, \Pt_{22}-\Pt_{11} \}                    & \{\Pt_{11}+\Pt_{22} \}                              \\
     \text{-space}  & 4b  & \{\Jt,\Ht,\Gt_{12}, \Gt_{22}-\Gt_{11}, \Pt_{ab} \}                   & \{\Gt_{11}+\Gt_{22} \}                              \\
                    & 4c  & \{\Jt,\Ht,\Gt_{12}, \Gt_{22}-\Gt_{11},\Pt_{12}, \Pt_{22}-\Pt_{11} \} & \{\Gt_{11}+\Gt_{22}, \Pt_{11}+\Pt_{22} \}           \\ \midrule
                    & 8a  & \{\Jt,\Gt_{12}, \Gt_{22}-\Gt_{11} \}                                 & \{ \Gt_{11}+\Gt_{22},\Pt_{ab} \}                    \\
     \text{General} & 10a & \{\Jt,\Pt_{12}, \Pt_{22}-\Pt_{11} \}                                 & \{\Gt_{ab},\Pt_{11}+\Pt_{22} \}                     \\
                    & 8b  & \{\Jt,\Gt_{12}, \Gt_{22}-\Gt_{11},\Pt_{11}+\Pt_{22} \}               & \{ \Gt_{11}+\Gt_{22},\Pt_{12}, \Pt_{22}-\Pt_{11} \} \\
                    & 10b & \{\Jt,\Pt_{12}, \Pt_{22}-\Pt_{11}, \Gt_{11}+\Gt_{22}\}               & \{\Gt_{12}, \Gt_{22}-\Gt_{11},\Pt_{11}+\Pt_{22} \}  \\ \bottomrule
  \end{array}
$
\caption{All traceless sIW-contractions, where we have to add in the $\mathfrak{i}$ column
for the speed-space sIW-contractions $\{\Gt_{a},\Pt_{a},\Jt_{a}, \Ht_{a}\}$ and for the general sIW-contractions
$\{\Ht, \Pt_{a}, \Gt_{a},\Jt_{a},\Ht_{a}\}$.}
  \label{tab:excontr}
\end{table}
\end{theorem}

The complete proof of this theorem is given in the Appendix of \cite{Bergshoeff:2016soe}.
For now, let us suffice by saying that the proof starts
by noting that each of the subalgebras
$\mathfrak{h}$ in Table \ref{tab:spin2contr}
needs to be supplemented with spin-3 generators,
in order to have a contraction with a nonabelian spin-3 part.
The proof then proceeds by enumerating,
for each of the contractions of Table \ref{tab:spin2contr},
all possibilities in which spin-3 generators
can be added to $\mathfrak{h}$ such that one still obtains a subalgebra, that leads to a contraction with a nonabelian spin-3 part. We refer to appendix \ref{app:all-expl-contr} for the explicit Lie algebras of the contracted Lie algebra given in Table \ref{tab:contr}.

Finally, let us  comment on the terminology ``democratic'' and ``traceless''.
This terminology stems from the fact that the three independent
generators contained in $\Pt_{ab}$ ($\Gt_{ab}$) form a real,
reducible representation of $\Jt$, that can be split into a
tracefree symmetric part consisting of the generators
$\{\Pt_{12}, \Pt_{22}-\Pt_{11}\}$ ($\{\Gt_{12}, \Gt_{22}-\Gt_{11}\}$)
and a trace part $\Pt_{11} + \Pt_{11}$ ($\Gt_{11} + \Gt_{22}$).
The democratic contractions are such that the subalgebra
$\mathfrak{h}$ contains both tracefree symmetric and
trace components of $\Pt_{ab}$ ($\Gt_{ab}$), if present.
In some cases, it is not necessary to include the trace
component in $\mathfrak{h}$ in order to obtain a valid
subalgebra. This is the case for the democratic contractions,
numbered 4, 8 and 10 in Table \ref{tab:contr}.
Moving the trace component from $\mathfrak{h}$ to $\mathfrak{i}$
leads to the traceless cases $4a$, $4b$, $4c$, $8a$ and $10a$
in Table \ref{tab:excontr}.
In the last two remaining cases both the tracefree symmetric
part of $\Gt_{ab}$ ($\Pt_{ab}$) and the trace part
of $\Pt_{ab}$ ($\Gt_{ab}$) belong to the subalgebra  $\mathfrak{h}$.
Doing this leads to the traceless cases $8b$ and $10b$.

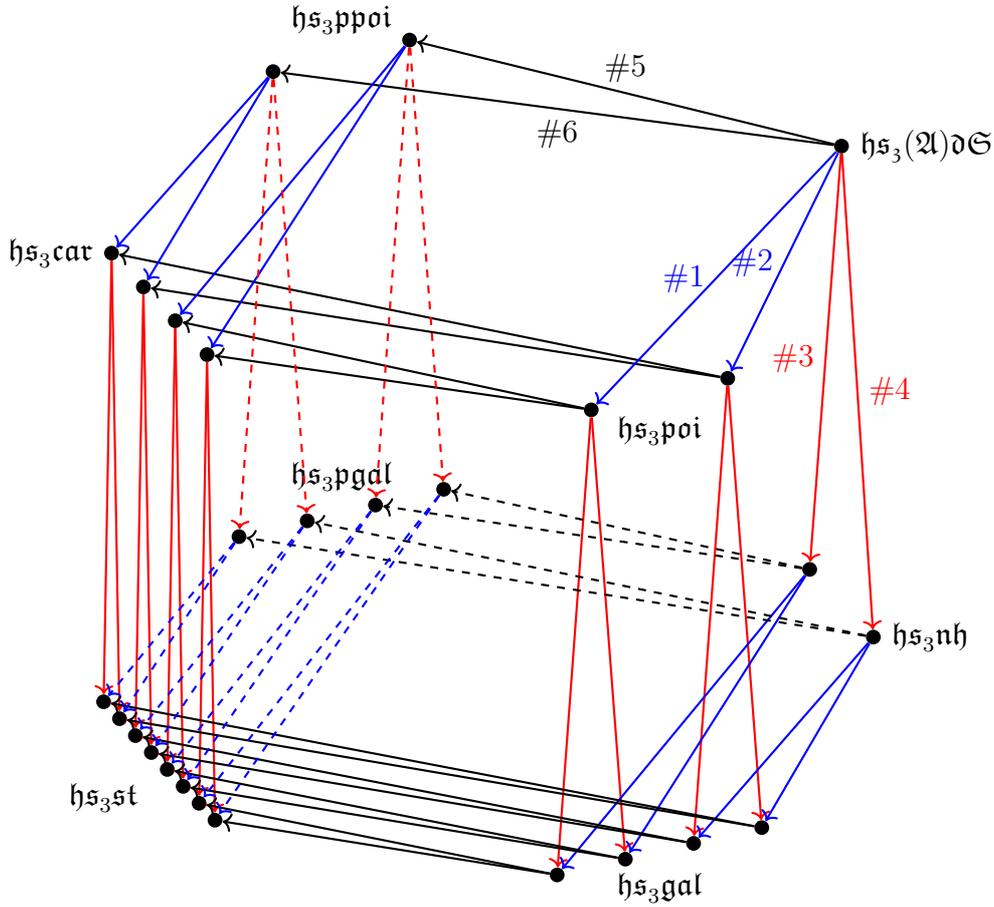
\begin{figure}[h]
  \centering
\tdplotsetmaincoords{60}{110}
\begin{tikzpicture}[
tdplot_main_coords,
dot/.style={circle,fill,scale=0.5},
linf/.style={thick,->,blue},
cinf/.style={thick,->,red},
tinf/.style={thick,->},
stinf/.style={ultra thick,->,gray},
scale=0.7
]


\node (ads) at (0,10,10) [dot, label=right:$\mathfrak{hs_3(A)}\mathfrak{dS}$] {};

\node (p) at (10,10,10) [label=below:$\mathfrak{hs}_{3}\mathfrak{poi}$] {};

\node (p1) at (10+\shi,10-\shi,10) [dot] {};
\node (p2) at (10-\shi,10+\shi,10) [dot] {};


\node (nh1) at (-\shi,10-\shi,0) [dot] {};
\node (nh2) at (\shi,10+\shi,0) [dot, label=right:$\mathfrak{hs}_{3}\mathfrak{nh}$] {};

\node (pp) at (0,0,10) [label=above:$\mathfrak{hs}_{3}\mathfrak{ppoi}$] {};

\node (pp1) at (-\shi,\shi,10) [dot] {};
\node (pp2) at (\shi,-\shi,10) [dot] {};

\node (g) at (10,10,0) [label=below:$\mathfrak{hs}_{3}\mathfrak{gal}$] {};

\node (g3) at (10-\shi-\shih,10+\shi+\shih,0) [dot] {};
\node (g4) at (10-\shi+\shih,10+\shi-\shih,0) [dot] {};

\node (g1) at (10+\shi+\shih,10-\shi-\shih,0) [dot] {};
\node (g2) at (10+\shi-\shih,10-\shi+\shih,0) [dot] {};

\node (pg) at (0,0,0) [label=above:$\mathfrak{hs}_{3}\mathfrak{pgal}$] {};


\node (pg1) at (-\shi+\shih,\shi-\shih,0) [dot] {};
\node (pg2) at (-\shi-\shih,\shi+\shih,0) [dot] {};


\node (pg3) at (\shi+\shih,-\shi-\shih,0) [dot] {};
\node (pg4) at (\shi-\shih,-\shi+\shih,0) [dot] {};


\node (car1) at (10+\shi-\shih,\shi-\shih,10) [dot] {};
\node (car2) at (10+\shi+\shih,\shi+\shih,10) [dot] {};

\node (car3) at (10-\shi-\shih,-\shi-\shih,10) [dot, label=left:$\mathfrak{hs}_{3}\mathfrak{car}$] {};
\node (car4) at (10-\shi+\shih,-\shi+\shih,10) [dot] {};

\node (st) at (10,0,0) [label=below left:$\mathfrak{hs}_{3}\mathfrak{st}$] {};

\node (st3) at (10-\shi-\shih-\shihh,-\shi-\shih-\shihh,0) [dot] {};
\node (st4) at (10-\shi-\shih+\shihh,-\shi-\shih+\shihh,0) [dot] {};
\node (st7) at (10+\shi+\shih+\shihh,\shi+\shih+\shihh,0) [dot] {};
\node (st8) at (10+\shi+\shih-\shihh,\shi+\shih-\shihh,0) [dot] {};

\node (st5) at (10+\shi-\shih+\shihh,\shi-\shih+\shihh,0) [dot] {};
\node (st6) at (10+\shi-\shih-\shihh,\shi-\shih-\shihh,0) [dot] {};

\node (st1) at (10-\shi+\shih+\shihh,-\shi+\shih+\shihh,0) [dot] {};
\node (st2) at (10-\shi+\shih-\shihh,-\shi+\shih-\shihh,0) [dot] {};

\draw[tinf] (p1) -- (car2);
\draw[tinf] (p1) -- (car1);
\draw[tinf] (p2) -- (car3);
\draw[tinf] (p2) -- (car4);

\draw[cinf] (ads) -- node [left] {\#3} (nh1);
\draw[cinf] (ads) -- node [right] {\#4}(nh2);

\draw[tinf] (ads) -- node [above] {\#5}(pp1);
\draw[tinf] (ads) -- node [below] {\#6}(pp2);

\draw[cinf] (p1) -- (g1);
\draw[cinf] (p1) -- (g2);
\draw[cinf] (p2) -- (g3);
\draw[cinf] (p2) -- (g4);

\draw[linf] (ads) -- node [left] {\#1} (p1);
\draw[linf] (ads) -- node [left] {\#2} (p2);

\draw[linf] (nh1) -- (g1);
\draw[linf] (nh1) -- (g2);
\draw[linf] (nh2) -- (g3);
\draw[linf] (nh2) -- (g4);

\draw[linf] (pp1) -- (car1);
\draw[linf] (pp1) -- (car2);
\draw[linf] (pp2) -- (car3);
\draw[linf] (pp2) -- (car4);

\draw[cinf] (car2) -- (st8);
\draw[cinf] (car2) -- (st7);

\draw[cinf] (car1) -- (st6);
\draw[cinf] (car1) -- (st5);

\draw[cinf] (car3) -- (st4);
\draw[cinf] (car3) -- (st3);
\draw[cinf] (car4) -- (st2);
\draw[cinf] (car4) -- (st1);

\draw[tinf,dashed] (nh1) -- (pg1);
\draw[tinf,dashed] (nh1) -- (pg2);
\draw[tinf,dashed] (nh2) -- (pg3);
\draw[tinf,dashed] (nh2) -- (pg4);

\draw[cinf,dashed] (pp1) -- (pg1);
\draw[cinf,dashed] (pp1) -- (pg2);
\draw[cinf,dashed] (pp2) -- (pg3);
\draw[cinf,dashed] (pp2) -- (pg4);

\draw[tinf] (g1) -- (st8);
\draw[tinf] (g1) -- (st7);
\draw[tinf] (g2) -- (st6);
\draw[tinf] (g2) -- (st5);
\draw[tinf] (g3) -- (st4);
\draw[tinf] (g3) -- (st3);
\draw[tinf] (g4) -- (st2);
\draw[tinf] (g4) -- (st1);

\draw[linf,dashed] (pg1) -- (st6);
\draw[linf,dashed] (pg1) -- (st5);
\draw[linf,dashed] (pg2) -- (st8);
\draw[linf,dashed] (pg2) -- (st7);

\draw[linf,dashed] (pg3) -- (st3);
\draw[linf,dashed] (pg3) -- (st4);
\draw[linf,dashed] (pg4) -- (st1);
\draw[linf,dashed] (pg4) -- (st2);

\end{tikzpicture}
  \caption{This figure summarizes the sequential democratic contractions of Table \ref{tab:contr}. There are 2 space-time (blue; \#1,\#2), 2 speed-space (red; \#3,\#4) and 2 speed-time (black; \#5,\#6) contractions and combining them leads to the full cube. The commutators of the algebras corresponding to the dots are given in Table \ref{tab:adspoin}-\ref{tab:hsstat2}. In comparison to Figure \ref{fig:cube}, we have for clarity omitted the double lines and the diagonal lines that indicate the direct sIW-contractions to the static algebras.}
  \label{fig:hscube}
\end{figure}

The democratic contractions can again be summarized as a cube, see Figure \ref{fig:hscube}.

\section{Carroll, Galilei and Extended Bargmann Theories}
\label{sec:spin-3}

In the previous section, we have classified all
possible (sIW-)contractions of the spin-3 AdS$_3$ and dS$_3$ algebras. Combining some of these contractions can lead to algebras whose spin-2 part corresponds to the Carroll or Galilei algebra. Here, we will study these cases in more detail. In particular, we will be concerned with constructing Chern--Simons theories for these spin-3 algebras, or suitable extensions thereof.
This extends \cite{Bergshoeff:2017btm} where the case of spin-2 Carroll and spin-2 Galilei gravity is discussed. 


In order to construct Chern--Simons actions for Carroll and Galilei spin-3 algebras, one therefore needs to know whether these algebras can be equipped with an invariant metric.
We have already seen in Section \ref{cha:kinem-chern-simons}
that this is not even for the spin-2 algebras always possible.
In this respect, it is useful to remember that it is not always true that the contraction of a Lie algebra equipped with an invariant metric, also admits one.
A counter-example was provided by the three-dimensional spin-2 Galilei algebra
which arises as sIW contractions of the Poincar\'e algebra, that in three dimensions has an invariant metric.
Naively, one can thus not construct a Chern--Simons action for the Galilei algebra how ever as shown, there exists an extension of the Galilei algebra, the so-called Extended Bargmann algebra, that can be equipped with an invariant metric and for which a Chern--Simons action can be constructed.

In this section, we will show that similar results hold in the spin-3 case. In particular, we will see that the spin-3 versions of the Carroll algebra admit an invariant metric and that a Chern--Simons action can be straightforwardly constructed.
The spin-3 versions of the Galilei algebra, like their spin-2 versions, do not have an invariant metric.
However, using double extensions we can extended them to Lie algebras with an invariant metric.
In contrast to the spin-2 case this double extension is not just given
by nontrivial central extensions.
We will explicitly construct these ``spin-3 Extended Bargmann'' algebras and their associated Chern--Simons actions. In this way, we will obtain spin-3 versions of Carroll gravity \cite{Hartong:2015xda,Bergshoeff:2017btm} and Extended Bargmann gravity \cite{Papageorgiou:2009zc,Papageorgiou:2010ud,Bergshoeff:2016lwr,Hartong:2016yrf}.

We will first treat the case of spin-3 Carroll gravity, while the spin-3 Extended Bargmann gravity case will be discussed afterwards. In both cases, we will also study the equations of motion, at the linearized level. This will allow us to interpret the Chern-Simons actions for these theories as suitable spin-3 generalizations of the actions of Carroll and Extended Bargmann gravity, in a first order formulation. In particular, this linearized analysis will show that some of the gauge fields appearing in these actions can be interpreted as generalized vielbeine, while others can be viewed as generalized spin connections. The latter in particular appear only algebraically in the equations of motion and are therefore dependent fields that can be expressed in terms of other fields. We will give these expressions. In some cases, we will see that not all spin connection components become dependent. We will argue that the remaining independent spin connection components can be viewed as Lagrange multipliers that implement certain constraints on the geometry. For simplicity, we will restrict ourselves to Carroll and Galilei spin-3 gravity theories. The analysis provided here can be straightforwardly extended to include a cosmological constant.

\subsection{Spin-3 Carroll Gravity}
\label{ssec:CarrollGravity}

There are four distinct ways of contracting $\mathfrak{hs}_3\mathfrak{(A)dS}$, such that a spin-3 algebra whose spin-2 part coincides with the Carroll algebra is obtained. These four ways correspond to combining the contractions 1 and 5, 1 and 6, 2 and 5 or 2 and 6 of Table \ref{tab:contr}, respectively.
We will denote the resulting algebras as $\mathfrak{hs}_3\mathfrak{car1}$, $\mathfrak{hs}_3\mathfrak{car2}$, $\mathfrak{hs}_3\mathfrak{car3}$ and $\mathfrak{hs}_3\mathfrak{car4}$.
Their structure constants are summarized in Table \ref{tab:hscar}. Note that $\mathfrak{hs}_3\mathfrak{car3}$ and $\mathfrak{hs}_3\mathfrak{car4}$ each come in two versions, since we apply the sIW-contractions to AdS and dS simultaneously. These versions differ in the signs of some of their structure constants, as can be seen from Table \ref{tab:hscar}. The existence of these different versions when applying the contractions 2 and 5 (or 2 and 6) stems from the fact that the combination of these contractions leads to different algebras, depending on whether one starts from $\mathfrak{hs}_3\mathfrak{AdS}$ or from $\mathfrak{hs}_3\mathfrak{dS}$.
By contrast, applying contraction  1 and 5 (or 1 and 6) on $\mathfrak{hs}_3\mathfrak{AdS}$ and $\mathfrak{hs}_3\mathfrak{dS}$ leads to the same result, namely $\mathfrak{hs}_3\mathfrak{car1}$ (or $\mathfrak{hs}_3\mathfrak{car2}$).

All these spin-3 algebras
have an invariant metric.
This can either be seen using the just mentioned contractions
or using the invariant metric preserving contractions
discussed in Section \ref{sec:double-extens-pres}.
The invariant metric preserving contractions
are specified by just a subalgebra $\mathfrak{h}$ of the original algebra
and for these cases are given by:
\begin{itemize}
\item $\mathfrak{hs}_{3}\mathfrak{poi1} \to \mathfrak{hs}_{3}\mathfrak{car1}$: $\mathfrak{h}=\{\Jt, \Jt_{a}\}$,
\item   $\mathfrak{hs}_{3}\mathfrak{poi1} \to \mathfrak{hs}_{3}\mathfrak{car2}$: $\mathfrak{h}=\{\Jt, \Gt_{ab}\}$,
\item   $\mathfrak{hs}_{3}\mathfrak{poi2} \to \mathfrak{hs}_{3}\mathfrak{car3}$: $\mathfrak{h}=\{\Jt, \Pt_{ab}\}$,
\item   $\mathfrak{hs}_{3}\mathfrak{poi2} \to \mathfrak{hs}_{3}\mathfrak{car4}$: $\mathfrak{h}=\{\Jt, \Ht_{a}\}$.
\end{itemize}

By examining the structure constants of Table \ref{tab:hscar} and \ref{tab:hscar2}, one can see that $\mathfrak{hs}_3\mathfrak{car1}$ ($\mathfrak{hs}_3\mathfrak{car2}$) and $\mathfrak{hs}_3\mathfrak{car3}$ ($\mathfrak{hs}_3\mathfrak{car4}$) are related via the following interchange of generators
\begin{equation}
  \Ht^a \leftrightarrow \Jt^a  \qquad \qquad \Pt^{ab} \leftrightarrow \Gt^{ab}
\end{equation}
plus potentially some sign changes in structure constants, as mentioned in the previous paragraph.
The structure of the Chern--Simons theories will therefore be very similar for $\mathfrak{hs}_3\mathfrak{car1}$ ($\mathfrak{hs}_3\mathfrak{car2}$) and $\mathfrak{hs}_3\mathfrak{car3}$ ($\mathfrak{hs}_3\mathfrak{car4}$). In the following, we will restrict to the case of $\mathfrak{hs}_3\mathfrak{car1}$.
The CS theory based on  $\mathfrak{hs}_3\mathfrak{car2}$ is explicitly treated
in \cite{Bergshoeff:2016soe}.

\subsubsection{Chern--Simons Theory for $\mathfrak{hs}_3\mathfrak{car1}$}
\label{sssec:hscar1}

The commutation relations of $\mathfrak{hs}_3\mathfrak{car1}$ are summarized in the first column of Table \ref{tab:hscar}. This algebra admits the following invariant metric
\begin{alignat}{2}
  \label{eq:invmetrichscar1}
  & \langle \Ht\,, \Jt \rangle = - 1  \qquad \qquad & & \langle\Pt_a  \Gt_b\rangle = \delta_{ab}  \\
  & \langle \Ht_a\,, \Jt_b \rangle = - \delta_{ab} \qquad \qquad & & \langle\Pt_{ab} \,, \Gt_{cd} \rangle = \delta_{a(c} \delta_{d)b} - \frac23 \delta_{ab} \delta_{cd} \,.
\end{alignat}
Using the commutation relations of $\mathfrak{hs}_3\mathfrak{car1}$ and the invariant metric (\ref{eq:invmetrichscar1}), the Chern--Simons action (~\eqref{eq:CSAction}) and its equations of motion can be explicitly written down. Here, we will be interested in studying the action and equations of motion, linearized around a flat background solution\footnote{For fields in this flat background solution, the curved $\mu$ index becomes equivalent to a flat one. In the following, we will therefore denote the time-like and spatial values of the $\mu$ index by 0 and $a$. The $a$ index can moreover be freely raised and lowered using a Kronecker delta. We will often raise or lower spatial $a$ indices on field components (even if it leads to equations with non-matching index positions on the left- and right-hand-sides), to make more clear which field components are being meant.} given by
\begin{equation}
  \label{eq:flatsol}
  \bar{A}_\mu = \delta_\mu^0 \, \Ht + \delta_\mu^a \, \Pt_a \,.
\end{equation}
We will therefore assume that the gauge field is given by this background solution $\bar{A}_\mu$, plus fluctuations around this background

\begin{align}
  A_\mu &= \left(\delta_\mu^0 + \tau_\mu \right) \, \Ht + \left( \delta_\mu^a + e_\mu{}^a \right) \, \Pt_a + \omega_\mu \, \Jt + B_\mu{}^a \, \Gt_a
  \nonumber\\
  &\quad + \tau_\mu{}^a \, \Ht_a
  + e_\mu{}^{ab} \, \Pt_{ab} + \omega_\mu{}^a \, \Jt_a + B_\mu{}^{ab} \, \Gt_{ab} \,.
      \label{eq:gaugefieldlin}
\end{align}
Here, $\tau_\mu$ can be interpreted as a linearized time-like vielbein, $e_\mu{}^a$ as a linearized spatial vielbein, while $\omega_\mu$ and $B_\mu{}^a$ can be viewed as linearized spin connections for spatial rotations and boosts respectively. Similarly, $\tau_\mu{}^a$, $e_\mu{}^{ab}$, $\omega_\mu{}^a$ and $B_\mu{}^{ab}$ can be interpreted as spin-3 versions of these linearized vielbeine and spin connections.

Using the expansion (\ref{eq:gaugefieldlin}) in the Chern--Simons action and keeping only the terms quadratic in the fluctuations, one finds the following linearized action:
\begin{align}
  \label{eq:linScar1}
  S_{\mathfrak{hs}_3\mathfrak{car1}} &= \int\, d^3 x \, \epsilon^{\mu\nu\rho} \Big( -2 \tau_\mu \partial_\nu \omega_\rho + 2 e_\mu{}^a \partial_\nu B_{\rho}{}^a - 2 \tau_\mu{}^a \partial_\nu \omega_\rho{}^a + 4 e_\mu{}^{ab} \partial_\nu B_{\rho}{}^{ab} \nonumber \\
  &\quad - \frac43 e_{\mu}{}^{aa} \partial_\nu B_\rho{}^{bb} - \delta_\mu^0 \omega_\nu{}^a \omega_\rho{}^b \epsilon_{ab} - 2 \delta_\mu^a \omega_\nu B_\rho{}^b \epsilon_{ab} - 4 \delta_\mu^a \omega_\nu{}^c B_\rho{}^{cb} \epsilon_{ab} \Big) \,.
\end{align}
The linearized equations of motion corresponding to this action are given by
\begin{align}
  \label{eq:lineomscar1}
  & 0 = R_{\mu\nu}(\Ht) \equiv  \partial_\mu \tau_\nu - \partial_\nu \tau_\mu - \delta_\mu^a B_\nu{}^b \epsilon_{ab} + \delta_\nu^a B_\mu{}^b \epsilon_{ab}  \nonumber \\
  & 0 = R_{\mu\nu}(\Pt^a) \equiv  \partial_\mu e_\nu{}^a - \partial_\nu e_\mu{}^a + \epsilon^{ab} \delta_\mu^b \omega_\nu - \epsilon^{ab} \delta_\nu^b \omega_\mu  \nonumber \\
  & 0 = R_{\mu\nu}(\Jt) \equiv  \partial_\mu \omega_\nu - \partial_\nu \omega_\mu  \nonumber \\
  & 0 = R_{\mu\nu}(\Gt^a) \equiv  \partial_\mu B_\nu{}^a - \partial_\nu B_\mu{}^a  \nonumber \\
  & 0 = R_{\mu\nu}(\Ht^a) \equiv  \partial_{[\mu} \tau_{\nu]}{}^a  - \epsilon^{ab} \delta_\mu^0 \omega_\nu{}^b + \epsilon^{ab} \delta_\nu^0 \omega_\mu{}^b - 2 \delta_\mu^b B_\nu{}^{ac} \epsilon_{bc} + 2 \delta_\nu^b B_\mu{}^{ac} \epsilon_{bc}  \nonumber \\
  & 0 = R_{\mu\nu}(\Pt^{ab}) \equiv \partial_{[\mu} e_{\nu]}{}^{ab}  + \frac12 \delta_{[\mu}^c \omega_{\nu]}{}^{(a} \epsilon^{b)c} - \delta_\mu^c \omega_\nu{}^d \epsilon_{cd} \delta^{ab} + \delta_\nu^c \omega_\mu{}^d \epsilon_{cd} \delta^{ab}  \nonumber \\
  & 0 = R_{\mu\nu}(\Jt^a) \equiv \partial_\mu \omega_\nu{}^a - \partial_\nu \omega_\mu{}^a  \nonumber \\
  & 0 = R_{\mu\nu}(\Gt^{ab}) \equiv \partial_\mu B_\nu{}^{ab} - \partial_\nu B_\mu{}^{ab} 
\end{align}
The equations
\begin{equation}
  R_{\mu\nu}(\Ht) = 0  \quad R_{\mu\nu}(\Pt^a) = 0  \quad R_{\mu\nu}(\Ht^a) = 0  \quad R_{\mu\nu}(\Pt^{ab}) = 0
\end{equation}
 contain the spin connections $\omega_\mu$, $B_\mu{}^a$, $\omega_\mu{}^a$ and $B_\mu{}^{ab}$ only in an algebraic way. These equations can thus be solved to yield expressions for some of the spin connection components in terms of the vielbeine and their derivatives.

Let us first see how this works for the spin-2 spin connections $\omega_\mu$ and $B_\mu{}^a$. The equation $R_{0a}(\Ht) = 0$ can be straightforwardly solved for $B_0{}^a$:
\begin{equation}
  B_0{}^a = \epsilon^{ab} \left( \partial_0 \tau_b - \partial_b \tau_0 \right) \,.
\end{equation}
Similarly, the equation $R_{ab}(\Ht) = 0$ (or equivalently $\epsilon^{ab} R_{ab}(\Ht) = 0$) can be solved for $B_c{}^c$ (the spatial trace of $B_\mu{}^a$):
\begin{equation}
  B_c{}^c = \frac12 \epsilon^{ab} \left(\partial_a \tau_b - \partial_b \tau_a \right) \,.
\end{equation}
From $R_{ab}(\Pt^c) = 0$ (or equivalently $\epsilon^{ab} R_{ab}(\Pt^c) = 0$) one finds the spatial part of $\omega_\mu$:
\begin{equation}
  \omega_a = \frac12 \epsilon^{bc} \left( \partial_b e_{ca} - \partial_c e_{ba} \right) \,.
\end{equation}
Finally, let us consider the equation $R_{0a}(\Pt_b) = 0$. The anti-symmetric part of this equation $\epsilon^{ab} R_{0a}(\Pt_b) = 0$ can be solved for the time-like part of $\omega_\mu$:
\begin{equation}
  \omega_0 = \frac12 \epsilon^{ab} \left(\partial_a e_{0b} - \partial_0 e_{ab} \right) \,.
\end{equation}
The symmetric part $R_{0(a}(\Pt_{b)}) = 0$ does not contain any spin connection and can be viewed as a constraint on the geometry
\begin{equation}
  \label{eq:constraintcar1}
  \partial_0 e_{(ab)} - \partial_{(a} e_{|0|b)} = 0 \,.
\end{equation}
In summary, we find that $R_{\mu\nu}(\Ht) = 0$ and $R_{\mu\nu}(\Pt^a) = 0$ lead to the constraint (\ref{eq:constraintcar1}) as well as the following solutions for $\omega_\mu$ and $B_\mu{}^a$
\begin{align}
  \label{eq:solspin2conncar1}
  \omega_\mu &= \frac12 \delta_\mu^0 \, \epsilon^{ab} \left(\partial_a e_{0b} - \partial_0 e_{ab} \right) + \frac12 \delta_\mu^a\, \epsilon^{bc} \left( \partial_b e_{ca} - \partial_c e_{ba} \right) \,, \nonumber \\
  B_\mu{}^a &= \delta_\mu^0 \, \epsilon^{ab} \left( \partial_0 \tau_b - \partial_b \tau_0 \right) + \frac14 \delta_\mu^a \, \epsilon^{bc} \left (\partial_b \tau_c - \partial_c \tau_b \right) + \delta_\mu^b \, \tilde{B}_b{}^a \,,
\end{align}
where $\tilde{B}_b{}^a$ is an undetermined traceless tensor. The boost connection $B_\mu{}^a$ is thus not fully determined in terms of $\tau_\mu$ and $e_\mu{}^a$.

A similar reasoning allows one to solve for certain components of the spin-3 connections $\omega_\mu{}^a$ and $B_\mu{}^{ab}$. In particular, the equation $R_{ab}(\Ht^c) = 0$ can be solved for $B_d{}^{da}$, a spatial trace of $B_\mu{}^{ab}$:
\begin{equation}
  B_d{}^{da} = \frac14 \epsilon^{bc} \left(\partial_b \tau_c{}^a - \partial_c \tau_b{}^a\right) \,.
\end{equation}
The equation $R_{ab}(\Pt^{cd}) = 0$ can be solved for the symmetric, spatial part of $\omega_\mu{}^a$:
\begin{equation}
  \omega^{(ab)} = \epsilon^{cd} \left(\partial_c e_d{}^{ab} - \partial_d e_c{}^{ab}\right) - \frac13 \delta^{ab} \epsilon^{cd} \left(\partial_c e_d{}^{ee} - \partial_d e_c{}^{ee} \right) \,.
\end{equation}
The anti-symmetric, spatial part of $\omega_\mu{}^a$ can be found from $R_{0a}(\Ht^a) = 0$:
\begin{equation}
  \epsilon^{ab} \omega_{ab} = \partial_0 \tau_a{}^a - \partial_a \tau_0{}^a \,.
\end{equation}
From the other equations contained in $R_{0b}(\Ht^a) = 0$ one then finds
\begin{equation}
  B_0{}^{ab} = \frac14 \epsilon^{(a|c|} \left( \partial_0 \tau_c{}^{b)} - \partial_c \tau_0{}^{b)} \right)
  + \frac14 \epsilon^{cd}  \partial_{[c} e_{d]}{}^{ab} 
  - \frac16 \delta^{ab} \epsilon^{cd} \left( \partial_c e_d{}^{ee} - \partial_d e_c{}^{ee} \right) \,.
\end{equation}
The equation $R_{0a}(\Pt_{bc}) = 0$ can be divided into a part that is fully symmetric in the indices $a$, $b$, $c$ and a part that is of mixed symmetry:
\begin{equation}
  R_{0a}(\Pt_{bc}) = 0 \qquad \Leftrightarrow \qquad R^{\mathrm{S}}_{0a}(\Pt_{bc}) = 0 \quad \mathrm{and} \quad R^{\mathrm{MS}}_{0a}(\Pt_{bc}) = 0 \,,
\end{equation}
where
\begin{align}
  R^{\mathrm{S}}_{0a}(\Pt_{bc}) & = \frac13 \left(R_{0a}(\Pt_{bc}) + R_{0c}(\Pt_{ab}) + R_{0b}(\Pt_{ca})\right) \,, \nonumber \\
  R^{\mathrm{MS}}_{0a}(\Pt_{bc}) &= \frac13 \left(2 R_{0a}(\Pt_{bc}) - R_{0c}(\Pt_{ab}) - R_{0b}(\Pt_{ca})\right) \,.
\end{align}
The equation $R^{\mathrm{MS}}_{0a}(\Pt_{bc}) = 0$ can be solved for $\omega_0{}^a$, by noting that
\begin{equation}
  R^{\mathrm{MS}}_{0a}(\Pt_{bc}) = 0 \qquad \Leftrightarrow \qquad \epsilon^{ab} R_{0a}(\Pt_{bc}) = 0 \,.
\end{equation}
The solution one finds is given by
\begin{equation}
  \omega_0{}^a = \frac25 \epsilon^{bc} \left( \partial_b e_0{}^{ca} - \partial_0 e_b{}^{ca} \right) \,.
\end{equation}
The fully symmetric part $R^{\mathrm{S}}_{0a}(\Pt_{bc}) = 0$ can not be used to solve for other spin connection components. Rather, it should be viewed as a constraint on the geometry:
\begin{align}
  \label{eq:constraint2car1}
& \partial_0 e_b{}^{ac} - \partial_b e_0{}^{ac} + \partial_0 e_a{}^{bc} - \partial_a e_0{}^{bc} + \partial_0 e_c{}^{ab} - \partial_c e_0{}^{ab} \nonumber \\
&\quad  + \frac25 \delta_{ac} \left( \partial_b e_0{}^{dd} - \partial_d e_0{}^{bd} + \partial_0 e_d{}^{bd} - \partial_0 e_b{}^{dd} \right) \nonumber \\
&\quad + \frac25 \delta_{bc} \left( \partial_a e_0{}^{dd} - \partial_d e_0{}^{ad} + \partial_0 e_d{}^{ad} - \partial_0 e_a{}^{dd} \right) \nonumber \\
&\quad + \frac25 \delta_{ab} \left( \partial_c e_0{}^{dd} - \partial_d e_0{}^{cd} + \partial_0 e_d{}^{cd} - \partial_0 e_c{}^{dd} \right) = 0 \,.
\end{align}
This constraint can be slightly simplified. By contracting it with $\delta^{bc}$, one finds that
\begin{equation}
  \partial_a e_0{}^{bb} - \partial_0 e_a{}^{bb} = 6 \left(\partial_b e_0{}^{ab} - \partial_0 e_b{}^{ab} \right) \,.
\end{equation}
Using this, one finds that (\ref{eq:constraint2car1}) simplifies to
\begin{align}
  \label{eq:constraint2car1v2}
& \partial_0 e_b{}^{ac} - \partial_b e_0{}^{ac} + \partial_0 e_a{}^{bc} - \partial_a e_0{}^{bc} + \partial_0 e_c{}^{ab} - \partial_c e_0{}^{ab}  + \frac13 \delta_{bc} \left( \partial_a e_0{}^{dd} - \partial_0 e_a{}^{dd} \right) \nonumber \\ & \quad + \frac13 \delta_{ac} \left( \partial_b e_0{}^{dd} - \partial_0 e_b{}^{dd} \right) + \frac13 \delta_{ab} \left( \partial_c e_0{}^{dd} - \partial_0 e_c{}^{dd} \right) = 0 \,.
\end{align}
One thus finds for the spin-3 sector, that the equations $R_{\mu\nu}(\Ht^a) = 0$ and $R_{\mu\nu}(\Pt^{ab}) = 0$ lead to the constraint (\ref{eq:constraint2car1v2}) and the following solutions for $\omega_\mu{}^a$ and $B_\mu{}^{ab}$:
\begin{align}
  \label{eq:solspin3conncar1}
  \omega_\mu{}^a &= \frac25 \delta_\mu^0 \epsilon^{bc} \left( \partial_b e_0{}^{ca} - \partial_0 e_b{}^{ca} \right) + \frac12 \delta_\mu^b \Big( \epsilon^{cd} \left( \partial_c e_d{}^{ba} - \partial_d e_c{}^{ba} \right) \nonumber \\ & \qquad - \frac13 \delta_b^a \epsilon^{cd} \left( \partial_c e_d{}^{ee} - \partial_d e_c{}^{ee} \right) + \epsilon_{ba} \left(\partial_0 \tau_c{}^c - \partial_c \tau_0{}^c \right)\Big) \,, \nonumber \\
  B_\mu{}^{ab} &= \frac14 \delta_\mu^0 \left(\epsilon^{(a|c|} \left( \partial_0 \tau_c{}^{b)} - \partial_c \tau_0{}^{b)} \right) + \epsilon^{cd}
                 \partial_{[c} e_{d]}{}^{ab} 
                 - \frac23 \delta^{ab} \epsilon^{cd} \left( \partial_c e_d{}^{ee} - \partial_d e_c{}^{ee} \right) \right) \nonumber \\
& \qquad + \frac{1}{12}  \delta_\mu^{(a} \epsilon^{|de|} \left(\partial_d \tau_e{}^{b)} - \partial_e \tau_d{}^{b)}\right) + \delta_\mu^c \tilde{B}_c{}^{ab} \,,
\end{align}
where $\tilde{B}_c{}^{ab}$ is an arbitrary tensor obeying $\tilde{B}_b{}^{ba} = 0$. As for the spin-2 sector, one thus finds that the spin-3 boost connection $B_\mu{}^{ab}$ can not be fully determined in terms of $\tau_\mu{}^a$ and $e_\mu{}^{ab}$.

It is interesting to see what role the undetermined components $\tilde{B}_b{}^a$ and $\tilde{B}_c{}^{ab}$ play. In particular, one can check how these components appear in the Lagrangian and what their equations of motion are. Upon partial integration in the action (\ref{eq:linScar1}), one finds that the terms in the Lagrangian involving $B_\mu{}^a$ can be written as
\begin{equation}
  \epsilon^{\mu\nu\rho} R_{\mu\nu}(\Pt_a) B_\rho{}^a \,.
\end{equation}
The traceless spatial components $\tilde{B}_b{}^a$ of $B_\rho{}^a$ thus couple to
\begin{equation}
  \epsilon^{cb} R_{0c}(\Pt_a) - \frac12 \delta_a^{b} \epsilon^{cd} R_{0c}(\Pt_d) \,.
\end{equation}
This can however be rewritten as
\begin{equation}
 - \frac12 \epsilon^{cb} R_{0(a}(\Pt_{b)}) \,.
\end{equation}
One thus sees that $\tilde{B}_b{}^a$ acts as a Lagrange multiplier for $R_{0(a}(\Pt_{b)}) = 0$, which led to the constraint (\ref{eq:constraintcar1}). Similarly, one can check that $\tilde{B}_c{}^{ab}$ plays the role of Lagrange multiplier for the constraint (\ref{eq:constraint2car1v2}).

\subsection{Spin-3 Galilei and Extended Bargmann Gravity}
\label{ssec:ExtBargGravity}

In the previous section, we have studied Carroll spin-3 algebras, whose spin-2 part corresponds to the Carroll algebra.
Using the contractions of Table \ref{tab:contr}, one can also obtain nonrelativistic spin-3 algebras, that contain the Galilei algebra.
As in the Carroll case, there are four distinct ways of doing this, namely by successively applying the contractions 1 and 3, 1 and 4, 2 and 3 or 2 and 4 of Table \ref{tab:contr}.
We have called the resulting algebras $\mathfrak{hs}_3\mathfrak{gal1}$, $\mathfrak{hs}_3\mathfrak{gal2}$, $\mathfrak{hs}_3\mathfrak{gal3}$ and $\mathfrak{hs}_3\mathfrak{gal4}$ respectively and summarized their commutation relations in Table \ref{tab:hsgal} and \ref{tab:hsgal2}.
As in the Carroll case, $\mathfrak{hs}_3\mathfrak{gal3}$ and $\mathfrak{hs}_3\mathfrak{gal4}$ each come in two different versions, depending on whether one applies the combination of contraction on $\mathfrak{hs}_3\mathfrak{AdS}$ or $\mathfrak{hs}_3\mathfrak{dS}$. They are again structurally similar to $\mathfrak{hs}_3\mathfrak{gal1}$ and $\mathfrak{hs}_3\mathfrak{gal2}$. We will therefore restrict our discussion here to these two cases.

In contrast to the spin-3 Carroll algebras, whose invariant metrics arose from applying the relevant contraction on (\ref{eq:bilformAdSdS}), a similar reasoning for the spin-3 Galilei algebras leads to degenerate bilinear forms. One can in fact show by direct computation that they can not be equipped with a nondegenerate symmetric invariant bilinear form.
This is even true when one allows nontrivial central extensions. One algebra admits no nontrivial central extensions (the second cohomology group is trivial), whereas the other does admit three nontrivial extensions of which no combination of them can be used to define an invariant metric.
In this sense the spin-3 version differs from the spin-2 one, see Section \ref{cha:kinem-chern-simons}.
It could be interesting to investigate these algebras, given explicitly in Table \ref{tab:hsgal}, and their degenerate bilinear forms. For the spin-2 case, this has been done in \cite{Bergshoeff:2017btm}. Due to the degeneracy of the bilinear form, some of the fields appear without kinetic term  in the action (see the discussion in Section \ref{sec:non-degenerate} concerning non-degeneracy) and are therefore not dynamical. In the spin-2 case, one can nevertheless interpret these non-dynamical fields as Lagrange multipliers for geometrical constraints, similarly to what happens in the Carroll cases of the previous section. Although it would be interesting to see whether similar results hold for the higher spin case, we will not do this here and instead we will look at Chern--Simons theories where each field has a kinetic term. These can not be based on the spin-3 Galilei algebras, but interestingly, double extensions help to find Lie algebras that admit an invariant metric, i.e. a nondegenerate invariant symmetric bilinear form. Remarkably, in this way  one ends up with a spin-3 version of the Extended Bargmann algebra,
it the sense that the spin-2 subalgebra is the $\mathfrak{ebarg}$ discussed in Section \ref{cha:kinem-high-spin}.

Double extensions applied to the ordinary Galilei algebra in three dimensions and yields the so-called Extended Bargmann algebra \cite{Papageorgiou:2009zc,Papageorgiou:2010ud,Bergshoeff:2016lwr,Hartong:2016yrf}, that extends the Galilei algebra with two central extensions. Applying the theorem to $\mathfrak{hs}_3 \mathfrak{gal1}$ and $\mathfrak{hs}_3\mathfrak{gal2}$ yields two spin-3 algebras, that we will denote, in hindsight,  by $\mathfrak{hs}_3 \mathfrak{ebarg1}$ and $\mathfrak{hs}_3 \mathfrak{ebarg2}$ (since they have an Extended Bargmann spin-2 subalgebra).

The algebra $\mathfrak{hs}_3\mathfrak{ebarg1}$ can be obtained by looking
for double extension for $\mathfrak{hs}_3\mathfrak{gal1}$. Indeed, with the choices $\mathfrak{g} = \{\Pt_a, \Gt_a, \Pt_{ab}, \Gt_{ab}\}$, $\mathfrak{h} = \{\Ht, \Jt, \Ht_a, \Jt_a\}$ and
\begin{equation}
  \label{eq:trggal1}
  \langle \Pt_a \,, \Gt_b\rangle_{\mathfrak{g}} = \delta_{ab} \,, \qquad \langle\Pt_{ab}\,, \Gt_{cd}\rangle_{\mathfrak{g}} = \delta_{a(c} \delta_{d)b} - \frac23 \delta_{ab} \delta_{cd} \,,
\end{equation}
the assumptions of a double extension theorem are fulfilled and the algebra $\mathfrak{hs}_3\mathfrak{ebarg1}$ can be constructed. Denoting the generators of $\mathfrak{h}^*$ by $\{\Ht^*, \Jt^*, \Ht_a^*, \Jt_a^*\}$, the commutation relations of $\mathfrak{hs}_3\mathfrak{ebarg1}$ are given in Table \ref{tab:ebarg1}.
\begin{table}
  \centering
$
\begin{array}{l r r l r r r}
\toprule %
                                               & \mathfrak{hs}_3\mathfrak{ebarg1}                  & \quad \quad\quad\quad\quad     &                                              & \mathfrak{hs}_3\mathfrak{ebarg2}     \\ \midrule  
  \left[\,\Jt  \comma \Gt_{a} \,\right]        & \epsilon_{am}  \Gt_{m}                            &                                 & \left[\,\Jt  \comma \Gt_{a} \,\right]        & \epsilon_{am}  \Gt_{m}               \\           
  \left[\, \Jt \comma \Pt_{a} \,\right]        & \epsilon_{am}  \Pt_{m}                            &                                 & \left[\, \Jt \comma \Pt_{a} \,\right]        & \epsilon_{am}  \Pt_{m}               \\           
  \left[\, \Gt_{a} \comma \Ht \,\right]        & -\epsilon_{am}  \Pt_{m}                           &                                 & \left[\, \Gt_{a} \comma \Ht \,\right]        & -\epsilon_{am}  \Pt_{m}              \\           
    \left[\, \Gt_a \comma \Gt_b \, \right]     & \epsilon_{ab} \Ht^*                               &                                 & \left[\, \Gt_a\comma \Gt_b\,\right]          & \epsilon_{ab} \Ht^*                  \\           
  \left[\, \Pt_a \comma \Gt_b \, \right]       & \epsilon_{ab} \Jt^*                               &                                 & \left[\, \Gt_a\comma \Pt_b\,\right]          & \epsilon_{ab} \Jt^*                  \\\midrule   
  \left[\, \Jt \comma \Jt_{a} \,\right]        & \epsilon_{am} \Jt_{m}                             &                                 & \left[\, \Jt \comma \Jt_{a} \,\right]        & \epsilon_{am} \Jt_{m}                \\           
  \left[\, \Jt \comma \Gt_{ab} \,\right]       & -\epsilon_{m(a}\Gt_{b)m}                          &                                 & \left[\, \Jt \comma \Gt_{ab} \,\right]       & -\epsilon_{m(a}\Gt_{b)m}             \\           
  \left[\, \Jt \comma \Ht_{a} \,\right]        & \epsilon_{am} \Ht_{m}                             &                                 & \left[\, \Jt \comma \Ht_{a} \,\right]        & \epsilon_{am} \Ht_{m}                \\           
  \left[\, \Jt \comma \Pt_{ab} \,\right]       & -\epsilon_{m(a}\Pt_{b)m}                          &                                 & \left[\, \Jt \comma \Pt_{ab} \,\right]       & -\epsilon_{m(a}\Pt_{b)m}             \\           
  \left[\, \Gt_{a} \comma \Jt_{b}  \,\right]   & -(\epsilon_{am} \Gt_{bm}+ \epsilon_{ab} \Gt_{mm}) &                                 & \left[\, \Gt_{a} \comma \Gt_{bc}  \,\right]  & -\epsilon_{a(b}\Jt_{c)}              \\          
  \left[\, \Gt_{a} \comma \Ht_{b}  \,\right]   & -(\epsilon_{am} \Pt_{bm}+ \epsilon_{ab} \Pt_{mm}) &                                 & \left[\, \Gt_{a} \comma \Pt_{bc}  \,\right]  & -\epsilon_{a(b}\Ht_{c)}              \\          
  \left[\, \Ht \comma \Jt_{a} \,\right]        & \epsilon_{am} \Ht_{m}                             &                                 & \left[\, \Ht \comma \Jt_{a} \,\right]        & \epsilon_{am} \Ht_{m}                \\           
  \left[\, \Ht \comma \Gt_{ab} \,\right]       & -\epsilon_{m(a}\Pt_{b)m}                          &                                 & \left[\, \Ht \comma \Gt_{ab} \,\right]       & -\epsilon_{m(a}\Pt_{b)m}             \\           
  \left[\, \Pt_{a} \comma \Jt_{b}  \,\right]   & -(\epsilon_{am} \Pt_{bm}+ \epsilon_{ab} \Pt_{mm})  &                                 & \left[\, \Pt_{a} \comma \Gt_{bc}  \,\right]  & -\epsilon_{a(b}\Ht_{c)}              \\ \midrule 
  \left[\, \Jt_{a} \comma \Jt_{b} \,\right]    & \epsilon_{ab} \Jt                                 &                                 & \left[\, \Jt_{a} \comma \Gt_{bc} \,\right]   & \delta_{a(b}\epsilon_{c)m} \Gt_{m}   \\           
  \left[\, \Jt_{a} \comma \Gt_{bc} \,\right]   & \delta_{a(b}\epsilon_{c)m} \Gt_{m}                &                                 & \left[\, \Jt_{a} \comma \Pt_{bc} \,\right]   & \delta_{a(b}\epsilon_{c)m} \Pt_{m}   \\           
  \left[\, \Jt_{a} \comma \Ht_{b} \,\right]    & \epsilon_{ab} \Ht                                 &                                 & \left[\, \Gt_{ab} \comma \Gt_{cd}  \,\right] & \delta_{(a(c}\epsilon_{d)b)}\Jt      \\           
  \left[\, \Jt_{a} \comma \Pt_{bc} \,\right]   & \delta_{a(b}\epsilon_{c)m} \Pt_{m}                &                                 & \left[\, \Gt_{ab} \comma \Ht_{c} \,\right]   & - \delta_{c(a}\epsilon_{b)m} \Pt_{m} \\           
  \left[\, \Gt_{ab} \comma \Ht_{c} \,\right]   & - \delta_{c(a}\epsilon_{b)m} \Pt_{m}              &                                 & \left[\, \Gt_{ab} \comma \Pt_{cd}  \,\right] & \delta_{(a(c}\epsilon_{d)b)}\Ht      \\ \midrule  
  \left[\, \Gt_{ab} \comma \Gt_{cd}  \,\right] & \epsilon_{(a(c}\delta_{d)b)} \Ht^{*}              &                                 & \left[\, \Gt_a\comma \Jt_b\,\right]          & -\epsilon_{am} \Pt^*_{mb}            \\           
  \left[\, \Pt_{ab} \comma \Gt_{cd}  \,\right] & \epsilon_{(a(c}\delta_{d)b)} \Jt^{*}              &                                 & \left[\, \Gt_a\comma \Ht_b\,\right]          & - \epsilon_{am} \Gt^*_{mb}           \\           
  \left[\, \Pt_{a} \comma \Gt_{bc}  \,\right]  & \epsilon_{a(b} \Jt_{c)}^{*}                       &                                 & \left[\, \Pt_a\comma \Jt_b\,\right]          & - \epsilon_{am} \Gt^*_{mb}           \\           
  \left[\, \Gt_{a} \comma \Gt_{bc}  \,\right]  & \epsilon_{a(b} \Ht_{c)}^{*}                       &                                 & \left[\, \Jt_a\comma \Jt_b\,\right]          & - \epsilon_{ab} \Ht^*                \\           
  \left[\, \Gt_{a} \comma \Pt_{bc}  \,\right]  & \epsilon_{a(b} \Jt_{c)}^{*}                       &                                 & \left[\, \Jt_a\comma \Ht_b\,\right]          & - \epsilon_{ab} \Jt^*                \\ \midrule   
  \left[\, \Jt\comma \Ht^*_a\,\right]          & \epsilon_{am} \Ht^*_m                             &                                 & \left[\,\Jt\comma \Pt^*_{ab}\,\right]        & -\epsilon_{m(a} \Pt^*_{b)m}          \\           
  \left[\, \Jt\comma \Jt^*_a \,\right]         & \epsilon_{am} \Jt^*_m                             &                                 & \left[\,\Jt\comma \Gt^*_{ab}\,\right]        & -\epsilon_{m(a} \Gt^*_{b)m}          \\           
  \left[\, \Ht\comma \Ht^*_a \,\right]         & \epsilon_{am} \Jt^*_m                             &                                 & \left[\, \Ht \comma \Pt^*_{ab}\,\right]      & -\epsilon_{m(a} \Gt_{b)m}^*          \\           
  \left[\, \Jt_a\comma \Jt^* \,\right]         & - \epsilon_{am} \Jt^*_m                           &                                 & \left[\,\Gt_{ab}\comma\Jt^*\,\right]         & - \epsilon_{m(a} \Gt^*_{b)m}         \\           
  \left[\,\Jt_a\comma \Ht^*\,\right]           & - \epsilon_{am} \Ht^*_m                           &                                 & \left[\,\Gt_{ab}\comma\Ht^*\,\right]         & - \epsilon_{m(a} \Pt^*_{b)m}         \\           
  \left[\, \Jt_a\comma \Jt^*_b \,\right]       & \epsilon_{ab} \Jt^*                               &                                 & \left[\,\Gt_{ab}\comma \Gt^*_{cd}\,\right]   & \epsilon_{(a(c} \delta_{d)b)}  \Jt^* \\           
  \left[\,\Jt_a\comma \Ht^*_b\,\right]         & \epsilon_{ab} \Ht^*                               &                                 & \left[\,\Gt_{ab}\comma \Pt^*_{cd}\,\right]   & \epsilon_{(a(c} \delta_{d)b)} \Ht^*  \\           
  \left[\,\Ht_a\comma \Ht^*\,\right]           & - \epsilon_{am} \Jt^*_m                           &                                 & \left[\,\Pt_{ab}\comma\Ht^*\,\right]         & - \epsilon_{m(a} \Gt^*_{b)m}         \\           
  \left[\, \Ht_a\comma \Ht^*_b \,\right]       & \epsilon_{ab} \Jt^*                               &                                 & \left[\,\Pt_{ab}\comma \Pt^*_{cd}\,\right]   & \epsilon_{(a(c} \delta_{d)b)} \Jt^*  \\           
\bottomrule
\end{array}
$
\caption{Nonzero commutators of $\mathfrak{hs}_3\mathfrak{ebarg1}$ and $\mathfrak{hs}_3\mathfrak{ebarg2}$. This algebras admit an invariant metric, given by equation \eqref{eq:trebarg1} and \eqref{eq:trebarg2}, respectively.}
\label{tab:ebarg1}
\end{table}
The invariant metric of $\mathfrak{hs}_3\mathfrak{ebarg1}$ is explicitly given by
\begin{alignat}{2} \label{eq:trebarg1}
 \langle \Pt_a\,, \Gt_b\rangle                  & =  \delta_{ab} \,, \qquad                                                        &
\langle \Pt_{ab}\,, \Gt_{cd}\rangle              & = \delta_{a(c} \delta_{d)b} - \frac23 \delta_{ab}\delta_{cd} \,, \nonumber                                                                                                                              \\
  \langle \Ht\,, \Ht^*\rangle                    & = 1 \,, \qquad                                                                   & \langle \Jt\,, \Jt^*\rangle     & = 1 \,, \nonumber                                                                                         \\
  \langle \Ht_a\,, \Ht^*_b\rangle                & = \delta_{ab} \,, \qquad                                                         & \langle \Jt_a\,, \Jt^*_b\rangle & = \delta_{ab} \,.
\end{alignat}
Similarly, starting from $\mathfrak{hs}_3\mathfrak{gal2}$ and
double extending $\mathfrak{g} = \{\Pt_a, \Gt_a, \Ht_a, \Jt_a\}$
by
$\mathfrak{h} = \{\Ht, \Jt, \Pt_{ab},\Gt_{ab}\}$ and
\begin{equation}
  \langle \Pt_a \,, \Gt_b \rangle_{\mathfrak{g}} = \delta_{ab} \,, \qquad \langle \Ht_a\,, \Jt_b \rangle_{\mathfrak{g}} = - \delta_{ab} \,,
\end{equation}
the algebra $\mathfrak{hs}_3 \mathfrak{ebarg2}$ can be constructed. Denoting the generators of $\mathfrak{h}^*$ by $\{\Ht^*, \Jt^*, \Pt^*_{ab}, \Gt^*_{ab}\}$, its commutation relations are given in Table \ref{tab:ebarg1}. 

This algebra admits the following invariant metric
\begin{alignat}{2} \label{eq:trebarg2}
  \langle \Pt_a, \Gt_b\rangle                  & =  \delta_{ab} \,, \qquad                                                                            &
\langle \Ht_{a}, \Jt_{b}\rangle                & = - \delta_{ab} \,, \nonumber                                                                                                                                                                                                                     \\
  \langle \Ht, \Ht^*\rangle                    & = 1 \,, \qquad                                                                                       & \langle \Jt, \Jt^*\rangle           & = 1 \,, \nonumber                                                                                         \\
  \langle \Pt_{ab}, \Pt^*_{cd}\rangle          & =  \delta_{a(c} \delta_{d)b} \,, \qquad                                                           & \langle \Gt_{ab}, \Gt^*_{cd}\rangle & =  \delta_{a(c}\delta_{d)b} \,.
\end{alignat}
Note that for both $\mathfrak{hs}_3\mathfrak{ebarg1}$ and $\mathfrak{hs}_3\mathfrak{ebarg2}$ the generators $\{\Ht, \Jt, \Pt_a, \Gt_a, \Ht^*, \Jt^*\}$ form a subalgebra that coincides with the Extended Bargmann algebra. The Chern--Simons theories based on these algebras can therefore be viewed as spin-3 extensions of Extended Bargmann gravity, studied in \cite{Papageorgiou:2009zc,Papageorgiou:2010ud,Bergshoeff:2016lwr,Hartong:2016yrf}. In \cite{Bergshoeff:2016soe} these spin-3 Extended Bargmann gravity theories
were studied in detail at the linearized level.

\chapter{Conclusions}
\label{cha:conclusions}

We will summarize the accomplished 
results and
highlight areas
that
permit
further investigations.

\subsection{Algebraic Tools for CS Theories}
\label{sec:algebraic-tools-cs}

In Chapter \ref{cha:chern-simons-theory}
we established that
the natural set-up for
CS theories 
is based on gauge algebras
admitting an invariant metric.
Besides the well known direct sum
of abelian and simple Lie algebras
which lead to reductive Lie algebras
another construction needs to be added.
With the addition of double extensions, see Definition \ref{def:double-ext},
one fully exhausts the possible 
symmetric self-dual Lie algebras.
This is due to the remarkable Theorem \ref{thm:MR} of Medina and Revoy
which states how every such indecomposable Lie algebra has to look like.

With this knowledge we reviewed Lie algebra contractions
whose physical interpretation is that of an approximation.
Therefore not only a lot can be learned from the original algebra,
but they can also be used to classify possible
physical systems in various interesting limits.

The combination of invariant metrics with contractions,
see Chapter \ref{sec:contr-invar-metr},
paired with the knowledge of double extensions
is the ideal set-up for investigations of (approximate) CS theories.
A new type of contraction, which can be seen as
a natural generalization of (simple) Inönü--Wigner contractions,
is presented in Theorem \ref{thm:invar-metr-pres} of this work.
The special feature
that it preserves the invariant metric
explains why (higher spin) Carroll algebras in $2+1$
dimensions stay equipped with an invariant
metric after the limit from Poincar\'e.

The generalization to Lie superalgebras seems like a
fruitful endeavor,
especially since 
double extensions generalize~\cite{FigueroaO'Farrill:1995cy}.

Even for Lie algebras a more systematic study
of contractions of the various types of symmetric self-dual
Lie algebras seems of interest.
Especially since the importance of
invariant metrics are not restricted to CS gauge theories.

From an algebraic point of view it might be of interest
which (simple) Lie algebras contract
to Lie algebras that are double extensions.
This could for example explain from which (simple) Lie algebra
one could arrive at the spin-3 Extended Bargmann algebras.
Notice that this is different to the spin-2 case since we
needed more than just central extensions.
Another point for why this might be of importance is
that the ``inverse'' of a contraction might lead,
analog to the deformation from Galilei to Poincar\'e algebras,
to more fundamental theories.

\subsection{Boundary Conditions}
\label{sec:boundary-conditions-2}

In Chapter \ref{sec:boundary-conditions} the
concepts of global charge and boundary conditions was reviewed,
and afterwards applied to AdS (Chapter \ref{cha:ads-spin-3}),
in the form of the $\mathfrak{u}(1)$ boundary conditions~\cite{Grumiller:2016kcp},
as well as to Lifshitz~ and null warped AdS (Chapter \ref{cha:non-ads-cs}).

It might be interesting to re-investigate
possibilities to consistently  break the boundary
conditions of the $\mathcal{W}$ algebras.
Maybe contractions are useful for this task.
On the more speculative side one might try to
restrict boundary conditions
mode wise.

In Section~\ref{ssec:CarrollGravity} consistent
boundary conditions for Carroll Gravity were found.
As discussed, the spin-3 Carroll algebras 
also permit an invariant metric, but the generalization
of the boundary conditions to the higher spin case 
has so far not been done.
Actually, even for some of the extended kinematical spin-2 algebras
(see Section \ref{sec:extend-kinem})
consistent boundary conditions have not been established.
See also the algebras proposed in \cite{Hartong:2016yrf}.
Extended Newton--Hooke might provide an interesting
intermediate step since it might have more ``box-like'' properties due to non-vanishing cosmological constant. 
This means it might be closer to AdS than, e.g.,
Poincar\'e, and for  AdS/CFT generalizations better suited.

\subsection{Kinematical Chern--Simons Theories}
\label{sec:spin-3-kinematical}

In Chapter \ref{cha:kinem-high-spin} we have extended the work of Bacry and L\'evy-Leblond \cite{Bacry:1968zf} by classifying all possible kinematical algebras of three-dimensional theories of a spin-3 field coupled to gravity, that can be obtained via (sequential) simple In\"on\"u-Wigner contractions of the algebras of spin-3 gravity in (A)dS. This classification can be found in Section \ref{sec:algebras} and the resulting possible kinematical algebras, along with their origin via contraction, are  summarized in Figure  \ref{fig:hscube}. We have summarized the commutation relations of the algebras in Tables \ref{tab:adspoin}-\ref{tab:hsstat4}. The algebras of Tables \ref{tab:hscar} to \ref{tab:hsgal2} are suitable generalizations of the Carroll and Galilei algebras, that correspond to the ultra-relativistic and nonrelativistic limits of the Poincar\'e algebra. 
We have argued that one can easily construct a Chern-Simons action for the spin-3 Carroll algebras (here invariant metric preserving contractions were useful),
that leads to a spin-3 generalization of Carroll gravity. We have moreover shown that Chern-Simons actions can be written down for suitable extensions of the spin-3 Galilei algebras, that lead to spin-3 generalizations of Extended Bargmann gravity (see Table \ref{tab:ebarg1}).

The constructed kinematical algebras are finite-dimensional.
We have shown in Section \ref{sec:4} that the three-dimensional Carroll algebra admits an infinite-dimensional extension, that is the asymptotic symmetry algebra of Carroll gravity with suitable boundary conditions. This can be taken as a hint that similar results hold for the higher spin non- and ultra-relativistic algebras as well as for the spin-2 algebras whose infinite-dimensional extensions have not been addressed in the literature yet.

There are several questions that are worthwhile for future study. The non- and ultra-relativistic spin-3 gravity theories constructed here, are given in the Chern-Simons (i.e.~first order `zuvielbein') formulation. It is interesting to see whether a metric-like~\cite{Campoleoni:2012hp} formulation can be constructed and whether the linearized field equations can be rewritten as Fronsdal-like equations. The results for the linearized spin connections given in Section \ref{sec:spin-3}
and more exhaustively in \cite{Bergshoeff:2016soe} should be useful in this regard.

We have restricted our investigations to spin-3 theories.
This analysis can be extended to theories with fields up to spin $N$, by considering sIW-contractions of $\mathfrak{sl}(N,\mathbb{R}) \oplus \mathfrak{sl}(N,\mathbb{R})$ or $\mathfrak{sl}(N,\mathbb{C})$~\cite{Campoleoni:2011hg}. One can then study the non- and ultra-relativistic gravity theories that arise in this way and in particular investigate the types of boundary conditions that lead to interesting asymptotic symmetry algebras. It would be particularly interesting to see whether it is possible to construct non- and ultra-relativistic versions of non-linear $\mathcal{W}$-algebras.

Another research direction concerns the inclusion of fermionic fields with spins higher than or equal to 3/2. This will require a classification of contractions of Lie superalgebras and can lead to higher spin generalizations of three-dimensional Extended Bargmann supergravity \cite{Bergshoeff:2016lwr}.

Some of the results presented in this thesis are also useful for studies of Ho\v{r}ava--Lifshitz gravity, that has been proposed as a new framework for Lifshitz holography \cite{Griffin:2011xs,Griffin:2012qx,Kiritsis:2012ta,Janiszewski:2012nf,Janiszewski:2012nb,Wu:2014dha,Hartong:2015zia,Hartong:2016yrf}. Extended Bargmann gravity has been argued to correspond to a special case of Ho\v{r}ava--Lifshitz gravity \cite{Hartong:2016yrf}. In this paper, we have constructed spin-3 generalizations of Extended Bargmann gravity. It is conceivable that these can be interpreted as suitable spin-3 generalizations of Ho\v{r}ava--Lifshitz gravity. It would be interesting to check whether this is indeed the case and whether the construction presented here can be generalized to yield spin-3 generalizations of generic Ho\v{r}ava--Lifshitz gravity theories.

Finally, higher spin theory has recently been argued to describe some of the excitations in fractional quantum Hall liquids \cite{Golkar:2016thq}. Newton--Cartan geometry and gravity, that are based on extensions of the Galilei algebra, have been very useful in constructing effective actions that can capture transport properties in studies of the fractional quantum Hall effect. It would be interesting to investigate whether the nonrelativistic higher spin gravity theories that can be constructed using the results of this paper, can play a similar role.

\appendix

\chapter{Conventions}
\label{cha:conventions}

\section{Symmetrization and Indices}
\label{sec:symm-indic}

We adopt the convention that the symmetrization of a pair of indices $a$, $b$ are denoted with parentheses $(a b)$, while anti-symmetrization is denoted with square brackets $[a b]$. Symmetrization and anti-symmetrization is performed without normalization factor, i.e.,
\begin{align}
  T_{(ab)}= T_{ab}+T_{ba} \qquad T_{[ab]}= T_{ab}-T_{ba} \,.
\end{align}
Nested (anti-)symmetrizations are understood to be taken from the outermost ones to the innermost ones, e.g.
\begin{equation}
  T_{(a(bc)d)}= T_{a(bc)d}+ T_{d(bc)a}= T_{abcd}+T_{acbd}+ T_{dbca}+ T_{dcba} \,.
\end{equation}
Vertical bars denote that the (anti-)symmetrization does not affect the enclosed indices, e.g.,
\begin{equation}
  T_{[a|bc|d]}= T_{abcd}-T_{dbca} \,.
\end{equation}
With our conventions this means that $T_{(a|(bc)|d)}= T_{(a(bc)d)}$.

Upper case Latin indices denote spacetime indices, while lower case ones denote spatial indices:
\begin{align}
A,B,C,M,\dots&=0,1,2\,, & a,b,c,m,\dots&=1,2 \,.
\end{align}
We take the following conventions for the metric
\begin{align}
  \eta_{AB}=\mathrm{diag}(-,+,+) \qquad  \eta_{ab}=\delta_{ab}=\mathrm{diag}(+,+)\,.
\end{align}
For the Levi-Civita symbol, we adopt the following convention:
 \begin{align}
 \epsilon_{012}=\epsilon_{12}=1\,, \qquad \epsilon_{0ab}=\epsilon_{ab}\,, \qquad \epsilon^{ab} = \epsilon_{ab} \,.
 \end{align}

 Any convention concerning Lie algebras and vector spaces is given in Appendix \ref{cha:lie-algebras}.
 Definitions of various symbols can also be found using the Index at
 the end of the document.

\section{Differential Forms}
\label{sec:differentialforms}

These useful identities for the $a$-form $\alpha$ and the $b$-form $\beta$
with the normalization
\begin{align}
  \alpha&=\frac{1}{a!}\, \alpha_{\mu_{1}\dotsm \mu_{a}} \dd x^{\mu_{1}}\wedge \dotsm \wedge \dd x^{\mu_{a}}
  \\
  \dd x^{\mu} \wedge \dd x^{\nu} &= \dd x^{\mu} \otimes \dd x^{\nu}
  - \dd x^{\nu} \otimes \dd x^{\mu}
\end{align}
are taken from \cite{Nakahara:2003nw}, \cite{WoodhouseGeoQu} and \cite{Lee2012a}.
We denote the exterior product by $\alpha  \wedge \beta$,
the Lie derivative by $\Ld_{X}\alpha$
and the contraction of the vector field $X$ with $\alpha$ by $\ic_{X}\alpha$.
\begin{align}
  \alpha \wedge \beta = (-1)^{ab} \beta \wedge \alpha
\end{align}
\begin{align}
  \dd(\alpha \wedge \beta)&= \dd \alpha  \wedge \beta + (-1)^{a} \alpha \wedge \dd \beta
  \\
  \dd^{2}&=0
  \\
  \dd \Ld_{X}&= \Ld_{X} \dd
\end{align}
\begin{align}
  \ic_{X}(\alpha \wedge \beta)&= \ic_{X} \alpha  \wedge \beta + (-1)^{a} \alpha \wedge \ic_{X} \beta
  \\
  \ic_{X}^{2}&=0
  \\
  \ic_{X} \Ld_{X}&= \Ld_{X} \ic_{X}
\end{align}
\begin{align}
  \Ld_{X}(\alpha \wedge \beta)
  &= \Ld_{X} \alpha  \wedge \beta +  \alpha \wedge \Ld_{X} \beta 
  \\
  \Ld_{X}
  &= \dd \circ \ic_{X} + \ic_{X} \circ \dd
  \\
  [\Ld_{X},\ic_{Y}]\alpha
  &= \ic_{[X,Y]} \alpha  
  \\
  [\Ld_{X},\Ld_{Y}]\alpha
  &=\Ld_{[X,Y]}\alpha
\end{align}

\section{\texorpdfstring{$2+1$ Decomposition}{2+1 Decomposition}}
\label{sec:2+1-decomposition}

\begin{align}
  \label{eq:Asplit}
  A&=A_{N}+\At=A_{t} \dd t + A_{i} \dd x^{i}\\
  \dd&=\dd_{N}+\ddt\\
  \widetilde{F}&=\widetilde{\dd} \At + \At \wedge \At=\frac{1}{2}F_{ij} \dd x^{i}\wedge \dd x^{j}
\end{align}
\begin{align}
  \label{eq:splituse}
  A_{N}^{2}&=\At^{3}=0\\
  \dd (A_{N}+\At)&=\dd_{N} \At+\ddt A_{N}+\ddt \At\\
  \ddt(A_{N}\wedge \At)&=\ddt A_{N} \wedge \At-A_{N}\wedge\ddt \At
\end{align}

\chapter{Lie Algebras}
\label{cha:lie-algebras}

This appendix provides further introductory material for Lie algebras
and fixes the notation that is used in the main sections.
Since it is standard material
this section is neither complete nor are all necessary details provided.
The following references where used and provide further information concerning 
Lie algebra concepts~\cite{Barut:1986dd,HallLie2nd,KnappLieGroups},
cohomology \cite{Azcarraga:2011hqa},
abelian~\cite{Azcarraga:2011hqa}
and nonabelian extensions~\cite{Alekseevsky:2000ext}.

Any Lie algebra, if not mentioned otherwise is assumed to be
real and finite-dimensional.
Furthermore, if Lie algebra brackets or invariant metric components are not
explicitly mentioned they are vanishing.

\section{Basic Concepts of Lie Algebras}
\label{sec:basic-concepts}
\begin{definition}
  A real or complex \textbf{Lie algebra}
  \index{Lie algebra}
  is a real or complex vector space with a map $[\cdot ,\cdot ]:\mathfrak{g} \times \mathfrak{g} \to \mathfrak{g}$ with the following properties:
  \begin{enumerate}
  \item $[\cdot , \cdot]$ is bilinear.
  \item $[\cdot , \cdot]$ is skew-symmetric: $[X,Y]=-[Y,X]$ for all $X,Y \in \mathfrak{g}$.
  \item The Jacobi identity holds
    \begin{equation}
      \label{eq:Jacobiid}
    \underset{XYZ}{\circlearrowleft}[X,[Y,Z]] \equiv [X,[Y,Z]]+[Z,[X,Y]]+[Y,[Z,X]]=0
  \end{equation}
  for all $X,Y,Z \in \mathfrak{g}$.
  \end{enumerate}
\end{definition}
If we choose a basis $\Tb_{a} \in \mathfrak{g}$, where $a=1, \dotsc, \dim \mathfrak{g}$, and use bilinearity the Lie algebra can be written as
\begin{align}
  [\Tb_{a},\Tb_{b}]=f\indices{_{ab}^{c}} \Tb_{c}
\end{align}
where $f\indices{_{ab}^{c}}$ are the structure constants of the Lie algebra $\mathfrak{g}$.
They fully specify a Lie algebra.
Skew-symmetry and the Jacobi identity yield
\begin{align}
  f\indices{_{ab}^{c}}
  &=-f\indices{_{ba}^{c}}
  \\
  \underset{abc}{\circlearrowleft} f\indices{_{ab}^{d}} f\indices{_{cd}^{e}}
   &=0  \,.
\end{align}

A \textbf{homomorphism}
\index{Lie algebra!Homomorphism}
is a linear map $\phi: \mathfrak{g} \to \mathfrak{h}$ with
\begin{align}
  \phi([X,Y]_{\mathfrak{g}})=[\phi(X),\phi(Y)]_{\mathfrak{h}}
  \quad
  \text{ for all }
  \quad
  X, Y \in \mathfrak{g}\,.
\end{align}
An \textbf{isomorphism} between the two Lie algebras is an injective and surjective homomorphism.
\index{Lie algebra!Isomorphism}
For explicit calculations we fix for the Lie algebra $\mathfrak{g}$
the basis and the structure constants by
$[\Gb_{a},\Gb_{b}]=g\indices{_{ab}^{c}} \Gb_c$
and for $\mathfrak{h}$ by 
$[\Hb_{i},\Hb_{j}]=h\indices{_{ij}^{k}} \Hb_{k}$.
The linear map $\phi(\Gb_{a}) = T\indices{_{a}^{i}} \Hb_{i}$ is then a
homomorphism if
\begin{align}
  g\indices{_{ab}^{c}} T\indices{_{c}^{k}}
  =
   T\indices{_{a}^{i}}  T\indices{_{b}^{j}} h\indices{_{ij}^{k}} \,.
\end{align}
For an isomorphism invertibility  leads to
$(T^{-1})\indices{_{i}^{a}} T\indices{_{a}^{j}} = \delta^{j}_{i}$
and therefore 
\begin{align}
h\indices{_{ij}^{k}}
=
(T^{-1})\indices{_{i}^{a}} (T^{-1})\indices{_{j}^{b}}
  g\indices{_{ab}^{c}} T\indices{_{c}^{k}} \,.
\end{align}

When $\mathfrak{a}$ and $\mathfrak{b}$ are subsets of $\mathfrak{g}$, we write
\begin{align}
  [\mathfrak{a},\mathfrak{b}]
  \equiv
  \spn
  \{
  [X,Y]\, | \, X \in \mathfrak{a}, Y \in \mathfrak{b}
  \} \,.
\end{align}
Given a Lie algebra $\mathfrak{g}$ a subspace $\mathfrak{h}$ is a \textbf{subalgebra},
\index{Lie algebra!Subalgebra}
if $[\mathfrak{h},\mathfrak{h}] \subset \mathfrak{h}$,
and an \textbf{ideal} if $[\mathfrak{g},\mathfrak{h}] \subset \mathfrak{h}$.
\index{Lie algebra!Ideal}
If the commutator of all elements of the Lie algebra vanishes,
$[\mathfrak{g},\mathfrak{g}]=0$, then it is called \textbf{abelian}.
\index{Lie algebra!Abelian}
The maximal ideal $\mathfrak{z}$ for which $[\mathfrak{g},\mathfrak{z}] = 0$
is called the \textbf{center} of the Lie algebra.
\index{Lie algebra!Center}

Given an ideal $\mathfrak{h}$ of the Lie algebra $\mathfrak{g}$ the \textbf{quotient algebra} $\mathfrak{g}/\mathfrak{h}$
\index{Lie algebra!Quotient algebra}
is the vector space quotient $\mathfrak{g}/\mathfrak{h}$ with the definition
\begin{align}
  [X +\mathfrak{h}, Y + \mathfrak{h}] = [X,Y] + \mathfrak{h}
  \quad \text{for all} \quad
  X,Y \in \mathfrak{g}\,.
\end{align}

A Lie algebra $\mathfrak{g}$ is a \textbf{direct sum} of Lie algebras,
\index{Lie algebra!Direct sum}
denoted by $\mathfrak{g}= \mathfrak{g}_{1} \dis \mathfrak{g}_{2}$,
if it is a direct sum of vector spaces,
denoted by  $\mathfrak{g}=\mathfrak{g}_{1} \vdis \mathfrak{g}_{2} $
\index{$\dis$}%
\index{$\vdis$}%
and fulfills
\begin{align}
  [\mathfrak{g}_{i},\mathfrak{g}_{i}] \subset \mathfrak{g}_{i} \quad \text{ and } \quad
  [\mathfrak{g}_{i},\mathfrak{g}_{j}] =0 \quad \text{ for } \quad i,j = 1,2 \,.
\end{align}
\textbf{Semidirect sums} are denoted by $\mathfrak{i} \sdis \mathfrak{g}$
\index{Lie algebra!Semidirect sum}
\index{$\sdis$}%
where $\mathfrak{i}$ is an ideal and $\mathfrak{g}$ is a subalgebra, see Appendix \ref{sec:sequences}.

A Lie algebra is \textbf{semisimple} if it has no non-zero commutative ideals
\index{Lie algebra!Semisimple}
and \textbf{simple} if it has dimension bigger than one and no ideals other than $\{0\}$ and the Lie algebra itself.
\index{Lie algebra!Simple}
Semisimple Lie algebras are direct sums of simple ones.
They are \textbf{perfect}, which means that they obey
$[\mathfrak{g},\mathfrak{g}]=\mathfrak{g}$. 
\index{Lie algebra!Perfect}
Not all perfect Lie algebras are semisimple, e.g., there exist semidirect sums that are perfect.

The \textbf{adjoint representation} is given by $\ad : X \in \mathfrak{g} \mapsto  \ad_{X} \in \End (\mathfrak{g})$\footnote{$\End(V)$ denote the endomorphisms of $V$.}
\index{Lie algebra!Adjoint representation}
where $\ad_{X}Y \equiv [X,Y]$
or in a basis $(\ad_{\Tb_{a}})\indices{^{c}_{b}}=f\indices{_{ab}^{c}}$.
On the vector space dual $\mathfrak{g}^{*}$ the \textbf{coadjoint
  representation} is defined by
\index{Lie algebra!Coadjoint representation}
\begin{align}
  \langle \ad^{*}_{X} \alpha , Y \rangle
  \equiv
  -\langle
  \alpha , \ad_{X} Y
  \rangle
\end{align}
where $\alpha \in \mathfrak{g}^{*}$ and $X \in \mathfrak{g}$ and
$\langle \alpha ,X \rangle$ is the value of the linear functional
$\alpha$ evaluated on the vector $X$. The representation can be
written in a basis as
$(\ad^{*}_{\Tb_{a}})\indices{^{c}_{b}}=-f\indices{_{ab}^{c}}$.

Any linear mapping $D: \mathfrak{g} \to \mathfrak{g}$ for which
\begin{align}
  D[X,Y]= [DX,Y] + [X,DY]
\end{align}
is a \textbf{derivation} and is an element of the space of derivations $\der(\mathfrak{g})$.
\index{Derivation}
\index{Lie algebra!Derivation|see {Derivation}}
\index{$\der(\mathfrak{g})$|see {Derivation}}%
An example of a derivation is $\ad_{X}$.
An \textbf{inner derivation} can be written in this
\index{Derivation!Inner derivation}
form, i.e., $D=\ad_{X}$ for some $X \in \mathfrak{g}$.
Derivations for which this is not possible are called
\textbf{outer derivations}.
\index{Derivation!Outer derivation}

\section{Sequences}
\label{sec:sequences-1}

A \textbf{sequence} consists of objects $O_{n}$ and homomorphisms $f_{n}$ between them
\index{Sequence}
\begin{align}
  \dotsb  \rightarrow O_{n} \overset{f_n}{\rightarrow} O_{n+1} \overset{f_{n+1}}{\rightarrow} O_{n+2} \rightarrow \dotsb \,.
\end{align}
The sequence is \textbf{exact} if the image of each homomorphism is
equal to the
\index{Sequence!Exact}
kernel of the next, i.e.,
\begin{align}
  \im(f_{n})= \ker(f_{n+1}) \text{ for all } n \,.
\end{align}
A \textbf{short} exact sequence is an exact sequence with
\index{Sequence!Short exact}
\begin{align}
  0 \rightarrow A \rightarrow B \rightarrow C \rightarrow 0 \,.
\end{align}

\section{Lie Algebra Cohomology}
\label{sec:lie-algebra-cohomology}

Suppose we have a Lie algebra $\mathfrak{g}$
and a vector space $V$ which is an
$\alpha_{\mathfrak{g}}$-module\footnote{
  This means
  $\mathfrak{g}\times V \to V:(X,v)\mapsto \alpha_{X} v$ which satisfies
  $\alpha_{X}(v_{1}+v_{2})=\alpha_{X}v_{1}+\alpha_{X}v_{2}$;
  $\alpha_{X_{1}+X_{2}}v=\alpha_{X_{1}}v+\alpha_{X_{2}}v$
  and
  $\alpha_{[X_{1},X_{2}]}v=[\alpha_{X_{1}},\alpha_{X_{2}}]v$.
}.
An $n$-dimensional $V$-\textbf{cochain} $\omega_{n}$ for
\index{Cochain}
the Lie algebra $\mathfrak{g}$ is a skew-symmetric $n$-linear mapping
  \begin{align}
    \omega_{n}: \underbrace{\mathfrak{g}\times \dots \times \mathfrak{g}}_{n} \to V \,.
  \end{align}
The (abelian) group of all $n$-cochains will be denoted $C^{n}(\mathfrak{g},V)$.

The \textbf{coboundary operator}  $\dco_{n}:C^{n}(\mathfrak{g},V)\to C^{n+1}(\mathfrak{g},V)$ is defined by its action on the cochains by
\index{Coboundary operator}%
\begin{align}
    (\dco \omega_{n})(X_{1},\dots,X_{n+1})
    \equiv \sum_{i=1}^{n+1}(-)^{i+1} \alpha_{X_{i}}(\omega(X_{1},\dotsc, \hat X_{i},\dotsc X_{n+1}))
    \nonumber\\
    \quad+ \sum_{\substack{j,k=1\\ j<k}}^{n+1}(-)^{j+k} \omega([X_{j},X_{k}],X_{1},\dotsc, \hat X_{j}, \dotsc, \hat X_{k}, \dotsc, X_{n+1})
\end{align}
where the hat above the Lie algebra elements means that this element should be omitted.
The coboundary operator has the property that $\dco^{2}=0$.
This can be checked explicitly for the first few cases
\begin{align}
    (\dco \omega_{0})(X)
    &= \alpha_{X}\omega_{0}
    \\
    (\dco \omega_{1})(X_{1},X_{2})
    &=\alpha_{X_{1}} \omega_{1}(X_{2})- \alpha_{X_{2}}\omega_{1}(X_{1})-\omega_{1}([X_{1},X_{2}])
    \\
  (\dco \omega_{2})(X_{1},X_{2},X_{3})
    &=\alpha_{X_{1}} \omega_{2}(X_{2},X_{3})+\alpha_{X_{3}} \omega_{2}(X_{1},X_{2})+\alpha_{X_{2}} \omega_{2}(X_{3},X_{1})
   \nonumber \\
    &- \omega_{2}([X_{1},X_{2}],X_{3})-\omega_{2}([X_{3},X_{1}],X_{2})-\omega_{2}([X_{2},X_{3}],X_{1})
    \nonumber \\
    &=
      \underset{X_{1}X_{2}X_{3}}{\circlearrowleft}
    \left(
    \alpha_{X_{1}} \omega_{2}(X_{2},X_{3})
    -
    \omega_{2}([X_{1},X_{2}],X_{3})
    \right) \,.
\end{align}
Using the coboundary operator one can define the following sequence
\begin{align}
  0 \overset{\dco_{-1}}{\to} C^{0}(\mathfrak{g},V) \overset{\dco_{0}}{\to} C^{1}(\mathfrak{g},V) \overset{\dco_{1}}{\to} \dotsb
\end{align}
and 
furthermore the quotient group $H^{n}_{\alpha}(\mathfrak{g},V)$, called the \textbf{$n$-th cohomology group},
\index{Cohomology group}
\index{Cocyle}
\index{Coboundary}
by
\begin{align}
  H^{n}_{\alpha}(\mathfrak{g},V)
  \equiv \frac{\ker{\dco_{n}}}{\im\dco_{n-1}}
  =\frac{\{n-\text{cocycles}\}}{\{n-\text{coboundarys}\}} \,.
\end{align}
The cohomology group ``measures'' the amount at which the sequence fails to be exact.
When $\alpha$ is trivial we will sometimes omit it and write $H^{n}
\equiv H^{n}_{0}$.

\section{A Sketch of Lie Algebra Extensions}
\label{sec:sequences}

We now use sequences between Lie algebras to define Lie algebra extensions.
\begin{definition}
  The Lie algebra $\mathfrak{e}$ is a  Lie algebra \textbf{extension}
  of $\mathfrak{g}$ by $\mathfrak{h}$ if
  \index{Extension}
  \begin{equation}
    0 \to \mathfrak{h} \overset{i}{\inj} \mathfrak{e}  \overset{\pi}{\surj} \mathfrak{g} \to 0
  \end{equation}
  is a short exact sequence.
  Two extensions are equivalent
  \index{Extension!Equivalent}
  if there exists a Lie algebra isomorphism $\phi$ and the following diagram commutes
  \begin{equation*}
  \begin{tikzcd}
                &  & \mathfrak{e} \arrow[dd,"\phi"]\arrow[dr] &
 \\
    0 \arrow[r] & \mathfrak{h} \arrow[rd] \arrow[ru] & &\mathfrak{g} \arrow[r]& 0
 \\
                &  & \mathfrak{e}' \arrow[ur]&
  \end{tikzcd}\, .
\end{equation*}
\end{definition}
Since the homomorphism $i$ has a trivial kernel it is an injective map.
Furthermore, is the image  isomorphic to the original algebra,
so $\im(i) \simeq \mathfrak{h}$.
When the context is clear and since they are isomorphic
we will often use $\mathfrak{h}$ instead of $\im(i)$.
Another consequence of the definition is that $\pi$ is surjective
and therefore (the image of) $\mathfrak{h}$ is an ideal in $\mathfrak{e}$.
On the other hand there might not exist a subalgebra of $\mathfrak{e}$ that is isomorphic to $\mathfrak{g}$.
But there exists a quotient that leads to  $\mathfrak{e}/\mathfrak{h} \simeq \mathfrak{g}$.
The linear mapping $\tau:\mathfrak{g} \to \mathfrak{e}$ with $\pi \circ \tau= \mathrm{id}_{\mathfrak{g}}$ induces the mappings
\begin{align}
  \label{eq:alphasketch}
  \alpha&:
          \mathfrak{g} \to \der(\mathfrak{h}) \, ,
  &
    \alpha_{X}(H)&= [\tau(X),H]
  \\
  \label{eq:omegtau}
  \omega&:
          \mathfrak{g} \times \mathfrak{g} \to \mathfrak{h} \, ,
  &
    \omega(X,Y)&=[\tau(X),\tau(Y)]- \tau([X,Y])
\end{align}
where $\omega$ is skew-symmetric
and which satisfy
\begin{align}
  \label{eq:ext1}
  [\alpha_{X}, \alpha_{Y}]- \alpha_{[X,Y]}&= \ad_{\omega(X,Y)}
  \\
    \label{eq:ext2}
  \underset{XYZ}{\circlearrowleft}
  \left(
  \alpha_{X}\omega(Y,Z)
  -
  \omega([X,Y],Z)
  \right)
  &=0 \, .
\end{align}
They describe the Lie algebra structure on $\mathfrak{e}= \mathfrak{h} \vdis \tau(\mathfrak{g})$ as
\begin{align}
  [H_{1}+\tau(X_{1}),H_{2}+\tau(X_{2})]
  &=[H_{1},H_{2}] + \alpha_{X_{1}}H_{2} - \alpha_{X_{2}}H_{1}
  \nonumber \\
  &\quad + \tau([X_{1},X_{2}]) + \omega(X_{1},X_{2})  \,.
\end{align}

On the other hand, we can start with two Lie algebras $\mathfrak{g}$ and $\mathfrak{h}$ and maps
$\alpha:\mathfrak{g} \to \der(\mathfrak{h})$ 
and
skew-symmetric $\omega:\mathfrak{g} \times \mathfrak{g} \to \mathfrak{h}$
fulfilling \eqref{eq:ext1} and \eqref{eq:ext2}.
Then on the vector space $\mathfrak{e}= \mathfrak{h} \vdis \mathfrak{g}$ a Lie algebra structure is given by
\begin{align}
  [H_{1}+X_{1},H_{2}+X_{2}]_{\mathfrak{e}}
  &= [H_{1},H_{2}]_{\mathfrak{h}} + \alpha_{X_{1}}H_{2} - \alpha_{X_{2}}H_{1}
  \nonumber \\
  &\quad + [X_{1},X_{2}]_{\mathfrak{g}} + \omega(X_{1},X_{2})  \,.
\end{align}
So a general Lie algebra extension schematically has the form
\begin{align}
    [\mathfrak{g},\mathfrak{g}] &\subset \mathfrak{g} \vdis  \mathfrak{h}
&
    [\mathfrak{h},\mathfrak{h}] &\subset \mathfrak{h}
&
    [\mathfrak{g},\mathfrak{h}] &\subset \mathfrak{h} \,.
\end{align}
An extension is \textbf{trivial} if $\mathfrak{e} \simeq \mathfrak{h}\oplus \mathfrak{g}$,
\index{Extension!Trivial}
which means that it is just the direct sum discussed in Section \ref{sec:basic-concepts},
\begin{align}
    [\mathfrak{g},\mathfrak{g}] &\subset \mathfrak{g} 
&
    [\mathfrak{h},\mathfrak{h}] &\subset \mathfrak{h}
&
    [\mathfrak{g},\mathfrak{h}] &= 0 \,.
\end{align}
Equivalently, this means that $\alpha=\omega=0$.
A \textbf{split} extension is a Lie algebra extension
\index{Extension!Split}
with a homomorphism $\tau:\mathfrak{g} \to \mathfrak{e}$
and $\pi \circ \tau=\mathrm{id}_{\mathfrak{g}}$.
Since $\tau$ is a homomorphism it follows from \eqref{eq:omegtau} that $\omega=0$,
so this extension can be written as
\begin{align}
    [\mathfrak{g},\mathfrak{g}] &\subset \mathfrak{g}
&
    [\mathfrak{h},\mathfrak{h}] &\subset \mathfrak{h}
&
    [\mathfrak{g},\mathfrak{h}] &\subset \mathfrak{h} \,.
\end{align}
Since it is a semidirect sum it will be denoted by $\mathfrak{e} \simeq \mathfrak{h} \sdis \mathfrak{g}$.
The following theorem characterizes
the extensions of simple or one-dimensional Lie algebras.
\begin{theorem}
  \label{thm:split}
  If $\mathfrak{g}$ is simple or one-dimensional, every Lie algebra extension
  \begin{equation}
    0 \to \mathfrak{h} \inj \mathfrak{e} \surj \mathfrak{g} \to 0
  \end{equation}
splits~\cite[Prop. A.1]{FigueroaO'Farrill:1995cy}.
\end{theorem}

A \textbf{central} extension is a Lie algebra extension where
\index{Extension!Central}
$\mathfrak{h}$ is in the center of $\mathfrak{e}$.
It follows that $\mathfrak{h}$ is abelian
and that $\alpha=0$ (see equation \eqref{eq:alphasketch}).
It can be written as
\begin{align}
    &[\mathfrak{g},\mathfrak{g}] \subset \mathfrak{g} \dot + \mathfrak{h}
&
    &[\mathfrak{h},\mathfrak{h}] =0
&
    &[\mathfrak{g},\mathfrak{h}] =0 
\end{align}
and we will denote it by $\mathfrak{g} \dis_{c} \mathfrak{h}$.
\index{$\dis_{c}$}
By definition a split central extension is a trivial extension.
Therefore, as a consequence of Theorem \ref{thm:split},
we have the well known result
(part of  Whitehead's lemma) that
a simple Lie algebra has no nontrivial central extension.

\section{Abelian Lie Algebra Extension}
\label{sec:lie-algebra-extens}
\index{Extension!Abelian}
For a Lie algebra extension by an abelian Lie algebra $\mathfrak{a}$,
i.e., for the short exact sequence
\begin{equation}
  0 \to \mathfrak{a} \overset{i}{\inj} \mathfrak{e}  \overset{\pi}{\surj} \mathfrak{g} \to 0 
\end{equation}
we can make contact with Lie algebra cohomology discussed in  Appendix \ref{sec:lie-algebra-cohomology}.
Because for abelian extensions
$\mathfrak{a}$ is an $\alpha_{\mathfrak{g}}$-module
and therefore
\begin{equation}
[\alpha_{X}, \alpha_{Y}]H = \alpha_{[X,Y]}H \,.
\end{equation}

The coboundary operator acting on $\omega$ vanishes ($\vd\omega=0$)
and therefore $\omega$ is a $2$-cocycle.
Inequivalent extensions differ by $2$-coboundaries and
we obtain the following theorem.
\begin{theorem}
  For a given $\alpha$,
  the classes of equivalent extensions $\mathfrak{e}$ of $\mathfrak{g}$
  by the abelian algebra $\mathfrak{a}$
  are in one-to-one correspondence with the elements of the second cohomology group $H_{\alpha}^{2}(\mathfrak{g},\mathfrak{a})$.
\end{theorem}






\section{Central Extensions}
\label{sec:central-extension-1}
\index{Extension!Central}
A special class of Lie algebra extensions are the central extensions.
For central extensions the Lie algebra structure simplifies,
and can be written as
\begin{align}
  [H_{1}+X_{1},H_{2}+X_{2}]_{\mathfrak{e}}
  = [X_{1},X_{2}]_{\mathfrak{g}} + \omega(X_{1},X_{2})  \,.
\end{align}
Choosing the basis $\Tb_{a}$ for $\mathfrak{g}$
and the basis $\Zb_{\alpha}$ for $\mathfrak{a}$
we can write the commutation relations in form
\begin{align}
  [\Tb_{a},\Tb_{b}]&=f\indices{_{ab}^{c}} \Tb_{c} + \omega\indices{_{ab}^{\alpha}} \Zb_{\alpha}
  &
    [\Tb_{a},\Zb_{\alpha}]&= [\Zb_{\alpha},\Zb_{\beta}]=0
\end{align}
The inequivalent central extensions are given by the
second cohomology group $H^{2}_{0}(\mathfrak{g},\mathfrak{a})$.
Therefore, $\omega$ is a $2$-cocycle which means that it is
\index{Extension!Central!Cocycle}
\index{Extension!Central!Coboundary}
antisymmetric
  \begin{align}
      \omega(X,Y)&=-\omega(Y,X)  & \omega\indices{_{ab}^{\alpha}}&= -\omega\indices{_{ba}^{\alpha}}
  \end{align}
and that $\dco \omega=0$, which leads to
\begin{align}
  \label{eq:cent-cocy}
  \underset{XYZ}{\circlearrowleft}
  \omega([X,Y],Z)&=0
                             &
  \underset{abc}{\circlearrowleft}
  f\indices{_{ab}^{d}}\omega\indices{_{dc}^{\alpha}}&=0 \,.
\end{align}
The last condition also ensures that the Jacobi identities of the whole Lie algebra are satisfied.

Central extensions are seen as equivalent if they differ by a $2$-coboundary
which is given by
\begin{align}
  \label{eq:cent-cob}
  \dco \eta(X,Y)&= - \eta([X,Y]) & \dco \eta\indices{_{ab}^{\alpha}}&=-f\indices{_{ab}^{c}} \eta\indices{_{c}^{\alpha}} \,.
\end{align}
So for a nontrivial central extension necessarily the cocycle should not be given by a coboundary,
i.e., $\omega \neq \dco \eta$.

\subsection{Example:  Canonical Commutation Relations}
\label{sec:example}

We start with an abelian algebra $\mathfrak{g}_{d}$ with the basis
\begin{equation}
  \label{eq:tbasis}
 \Tb_{a}=(q_{1}, \dotsc, q_{d},p_{1},\dotsc, p_{d}).
\end{equation}
So we have the commutation relations
\begin{align}
  [q_{i},p_{j}]=[q_{i},q_{j}]=[p_{i},p_{j}]=0
\end{align}
or equivalently  $f\indices{_{ab}^{c}}=0$.
This means that every skew-symmetric $\omega$ leads to a $2$-cocycle, see equation \eqref{eq:cent-cocy}.
Since all the $2$-coboundaries are trivial, see equation \eqref{eq:cent-cob}, they are all inequivalent.

For the case of $d=1$, spanned by $q$ and $p$, the cohomology group is one-dimensional, $\dim H_{0}^{2}(\mathfrak{g}_{1},\R)=1$.
And the commutation relations of the nontrivial central extension are the canonical commutation relations
\begin{align}
  [q,p]&=\omega Z & [q,q]&=[p,p]=0
\end{align}
where $\omega \neq 0$.
For arbitrary dimension $d$ every skew symmetric $\omega\indices{_{ab}^{\alpha}}$ is possible.
So $\dim H^{2}_{0}(\mathfrak{g}_{d},\R)=d(2 d-1)$.

\chapter{Useful Formulas}
\label{sec:useful}

\section{\texorpdfstring{Details: Solutions of $F=0$}{Details: Solutions of F=0}}
\label{sec:deta-solut-f=0}

To show that $F=\dd A+ A \wedge A=0$ is solved by $A= g^{-1} \dd g$
one uses $\dd (g^{-1} g)=\dd g^{-1} g+ g^{-1} \dd g=\dd (1) = 0$ to derive
\begin{align}
  \dd g^{-1} =- g^{-1} \dd g g^{-1} \,.
\end{align}
Then we just insert it in the third line of
\begin{align}
  \dd A &= \dd (g^{-1} \dd g )
  \\
        &= \dd g^{-1} \wedge \dd g
  \\
        &= - g^{-1} \dd g \wedge g^{-1} \dd g
  \\
        &= - A \wedge A
\end{align}
from which the flatness condition on the connection can be read of.




\section{Finite Gauge Transformation}
\label{sec:finite-gauge-transf-1}

For finite gauge transformations
\begin{equation}
  \label{eq:gaugetrafa}
  A \to g^{-1}A g + g^{-1}\dd g
\end{equation}
the action transforms as
\begin{align}
  \label{eq:CSfinitegauge}
  \CS[A]&\rightarrow \CS[A]-\frac{1}{3} \langle (g^{-1} \dd g)^{3}\rangle -\dd \langle A \wedge g\dd g^{-1}\rangle \,.
\end{align}
This can be seen using the following useful formulas.
As already seen in Section \ref{sec:what-chern-simons} we define
\begin{align}
  \label{eq:ABCdef}
  \langle A \wedge B \wedge C \rangle \equiv  \frac{1}{2} \langle[A,B]\wedge C\rangle \, ,
\end{align}
which satisfies
\begin{align}
  \langle[A,B]\wedge C\rangle=\langle[B,A]\wedge C\rangle=\langle B\wedge [A,C]\rangle= \langle A \wedge [B,C] \rangle \,.
\end{align}
Since \eqref{eq:ABCdef} is symmetric under any permutation of $A$, $B$
and $C$ we will, if convenient, omit the wedge product. Using
\begin{align}
  \label{eq:gdgm1}
  \dd(g g^{-1}) &= \dd g g^{-1}+ g \dd g^{-1}=0
  \\
  \alpha &= \dd g g^{-1}=-g \dd g^{-1}
  \\
  \dd g&=\alpha g \quad \dd g^{-1}=-g^{-1}\alpha
\end{align}
and
\begin{align}
  \label{eq:62}
  A &\rightarrow g^{-1}(A+\dd)g=g^{-1}(A+\alpha)g
  \\
  \dd A &\rightarrow g^{-1}(\dd A-\alpha^{2}-\alpha A-A \alpha )g
  \\
  \langle \dd A \wedge A \rangle
    & \to \langle  \dd A A + \dd A \alpha - \alpha^{3} - 3 A \alpha^{2} - 2 A^{2} \alpha \rangle
  \\
  \langle A^{3} \rangle &\rightarrow \langle A^{3}+\alpha^{3}+3 A\alpha^{2}+ 3 A^{2}\alpha \rangle
\end{align}
leads to \eqref{eq:CSfinitegauge}.

Equivalently, one can express the CS Lagrangian in terms of its curvature
\begin{align}
  \label{eq:CScurvact}
  \CS[A]
  &= \langle F  \wedge A - \frac{1}{6} [A,A] \wedge A \rangle  
\end{align}
where one can use that curvature transforms as 
\begin{align}
  F \to g^{-1} F g \,.
\end{align}

\section{Infinitesimal Gauge Transformations}
\label{sec:deta-infin-gauge}

The infinitesimal gauge (like) transformation
\begin{align}
 \vd_{\lambda}A= \Dd \lambda \equiv \dd \lambda+[A,\lambda]
\end{align}
is an infinitesimal divergence symmetry of $I_{\CS}$
\begin{align}
  \vd_{\lambda} \CS[A]
  =\dd
    \langle 
      \lambda \wedge \dd A 
    \rangle  \,.
\end{align}

The explicit calculation is given by
\begin{align}
  \vd_{\lambda} \CS [A]
  &=\langle \dd\! \Dd \lambda \wedge A + \dd A \wedge \Dd \lambda + 2 A \wedge A \wedge \Dd \lambda \rangle
  \\
  &=\langle ([\dd A, \lambda] - [A,\dd \lambda])\wedge A
    + \dd A \wedge \dd \lambda+\dd A \wedge [A , \lambda]
  \nonumber\\
  & \quad + [A ,A] \wedge \dd \lambda+[A ,A] \wedge [A,\lambda] \rangle
   \\
  &=\langle - \dd A \wedge [A,\lambda]- [A,A]\wedge  \dd \lambda
    + \dd A \wedge \dd \lambda+\dd A \wedge [A , \lambda]
  \nonumber  \\
  & \quad + [A ,A] \wedge \dd \lambda-\lambda \wedge [A ,[A,A]] \rangle
  \\
  &=\dd
    \langle 
      \lambda \wedge \dd A 
    \rangle
\end{align}
where $[A ,[A,A]]=0$ using the Jacobi identity.

\section{Infinitesimal Diffeomorphisms}
\label{sec:diffeomorphisms}

That the CS action is invariant under diffeomorphisms
is evident from the fact that it is a (covariant) differential form.
Infinitesimal diffeomorphisms are given by the Lie derivative
\begin{align}
  \vd_{\xi}A= \Ld_{\xi} A= \ic_{\xi}(\dd A) + \dd (\ic_{\xi}A)
\end{align}
and lead to an infinitesimal divergence symmetry
\begin{align}
  \vd_{\xi} \CS[A]
  &=\Ld_{\xi}\CS[A]= \dd (\ic_{\xi} \CS[A]) \,.
\end{align}




\section{Diffeomorphisms as Gauge Transformations}
\label{sec:diff-as-gauge}

On-shell an  infinitesimal diffeomorphism generated by $\xi$ can be written as a gauge transformation~\cite{Witten:1988hc} defined by
\begin{align}
 \vd_{\lambda}A= \Dd \lambda \equiv \dd \lambda+[A,\lambda]
\end{align}
since with the gauge parameter given by $\lambda=\ic_{\xi} A = \xi^{\mu} A_{\mu}$
we get
\begin{align}
  \vd_{\xi}A
  &=\dd \ic_{\xi}A + \ic_{\xi}\dd A
  \\
  &= \dd \ic_{\xi}A - \ic_{\xi}(A \wedge A) + \ic_{\xi} F
  \\
  &=\dd \ic_{\xi} A + [A, \ic_{\xi}A ]+\ic_{\xi} F
  \\
  &\os \Dd (\ic_{\xi} A) \,.
\end{align}
Or said differently, gauge transformations with the gauge parameter $\lambda=\ic_{\xi} A$
should be regarded as diffeomorphisms.



\chapter{Explicit Lie Algebra Relations}
\label{cha:explicit-lie-algebra}

\section{\texorpdfstring{$\mathfrak{sl}(2,\R) \simeq
    \mathfrak{so}(2,1)$}{sl(2,R) = sl(2,1)}}
\label{sec:sl2-r}

The simple real Lie algebra $\mathfrak{sl}(2,\R)$ is given by the
commutation relations
\begin{align}
  \label{eq:sl2app}
  [\Lt_{a},\Lt_{b}]=(a-b) \Lt_{a+b}
\end{align}
where $a,b={-1,0,+1}$.
A defining representation are tracefree $2 \times 2$ matrices
\begin{align}
  \label{eq:sl2}
  \Lt_{-1}&=
  \begin{pmatrix}
    0 & 1\\
    0 & 0    
  \end{pmatrix}
&  \Lt_{0}&=\frac{1}{2}
  \begin{pmatrix}
    1&0 \\
    0 & -1    
  \end{pmatrix}
&  \Lt_{+1}&=
  \begin{pmatrix}
    0 & 0\\
    -1 & 0    
  \end{pmatrix}
\end{align}
for which the trace defines an invariant metric
\begin{align}
  \label{eq:sl2tr}
    \tr (\Lt_{a}\,\Lt_{b})=
    \left(
    \begin{array}{c|ccc}
                 & \Lt_{-1}       & \Lt_0       & \Lt_{+1}    \\
      \hline
      \Lt_{-1}   & 0              & 0           & -1          \\
      \Lt_0      & 0              & \frac{1}{2} & 0           \\
      \Lt_{+1}   & -1             & 0           & 0           \\
    \end{array}
  \right)\,.
\end{align} 
This metric is, like every invariant metric of a simple Lie algebra, proportional to the
Killing form, $\kappa_{ab}=4 \tr (\Lt_{a}\,\Lt_{b})$.
This can be verified by using the  
adjoint representation
\begin{align}
  \label{eq:30}
  \ad_{\Lt_{-1}} & =
  \begin{pmatrix}
 0               & -1             & 0                         \\
 0               & 0              & -2                        \\
 0               & 0              & 0    
  \end{pmatrix}  & 
\ad_{\Lt_{0}}    & =
\begin{pmatrix}
 1               & 0              & 0                         \\
 0               & 0              & 0                         \\
 0               & 0              & -1 
\end{pmatrix}    & 
\ad_{\Lt_{+1}}   & =
\begin{pmatrix}
 0               & 0              & 0                         \\
 2               & 0              & 0                         \\
 0               & 1              & 0
\end{pmatrix}
\end{align}
and the definition of the Killing form
$\kappa_{ab}=\tr(\ad_{\Lt_{a}} \ad_{\Lt_{b}})$.
Since the Killing form is nondegenerate this algebra is simple.

The Lie algebra $\mathfrak{so}(2,1)$ is defined by $3\times 3$
matrices $M$, which have to satisfy $M=- \eta \cdot M^{T} \cdot \eta$ where the
superscript $T$ denotes transpose and
$\eta=\mathrm{diag}(-1,1,1)$.
We use $\eta=\mathrm{diag}(-1,1,1)$ and $\epsilon_{012}=1$ and simplify the above expressions to get
\begin{align}
      [\Jt_{A},\Jt_{B}]=\epsilon\indices{_{ABC}}\eta^{CD}\Jt_{D}=\epsilon\indices{_{AB}^{C}}\Jt_{C}
\end{align}
and
\begin{align}
  \label{eq:30a}
  \Jt_{0}       & =
  \begin{pmatrix}
 0              & 0 & 0  \\
 0              & 0 & -1 \\
 0              & 1 & 0    
  \end{pmatrix} & 
\Jt_{1}         & =
\begin{pmatrix}
 0              & 0 & -1 \\
 0              & 0 & 0  \\
 -1             & 0 & 0    
\end{pmatrix}   & 
\Jt_{2}         & =
\begin{pmatrix}
 0              & 1 & 0  \\
 1              & 0 & 0  \\
 0              & 0 & 0    
\end{pmatrix} \,.
\end{align}
Again the Lie algebra permits an invariant metric given by
\begin{align}
  \langle \Jt_{A},\Jt_{B} \rangle = \eta_{AB}
\end{align}
which is related to the trace and the Killing form by
\begin{equation}
\langle
\Jt_{A},\Jt_{B} \rangle = 2 \tr (\Jt_{A},\Jt_{B}) = 2 \kappa_{AB} \,.
\end{equation}

The isomorphism between $\mathfrak{sl}(2,\R)$ and $\mathfrak{so}(2,1)$
is given by
\begin{align}
  \label{eq:isoslso}
  \Jt_{0}&=-\frac{1}{2} (\Lt_{+1}+ \Lt_{-1})
  &
    \Jt_{1}&=-\frac{1}{2} (\Lt_{+1}- \Lt_{-1})
  &
    \Jt_{2}&=-\Lt_{0} \,.
\end{align}
We should note that the invariant metric given using the isomorphism
and the invariant metric \eqref{eq:sl2tr} are related by
$\langle \Jt_{a},\Jt_{b} \rangle= 2\langle \Jt_{a},\Jt_{b} \rangle_{\text{iso}}$.

\section{\texorpdfstring{$\mathfrak{sl}(3,\R)$}{sl(3,R)}}
\label{sec:sl3-r}

\begin{align}
  \left[\Lt_{i},\Lt_{j}\right]
  & =\left(i-j\right)\Lt_{i+j}
    \label{eq:LLsl3}  \\
  \left[\Lt_{i},\Wt_{m}\right]
  & =\left(2i-m\right)\Wt_{i+m}
    \label{eq:LWsl3}  \\
  \left[\Wt_{m},\Wt_{n}\right]
  &=-\frac{\sigma}{3}\left(m-n\right)\left(2m^{2}+2n^{2}-mn-8\right)\Lt_{m+n}.
    \label{eq:WWsl3}
\end{align}
with $i,j=-1,0,1$ and $m,n=-2,-1,0,1,2$. With our conventions
The constant $\sigma$ is restricted to be positive for
$\mathfrak{sl}(3,\R)$  while negative $\sigma$ would lead to $\mathfrak{su}(1,2)$.
With our conventions $\sigma=1$ for \cite{Bergshoeff:2016soe,Ammon:2017vwt,Grumiller:2016kcp,Bunster:2014mua,deBoer:2014fra},
and $\sigma_{\mathrm{here}}=-\sigma_{\mathrm{there}}$ for \cite{Campoleoni:2010zq}.

A matrix representation for is given by
\begin{align}
  \label{eq:sl2matr}
\Lt_{-1}&=\begin{pmatrix}
0 & \sqrt{2} & 0\\
0 & 0 & \sqrt{2}\\
0 & 0 & 0
\end{pmatrix}
             &
             \Lt_{0}=&\begin{pmatrix}
1 & 0 & 0\\
0 & 0 & 0\\
0 & 0 & -1
\end{pmatrix}
             &
             \Lt_{1}&=\begin{pmatrix}
0 & 0 & 0\\
-\sqrt{2} & 0 & 0\\
0 & -\sqrt{2} & 0
\end{pmatrix}
\end{align}
and
\begin{align}
                \Wt_{-2}&=\sqrt{4 \sigma}\begin{pmatrix}0 & 0 & 2\\
0 & 0 & 0\\
0 & 0 & 0
\end{pmatrix}
        &
          \Wt_{-1}&= \frac{\sqrt{4 \sigma}}{\sqrt{2}}
                    \begin{pmatrix}0 & 1 & 0\\
                      0 & 0 & -1 \\
                      0 & 0 & 0
                    \end{pmatrix}
             &
               \Wt_{0}&=\frac{\sqrt{4 \sigma}}{3}
                        \begin{pmatrix}1 & 0 & 0\\
                          0 & -2 & 0\\
                          0 & 0 & 1
                        \end{pmatrix}
\nonumber  \\
    \Wt_{1}&= \frac{\sqrt{4 \sigma}}{\sqrt{2}}
           \begin{pmatrix}0 & 0 & 0\\
             -1 & 0 & 0\\
             0 & 1 & 0
           \end{pmatrix}
                     &
        \Wt_{2}
   &=\sqrt{4 \sigma}
     \begin{pmatrix}0 & 0 & 0\\
       0 & 0 & 0\\
       2 & 0 & 0
     \end{pmatrix}\;.  
\end{align}
\begin{align}
  \label{eq:sl3inv}
 \langle \Tt_{a}\Tt_{b} \rangle=\left(
\begin{array}{c|ccc|ccccc}
  & \Lt_{-1} & \Lt_0 & \Lt_1 & \Wt_{-2} & \Wt_{-1} & \Wt_0 & \Wt_1 & \Wt_2 \\
\hline
 \Lt_{-1} & 0 & 0 & -1 & 0 & 0 & 0 & 0 & 0 \\
 \Lt_0 & 0 & \frac{1}{2} & 0 & 0 & 0 & 0 & 0 & 0 \\
 \Lt_1 & -1 & 0 & 0 & 0 & 0 & 0 & 0 & 0 \\
\hline
 \Wt_{-2} & 0 & 0 & 0 & 0 & 0 & 0 & 0 & 4  \sigma\\
 \Wt_{-1} & 0 & 0 & 0 & 0 & 0 & 0 & -\sigma& 0 \\
 \Wt_0 & 0 & 0 & 0 & 0 & 0 & \frac{2}{3} \sigma & 0 & 0 \\
 \Wt_1 & 0 & 0 & 0 & 0 & - \sigma& 0 & 0 & 0 \\
 \Wt_2 & 0 & 0 & 0 & 4 \sigma& 0 & 0 & 0 & 0 \\
\end{array}
\right)
\end{align}

The invariant metric is proportional to the trace and the Killing form
in the following form
$\langle \Tt_{a}\Tt_{b} \rangle = \frac{1}{4} \tr(\Tt_{a}\Tt_{b}) =
\frac{1}{24} \kappa_{ab}$.

There is another useful form to write $\mathfrak{sl}(3,\R)$
which makes its interpretation as spin-2 and spin-3 fields more obvious~\cite{Campoleoni:2010zq}.
One introduces symmetric and traceless generators $\Jt_{AB}$, i.e.,
\begin{align}
  & \Jt_{AB} =\Jt_{BA}\,,     & \eta^{AB} \Jt_{AB} = 0 
\end{align}
and defines the Lie algebra
\begin{align}
  [\Jt_{A},\Jt_{B}]
  &=
    \epsilon\indices{_{AB}^{C}}\Jt_{C}
  \\
  [ \Jt_A , \Jt_{BC}]
  & =
    \epsilon\indices{^M_{A(B}}\,  \Jt_{C)M}
  \\
  [ \Jt_{AB} , \Jt_{CD}]
  &=
    -\sigma \eta_{(A(C} \epsilon_{D)B)M}  \, \Jt^M \,,
\end{align}
It permits the invariant metric
\begin{align}
  \langle \Jt_A, \Jt_B\rangle &= \eta_{AB}
 \label{eq:sl3AA} \\
  \langle \Jt_A, \Jt_{BC} \rangle &= 0
 \label{eq:sl3ABC} \\
  \langle \Jt_{AB} , \Jt_{CD}\rangle& = \sigma (\eta_{A(C} \eta_{D)B} - \tfrac{2}{3} \eta_{AB} \eta_{CD}) \,.
  \label{eq:sl3ABCD}
\end{align}

The isomorphism to the basis given by \eqref{eq:LLsl3} to
\eqref{eq:WWsl3} is given by \eqref{eq:isoslso} combined with
\begin{align}
\Jt_{00} \, &= \, \frac{1}{4} \left( \Wt_2 + \Wt_{-2} + 2\, \Wt_0
              \right) \, , &  \Jt_{01} \, &= \,  \frac{1}{4} \left(
                                            \Wt_2 - \Wt_{-2} \right)
  \\
\Jt_{11} \, &= \, \frac{1}{4} \left( \Wt_2 + \Wt_{-2} - 2\, \Wt_0
              \right) \, , &  \Jt_{02} \, &= \,  \frac{1}{2} \left(
                                            \Wt_1 + \Wt_{-1} \right)
  \\
\Jt_{22} \, &= \, \Wt_0 \, , & \Jt_{12} \,& = \,  \frac{1}{2} \left( \Wt_1 - \Wt_{-1} \right) \, .
\end{align}
This transformation shows explicitly that $\Wt_{m}$ automatically satisfies the traceless condition
\begin{align}
  -\Jt_{00}+\Jt_{11}+\Jt_{22}=0 \,.
\end{align}
The invariant metric given by \eqref{eq:sl3AA} to
\eqref{eq:sl3ABCD} is rescaled by two with respect to the invariant
metric given by the isomorphism and using \eqref{eq:sl3inv}, e.g., 
$\langle \Jt_{AB} , \Jt_{CD}\rangle = 2 \langle \Jt_{AB} , \Jt_{CD}\rangle_{\text{iso}}$.

\section{\texorpdfstring{Principal $\mathfrak{sl}(N,\R)$}{Principal sl(N,R)}}
\label{sec:slN-r}

The conventions are the ones used in \cite{Castro:2011iw}
with the difference that a conventional positive constant $\sigma$ is introduced.

The $\mathfrak{sl}(N,\mathbb{R})$ with a principally embedded $\mathfrak{sl}(2,\mathbb{R})$ have generators of spin $s = 2, 3, \ldots, N$. The generators $\{\Lt_0, \Lt_{\pm 1}\}$ label the $sl(2,\mathbb{R})$ subalgebra, while the higher spin generators are denoted by $\Wt_m^{(s)}$ for $m = -(s-1),\ldots, 0 ,\ldots, s-1$. The algebra in this representation is
\begin{align}
  [ \Lt_i , \Lt_j]           & = (i-j) \Lt_{i+j}\,,
                                                                                                                       \\
  [\Lt_i, \Wt_m^{(s)}]       & = \left( i(s-1)- m\right) \Wt_{i+m}^{(s)}\,.
\end{align}
and additional commutators for $[\Wt_m^{(s)},\Wt_{n}^{(t)}]$.
We take the $N$-dimensional generators of the principally embedded
$sl(2,\mathbb{R})$, denoted as $\Lt_i$ to be
\begin{align}
  \left(\Lt_{1}\right)_{jk}  & =-\sqrt{ j\left(N-j\right)}\delta_{j+1,k}
                                                                                                                       \\
  \left(\Lt_{-1}\right)_{jk} & =\sqrt{ k\left(N-k\right)}\delta_{j,k+1}
                                                                                                                       \\
  \left(\Lt_{0}\right)_{jk}  & =\frac{1}{2}\left(N+1-2j\right)\delta_{j,k}\;,
\end{align}
or explicitly
\begin{align}
  \Lt_1  = -                 & 
                           \begin{pmatrix}
0                            &               & \cdots        &               &               &            & 0          \\
\sqrt{N-1}                   & 0	     &               &               &               &            &            \\
0                            & \sqrt{2(N-2)} & 0             &               &               &            &            \\
                             &               & \ddots        & \ddots        &               &            & \vdots     \\
\vdots                       &               &               & \sqrt{k(N-k)} & 0             &            &            \\
                             &               &               &               & \ddots        & \ddots     &            \\
0                            &               & \cdots        &               &               & \sqrt{N-1} & 0          \\
\end{pmatrix}\,,                                                                                                       \\
  \Lt_{-1}  =                & 
                           \begin{pmatrix}
0                            & \sqrt{N-1}    &               & \cdots        &               &            & 0          \\
                             & 0	     & \sqrt{2(N-2)} &               &               &            &            \\
                             &               & \ddots        & \ddots        &               &            &            \\
\vdots                       &               &               & 0             & \sqrt{k(N-k)} &            & \vdots     \\
                             &               &               &               & \ddots        & \ddots     &            \\
                             &               &               &               &               & 0	  & \sqrt{N-1} \\
0                            &               &               & \cdots        &               &            & 0          \\
\end{pmatrix} \,,                                                                                                      \\
  \Lt_0  = \frac12           & 
                           \begin{pmatrix}
(N-1)                        & 0             &               & \cdots        &               &            & 0          \\
0                            & (N-3)         &               &               &               &            &            \\
                             &               & \ddots        &               &               &            &            \\
\vdots                       &               &               & (N+1-2k)      &               &            & \vdots     \\
                             &               &               &               & \ddots        &            &            \\
                             &               &               &               &               & -(N-3)     & 0          \\
0                            &               &               & \cdots        &               & 0          & -(N-1)     \\
\end{pmatrix}\,.
\end{align}
The normalization from this choice of generators is
\begin{equation}
\tr (\Lt_0 \Lt_0) = \frac{1}{12}N(N^2-1)\,.
\end{equation}
The representation for the higher spin generators follows from
\begin{align}
\Lt_m^{(s)} &=(\sqrt{4 \sigma})^{1-\delta_{s,2}} (-1)^{s+m-1}\frac{(s+m-1)!}{(2s-2)!} \underbrace{[\Lt_{-1}, [\Lt_{-1}, \ldots [\Lt_{-1}}_\text{$s-m-1$ terms}, (\Lt_1)^{s-1}] \ldots ]] \,.
  \\
  &=(\sqrt{4 \sigma})^{1-\delta_{s,2}} (-1)^{s+m-1}\frac{(s+m-1)!}{(2s-2)!}  (\ad_{\Lt_{-1}})^{s-m-1} (\Lt_1)^{s-1} \,.
\end{align}
where the $(\sqrt{4 \sigma})^{1-\delta_{s,2}}$ term is added such
that the definitions are still true for $s=2$.
The matrices obey the hermiticity property
\begin{eqnarray}
\Lt_{i}^{\dagger} & = & \left(-1\right)^{i}\Lt_{-i},\\
(\Lt_{m}^{(s)})^{\dagger} & = & \left(-1\right)^{m}\Lt_{-m}^{(s)}.
\end{eqnarray}
The trace of the matrix representation given above is given by\footnote{
  It is called ``Killing Cartan form'' in \cite{Castro:2011iw},
  but this is not the Killing form as defined here.
}
\begin{equation}
\tr (\Lt_m^{(s)} \Lt_n^{(t)} ) =(4 \sigma)^{1-\delta_{s,2}} t_m^{(s)} \delta^{s,t} \delta_{m,-n}\,,
\end{equation}
with
\begin{equation}
t_m^{(s)} = (-1)^m \frac{(s-1)!^2(s+m-1)!(s-m-1)!}{(2s-1)!(2s-2)!} N \prod_{i=1}^{s-1} (N^2 - i^2)\,.
\end{equation}
The relationship between the Killing form $\kappa$ and the invariant
metric given by the trace in the fundamental $n \times n$ matrix
representation for $\mathfrak{sl}(N,\R)$ is
\begin{align}
  \kappa(x,y)= 2 N \tr (xy) \,.
\end{align}
A normalization where the $\mathfrak{sl}(2,\R)$ sector
is in agreement with \eqref{eq:sl2tr} is given by
\begin{align}
  \langle \Lt_m^{(s)} \Lt_n^{(t)} \rangle
  = \frac{24}{N(N^{2}-1)}\tr (\Lt_m^{(s)}\Lt_n^{(t)} )
\end{align}

\section{\texorpdfstring{$\mathfrak{hs}[\lambda]$}{hs[lambda]}}
\label{sec:hs}

We define here the infinite-dimensional Lie algebra $\mathfrak{hs}[\lambda]$.
The finite-dimensional algebra $\mathfrak{sl}(N,\R)$ is then given by a Lie algebra quotient thereof.
We will provide an invariant metric for both algebras as well as the commutators for spins $s \leq 4$ of $\mathfrak{hs}[\lambda]$.

The generators of $\mathfrak{hs}[\lambda]$ are given by 
\begin{align}
  \Lt_{n}^{(s)}, \quad  s\geq2 , \quad |n|<s \,.
\end{align}
With the notation used in the previous sections $\Lt_{n}^{(2)}=\Lt_{n}$ and $\Lt_{n}^{(3)}=\Wt_{n}$.
Using the contraction described in the preceding subsection we can use the commutation relations of $\mathfrak{hs}[\lambda]$~\cite{Feigin:1988,Bordemann:1989zi,Bergshoeff:1989ns,Pope:1989sr,Fradkin:1990ir}\footnote{%
The commutation relations were explicitly given in \cite{Pope:1989sr}. Our structure constants are divided by four with respect to the ones given in \cite{Gaberdiel:2011wb}, but we otherwise closely follow \cite{Gaberdiel:2011wb} (see also \cite{Henneaux:2010xg,Campoleoni:2011hg,Ammon:2011ua}).
}
  \begin{align}
    [\Lt_{n}^{(s)},\Lt_{m}^{(t)}]&=\sum_{ \stackrel{u=2}{\mathrm{ even}}}^{s+t-1}  g_u^{st}(n,m;\lambda) \,\Lt_{n+m}^{(s+t-u)}
  \end{align}
where
\begin{subequations}
  \begin{align}
    g_u^{st}(n,m;\lambda) &= {\frac{q^{u-2}}{2(u-1)!}} \phi_u^{st}(\lambda) N_u^{st}(n,m)
                            \\
    N_u^{st}(n,m) &=  \sum_{k=0}^{u-1}(-1)^k
                    \binom{u-1}{k}
                    [s-1+n]_{u-1-k}[s-1-n]_k
                    \nonumber \\
                    & \phantom{\sum_{k=0}^{u-1}}\times [t-1+m]_k[t-1-m]_{u-1-k}
    \\
    \phi_u^{st}(\lambda) &= \ _4F_3\left[
                           \begin{array}{c|}
                             \frac{1}{2} + \lambda \ ,  \ \frac{1}{2} - \lambda \ , {\frac{2-u}{2}} \ , {\frac{1-u}{2}}\\
                             {\frac{3}{2}}-s \ , \ {\frac{3}{2}} -t\ , \ \frac{1}{2} + s+t-u
                           \end{array}  \ 1\right] \, .  
  \end{align}
\end{subequations}
The number $q$ is a normalization factor that can be set to any fixed value (for more details see Appendix A in \cite{Gaberdiel:2011wb}).
The falling factorial or Pochhammer symbol is given by 
\begin{align}
[a]_{n}= a (a-1) (a-2) \cdots (a-n+1) =\frac{a!}{(a-n)!} = \frac{\Gamma(a+1)}{\Gamma(a+1-n)}
\end{align}
the rising factorial or Pochhammer symbol is given by 
\begin{align}
  (a)_{n}= a (a+1) \cdots (a+n-1)= \frac{(a+n-1)!}{(a-1)!}= \frac{\Gamma(a+n)}{\Gamma(a)}
\end{align}
with $(a)_{0}=[a]_{0}=1$.
The generalized hypergeometric function $_{m}F_{n}(z)$ is defined by 
\begin{align}
  _{m}F_{n}
  \left[
  \begin{array}{c|}
    a_{1},\dots,a_{m} \\
    b_{1},\dots,b_{n}
  \end{array} \  z
  \right]
  = \sum_{k=0}^{\infty}\frac{(a_{1})_{k} (a_{2})_{k} \dots (a_{m})_{k}}{(b_{1})_{k} (b_{2})_{k} \dots (b_{n})_{k}} \frac{z^{k}}{k!} \,.
\end{align}
The infinite-dimensional Lie algebra $\mathfrak{hs}[\lambda]$ possesses an invariant metric given by 
\begin{subequations}
\label{eq:invmeihs}
  \begin{align}
    \langle \Lt^{(s)}_n \Lt^{(t)}_m \rangle &\equiv \frac{3}{4 q (\lambda^2-1)} g^{s t}_{s+t-1}(n,m,\lambda)\\
                                            &=N_s  \frac{(-1)^{s-n-1}}{4 (2s-2)!}\Gamma(s+n)\Gamma(s-n)  \delta^{st}\delta_{n,-m} \notag
  \end{align}
\end{subequations}
with
\begin{align}
    N_s &\equiv {3 \cdot 4^{s-3}\sqrt{\pi}q^{2s-4}\Gamma(s)\over (\lambda^2-1)
          \Gamma(s+\frac{1}{2})} (1-\lambda)_{s-1} (1+\lambda)_{s-1} \ . \label{appN}
\end{align}
The overall constant has been chosen so that
\begin{align}
\langle \Lt^{(2)}_{0} \Lt^{(2)}_{0} \rangle= \frac{1}{2}
\end{align}
which ensures that the $\mathfrak{sl}(2,\R)$ sector agrees with
\eqref{eq:sl2tr}.

\subsection{\texorpdfstring{From $\mathfrak{hs}[\lambda]$ to $\mathfrak{sl}(N,\R)$}{From hs[lambda] to sl(N,R)}}
\label{sec:hstosl}

Using $\mathfrak{hs}[\lambda]$ one can define $\mathfrak{sl}(N,\R)$ as a Lie algebra quotient. This is only possible for $\lambda=N$ since this leads to an ideal $\chi_{N}$~\cite{Feigin:1988,Vasiliev:1989re,Fradkin:1990qk} spanned by $\Lt^{(s)}_{n}$ with $s>N$. Using this ideal we can then define the finite-dimensional algebra $\mathfrak{sl}(N,\R)$ by the quotient
\begin{align}
  \mathfrak{sl}(N,\R)=\mathfrak{hs}[N]/\chi_{N} \,.
\end{align}
The invariant metric, equation \eqref{eq:invmeihs} with $\lambda=N$, stays an invariant metric for $\mathfrak{sl}(N,\R)$. It is  zero for higher spins. In the next section this can be seen explicitly.

\subsection{\texorpdfstring{Commutators of $\mathfrak{hs}[\lambda]$ for $s \leq 4$}{Commutators of hs[lambda] for s >= 4}}
\label{sec:first-commutators}

We list here the commutators for $s \leq 4$ of $\mathfrak{hs}[\lambda]$ (with $q=1/4$)\footnote{%
A \texttt{Mathematica} workbook that reproduces the commutation relations and might be useful for further checks is uploaded with \cite{Ammon:2017vwt} as an ancillary file on the \texttt{arxiv} server.
}
\begin{subequations}
  \begin{align}
    [\Lt_{n}^{(2)},\Lt_{m}^{(2)}]&=(n-m) \Lt^{(2)}_{n+m}
    \\
    [\Lt_{n}^{(2)},\Lt_{m}^{(3)}]&=(2n-m) \Lt^{(3)}_{n+m}
    \\
    [\Lt_{n}^{(3)},\Lt_{m}^{(3)}]&=-\frac{1}{60}(\lambda^{2}-4) (n-m) (2 n^{2}-n m + 2m^{2} - 8 )  \Lt^{(2)}_{n+m}
                                   \nonumber \\
                                 &\quad + 2 (n-m) \Lt^{(4)}_{n+m}
    \\
    [\Lt_{n}^{(2)},\Lt_{m}^{(4)}]&=(3n-m) \Lt^{(4)}_{n+m} 
    \\
    [\Lt_{n}^{(3)},\Lt_{m}^{(4)}]&=-\frac{1}{70} (\lambda^{2}-9) (5 n^{3}-5n^{2}m -17 n + 3n m^{2} + 9 m - m^{3}) \Lt^{(3)}_{n+m} 
                                   \nonumber \\
                                 & \quad + (3 n - 2m) \Lt^{(5)}_{n+m}
    \\
    [\Lt_{n}^{(4)},\Lt_{m}^{(4)}]&=(\lambda^{2}-4)(\lambda^{2}-9) (n-m) f(n,m) \Lt^{(2)}_{n+m}
                                   \nonumber  \\
                                 & \quad -\frac{1}{30}(\lambda^{2}-19) (n-m) (n^{2}-nm + m^{2}-7) \Lt^{(4)}_{n+m}
                                   \nonumber  \\
                                 & \quad +3 (n-m) \Lt^{(6)}_{n+m}
  \end{align}
\end{subequations}
with
\begin{equation}
 f(n,m)=\tfrac{1}{8400}\left[3n^4 + 3m^4 -2 n m (n-m)^2  - 39(n^2 + m^2) + 20 n m + 108\right] .
\end{equation}
The invariant metric for $s \leq 4$ is given by the anti-diagonal matrices
\begin{subequations}
  \begin{align}
    \langle \Lt^{(2)}_{n} \, \Lt^{(2)}_{m} \rangle &= \mathrm{adiag}(-1,\tfrac{1}{2},-1)
    \\
    \langle \Lt^{(3)}_{n} \, \Lt^{(3)}_{m} \rangle &= \frac{1}{20} (\lambda^{2}-4)\cdot \mathrm{adiag}(4,-1,\tfrac{2}{3},-1,4)
    \\
    \langle \Lt^{(4)}_{n} \, \Lt^{(4)}_{m} \rangle &= \frac{1}{140} (\lambda^{2}-4)(\lambda^{2}-9) \cdot \mathrm{adiag}(-6,1,\tfrac{2}{5}, \tfrac{3}{10}, \tfrac{2}{5},1,-6) \,.
  \end{align}
\end{subequations}
So the quotient agrees with $\mathfrak{sl}(2,\R)$ and
$\mathfrak{sl}(3,\R)$ with $\sigma=1/4$.

\section{\texorpdfstring{Virasoro and $\W_3$ Algebra}{Virasoro and W3 Algebra}}
\label{app:Walg}

The $\W_3$ algebra at finite central charge, first introduced in \cite{Zamolodchikov:1985wn} and reviewed in \cite{Bouwknegt:1992wg},
is explicitly given by 
\begin{subequations}
   \label{eq:W3}
\begin{align}
 \label{eq:Walg}
 [\L_n,\,\L_m] &= (n-m)\,\L_{n+m} + \frac{c}{12}\,(n^3-n)\,\delta_{n+m,\,0} \\
 [\L_n,\,\W_m] &= (2n-m)\,\W_{n+m} \label{eq:Wal1}\\
 [\W_n,\,\W_m] &= (n-m)(2n^2+2m^2-nm-8)\,\L_{n+m}    \label{eq:Wal2}
\\
&\quad +\frac{c}{12}\,(n^2-4)(n^3-n)\,\delta_{n+m,\,0}  + \frac{96}{c+\tfrac{22}{5}}\,(n-m) \, \Lambda_{n+m} \nonumber
\label{eq:quadratic}
\end{align}
\end{subequations}
where
\begin{equation}
  \label{eq:lambda}
\Lambda_{n}=  \sum_{p\in\mathbb{Z}} :(\L_{n-p}\L_p): -\frac{3}{10} (n+3) (n+2) \L_{n} \ .
\end{equation}
The generators split into the Virasoro generators $\L_n$
and of spin-3 generators $\W_n$ both with integer $n$.

\section{Kinematical Spin-2 Algebras}
\label{cha:kinematical-spin-2}

\begin{table}[H]
  \centering
$
\begin{array}{l r r r r r}
\toprule %
                                            & \mathfrak{(A)dS}_{\parmp} & \mathfrak{poi}          & \mathfrak{nh}             & \mathfrak{ppoi}           \\ \midrule
  \left[\,\Jt  \comma \Jt \,\right]        & 0                         & 0                       & 0                         & 0                         \\
  \left[\,\Jt  \comma \Gt_{a} \,\right]     & \epsilon_{am}  \Gt_{m}    & \epsilon_{am}  \Gt_{m}  & \epsilon_{am}  \Gt_{m}    & \epsilon_{am}  \Gt_{m}    \\
  \left[\, \Jt\comma \Ht \,\right]          & 0                         & 0                       & 0                         & 0                         \\
  \left[\, \Jt \comma \Pt_{a} \,\right]     & \epsilon_{am}  \Pt_{m}    & \epsilon_{am}  \Pt_{m}  & \epsilon_{am}  \Pt_{m}    & \epsilon_{am}  \Pt_{m}    \\ 
  \left[\,\Gt_{a}  \comma \Gt_{b} \,\right] & - \epsilon_{ab}  \Jt      & - \epsilon_{ab}  \Jt    & 0                         & 0                         \\
  \left[\, \Gt_{a} \comma \Ht \,\right]     & -\epsilon_{am}  \Pt_{m}   & -\epsilon_{am}  \Pt_{m} & -\epsilon_{am}  \Pt_{m}   & 0                         \\
  \left[\, \Gt_{a} \comma \Pt_{b} \,\right] & -\epsilon_{ab} \Ht        & -\epsilon_{ab} \Ht      & 0                         & -\epsilon_{ab} \Ht        \\
  \left[\,\Ht  \comma \Pt_{a} \,\right]     & \pm \epsilon_{am} \Gt_{m} & 0                       & \pm \epsilon_{am} \Gt_{m} & \pm \epsilon_{am} \Gt_{m} \\
  \left[\,\Pt_{a}  \comma \Pt_{b} \,\right] & \mp  \epsilon_{ab}\Jt     & 0                       & 0                         & \mp  \epsilon_{ab}\Jt     \\ \midrule
\end{array}
$
\caption{(Anti-)de Sitter, Poincar\'e, Newton--Hooke and para-Poincar\'e algebras. The upper sign is for AdS (and contractions thereof) and the lower sign for dS (and contractions thereof).}
\label{tab:KinSpin2AdS}
\end{table}

\begin{table}[H]
  \centering
$
\begin{array}{l r r r r r}
\toprule %
                                            & \mathfrak{car}         & \mathfrak{gal}          & \mathfrak{pgal}           & \mathfrak{st}          \\ \midrule
 \left[\,\Jt  \comma \Jt \,\right]          & 0                      & 0                       & 0                         & 0                      \\
  \left[\,\Jt  \comma \Gt_{a} \,\right]     & \epsilon_{am}  \Gt_{m} & \epsilon_{am}  \Gt_{m}  & \epsilon_{am}  \Gt_{m}    & \epsilon_{am}  \Gt_{m} \\
  \left[\, \Jt\comma \Ht \,\right]          & 0                      & 0                       & 0                         & 0                      \\
  \left[\, \Jt \comma \Pt_{a} \,\right]     & \epsilon_{am}  \Pt_{m} & \epsilon_{am}  \Pt_{m}  & \epsilon_{am}  \Pt_{m}    & \epsilon_{am}  \Pt_{m} \\ 
  \left[\,\Gt_{a}  \comma \Gt_{b} \,\right] & 0                      & 0                       & 0                         & 0                      \\
  \left[\, \Gt_{a} \comma \Ht \,\right]     & 0                      & -\epsilon_{am}  \Pt_{m} & 0                         & 0                      \\
  \left[\, \Gt_{a} \comma \Pt_{b} \,\right] & -\epsilon_{ab} \Ht     & 0                       & 0                         & 0                      \\
  \left[\,\Ht  \comma \Pt_{a} \,\right]     & 0                      & 0                       & \pm \epsilon_{am} \Gt_{m} & 0                      \\
  \left[\,\Pt_{a}  \comma \Pt_{b} \,\right] & 0                      & 0                       & 0                         & 0                      \\ \midrule
\end{array}
$
\caption{Carroll, Galilei, para-Galilei and static algebra. The upper sign is for AdS (and contractions thereof) and the lower sign for dS (and contractions thereof).}
\label{tab:KinSpin2Car}
\end{table}


The most general invariant metric for the $\mathfrak{(A)dS}_{\parmp}$
algebra is given by
\begin{align}
  \langle \Ht \,, \Jt \rangle           & = -\mu^{-}
 &
    \langle \Pt_{a} \,, \Gt_{b} \rangle & = \mu^{-} \delta_{ab}
 \\
  \langle \Jt \,, \Jt \rangle           & = -\mu^{+}
 &
    \langle \Gt_{a} \,, \Gt_{b} \rangle & = \mu^{+} \delta_{ab}
 \\
  \langle \Ht \,, \Ht \rangle
                                        & = \mp \mu^{+}
 &
    \langle \Pt_{a} \,, \Pt_{b} \rangle
                                        & = \pm \mu^{+} \delta_{ab} \,.
\end{align}
The two real constants need to satisfy
$\mu^{+} \neq \pm\mu^{-}$ for the metric to be
nondegenerate, see Section \ref{sec:cont-inh}.

\section{Democratic Spin-3 Algebras}
\label{app:all-expl-contr}

This appendix contains tables with all the commutation relations of the spin-3 algebras that can be obtained via sequential application of the ``democratic'' sIW-contractions. We start each  table with the spin-2 commutation relations, then proceed with the mixed spin commutation relations and conclude with the spin-3 commutation relations. The table caption contains information about what type of higher spin version we are dealing with (e.g.~higher spin version of Poincar\'e, Galilei or Carroll). Under the heading `Contraction \#', we have indicated one possibility of obtaining the corresponding algebra as a sequential application of IW contraction procedures. The numbers in this heading refer to the contraction procedures of Table \ref{tab:contr}.

\newpage

\begin{table}[H]
  \centering
$
\begin{array}{l r r r r r}
\toprule %
                                               & \mathfrak{hs}_{3}\mathfrak{(A)dS}_{\parmp}          & \mathfrak{hs}_{3}\mathfrak{poi1}                  & \mathfrak{hs}_{3}\mathfrak{poi2}              \\
  \text{Contr. \#}                        &                                                     & 1                                                 & 2                                                   \\ \midrule
  \left[\,\Jt  \comma \Gt_{a} \,\right]        & \epsilon_{am}  \Gt_{m}                              & \epsilon_{am}  \Gt_{m}                            & \epsilon_{am}  \Gt_{m}                              \\
  \left[\, \Jt\comma \Ht \,\right]             & 0                                                   & 0                                                 & 0                                                   \\
  \left[\, \Jt \comma \Pt_{a} \,\right]        & \epsilon_{am}  \Pt_{m}                              & \epsilon_{am}  \Pt_{m}                            & \epsilon_{am}  \Pt_{m}                              \\ 
  \left[\,\Gt_{a}  \comma \Gt_{b} \,\right]    & - \epsilon_{ab}  \Jt                                & - \epsilon_{ab}  \Jt                              & - \epsilon_{ab}  \Jt                                \\
  \left[\, \Gt_{a} \comma \Ht \,\right]        & -\epsilon_{am}  \Pt_{m}                             & -\epsilon_{am}  \Pt_{m}                           & -\epsilon_{am}  \Pt_{m}                             \\
  \left[\, \Gt_{a} \comma \Pt_{b} \,\right]    & -\epsilon_{ab} \Ht                                  & -\epsilon_{ab} \Ht                                & -\epsilon_{ab} \Ht                                  \\
\end{array}
$
\caption{Higher spin versions of the (A)dS and Poincar\'e algebra. The upper sign is for AdS (and contractions thereof) and the lower sign for dS (and contractions thereof).}
\label{tab:adspoin}
\end{table}

\begin{table}[H]
  \centering
$
\begin{array}{l r r r r r}
\toprule %
                                               & \mathfrak{hs}_{3}\mathfrak{nh1}                          & \mathfrak{hs}_{3}\mathfrak{nh2}                    \\
  \text{Contraction \#}                           & 3                                                   & 4                                             \\ \midrule
  \left[\,\Jt  \comma \Gt_{a} \,\right]        & \epsilon_{am}  \Gt_{m}                              & \epsilon_{am}  \Gt_{m}                        \\
  \left[\, \Jt\comma \Ht \,\right]             & 0                                                   & 0                                             \\
  \left[\, \Jt \comma \Pt_{a} \,\right]        & \epsilon_{am}  \Pt_{m}                              & \epsilon_{am}  \Pt_{m}                        \\ 
  \left[\,\Gt_{a}  \comma \Gt_{b} \,\right]    & 0                                                   & 0                                             \\
  \left[\, \Gt_{a} \comma \Ht \,\right]        & -\epsilon_{am}  \Pt_{m}                             & -\epsilon_{am}  \Pt_{m}                       \\
  \left[\, \Gt_{a} \comma \Pt_{b} \,\right]    & 0                                                   & 0                                             \\
\end{array}
$
\caption{Higher spin versions of the Newton--Hooke algebra. The upper sign is for contractions of AdS and the lower sign for contractions of dS.}
\label{tab:nh}
\end{table}

\begin{table}[H]
  \centering
$
\begin{array}{l r r r r r}
\toprule %
                                               & \mathfrak{hs}_{3}\mathfrak{ppoi1}                        & \mathfrak{hs}_{3}\mathfrak{ppoi2}                        \\
  \text{Contraction \#}                        & 5                                                   & 6                                                   \\ \midrule
  \left[\,\Jt  \comma \Gt_{a} \,\right]        & \epsilon_{am}  \Gt_{m}                              & \epsilon_{am}  \Gt_{m}                              \\
  \left[\, \Jt\comma \Ht \,\right]             & 0                                                   & 0                                                   \\
  \left[\, \Jt \comma \Pt_{a} \,\right]        & \epsilon_{am}  \Pt_{m}                              & \epsilon_{am}  \Pt_{m}                              \\ 
  \left[\,\Gt_{a}  \comma \Gt_{b} \,\right]    & 0                                                   & 0                                                   \\
  \left[\, \Gt_{a} \comma \Ht \,\right]        & 0                                                   & 0                                                   \\
  \left[\, \Gt_{a} \comma \Pt_{b} \,\right]    & -\epsilon_{ab} \Ht                                  & -\epsilon_{ab} \Ht                                  \\
\end{array}
$
\caption{Higher spin versions of para-Poincar\'e algebra. The upper sign is for contractions of AdS and the lower sign for contractions of dS.}
\label{tab:poins}
\end{table}

\begin{table}[H]
  \centering
$
\begin{array}{l r r r r}
\toprule %
                                               & \mathfrak{hs}_{3}\mathfrak{car1}                  & \mathfrak{hs}_{3}\mathfrak{car2}                  \\
  \text{Contraction \#}                        & 1,5                                               & 1,6                                               \\ \midrule
  \left[\,\Jt  \comma \Gt_{a} \,\right]        & \epsilon_{am}  \Gt_{m}                            & \epsilon_{am}  \Gt_{m}                            \\
  \left[\, \Jt\comma \Ht \,\right]             & 0                                                 & 0                                                 \\
  \left[\, \Jt \comma \Pt_{a} \,\right]        & \epsilon_{am}  \Pt_{m}                            & \epsilon_{am}  \Pt_{m}                            \\ 
  \left[\,\Gt_{a}  \comma \Gt_{b} \,\right]    & 0                                                 & 0                                                 \\
  \left[\, \Gt_{a} \comma \Ht \,\right]        & 0                                                 & 0                                                 \\
  \left[\, \Gt_{a} \comma \Pt_{b} \,\right]    & -\epsilon_{ab} \Ht                                & -\epsilon_{ab} \Ht                                \\
\end{array}
$
\caption{Higher spin versions of the Carroll algebra.}
\label{tab:hscar}
\end{table}

\begin{table}[H]
  \centering
$
\begin{array}{l r r r r}
\toprule %
                                               & \mathfrak{hs}_{3}\mathfrak{car3}                    & \mathfrak{hs}_{3}\mathfrak{car4}                    \\
  \text{Contraction \#}                        & 5,2                                                 & 6,2                                                 \\ \midrule
  \left[\,\Jt  \comma \Gt_{a} \,\right]        & \epsilon_{am}  \Gt_{m}                              & \epsilon_{am}  \Gt_{m}                              \\
  \left[\, \Jt\comma \Ht \,\right]             & 0                                                   & 0                                                   \\
  \left[\, \Jt \comma \Pt_{a} \,\right]        & \epsilon_{am}  \Pt_{m}                              & \epsilon_{am}  \Pt_{m}                              \\ 
  \left[\,\Gt_{a}  \comma \Gt_{b} \,\right]    & 0                                                   & 0                                                   \\
  \left[\, \Gt_{a} \comma \Ht \,\right]        & 0                                                   & 0                                                   \\
  \left[\, \Gt_{a} \comma \Pt_{b} \,\right]    & -\epsilon_{ab} \Ht                                  & -\epsilon_{ab} \Ht                                  \\
\end{array}
$
\caption{Higher spin versions of the Carroll algebra. The upper sign is for contractions of AdS and the lower sign for contractions of dS.}
\label{tab:hscar2}
\end{table}

\begin{table}[H]
  \centering
$
\begin{array}{l r r r r r}
\toprule %
                                               & \mathfrak{hs}_{3}\mathfrak{gal1}                  & \mathfrak{hs}_{3}\mathfrak{gal2}     \\
  \text{Contraction \#}                        & 1,3                                               & 1,4                                  \\ \midrule
  \left[\,\Jt  \comma \Gt_{a} \,\right]        & \epsilon_{am}  \Gt_{m}                            & \epsilon_{am}  \Gt_{m}               \\
  \left[\, \Jt\comma \Ht \,\right]             & 0                                                 & 0                                    \\
  \left[\, \Jt \comma \Pt_{a} \,\right]        & \epsilon_{am}  \Pt_{m}                            & \epsilon_{am}  \Pt_{m}               \\ 
  \left[\,\Gt_{a}  \comma \Gt_{b} \,\right]    & 0                                                 & 0                                    \\
  \left[\, \Gt_{a} \comma \Ht \,\right]        & -\epsilon_{am}  \Pt_{m}                           & -\epsilon_{am}  \Pt_{m}              \\
  \left[\, \Gt_{a} \comma \Pt_{b} \,\right]    & 0                                                 & 0                                    \\
\end{array}
$
\caption{Higher spin versions of the Galilei algebra.}
\label{tab:hsgal}
\end{table}

\begin{table}[H]
  \centering
$
\begin{array}{l r r r r r}
\toprule %
                                               & \mathfrak{hs}_{3}\mathfrak{gal3}                    & \mathfrak{hs}_{3}\mathfrak{gal4}        \\
  \text{Contraction \#}                        & 3,2                                                 & 4,2                                     \\ \midrule
  \left[\,\Jt  \comma \Gt_{a} \,\right]        & \epsilon_{am}  \Gt_{m}                              & \epsilon_{am}  \Gt_{m}                  \\
  \left[\, \Jt\comma \Ht \,\right]             & 0                                                   & 0                                       \\
  \left[\, \Jt \comma \Pt_{a} \,\right]        & \epsilon_{am}  \Pt_{m}                              & \epsilon_{am}  \Pt_{m}                  \\ 
  \left[\,\Gt_{a}  \comma \Gt_{b} \,\right]    & 0                                                   & 0                                       \\
  \left[\, \Gt_{a} \comma \Ht \,\right]        & -\epsilon_{am}  \Pt_{m}                             & -\epsilon_{am}  \Pt_{m}                 \\
  \left[\, \Gt_{a} \comma \Pt_{b} \,\right]    & 0                                                   & 0                                       \\
\end{array}
$
\caption{Higher spin versions of the Galilei algebra. The upper sign is for contractions of AdS and the lower sign for contractions of dS.}
\label{tab:hsgal2}
\end{table}

\begin{table}[H]
  \centering
$
\begin{array}{l r r r r r}
\toprule %
                                               & \mathfrak{hs}_{3}\mathfrak{pgal1}                   & \mathfrak{hs}_{3}\mathfrak{pgal2}                   \\
  \text{Contraction \#}                        & 3,5                                                 & 3,6                                                 \\ \midrule
  \left[\,\Jt  \comma \Gt_{a} \,\right]        & \epsilon_{am}  \Gt_{m}                              & \epsilon_{am}  \Gt_{m}                              \\
  \left[\, \Jt\comma \Ht \,\right]             & 0                                                   & 0                                                   \\
  \left[\, \Jt \comma \Pt_{a} \,\right]        & \epsilon_{am}  \Pt_{m}                              & \epsilon_{am}  \Pt_{m}                              \\ 
  \left[\,\Gt_{a}  \comma \Gt_{b} \,\right]    & 0                                                   & 0                                                   \\
  \left[\, \Gt_{a} \comma \Ht \,\right]        & 0                                                   & 0                                                   \\
  \left[\, \Gt_{a} \comma \Pt_{b} \,\right]    & 0                                                   & 0                                                   \\
\end{array}
$
\caption{Higher spin versions of the  para-Galilei algebra. The upper sign is for contractions of AdS and the lower sign for contractions of dS.}
\label{tab:pgal}
\end{table}

\begin{table}[H]
  \centering
$
\begin{array}{l r r r r r}
\toprule %
                                               & \mathfrak{hs}_{3}\mathfrak{pgal3}      & \mathfrak{hs}_{3}\mathfrak{pgal4}      \\
  \text{Contraction \#}                        & 4,5                                    & 4,6                                    \\ \midrule
  \left[\,\Jt  \comma \Gt_{a} \,\right]        & \epsilon_{am}  \Gt_{m}                 & \epsilon_{am}  \Gt_{m}                 \\
  \left[\, \Jt\comma \Ht \,\right]             & 0                                      & 0                                      \\
  \left[\, \Jt \comma \Pt_{a} \,\right]        & \epsilon_{am}  \Pt_{m}                 & \epsilon_{am}  \Pt_{m}                 \\ 
  \left[\,\Gt_{a}  \comma \Gt_{b} \,\right]    & 0                                      & 0                                      \\
  \left[\, \Gt_{a} \comma \Ht \,\right]        & 0                                      & 0                                      \\
  \left[\, \Gt_{a} \comma \Pt_{b} \,\right]    & 0                                      & 0                                      \\
\end{array}
$
\caption{Higher spin versions of the  para-Galilei algebra. The upper sign is for contractions of AdS and the lower sign for contractions of dS.}
\label{tab:pgal2}
\end{table}

\begin{table}[H]
  \centering
$
\begin{array}{l r r r r r}
\toprule %
                                               & \mathfrak{hs}_{3}\mathfrak{st1}                   & \mathfrak{hs}_{3}\mathfrak{st2}      \\
  \text{Contraction \#}                        & 1,3,5=7                                           & 1,4,6=8                              \\ \midrule
  \left[\,\Jt  \comma \Gt_{a} \,\right]        & \epsilon_{am}  \Gt_{m}                            & \epsilon_{am}  \Gt_{m}               \\
  \left[\, \Jt\comma \Ht \,\right]             & 0                                                 & 0                                    \\
  \left[\, \Jt \comma \Pt_{a} \,\right]        & 0                                                 & 0                                    \\ 
  \left[\,\Gt_{a}  \comma \Gt_{b} \,\right]    & 0                                                 & 0                                    \\
  \left[\, \Gt_{a} \comma \Ht \,\right]        & 0                                                 & 0                                    \\
  \left[\, \Gt_{a} \comma \Pt_{b} \,\right]    & 0                                                 & 0                                    \\
\end{array}
$
\caption{Higher spin versions of the static algebra. The upper sign is for contractions of AdS and the lower sign for contractions of dS.}
\label{tab:hsstat}
\end{table}

\begin{table}[H]
  \centering
$
\begin{array}{l r r r r r}
\toprule %
                                               & \mathfrak{hs}_{3}\mathfrak{st3}                     & \mathfrak{hs}_{3}\mathfrak{st4}         \\
  \text{Contraction \#}                        & 2,3,6=9                                             & 2,4,5=10                                \\ \midrule
  \left[\,\Jt  \comma \Gt_{a} \,\right]        & \epsilon_{am}  \Gt_{m}                              & \epsilon_{am}  \Gt_{m}                  \\
  \left[\, \Jt\comma \Ht \,\right]             & 0                                                   & 0                                       \\
  \left[\, \Jt \comma \Pt_{a} \,\right]        & 0                                                   & 0                                       \\ 
  \left[\,\Gt_{a}  \comma \Gt_{b} \,\right]    & 0                                                   & 0                                       \\
  \left[\, \Gt_{a} \comma \Ht \,\right]        & 0                                                   & 0                                       \\
  \left[\, \Gt_{a} \comma \Pt_{b} \,\right]    & 0                                                   & 0                                       \\
\end{array}
$
\caption{Higher spin versions of the static algebra. The upper sign is for contractions of AdS and the lower sign for contractions of dS.}
\label{tab:hsstat2}
\end{table}

\begin{table}[H]
  \centering
$
\begin{array}{l r r r r r}
\toprule %
                                               & \mathfrak{hs}_{3}\mathfrak{st5}                   & \mathfrak{hs}_{3}\mathfrak{st6}    \\
  \text{Contraction \#}                        & 3,1,6                                             & 1,4,5                              \\ \midrule
  \left[\,\Jt  \comma \Gt_{a} \,\right]        & \epsilon_{am}  \Gt_{m}                            & \epsilon_{am}  \Gt_{m}             \\
  \left[\, \Jt\comma \Ht \,\right]             & 0                                                 & 0                                  \\
  \left[\, \Jt \comma \Pt_{a} \,\right]        & \epsilon_{am}  \Pt_{m}                            & \epsilon_{am}  \Pt_{m}             \\ 
  \left[\,\Gt_{a}  \comma \Gt_{b} \,\right]    & 0                                                 & 0                                  \\
  \left[\, \Gt_{a} \comma \Ht \,\right]        & 0                                                 & 0                                  \\
  \left[\, \Gt_{a} \comma \Pt_{b} \,\right]    & 0                                                 & 0                                  \\
\end{array}
$
\caption{Higher spin versions of the static algebra which can not be directly contracted.}
\label{tab:hsstat3}
\end{table}

\begin{table}[H]
  \centering
$
\begin{array}{l r r r r r}
\toprule %
                                               & \mathfrak{hs}_{3}\mathfrak{st7}                     & \mathfrak{hs}_{3}\mathfrak{st8}         \\
  \text{Contraction \#}                        & 3,2,5                                               & 4,2,6                                   \\ \midrule
  \left[\,\Jt  \comma \Gt_{a} \,\right]        & \epsilon_{am}  \Gt_{m}                              & \epsilon_{am}  \Gt_{m}                  \\
  \left[\, \Jt\comma \Ht \,\right]             & 0                                                   & 0                                       \\
  \left[\, \Jt \comma \Pt_{a} \,\right]        & \epsilon_{am}  \Pt_{m}                              & \epsilon_{am}  \Pt_{m}                  \\ 
  \left[\,\Gt_{a}  \comma \Gt_{b} \,\right]    & 0                                                   & 0                                       \\
  \left[\, \Gt_{a} \comma \Ht \,\right]        & 0                                                   & 0                                       \\
  \left[\, \Gt_{a} \comma \Pt_{b} \,\right]    & 0                                                   & 0                                       \\
\end{array}
$
\caption{Higher spin versions of the static algebra which can not be directly contracted. The upper sign is for contractions of AdS and the lower sign for contractions of dS.}
\label{tab:hsstat4}
\end{table}

\subsection{Invariant Metric of $\mathfrak{hs_{3}(A)dS}$}
\label{cha:invar-metr}

The most general invariant metric for both
$\mathfrak{hs}_{3}\mathfrak{AdS}$ and $\mathfrak{hs}_{3}\mathfrak{dS}$,
as well as their subalgebras $\mathfrak{AdS}$ and $\mathfrak{dS}$ in the
notation given in \eqref{eq:AdSdSalgebra} is
\begin{align}
  \langle \hP_A\,, \hJ_B\rangle &= \mu^{-}\eta_{AB}
  &
  \langle \hP_{AB}\,, \hJ_{CD}\rangle &= \mu^{-} (\eta_{A(C} \eta_{D)B} - \tfrac{2}{3} \eta_{AB} \eta_{CD})  
\end{align}
and additionally
\begin{align}
  \langle \hJ_A\,, \hJ_B\rangle &= \mu^{+}\eta_{AB}
  &
  \langle \hJ_{AB}\,, \hJ_{CD}\rangle &= \mu^{+} (\eta_{A(C} \eta_{D)B} - \tfrac{2}{3} \eta_{AB} \eta_{CD})  
  \\
  \langle \hP_A\,, \hP_B\rangle &= \pm \mu^{+}\eta_{AB}
  &
  \langle \hP_{AB}\,, \hP_{CD}\rangle &= \pm \mu^{+} (\eta_{A(C} \eta_{D)B}
  - \tfrac{2}{3} \eta_{AB} \eta_{CD}) \,.
\end{align}
where the upper sign is for the $\mathfrak{AdS}$ case.
While for $\mathfrak{AdS}$ non-degeneracy requires $\mu^{+} \neq \pm
\mu^{-}$
the $\mathfrak{dS}$ case requires only that not both $\mu^{\pm}$ vanish,.
The remaining products like, e.g., $\langle \hP_A\,, \hJ_{BC}\rangle$
are vanishing.

Using the decomposition \eqref{eq:nothsalgebras} leads to
\begin{align}
  \langle \Ht \,, \Jt \rangle                                                            & = -\mu^{-}
                                                                                         & 
                                    \langle \Ht_{a} \,, \Jt_{b}\rangle                   & = - \mu^{-} \delta_{ab}
 \\
    \langle \Pt_{a} \,, \Gt_{b} \rangle                                                  & = \mu^{-} \delta_{ab}
                                                                                         & 
                                    \langle \Pt_{ab} \,,
                                                                         \Gt_{cd}\rangle & = \mu^{-} (\delta_{a(c} \delta_{d)b} - \tfrac{2}{3} \delta_{ab} \delta_{cd})
 \\
  \langle \Jt \,, \Jt \rangle                                                            & = -\mu^{+}
                                                                                         & 
                                    \langle \Jt_{a} \,, \Jt_{b}\rangle                   & = - \mu^{+} \delta_{ab}
 \\
    \langle \Gt_{a} \,, \Gt_{b} \rangle                                                  & = \mu^{+} \delta_{ab}
                                                                                         & 
                                    \langle \Gt_{ab} \,, \Gt_{cd}\rangle                 & = \mu^{+} (\delta_{a(c} \delta_{d)b} - \tfrac{2}{3} \delta_{ab} \delta_{cd})  
 \\
  \langle \Ht \,, \Ht \rangle
                                                                                         & = \mp \mu^{+}
                                                                                         & 
                                    \langle \Ht_{a} \,, \Ht_{b}\rangle                   & = \mp \mu^{+} \delta_{ab}
 \\
    \langle \Pt_{a} \,, \Pt_{b} \rangle
                                                                                         & = \pm \mu^{+} \delta_{ab}
                                                                                         & 
                                    \langle \Pt_{ab} \,,
                                                                                           \Pt_{cd}\rangle                 & = \pm \mu^{+} (\delta_{a(c} \delta_{d)b} - \tfrac{2}{3} \delta_{ab} \delta_{cd})  
\end{align}
Again, only nonzero elements are displayed.

\bibliographystyle{utphys} 
\bibliography{bibl} 


\printindex

\end{document}